# Estimating Operational Risk Capital with Greater Accuracy, Precision, and Robustness

**J.D. Opdyke**[1]
**Head of Operational Risk Modeling, Quantitative Methods Group, GE Capital**

The largest US banks and Systemically Important Financial Institutions are required by regulatory mandate to estimate the operational risk capital they must hold using an Advanced Measurement Approach (AMA) as defined by the Basel II/III Accords. Most of these institutions use the Loss Distribution Approach (LDA) which defines the aggregate loss distribution as the convolution of a frequency distribution and a severity distribution representing the number and magnitude of losses, respectively. Capital is a Value-at-Risk estimate of this annual loss distribution (i.e. the quantile corresponding to the 99.9%tile, representing a one-in-a-thousand-year loss, on average). In practice, the severity distribution drives the capital estimate, which is essentially a very large quantile of the estimated severity distribution. Unfortunately, when using LDA with any of the widely used severity distributions (i.e. heavy-tailed, skewed distributions), all unbiased estimators of severity distribution parameters appear to generate biased capital estimates due to Jensen's Inequality: VaR always appears to be a convex function of these severities' parameter estimates because the (severity) quantile being estimated is so large *and* the severities are heavy-tailed. The resulting bias means that capital requirements always will be overstated, and this inflation is sometimes enormous (sometimes even billions of dollars at the unit-of-measure level). Herein I present an estimator of capital that essentially eliminates this upward bias when used with any commonly used severity parameter estimator. The Reduced-bias Capital Estimator (RCE), consequently, is more consistent with regulatory intent regarding the responsible implementation of the LDA framework than other implementations that fail to mitigate, if not eliminate this bias. RCE also notably increases the precision of the capital estimate and consistently increases its robustness to violations of the i.i.d. data presumption (which are endemic to operational risk loss event data). So with greater capital accuracy, precision, and robustness, RCE lowers capital requirements at both the unit-of-measure and enterprise levels, increases capital stability from quarter to quarter, ceteris paribus, and does both while more accurately and precisely reflecting regulatory intent. RCE is straightforward to explain, understand, and implement using any major statistical software package.

---

[1] J.D. Opdyke is Head of Operational Risk Modeling, Quantitative Methods Group at GE Capital where he is heading up all operational risk modeling and quantification efforts including AMA capital estimation, CCAR, and all related modeling (e.g. ECap) and reporting. J.D. has over 20 years of experience as a quantitative consultant, most of this in the banking and credit sectors where his clients have included multiple Fortune and Global 50 banks and financial credit organizations. J.D.'s recent Journal of Operational Risk paper (2012) was voted "Paper of the Year" by Operational Risk & Regulation staff in consultation with industry experts, and he has been invited to present his work at the American Bankers Association Operational Risk Forum, the Operational Risk Exchange Analytics Forum, and OpRisk North America 2014. J.D.'s other publications span statistical finance, computational statistics (solving "Big Data" problems using SAS®), number theory/combinatorics, and applied econometrics. J.D. earned his undergraduate degree, with honors, from Yale University, his Master's degree from Harvard University where he was awarded both Kennedy and Social Policy Research Fellowships, and he completed post-graduate statistics work as an ASP Fellow in the graduate mathematics department at MIT.

The views expressed in this paper are the views of the sole author and are not necessarily those of GE Capital or any other institution. The author may be contacted at J.D.Opdyke@GE.com.

The author extends his sincere appreciation to Toyo R. Johnson, Nicole Opdyke, and Ryan Opdyke for their thoughtful insights, and to Kirill Mayorov for a thorough review and for spotting an error in an earlier draft.

"Measurement is the first step that leads to control and eventually to improvement. If you can't measure something, you can't understand it. If you can't understand it, you can't control it. If you can't control it, you can't improve it." - H. J. Harrington

**Background, Introduction, and Objectives**

In the United States, regulatory mandate is compelling the larger banks[2] and companies designated as Systemically Important Financial Institutions ("SIFIs," both bank and non-bank)[3] to use an Advanced Measurement Approach (AMA) framework to estimate the operational risk capital they must hold in reserve.[4] Both industry practice and regulatory guidance have converged over the past decade[5] on the Loss Distribution Approach (LDA)[6] as the most widely used AMA framework. Under this approach, data on operational risk loss events[7] is used to estimate a frequency distribution, representing the *number* of loss events that could occur over a given time period (typically a year), and to estimate a severity distribution, representing the *magnitude* of those loss events. These two distributions are then combined via convolution to obtain an annual aggregate loss distribution. Operational risk regulatory capital (RCap) is the dollar amount associated with the 99.9%tile of this estimated loss distribution. Operational risk economic capital (ECap) is the quantile associated with, typically, the 99.97%tile of the aggregate loss distribution, depending on the institution's credit rating.[8]

The frequency, severity, and capital estimations take place at the level of the Unit-of-Measure (UoM). UoM's simply are the groups into which operational risk loss events are categorized, generally under the competing objectives of homogeneity and (larger) sample size. Basel II identifies eight business lines and seven event types that together comprise fifty-six UoM's. Individual institutions either use some or all of these UoM's as is, define their UoM's empirically, or use some combination of these two approaches. Capital estimated at the UoM level then must be aggregated to a single estimate at the enterprise level, and under the conservative (and unrealistic) assumption of perfect dependence, capital is simply summed across all UoM's. In reality, however, losses do not occur in perfect lockstep across UoM's no matter how they are defined, and so this imperfect dependence in the occurrence of loss events can be estimated and simulated, typically via copula models.[9] This potentially can provide an enormous diversification benefit to the banks/SIFIs,[10] and along with LDA's risk-sensitive nature generally, is the major 'carrot' that counterbalances the 'stick' that is the regulatory requirement of an AMA implementation. These potential benefits also have been a major motivation for LDA's adoption by many institutions beyond the US. For a more extensive and detailed background on the LDA and its widespread use for operational risk capital estimation, see Opdyke and Cavallo (2012a and 2012b).[11]

---

[2] These include banks and systemically important financial institutions ("SIFIs") with over $250 billion in total consolidated assets, or over $10 billion in total on-balance sheet foreign exposure (and includes the depository institution subsidiaries of these firms). See Federal Register (2007).

[3] On July 8, 2013, the Financial Stability Oversight Council of the U.S. Department of the Treasury, as authorized by Section 113 of the Dodd-Frank Act, voted to designate American International Group (AIG) and General Electric Capital Corporation, Inc. (GECC) as SIFIs. On September 19, 2013, the Council voted to designate Prudential Financial, Inc. a SIFI. See www.treasury.gov/initiatives/fsoc/designations/Pages/default.aspx

[4] See BCBS (2004). The other two, less empirically sophisticated methods – the Basic Indicator Approach and the Standardized Approach – are simple functions of gross income. As such, they are not risk sensitive and do not accurately reflect the complex risk profiles of these financial institutions.

[5] There have been no dramatic changes with respect to operational risk capital estimation under an AMA since the first comprehensive guidance was published in 2004 (see BCBS, 2004).

[6] This approach has a longer history of use within the insurance industry.

[7] An operational risk loss event only can result from an operational risk, which is defined by Basel II as, "the risk of loss resulting from inadequate or failed internal processes, people, and systems or from external events. This includes legal risk, but excludes strategic and reputational risk." See BCBS (2004).

[8] ECap is higher than RCap as it addresses the very solvency of the institution. The 99.97%tile is a typical value used for ECap (almost all are 99.95%tile or above), based on a firm's credit rating, since it reflects 100% minus the historical likelihood of an AA rated firm defaulting in a one-year period. See Hull (2012).

As described above, capital under the LDA is based on the convolution of the severity and frequency distributions. However, estimates of severity and frequency are exactly that: merely estimates based only on samples of operational risk loss event data. Their values will change from sample to sample, quarter to quarter, and because they are based directly on these varying estimates, the capital estimates, too, will vary from sample to sample, quarter to quarter. So it is essential to understand how this *distribution* of capital estimates is shaped if we are to attempt to make reliable inferences (about "true" capital numbers) based on it. Is it centered on "true" capital values (if we test it using known inputs with simulated data), or is it systematically biased? If biased, in what direction – upwards or downwards – and under what conditions is this bias material? Is the capital distribution reasonably precise, or do capital estimates vary so dramatically as to be completely unreliable and little better than a wild guess at what the true capital values really are? Is the distribution reasonably robust to real-world violations of the properties of the loss data assumed by the estimation methods, or do modest deviations from idealized, mathematically convenient textbook assumptions effectively distort the results in material ways, and arguably render them useless? These are questions that only can be answered via scrutiny of the entire *distribution* of capital estimates (say, at least one thousand estimates), as opposed to a few capital numbers that may or may not appear to be "reasonable" based on a few estimates of severity and frequency distribution parameters. And we should be ready for answers that may call into question the conceptual soundness of the LDA framework, or at least the manner in which major components of it are commonly implemented in this setting.[12]

[9] There are other approaches to estimating dependence structures and tail dependence in particular, such as mixture models (see Reshetar, 2008), but many are much newer and not yet tested extensively in practice (for example, see Arakelian and Karlis, 2014, Bernard and Vanduffel, 2014, Dhaene et al., 2013, and Polanski et al., 2013).

[10] See RMA (2011), OR&R (2009), and Haubenstock and Hardin (2003).

[11] A key point here that drives a focus of this paper is the fact that empirically, the severity distribution drives capital much more than does the frequency distribution – typically orders of magnitude more. This is true both from the perspective of the choice of which severity distribution is used vs. the choice of which frequency distribution is used (the latter changes capital very little compared to the former), as well as variance in the values of the severity parameters vs. variance in the values of the frequency parameter(s): a change of a standard deviation of the former typically has an enormous effect on estimated capital in both absolute and relative terms, while the same change in the latter has a much smaller, if not de minimis effect on estimated capital. This is well established in the literature (see Opdyke and Cavallo, 2012a and 2012b, and Ames et al., 2014), and the analytic reasons for this are demonstrated later in this paper. So while stochastic frequency parameter(s) always are and always should be included in operational risk capital estimation and simulation, the severity distribution typically (and rightly) is more of a focus of research on operational risk capital estimation than is the frequency distribution.

This paper addresses these issues directly by focusing on the capital distribution and what are arguably the three biggest challenges to LDA-based operational risk capital estimation: the fact that even under idealized data assumptions,[13] LDA-based capital estimates are i) systematically inflated (and sometimes grossly inflated by many hundreds of millions of dollars under conditions not uncommon for the largest, and even medium-sized banks),[14] ii) extremely imprecise by any reasonable measure (i.e. they are extremely variable from sample to sample – see Opdyke, 2013, Opdyke and Cavallo, 2012a, Cope et al., 2009, and OR&R, 2014, for more on this topic), and iii) extremely non-robust to violations of the (i.i.d.) data assumptions almost always made when implementing the LDA (and which are universally recognized as unrealistic; see, for example, Opdyke and Cavallo, 2012a, and Horbenko et al., 2011). Yet it is precisely these three factors – capital accuracy, capital precision, and capital robustness – that arguably are the only criteria that matter when assessing the efficacy of an operational risk (or any) capital estimation framework. Indeed, the stated requirement of the US Final Rule on the Advanced Measurement Approaches for Operational Risk (see US Final Rule, 2007, and Interagency Guidance, 2011) is for "credible, transparent, systematic, and verifiable processes that incorporate all four operational risk elements … [that should be combined] in a manner that most effectively enables it [the regulated bank/sifi] to quantify its exposure to operational risk." But can it even be seriously argued that an operational risk capital estimation framework that generates results consistent with i), ii), and iii) above could be deemed "credible"? Or even "verifiable" in the face of excessive variability in capital estimate outcomes? How could one even assess whether i), ii), and/or iii) are true without scrutinizing the distribution of capital estimates that the framework generates under controlled conditions (i.e. under well-specified and extensive loss data simulations)?

[12] One such example is the extremely large size of the quantile of the aggregate loss distribution – that corresponding to the 99.9%tile – that firms are required to estimate. See Degen and Embrechts (2011) and Nešlehová et al. (2006) for more details.

[13] The most sweeping, yet common assumption is that the loss data is "i.i.d." – independent and identically distributed. "Independent" means that the values of losses are unrelated across time periods, and "identically distributed" means that losses are generated from the same data generating process, typically characterized as a parametric statistical distribution (see Opdyke and Cavallo, 2012a, 2012b). The assumption that operational risk loss event data is "i.i.d" is widely recognized as unrealistic and made more for mathematical and statistical convenience than as a reflection of empirical reality (see Opdyke and Cavallo, 2012a, 2012b). The consequences of some violations of this assumption are examined later in this paper.

[14] This has been confirmed by empirical findings in the literature (see Opdyke and Cavallo, 2012a and 2012b, Opdyke, 2013, Joris, 2013, and Ergashev et al., 2014) as well as a recent position paper from AMAG ("AMA Group"), a professional association of major financial institutions subject to AMA requirements (see RMA, 2013, which cites the need for "Techniques to remove or mitigate the systematic overstatement (bias) of capital arising in the context of capital estimation with the LDA methodology").

Unfortunately, very little operational risk research tackles these three issues head-on through a systematic examination of the entire *distribution* of capital estimates, as opposed to simply presenting several capital estimates almost as an afterthought to an analysis that focuses primarily on severity parameter estimation (a few exceptions include Rozenfeld, 2011, Opdyke and Cavallo, 2012a and 2012b, Opdyke, 2013, Joris, 2013, and Zhou, 2013). However, cause for optimism lies in the fact that a single analytical source appears to account for much, if not most of the deleterious effect of these three issues on capital estimation. What has become known as Jensen's inequality – a time-tested analytical result first proven in 1906 (see Jensen, 1906) – appears to be the sole cause of i), as well as a major contributing factor to ii) and, to a lesser extent, iii). Yet this has been overlooked and virtually unmentioned in the operational risk quantification and capital estimation literature (see Opdyke and Cavallo, 2012a, b, Opdyke, 2013, and Joris, 2013 for the only known exceptions).[15] If a fraction of the effort that has gone into research on severity parameter estimation also is directed at capital estimation, and specifically on defining, confronting, and mitigating the biasing, imprecision, and non-robustness effects apparently caused by Jensen's inequality, then all in this space – practitioners, academics, regulated (and even non-regulated) financial institutions, and regulators – quickly will be much farther along the path toward making the existing LDA framework much more useable and valuable in practice.[16] It has been a decade since Basel II published comprehensive guidance on operational risk capital estimation,[17] and still these three issues remain to dog the industry's efforts at effectively utilizing the LDA framework to provide reasonably accurate, reasonably precise, and reasonably robust capital estimates. So we are long past due for a refocusing of our analytical lenses specifically on the capital distribution and on these three challenges to make some real strides towards providing measurable, implementable, and impactful solutions to them. The direct financial and risk mitigation stakes for getting these capital numbers "right" (according to these three criteria) are enormously high for individual financial institutions (especially the larger ones), as well as for the industry as a whole, so our best efforts as empirical researchers should require no less than this refocusing, if not complete problem resolution.

[15] Of course, Jensen's inequality has long been the subject of applied research in other areas of finance (see Fisher et al., 2009), applied econometrics (see Duan, 1983), and even bias in market risk VaR (see, for example, Liu and Luger, 2006). But proposed solutions to its deleterious effects on estimation have not been extended to operational risk capital, the literature for which has almost completely ignored it (with the exception of Opdyke and Cavallo, 2012a and 2012b, Opdyke, 2013, and Joris, 2013). Although it does not identify Jensen's inequality as the source, RMA (2013) does identify "the systematic overstatement (bias) of capital arising in the context of capital estimation with the LDA methodology," and Ergashev et al. (2014) present extensive empirical results exactly consistent with its effects and with the empirical results shown in this paper.

[16] Here, "useable" and "valuable" are based on assessments of the accuracy, precision, and robustness of the capital estimates that the framework generates. A realistic example, shown later in this paper, makes the point: when true capital is, say, $391m, but 1,000 LDA capital estimates (based on 1,000 i.i.d. simulated samples) average $640m with a standard deviation of over $834m, the framework generating the estimates, due to this large upward bias and gross imprecision, unarguably is not terribly useful or valuable to those needing to make business decisions based on its results. And this is under the most idealized i.i.d. data assumptions which are rarely, if ever, realized in actual practice.

[17] See BCBS (2004).

To this end, this paper has two main objectives: first, to identify and clearly demonstrate that Jensen's inequality is the most like source of the materially deleterious effects on LDA-based operational risk capital estimation, define the specific conditions under which these effects are material, and make the case for a shift in focus to the distribution of capital estimates, rather than focusing solely on the distribution of the severity parameter estimates. After all, capital estimation, not parameter estimation, is the endgame here. And secondly, to develop and propose a capital estimator – the Reduced-bias capital estimator (RCE) – that tackles all three of the major issues mentioned above – capital accuracy, capital precision, and capital robustness – and unambiguously improves capital estimates by all three criteria when compared to the most widely used implementations of LDA based on maximum likelihood estimation (MLE) (and a wide range of similar estimators). Requirements governing the development and design of RCE include:

- Its use and assumptions must not conflict with those supporting the LDA framework specifically,[18] and it must be entirely consistent with regulatory intent regarding this framework's responsible and prudent implementation generally (I argue below that RCE is *more* consistent with regulatory intent in the context of applying the LDA than most, if not all other known implementations of it).[19]
- It must utilize the same general methodological approach across sometimes very different severity distributions, including those that are truncated to account for data collection thresholds.
- It must "work" regardless of whether the mean of the severity distribution is infinite, or close to infinite.
- Its range of application must encompass most, if not all, of the commonly used estimators of the severity (and frequency) distributions.
- It must "work" regardless of the method used to approximate the VaR of the aggregate loss distribution.

[18] This is not to say that research that proposes changing the bounds or parameters of the framework is any less valuable per se, but rather, that this was a conscious choice made to maximize the range of application of the proposed solution (RCE). RCE is designed to be entirely consistent with the LDA framework specifically, and regulatory guidance and expectation generally so that an institution's policy decision to strictly adhere to all aspects of the framework would not preclude usage of RCE. In fact, RCE is arguably *more* consistent with regulatory guidance and expectation than are most, if not all other implementations of LDA, because its capital estimates are not systemically biased upwards: they are, on average, quite literally the expected values for capital, or very close, under the LDA framework (in other words, they are centered on true capital). So capital estimates based on RCE arguably are *most* consistent with regulatory intent regarding the responsible implementation of the LDA framework, as discussed below.

[19] It is important to note that regulatory guidance has avoided prescribing of any one AMA framework, including the LDA, even though the LDA has become the de facto choice among AMA institutions, including those that have recently exited parallel run.

- It must be easily understood and implemented using any widely available statistical software package.
- It must implementable using only a reasonably powerful desktop or laptop computer.
- It must provide unambiguous improvements over the most widely used implementations of LDA on all three of the key criteria for assessing the efficacy of an operational risk capital estimation framework: capital accuracy, capital precision, and capital robustness

The remainder of the paper is organized as follows. First, I define and discuss Jensen's inequality and its apparent effects on operational risk capital estimation under LDA, demonstrate the conditions under which these effects are material, and define the extremely wide range of (severity parameter) estimators for which these results are relevant. Next I develop and present the Reduced-bias Capital Estimator (RCE), discuss the details of its implementation, and present some new analytic derivations that assist in this implementation (as well as with the implementation of LDA generally). Thirdly, I conduct an extensive simulation study comparing RCE to the most widely used implementation of LDA as a benchmark (i.e. using maximum likelihood estimation (MLE)).[20] The study covers a number of very distinct severity distributions, both truncated and non-truncated, widely varying values for regulatory capital (RCap) and economic capital (ECap) at the unit-of-measure level (from \$38m to over \$10.6b), and wide ranges of severity parameter values that cover conditions of both finite and infinite severity mean (showing that RCE "works" even under the latter condition). The study also includes i) a new analytic derivation for the mean of a very commonly used severity distribution under truncation; ii) a very fast, efficient, and stable sampling (perturbation) method based on iso-densities; iii) an improved single loss approximation for estimating capital under conditions that may include infinite means; and iv) a new analytic approximation of the Fisher information a commonly used severity distribution under truncation (thus avoiding computationally expensive numeric integration). I discuss throughout how RCE is entirely consistent with the LDA framework specifically, and with regulatory intent and expectation generally regarding its responsible (i.e. unbiased, or close) implementation. I conclude with a summary and a discussion of areas for future research.

**Key Methodological Background**

Before discussing Jensen's inequality, I turn to a more recent result to provide some explanatory foundation for the relevance of the former in operational risk capital estimation. As mentioned above, under LDA the aggregate loss distribution is defined as the convolution of the frequency and severity distributions, and in

[20] For severity distribution estimation, AMAG (2013), in its range of practice survey from 2012, states "MLE is predominant, by far." This also is true for other components of the framework (e.g. dependence modeling across UoMs).

It is important to note that the MLE-based capital distributions do not dramatically differ from those of most other (severity) estimators in this setting, and so the sometimes enormous benefits of RCE over MLE also apply to most other implementations of LDA.

almost all cases no closed-form solutions exist to estimate the VaR of this compound distribution. Böcker and Klüppelberg (2005) and Böcker and Sprittulla (2006) were the first to provide an analytical approximation of this VaR in (1), and Degen (2010) refined this and expanded its application to include conditions of infinite mean in (2a,b,c).[21]

$$C_\alpha \approx F^{-1}\left(1-\frac{1-\alpha}{\lambda}\right)+(\lambda-1)\mu \qquad (1)$$

$$\text{if } \xi<1, \qquad C_\alpha \approx F^{-1}\left(1-\frac{1-\alpha}{\lambda}\right)+\lambda\mu \qquad (2.a)$$

$$\text{if } \xi=1, \qquad C_\alpha \approx F^{-1}\left(1-\frac{1-\alpha}{\lambda}\right)+c_\xi\lambda\mu_F\left[F^{-1}\left(1-\frac{1-\alpha}{\lambda}\right)\right] \qquad (2.b)$$

$$\text{if } 1<\xi<2, \quad C_\alpha \approx F^{-1}\left(1-\frac{1-\alpha}{\lambda}\right)-(1-\alpha)F^{-1}\left(1-\frac{1-\alpha}{\lambda}\right)\cdot\left(\frac{c_\xi}{1-1/\xi}\right) \qquad (2.c)$$

( $\xi \geq 2$ is so extreme as to not be relevant here) where, $c_\xi=(1-\xi)\dfrac{\Gamma^2(1-1/\xi)}{2\Gamma(1-2/\xi)}$ for $1<\xi<\infty$ and

$c_\xi=1$ for $\xi=1$, $\mu_F(x)=\int_0^x\left[1-F(s)\right]ds$ ; $C_\alpha$ = "capital"; $\alpha=$ "confidence level" (e.g. $\alpha=0.999$ for RCap);

$F^{-1}(\ )$ is the quantile function of the severity; $\lambda$ is the (typically Poisson) frequency parameter; $\mu$= mean of severity; $\xi$= the tail index; and $\Gamma(\ )$ is the gamma function.

I focus now on Degen's (2.a) to make the point that the first term, the severity quantile, is much larger – sometimes even orders of magnitude larger – than the second term (the "mean correction"), and so capital is essentially a very large quantile of the severity distribution (and this is consistent with the widely cited finding in the literature that severity, not frequency, is what really drives capital (see Opdyke and Cavallo, 2012a and 2012b)). But at least as important is the fact that the quantile of the severity distribution that must be estimated is much larger than that corresponding to the 99.9%tile – it actually corresponds to the [1 – (1-α)/λ] = 0.99997 = 99.997%tile (assuming λ=30), which is nearly two orders of magnitude larger (the corresponding percentiles for ECap are the 99.97%tile and 99.999%tile, respectively, assuming λ=30 and a good credit rating). So not only is capital essentially a quantile of the severity distribution, but this quantile also is extremely large. The essence of the problem, then, reduces to estimating an extremely large quantile of the severity distribution.[22] This fact, *combined* with the fact that the only severities used (and allowed) in operational risk capital estimation are

---

[21] Sahay et al. (2007) presented similar results a few years earlier.

[22] However, as noted above, estimation and simulation of the frequency parameter is never ignored in this paper. The purpose of making this point here is heuristic as it pertains to the explanation of the relevance of Jensen's inequality in this setting.

medium- to heavy-tailed, is the reason that Jensen's inequality apparently can so adversely and materially affect capital estimation, as described below.

## Jensen's Inequality

Jensen's Inequality Defined

In 1906, Johan Jensen proved that the (strictly) convex transformation of a mean is less than the mean after a (strictly) convex transformation (and that the opposite is true for strictly concave transformations). When applied to random variable $\beta$, this is shown in Figure 1 below as $E[g(\hat{\beta})] > g(E[\hat{\beta}])$, with a magnitude of

**FIGURE 1: Graph of Jensen's Inequality with Strictly Convex Function (right-skewed, heavy-tailed cdf)**

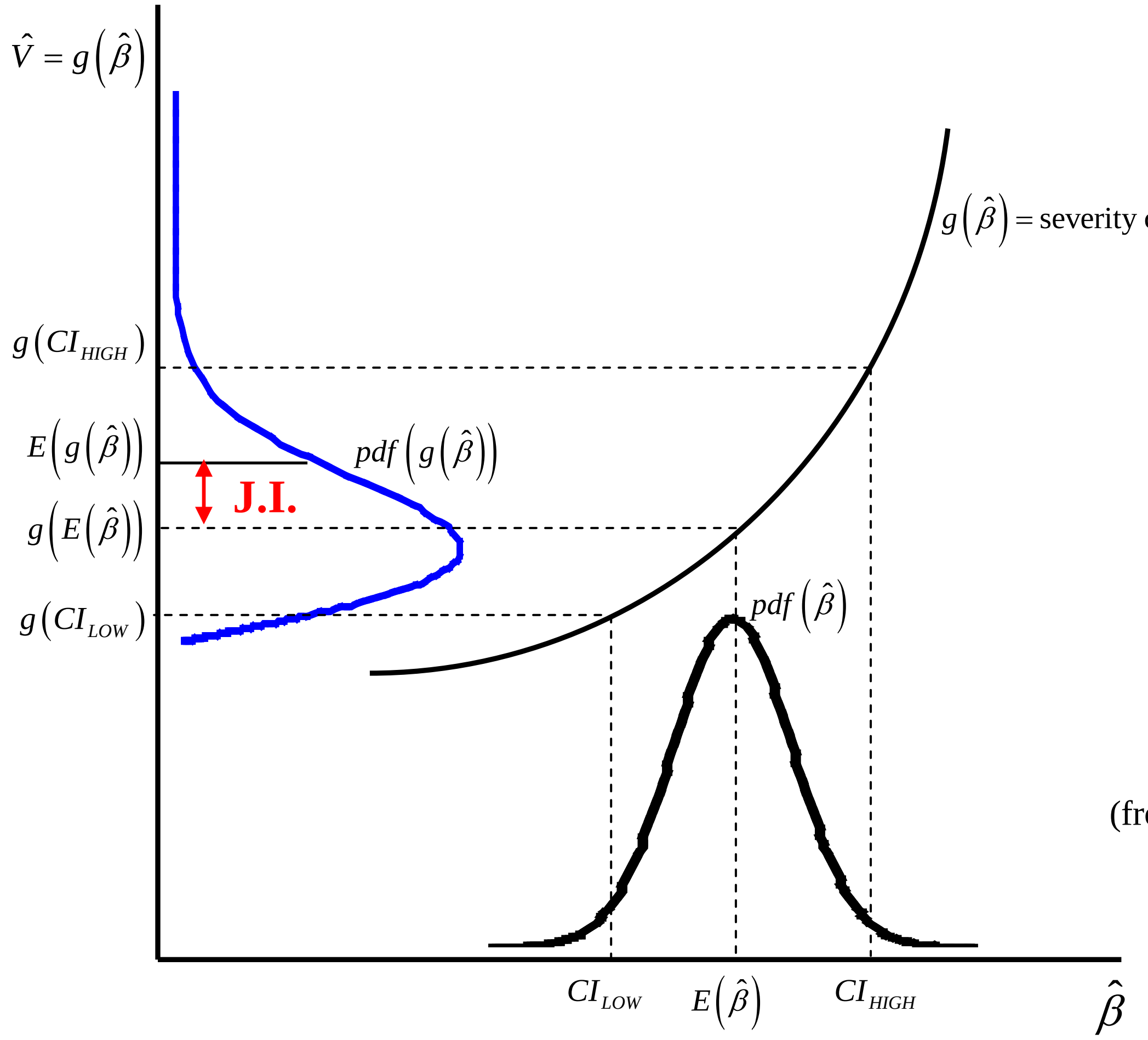

(from Kennedy, 1992. p.37)

Jensen's Inequality = J.I. = $E[g(\hat{\beta})] - g(E[\hat{\beta}])$.[23] An intuitive interpretation of Figure 1 is that the strictly convex function, g( ), "stretches out" the values of the random variable $\beta$ above its median more than it does below, thus positively skewing the distribution of $\hat{V} = g(\hat{\beta})$ and increasing its mean above what it would have been had the function g( ) been a linear function. In other words, $\hat{V}$ also would have been symmetric like $\hat{\beta}$, with a mean equal to its median, but because g( ) is convex, its upper tail is "stretched out" making its mean greater than its median.[24]

[23] Figure 1 shows VaR for a given cumulative probability, *p*. As *p* increases beyond some large level (e.g. *p* > 0.999), so does VaR's convexity in this setting, as discussed later in the paper.

Jensen's Inequality in Operational Risk Capital Estimation

The relevance of Jensen's inequality to operational risk capital estimation appears to be the joint fact that the only severities used (and permitted) in operational risk capital estimation are medium- to heavy-tailed, *and* the severity quantile being estimated is so extremely large: under these conditions, VaR appears to always be a convex function, like g( ), of the parameters of the severity distribution, which here is the vector $\beta$ (we can visualize $\beta$ as a single parameter without loss of generality as the multivariate case for Jensen's inequality is well established (see Schaefer, 1976). Consequently, the capital estimation, $\hat{V}$ = g( $\hat{\beta}$ ), will be biased upwards. In other words, its expected value, E[g( $\hat{\beta}$ )], will be larger than its true value, g(E[ $\hat{\beta}$ ]). Stated differently, if we generated 1,000 i.i.d. random samples of losses based on "true" severity parameter values = $\beta$, and for each of the 1,000 estimated $\hat{\beta}$'s we calculated capital $\hat{V}$ = g( $\hat{\beta}$ ), the average of these 1,000 capital estimates ( $\hat{V}$'s) will be larger than $V$ =g( $\beta$ ), which is "true" capital.

The above is straightforward, and the biasing effects of Jensen's inequality are very well established and not in doubt. The only question is whether VaR *always* is a strictly convex function of the estimators of the severity parameters. All of the estimators used in this setting are at least symmetrically distributed, and most are normally distributed, at least asymptotically.[25] So if VaR is a convex function of them, there is no doubt that capital will be systematically biased upwards (in addition to being, on average, more skewed, and with larger root mean squared error (RMSE)[26] and standard deviation, as shown empirically later in this paper). To check for this convexity, we can do several things: examine VaR as a function of each parameter individually (i.e.

---

[24] Importantly, note that the median of $\hat{V}$ actually is equal to the transformation of the original mean: g(E[ $\hat{\beta}$ ]) = g( $\beta$ ) . This is due to the fact that g( ) is a monotone transformation (here, of a symmetric, unbiased variable). This is shown below and exploited to our advantage later in this paper when designing a reduced-based capital estimator.

[25] All M-class estimators are asymptotically normal, and these include many of the most commonly used estimators in this setting (e.g. maximum likelihood estimation (MLE), many generalized method of moments (GMM) estimators, penalized maximum likelihood (PML), optimally bias-robust estimator (OBRE), Cramér von-Mises (CvM) estimator, and PITS estimator, among many others). See Hampel et al. (1986) and Huber and Ronchetti (2009) for more details.

[26] The MSE is the average of the squared deviations of a random variable from its true value. This is also equal to the variance of the random variable plus its bias squared $MSE = \frac{1}{n}\sum_{i=1}^{n}\left(\hat{V}_i - V\right)^2 = Var\left(\hat{V}\right) + \left[Bias\left(\hat{V}\right)\right]^2$ . The RMSE is the square root of MSE. So RMSE of the capital distribution = $RMSE = \sqrt{\frac{1}{n}\sum_{i=1}^{n}\left(\hat{V}_i - V\right)^2}$

check for marginal convexity); examine and attempt to define the multidimensional surface of VaR as a multivariate function of the severity parameters (i.e. check for multidimensional convexity (in three dimensional space for two-parameter severities)); and examine the behavior of VaR itself under straightforward i.i.d. Monte Carlo simulations to determine if it is consistent with the effects of Jensen's inequality as a convex, or at least "convex-dominant" function of the severity parameters.[27]

The check for marginal convexity has been performed graphically in Appendix A Figure A1, for three widely used severity distributions (7 others – the three-parameter Burr Type XII, the LogLogistic, and the truncated versions of all five – are available from the author upon request).[28] All demonstrate that for *sufficiently extreme* percentiles (e.g. $p > 0.999$), VaR is a convex function of either one or both of the severity parameters (and a linear function of the others). These results are summarized in Table 1.

One approach to checking for convexity (or convexity-dominance) in the multidimensional VaR surface is an examination of the signs and relative magnitudes of the eigenvalues of the shape operator (see Jiao and Zha, 2008). This turns out to be analytically nontrivial, if not intractable under truncation, and even numeric calculations for many of the relevant severities are nontrivial given the sizes of the severity percentiles (e.g. $p = 0.99999$) that must be used in this setting (because most of the gradients are exceedingly large for such high percentiles). So this research currently remains underway, and without this mathematical verification, attributions of capital inflation *apparently* due to Jensen's inequality and VaR's *apparent* convexity are based solely on empirical results, and conservatively and explicitly deemed "preliminary" or "presumed" herein.

However, arguably the most directly relevant of these three "checks" is the behavior of the capital estimate itself: if it consistently reflects what we would expect to see under Jensen's inequality, i.e. systematically inflated capital estimates under i.i.d. Monte Carlo simulations, then this, combined with consistent marginal convexity, would provide reasonably strong, if preliminary evidence that VaR is a convex (or convex-dominant) function of the entire vector of severity parameters. As such, severity parameter *estimates* that are subject to sampling variability will generate capital estimates that are, on average, inflated, as shown in Figure 1. And this is exactly what we observe: consistent marginal convexity, as shown in Appendix A, and consistent and strong capital inflation, as shown in the extensive simulation study presented below. But the broader question here is whether *all* severity distributions relevant to operational risk capital estimation can be so characterized.

---

[27] "Convex-dominant" is used here to indicate cases where VaR is not a convex function of each parameter individually, but may be a convex function of the entire vector of severity parameter estimates, given its variance-covariance matrix. For example, while VaR of GPD is marginally convex in $\xi$, it is marginally linear in $\theta$ (see Appendix A). Also, some areas of its multidimensional surface appear to indicate the existence of saddlepoints, i.e. surfaces with hyperbolic points. But if the convexity in one direction of such a surface is orders of magnitude larger than the concavity in the other, as measured by the relative sizes of their principal curvatures (i.e. the eigenvalues of its shape operator), then the net effect of sampling variation on VaR, under the estimator's variance-covariance matrix, would appear to be dominated by convexity rather than concavity.

[28] This is easily confirmed analytically for those distributions that have closed form representations of their inverse CDFs (i.e. VaR functions). For example, for the LogNormal, $VaR = \exp\left(\mu + \sigma\Phi^{-1}\left(p\right)\right)$; $\partial^2 VaR/\partial\mu^2 = VaR$; $\partial^2 VaR/\partial\sigma^2 = VaR \cdot \left[\Phi^{-1}\left(p\right)\right]^2$.

**TABLE 1: Marginal VaR Behavior OVER RELEVANT DOMAIN (p > 0.999) by Severity**

| Severity Distribution | VaR is a Convex/Linear Function of… | | | Relationship Between Parameters |
|---|---|---|---|---|
| | Parameter 1 | Parameter 2 | Parameter 3 | |
| **1) LogNormal ($\mu$, $\sigma$)** | Convex | Convex | | Independent |
| **2) LogLogistic ($\alpha$, $\beta$)** | Linear | Convex | | Independent |
| **3) LogGamma (a, b)** | Convex | Convex | | Dependent |
| **4) GPD ($\xi$, $\theta$)** | Convex | Linear | | Dependent |
| **5) Burr (type XII) ($\Upsilon$, $\alpha$, $\beta$)** | Convex | Convex | Linear | Dependent |
| **6) Truncated 1)** | Convex | Convex | | Dependent |
| **7) Truncated 2)** | Linear | Convex | | Dependent |
| **8) Truncated 3)** | Convex | Convex | | Dependent |
| **9) Truncated 4)** | Convex | Linear | | Dependent |
| **10) Truncated 5)** | Convex | Convex | Linear | Dependent |

Before answering this question, it should be noted here that convexity sometimes replaces subadditivity (as well as positive homogeneity; see Fölmer and Schied, 2002, and Frittelli and Gianin, 2002) as an axiom of coherent risk measures (see Artzner et al., 1999), and is only slightly less strong an axiom compared to subadditivity.[29] And while it is very well established that VaR is not *globally* subadditive across all quantiles for all parametric statistical distributions, for the specific group of medium- to heavy-tailed severities relevant to LDA-based operational risk capital estimation, *and* very extreme percentiles of those severities ($p > 0.999$), it appears that VaR may very well always be subadditive. Danielsson et al. (2005) proved that regularly-varying severities with finite means all were subadditive for sufficiently high percentiles (e.g. $p > 0.99$; for similar results, see also Embrechts and Neslehova, 2006, Ibragimov, 2008, and Hyung and de Vries, 2007). And the same result has been shown empirically in a number of publications (see, for example, Degen et al., 2007). Although supra-additivity has been proven for some families of extremely heavy-tailed severities with infinite mean, (see Embrechts and Nešlehová, 2006, Ibragimov, 2008, and Hyung and de Vries, 2007), and consequently strong caution has been urged when using such models for operational risk capital estimation (see Nešlehová et al., 2006), this does not cover all such severities. In fact, high VaR ($p > 0.999$) of the Generalized Pareto Distribution (GPD) with infinite mean ($\theta = 40{,}000$ and $\xi = 1.1$) is shown in Appendix B, Figure B1 to still be a convex function of $\xi$ and a linear function of $\theta$. And corresponding capital simulations in Appendix B (Table B1) demonstrate continued and notable capital bias consistent with Jensen's inequality, infinite mean notwithstanding (capital bias of more than 80% and more than 120% over true capital for RCap and ECap, respectively). These easily replicated results demonstrate that supra-additivity is not a given for very heavy-tailed severities with some infinite moments, at least for certain parameter values. What's more, many practitioners in this setting restrict severities, or severity parameter values, to those indicating finite mean, arguing that allowing expected losses to be infinite makes no sense for an operational risk capital framework. This would make moot the issue of the possible supra-additivity of the severity. Others counter that regulatory requirements dictate the estimation of quantiles, not moments, and that capital models, from a robustness perspective, should remain agnostic regarding the specific characteristics of a loss distribution's moments.

[29] In fact, for any normalized risk measure, the presence of any two of the three properties of convexity, subadditivity, and positive homogeneity implies remaining third (see Föllmer and Schied, 2011).

Regardless of the position one takes on this debate, a mathematical proof of VaR's subadditivity or convexity for all severities relevant to operational risk capital estimation (a group that is not strictly defined) is beyond the scope of this paper. However, while undoubtedly useful, this is not strictly necessary here, because the number of such severities in this setting is finite, and checking the subset of those in use by any given financial institution, one by one, is very simple to do graphically, as was done in Appendix A, Figure A1. Graphical checks can be complemented with a simple simulation study wherein capital is estimated, say, 1,000 times based on i.i.d. samples generated from a chosen severity. If the mean of these 1,000 capital estimates is noticeably larger than the "true" capital based on the "true" severity parameters (where the original parameter estimates are treated as "true"), and this is consistent with graphing VaR as a function of the parameter values, then attributing this systemic bias to Jensen's inequality remains the most plausible, if not highly probable explanation.[30]

Note that it is just as easy to demonstrate the opposite, too, for a given severity. For example, VaR of the Gaussian (Normal) distribution is a linear function of both of the distribution's parameters, $\mu$ and $\sigma$. These marginal results indicate that Jensen's inequality likely could never affect capital estimation based on this distribution. This is shown both graphically in Appendix B, Figure B1, as well as via capital simulation in Table B1 in Appendix B,[31] which shows no positive capital bias,[32] even for the extremely large quantiles that are estimated under LDA. Remember, however, that the normal distribution, whether truncated or not, is far too light-tailed to be considered for use in operational risk capital estimation. And this demonstrates that *both* characteristics – the medium- to heavy-tailed nature of the severity, *and* estimation of its very high percentiles (e.g. $p > 0.999$) – are required simultaneously for the presumed convexity of VaR to manifest, and thus, for Jensen's inequality to bias capital estimates.

[30] This assumes, of course, that any approximations used to estimate capital are correct and reasonably accurate, and that the simulated data is i.i.d. to remove any other potential source of bias. See discussion of the former point below.

[31] This simulation ignores the need for truncation of the normal distribution at zero as the findings do not change.

[32] Very slight negative capital bias due to the estimation of $\lambda$, the frequency parameter, is discussed below.

To conclude this section on the biasing effects on LDA-based capital most likely attributable to Jensen's inequality, we must address the effects of $\lambda$ on capital, both in the first terms of (2a,b,c) as well as the subsequent "correction" terms. Recall that $\lambda$ is the parameter of the frequency distribution, whose default is the Poisson distribution.[33] For the extremely wide range of severity and frequency parameter values examined in this paper, capital actually is a concave function of $\lambda$, but its (negative) biasing effects on capital estimation are very small, if not de minimis. This is shown in 216 simulation studies summarized in Table C1 in Appendix C wherein $\lambda$ is the only stochastic component of the capital estimate.[34] Bias due only to $\lambda$ always is negative, but rarely exceeds -1%, and then just barely. So for all practical purposes VaR is essentially a linear function of $\lambda$ in this setting, and any (negative) biasing effect on capital is swamped by the much larger (positive) biasing effect of the severity parameters on capital, as shown in the Results section below. And regardless, RCE takes the net effect of both sources of bias into account, as discussed below.

When Are the Presumed Effects of Jensen's Inequality Material?

When VaR is a convex function of the vector of severity parameters, capital estimates will be biased upwards – always. But when is this capital inflation material? The most straightforward and reasonable metric for materiality is the size of the bias, both relative to true capital and in absolute terms. A bias of, say, \$0.5m when true capital is \$250m arguably is not worth the concern of those estimating capital (especially if its standard deviation is, say, \$400m, which is actually somewhat conservative). However, it would be hard to argue that a bias of \$200m, \$75m, or even \$25m was not worth the trouble to address statistically and attempt to at least mitigate it, if not eliminate it. And in addition to bias that sometimes exceeds 100% of true capital, the dramatic increase in the skewness and spread of the distribution of capital estimates (as shown in the simulation study below) alone could be reason enough to justify the development and use of a statistical method to eliminate it, especially if its implementation is relatively straightforward and fast.

[33] Empirically there is rarely much difference in capital regardless of the frequency distribution chosen, and the Poisson is mathematically convenient as well, so it had become the widely used default. Also note that (2.a,b,c) require only slight modification to accommodate other reasonable, non-Poisson frequency distributions, such as the Negative Binomial.

[34] These simulations cover all severity conditions, and most sample sizes, under which LDA-MLE and RCE are tested later in the paper.

It turns out there are three factors that contribute to the size of the capital bias (and the other abovementioned effects on the capital distribution): a) the size of the variance of the severity parameter estimator; b) the heaviness of the tail of the given severity distribution; and c) the size of the quantile being estimated. Directionally, larger estimator variance is associated with larger bias; heavier tails are associated with larger bias, and more extreme quantiles are associated with larger bias. Typically a) is driven most by sample size, and because larger sample sizes are almost always associated with smaller estimator variance, larger samples are associated with smaller bias. The choice of severity, typically determined by goodness-of-fit tests,[35] along with the size of its estimated parameter values drive b). So for example, truncated distributions, all else equal, will exhibit more bias than their non-truncated counterparts (with the same parameter values). And the choice of quantile, c), is determined by $\alpha$ in formula (2.a), and $\alpha$ is set at 0.999 for regulatory capital (and typically $\alpha$ = 0.9997, or close, for economic capital, depending on the institution's credit rating). So ECap will exhibit larger capital bias than RCap, all else equal.

All of these factors, and the directions of their effects, are consistent with the effects of Jensen's inequality, and with VaR as a convex (or convex-dominant) function of the severity parameter estimates, All three, but particularly a), can be visualized with Figure 1. The smaller the variance of the estimator of the severity parameter, $\beta$, on the X-axis, the less the values of $g\left(\hat{\beta}\right)$ can be stretched out above the median, all else equal, and so the less capital estimates will exhibit bias. In the extreme, if there is no variance, then all we have is $\beta$, the true severity parameter, and there is no bias in our capital estimate (because it is no longer an estimate – it is true capital). For b) and c), heavier tails, and more extreme quantiles of those tails, both are associated with greater convexity as shown in Appendix A, Figure A1, so g( ) will be more curved and will "stretch out" the capital estimates more and increase bias, all else equal.

The effects of sample size on capital bias are shown empirically in Table 2 for sample sizes of approximately 150, 250, 500, 750, and 1,000,[36] corresponding to $\lambda$ = 15, 25, 50, 75, and 100, respectively, for a ten year period. The size of the bias relative to true capital is (almost) always greater when the number of operational risk loss events in the sample is smaller.[37] Unfortunately, UoM's with thousands of loss events are not nearly as common as those with a couple of hundred loss events, or less. So from an empirical perspective, we are squarely in the bias-zone: bias is material for many, if not most estimations of capital at the UoM level.[38] In fact, this is exactly what Ergashev et al. (2014) found in their study comparing capital based on shifted vs. truncated lognormal severity distributions. The latter exhibited notable bias that disappeared as sample sizes increased up to $n = 1{,}000$, exactly as in the simulation study in this paper. However, the authors did not attribute this empirical effect to an analytical result (i.e. Jensen's inequality), as is done here.

[35] In this setting these tests typically are empirical distribution function-based (EDF-based) statistics, based on the difference between the estimated cumulative distribution function (CDF) and the EDF. The most commonly used here are the Kolmogorov-Smirnov (KS), the Anderson-Darling (AD), and the Cramér-von Mises (CvM) tests.

[36] These are approximate sample sizes because the annual frequency, of course, is a random variable (i.e. $\lambda$ is stochastic). Because the Poisson distribution is used for this purpose, the standard deviation of the number of losses is, annually, StdDev = $\sqrt{\lambda}$ , and for a given number of years, StdDev = $\sqrt{\#\, years \cdot \lambda}$ .

It is important to explicitly note here the converse, that is, the conditions under which LDA-MLE-based capital bias apparently due to Jensen's inequality is *not* material. This is shown empirically in Table 2 and in the simulation study below, but general guidelines include a) *sample sizes*: sample sizes in the low hundreds, which are most common for operational risk loss event data, will exhibit notable bias, all else equal, while those in the thousands typically will exhibit much more modest, if any bias, depending on the severity (see Table 2 – three severities exhibit very little bias for $n \approx 1{,}000$ ($\lambda = 100$), while two others exhibit noticeable but arguably modest bias of around 20% over true capital, and the last exhibits 5%-20% bias, depending on the parameter values). b) *severities*: certain severities are more heavy-tailed than others (e.g. LogGamma is more heavy-tailed than LogNormal, and GPD is more heavy-tailed than LogGamma), and truncated severities, by definition, are heavier-tailed distributions than their non-truncated counterparts, all else equal (that is, with the same parameter values). c) *parameter values*: note that VaR sometimes is a convex function of only one of the parameters of the distribution (for example, as shown in Appendix A, Figure A1, for the GPD and Truncated GPD distributions VaR is linear in $\theta$ but convex in $\xi$), so the magnitude of capital bias, apparently, will hinge primarily on the magnitude of this parameter, all else equal. This can be seen for almost all cases of the GPD and Truncated GPD distributions in Table 2. Capital is approximately equal in the paired, adjacent rows for these severities, yet bias is larger for the second row of the pair, where $\xi$ is always larger. The only exception is where $\lambda = 15$ for the Truncated GPD, because sometimes the smaller number of losses decreases capital, on average, via the decrease in the quantile of the first term of (2.a) more than it increases capital, on average, due to an increase in parameter variance, so that on net, capital bias actually decreases even though $\xi$ is slightly larger.

---

[37] The one exception is the one case (LogNormal, $\mu = 10$, $\sigma = 2$) where the smaller sample size ($n \approx 150$) decreases capital, on average, via the decrease in the percentile of the first term of (2.a) more than it increases capital, on average, due to an increase in parameter variance, so that on net, capital bias actually decreases very slightly.

[38] Again, this is also confirmed in RMA (2013), which cites the need for "Techniques to remove or mitigate the systematic overstatement (bias) of capital arising in the context of capital estimation with the LDA methodology"

**TABLE 2: MLE Capital Bias Beyond True Capital by Sample Size by Severity by Parameter Values**

| Severity | | | + RCap % Bias | | | | + | + ECap % Bias | | | | + |
|---|---|---|---|---|---|---|---|---|---|---|---|---|
| Dist. | Parm1 | Parm2 | $\lambda = 15$ | $\lambda = 25$ | $\lambda = 50$ | $\lambda = 75$ | $\lambda = 100$ | $\lambda = 15$ | $\lambda = 25$ | $\lambda = 50$ | $\lambda = 75$ | $\lambda = 100$ |
| | $\mu$ | $\sigma$ | | | | | | | | | | |
| LogN | 10 | 2 | 6.0% | 6.7% | 3.0% | 1.5% | 1.5% | 7.3% | 7.8% | 3.5% | 1.8% | 1.8% |
| LogN | 7.7 | 2.55 | 11.9% | 11.5% | 5.4% | 3.0% | 2.8% | 14.2% | 13.2% | 6.2% | 3.4% | 3.3% |
| LogN | 10.4 | 2.5 | 11.3% | 11.0% | 5.1% | 2.8% | 2.7% | 13.5% | 12.7% | 5.9% | 3.2% | 3.1% |
| LogN | 9.27 | 2.77 | 14.9% | 13.8% | 6.5% | 3.7% | 3.4% | 17.6% | 15.8% | 7.5% | 4.2% | 3.9% |
| LogN | 10.75 | 2.7 | 13.9% | 13.1% | 6.2% | 3.4% | 3.2% | 16.5% | 15.0% | 7.1% | 3.9% | 3.7% |
| LogN | 9.63 | 2.97 | 17.9% | 16.1% | 7.7% | 4.4% | 4.0% | 21.1% | 18.5% | 8.8% | 5.0% | 4.6% |
| TLogN | 10.2 | 1.95 | 18.9% | 11.5% | 8.1% | 3.6% | 2.9% | 24.6% | 14.7% | 10.1% | 4.6% | 3.7% |
| TLogN | 9 | 2.2 | 52.0% | 26.5% | 13.9% | 7.3% | 5.3% | 76.8% | 35.0% | 17.7% | 9.5% | 6.9% |
| TLogN | 10.7 | 2.385 | 42.9% | 26.4% | 12.5% | 6.0% | 5.2% | 57.2% | 32.4% | 15.2% | 7.4% | 6.4% |
| TLogN | 9.4 | 2.65 | 64.2% | 39.1% | 20.0% | 13.9% | 8.4% | 87.8% | 51.6% | 24.8% | 17.0% | 10.3% |
| TLogN | 11 | 2.6 | 49.9% | 27.1% | 14.8% | 9.2% | 5.6% | 63.6% | 34.0% | 17.7% | 11.0% | 6.8% |
| TLogN | 10 | 2.8 | 90.9% | 40.2% | 17.1% | 13.2% | 8.8% | 127.3% | 51.5% | 21.1% | 16.1% | 10.8% |
| | **a** | **b** | | | | | | | | | | |
| Logg | 24 | 2.65 | 22.3% | 13.6% | 5.6% | 4.4% | 1.1% | 28.3% | 17.0% | 7.0% | 5.4% | 1.7% |
| Logg | 33 | 3.3 | 17.8% | 8.5% | 3.6% | 3.2% | 0.4% | 22.2% | 10.7% | 4.6% | 4.0% | 0.7% |
| Logg | 25 | 2.5 | 26.4% | 15.7% | 8.3% | 5.8% | 1.3% | 33.3% | 19.5% | 10.1% | 7.0% | 1.9% |
| Logg | 34.5 | 3.15 | 16.3% | 10.9% | 6.3% | 4.0% | 0.6% | 20.5% | 13.5% | 7.7% | 4.8% | 1.0% |
| Logg | 25.25 | 2.45 | 27.9% | 18.3% | 9.5% | 5.2% | 1.6% | 35.2% | 22.5% | 11.6% | 6.4% | 2.2% |
| Logg | 34.7 | 3.07 | 19.3% | 13.7% | 7.1% | 3.3% | 0.4% | 24.2% | 16.8% | 8.6% | 4.1% | 0.8% |
| TLogg | 23.5 | 2.65 | 166.7% | 56.1% | 31.7% | 14.6% | 13.5% | 329.3% | 83.1% | 45.0% | 20.1% | 18.5% |
| TLogg | 33 | 3.3 | 72.7% | 34.1% | 13.2% | 7.7% | 6.6% | 110.5% | 46.1% | 17.7% | 10.3% | 8.8% |
| TLogg | 24.5 | 2.5 | 110.2% | 60.4% | 25.8% | 16.9% | 9.9% | 169.5% | 84.9% | 34.2% | 22.4% | 13.3% |
| TLogg | 34.5 | 3.15 | 45.3% | 24.5% | 11.6% | 7.7% | 4.8% | 63.3% | 32.2% | 15.0% | 9.8% | 6.3% |
| TLogg | 24.75 | 2.45 | 102.1% | 62.9% | 23.4% | 16.0% | 9.9% | 152.3% | 87.6% | 31.2% | 20.6% | 13.2% |
| TLogg | 34.6 | 3.07 | 40.7% | 24.3% | 13.6% | 8.3% | 4.3% | 55.0% | 31.8% | 17.0% | 10.3% | 5.7% |
| | $\xi$ | $\vartheta$ | | | | | | | | | | |
| GPD | 0.8 | 35000 | 80.3% | 56.9% | 30.5% | 17.6% | 14.0% | 119.9% | 81.9% | 41.5% | 23.3% | 18.6% |
| GPD | 0.95 | 7500 | 108.8% | 75.6% | 39.8% | 23.0% | 18.2% | 163.4% | 109.2% | 54.0% | 30.2% | 23.9% |
| GPD | 0.875 | 47500 | 91.1% | 63.7% | 34.8% | 20.0% | 16.1% | 135.9% | 91.9% | 47.3% | 26.5% | 21.3% |
| GPD | 0.95 | 25000 | 105.7% | 73.2% | 39.7% | 22.8% | 18.3% | 158.8% | 105.9% | 53.8% | 30.0% | 24.1% |
| GPD | 0.925 | 50000 | 90.0% | 67.8% | 37.4% | 21.8% | 17.3% | 137.6% | 97.9% | 50.8% | 28.7% | 22.8% |
| GPD | 0.99 | 27500 | 109.5% | 76.4% | 41.6% | 24.3% | 19.3% | 164.9% | 110.7% | 56.5% | 31.9% | 25.3% |
| TGPD | 0.775 | 33500 | 81.6% | 52.0% | 25.3% | 17.7% | 14.4% | 127.8% | 75.7% | 34.7% | 23.9% | 19.1% |
| TGPD | 0.8 | 25000 | 71.3% | 56.9% | 28.3% | 19.6% | 16.0% | 108.5% | 82.9% | 38.8% | 26.5% | 20.9% |
| TGPD | 0.868 | 50000 | 101.2% | 63.0% | 33.1% | 20.6% | 15.8% | 154.8% | 92.0% | 45.5% | 27.7% | 20.7% |
| TGPD | 0.91 | 31000 | 93.8% | 68.6% | 34.1% | 23.2% | 17.8% | 146.7% | 100.4% | 46.3% | 30.9% | 23.2% |
| TGPD | 0.92 | 47500 | 115.9% | 64.7% | 35.7% | 24.0% | 17.1% | 176.7% | 93.9% | 48.6% | 32.0% | 22.5% |
| TGPD | 0.95 | 35000 | 105.6% | 68.2% | 39.0% | 24.6% | 19.1% | 168.6% | 100.8% | 53.7% | 32.8% | 25.1% |

*NOTE: #simulations = 1,000; time period = 10 years; for RCap and ECap, $\alpha$ = 0.999 and 0.9997, respectively.

Unfortunately there currently are no formulaic rules to determine whether LDA-MLE-based capital bias is material for a given sample of loss event data (and the best-fitting severity chosen), because all of these three factors – a), b), and c) – interact in ways that are not straightforward. And materiality is a subjective assessment as well. So the only way to answer this question of materiality is to conduct a simple simulation given the estimated values of the severity (and frequency) parameters: i) treat the estimated parameter values as "true" and calculate "true" capital; ii) use the "true" parameter values to simulate 1,000 i.i.d. data samples and for each of these samples, re-estimate the parameter values and calculate capital for each sample; iii) compare the mean of these 1,000 capital estimates to "true" capital, and if the (positive) difference is large or at least notable, then capital bias is material.[39] This is, in fact, exactly what was done for Table 2, which is taken from the simulation study presented later in this paper.

Estimators Apparently Affected by Jensen's Inequality

There are a wide range of estimators that have been brought to bear on the problem of estimating severity distribution parameters in this setting. Examples include maximum likelihood estimation (MLE; see Opdyke and Cavallo, 2012a and 2012b), penalized likelihood estimation (PLE; see Cope, 2011), Method of Moments (see Dutta and Perry, 2007) and Generalized Method of Moments (see RMA, 2013), Probability Weighted Moments (PWM – see BCBS, 2011), Bayesian estimators (with informative, non-informative, and flat priors; see Zhou et al., 2013), extreme value theory – peaks over threshold estimator (EVT-POT; see Chavez-Demoulez et al., 2013),[40] robust estimators such as the Quantile Distance estimator (QD; see Ergashev, 2008), Optimal Bias-Robust Estimator (OBRE; see Opdyke and Cavallo, 2012a), Cramér-von Mises estimator (CvM – not to be confused with the goodness-of-fit test by the same name; see Opdyke and Cavallo, 2012a), Generalized Median Estimator (see Serfling, 2002, and Wilde and Grimshaw, 2013), PITS Estimator (only for Pareto severity; see Finkelstein et al., 2006), and many others of the wide class of M-Class estimators (see Hampel et al., 1986, and Huber and Ronchetti, 2009). Which of these generate inflated capital estimates, apparently due to Jensen's inequality? The answer is any that would be represented as $\beta$ on Figure 1, which is to say, apparently all of them.[41] All the relevant estimators at least will be symmetrically distributed, and many, if not most, will be normally distributed, at least asymptotically (like all M-Class estimators). But normality most certainly is not a requirement for this bias to manifest (and even symmetry is not a requirement), and so capital based on all of these estimators will be subject to the same biasing effects outlined above. There is some evidence that robust estimators generate capital estimates that are less biased than their non-robust counterparts, and while this intuitively makes sense, unfortunately the mitigating effect on capital bias does not appear to be large (for some empirical results, see Opdyke and Cavallo, 2012a; Opdyke, 2013; and Joris, 2013). To the extent that there are differences in the size of the capital bias associated with each of these estimators, the size of the variance will be the main driver, but given the (maximal) efficiency of MLE,[42] it is safe to say that none of these other estimators will fare much better, if at all, regarding LDA-based capital bias, ceteris paribus.

[39] It is possible, of course, that the original estimated parameter values based on actual loss data are much larger than the "true" but unobservable parameter values due simply to random sampling error, in which case bias may not be material. But even in this case, the parameter values actually used to estimate capital will be the (high) estimates, because these are the best we have: we will never know the "true" values because we have only samples of loss data, not a population of loss data. And so bias will be material based on these estimated parameter values and the given sample of loss event data. Over time, unbiased estimates based on larger samples of data will converge (asymptotically) to true parameter values.

[40] Although estimating operational risk capital via EVT-POT was not explicitly tested in this paper for capital bias induced by VaR's apparent convexity, it would appear to be subject to the same effects. This approach relies on extreme value theory to estimate only the tail of the loss distribution which, beyond some high threshold, asymptotically converges to a GPD distribution (see Rocco, 2011, and Andreev et al., 2009). The estimated parameters of the GPD distribution, however, are generally unbiased (especially if specifically designed unbiased estimators are used in the case of very small samples; for example, see Pontines and Siregar, 2008). As such, they can be represented on Figure 1 as $\beta$, and thus would provide biased VaR estimates. See Chavez-Demoulez et al. (2013) for a rigorous application of EVT-POT to operational risk capital estimation.

<u>Severities Apparently Affected by Jensen's Inequality</u>

As discussed above, it appears that all severities commonly used in operational risk capital estimation satisfy the criteria of being heavy-tailed enough, and simultaneously the quantile being estimated is extreme enough, that the capital estimates they generate are upwardly biased. A number of papers have proposed using mixtures of severities in this setting, but as shown in Joris (2013), capital estimates based on these, too, appear to exhibit notable bias. Another common variant is to use spliced severities, wherein one distribution is used for the body of the losses and another is used for the right tail (see Ergashev, 2009, and RMA, 2013), and often the splice point is endogenized. Sometimes the empirical distribution is used for the body of the severity, and a parametric distribution is used for the tail. For these cases, too, we can say that as long as the ultimate estimates of the tail parameter can be represented as $\beta$ in Figure 1, the corresponding capital estimates also will exhibit the same systematic inflation. A simulation study testing the latter of these cases is beyond the scope of the current paper, but would be very useful to confirm results for spliced distributions similar to those of Joris (2013) for mixed distributions.

**Reduced-bias Capital Estimator**

Note that as mentioned above, the median of the capital distribution, *if sampled from a distribution centered on the true parameter values,* is an unbiased estimator of true capital, as shown below:

[41] One distinct approach proposed for operational risk capital estimation that may diverge from this paradigm is the semi-parametric kernel transformation (see Gustafsson and Nielson, 2008, Bolancé et al., 2012, and Bolancé et al., 2013). However, in a closely related paper, Alemany, Bolancé and Guillén (2012) discuss how variance reduction in their double transformation kernel estimation of VaR increases bias. In contrast, RCE simultaneously decreases both variance and bias in the VaR (capital) estimate.

[42] Of course, MLE achieves the maximally efficient Cramer-Rao lower bound only under i.i.d. data sample conditions.

From Figure 1, if $\hat{\beta}$ is symmetrically distributed and centered on true $\beta$ (that is, $\hat{\beta}$ is unbiased, as is the case, asymptotically, for MLE under i.i.d. data), then:

$E\left(\hat{\beta}\right) = F^{-1}\left(0.5\right)$, i.e. the mean equals the median, so

$$g\left[E\left(\hat{\beta}\right)\right] = g\left[F^{-1}\left(0.5\right)\right]$$

And as g[ ] is strictly convex and a monotonic transformation,

$$g\left[E\left(\hat{\beta}\right)\right] = g\left[F^{-1}\left(0.5\right)\right] = G^{-1}\left(0.5\right).$$

So as long as $\hat{\beta}$ is unbiased, the median of the capital distribution is an unbiased estimator of capital. In other words, given a monotonic and strictly convex transformation function (i.e. g( ), or VaR), the median of the transformed variable (i.e. capital estimates) is equal to the transformation of the original mean (i.e. $G^{-1}\left(0.5\right) = g\left(E\left[\hat{\beta}\right]\right) = g\left(\beta\right)$]) of a symmetric, unbiased variable (e.g. MLE estimates of severity parameters under i.i.d. data). However, this begs the question of unbiased capital estimation, because in reality we have only one sample and one corresponding vector of (estimated) parameter values, $\hat{\beta}$, and these will never exactly equal the true severity parameter values, $\beta$, of the underlying data generating process. So simply taking the median of the capital distribution will not work. But this relationship still can be exploited in constructing a reduced-bias capital estimator, as shown below.

The motivation behind the development of RCE is to design a simple scaler of capital that scales down capital inflated by Jensen's Inequality to remove its upward bias. And this scaling factor needs to be a function of the degree of convexity of VaR for a given severity, its parameter values, the percentile being estimated, and the sample size of the loss sample. The more apparently convex is VaR, the greater the downward scaling required to achieve an expected capital value centered on true capital. Both the magnitude of capital bias and the degree of presumed convexity of VaR, reflected in part in RCE's "*c*" parameter, are functions of i) the severity selected, ii) its estimated parameter values, and iii) the size of the quantile being estimated (e.g. for RCap vs. ECap). The magnitude of capital bias, although not the degree of presumed VaR convexity, also is a function of iv) the sample size of the loss dataset, as shown in Table 2 and Figure 1). However, in its current state of development, *c* is a function only of the severity selected and sample size, which appear to be the dominant drivers of capital bias. As shown in the Results section below, when using only sample size and the severity selected, RCE performs i) extremely well in terms of capital accuracy, eliminating virtually all capital bias except for a few cases under the smallest sample sizes $n \approx 150$, or $\lambda = 15$, ii) notably well in terms of capital precision, outperforming MLE by very wide and consistent margins, and iii) consistently, if less dramatically better than MLE in terms of capital robustness. If the size of the quantile mattered, we would see large differences, for a given value of *c*, in RCE's capital accuracy (and precision and robustness) for RCap vs. ECap, but that is not the case: there is negligible to very little difference (except for a few cases under the smallest sample sizes of $n \approx 150$, or $\lambda = 15$). Similarly for the parameter values: for a given value of *c*, but very different parameter values of the same severity, we would expect to see large differences in RCE's capital accuracy (and precision and robustness), but we do not: RCE's capital accuracy (and precision and robustness) is very similar across almost all parameter values of the same severity for a given value of *c*.

So while derivation of a fully analytic (yet practical) solution to estimating the degree of VaR's presumed convexity that relies on all four inputs may be very desirable, especially if it effectively addresses the few smaller-sample cases where RCE is not completely unbiased (although still much more accurate than MLE), it does not appear to be immediately essential: RCE effectively addresses MLE's deficiencies in terms of capital accuracy and capital precision, and to a lesser degree capital robustness, without identifiable areas in need of major improvements. So this analytic formula, if even possible to derive in tractable form,[43] is left for future research.

Finally, it is very important to note that all four of these inputs, i) – iv), and particularly the two currently used (i.e. the selected severity and sample size), are known ex ante, consistent with capital estimation under the LDA framework, and so they can be used as inputs to estimating capital using RCE without violating the ex ante nature of the estimation.

<u>RCE Conceptually Defined</u>: RCE is conceptually defined below in four straightforward steps.

**<u>Step 1</u>**: Estimate LDA-based capital using the chosen method (e.g. MLE).
**<u>Step 2</u>**: Use the severity (and frequency) parameter estimates from Step 1, treating them as reflecting the "true" data generating process, and simulate *K* data samples and estimate the severity (and frequency) parameter estimates of each.

[43] Note that for their fragility heuristic, a convexity metric in much simpler form than RCE and discussed later in this paper, Taleb et al. (2012) state: "Of course, ideally, losses would be derived in a closed-form expression that would allow the stress tester to trace out the complete arc of losses as a function of the state variables, but it is exceedingly unlikely that such a closed-form expression could be tractably derived, hence, the need for the simplifying heuristic." The excellent performance of RCE presented below in the Results section makes the need to derive undoubtedly complex, closed-form expressions for it much less pressing, or arguably even very useful, with the possible exception of its use under conditions of smaller sample sizes, as discussed below.

**Step 3**: For each of the $K$ samples in Step 2, simulate $M$ data samples using the estimated severity (and frequency) parameters as the data generation process, then estimate capital for each of the $M$ data samples, and calculate the median of the $M$ capital estimates, yielding $K$ medians of capital.

**Step 4**: Calculate the median of the $K$ medians of capital, calculate the mean of the $K$ medians of capital, and multiply the median of medians by the ratio of the two (median over mean) raised to the power "$c$":

$$\text{RCE} = \text{Median}(K \text{ capital medians}) * [\text{Median}(K \text{ capital medians}) \,/\, \text{Mean}(K \text{ capital medians})]^{c(sev,n)} \quad (3)$$

The first term of (3) can be viewed as close to the value that would be obtained using Step 1 alone, but it is more stable and thus, is preferable as it contributes to the stability of RCE. The second term – the ratio of median over mean – can be viewed as a measure of the apparent convexity of VaR, because the $K$ medians provide a stable trace of the VaR curve (g( ) in Figure 1), so the ratio of median to mean will decrease below 1.0 (one) as the apparent convexity of VaR increases. This ratio is augmented by $c(sev,n)$, which is a function of the sample size and the severity selected. Values for $c(sev,n)$ can be determined in one of two ways: i) using the values provided in Table E1 of Appendix E, by severity by sample size, or ii) using straightforward simulation study on a case by case basis (as was done to obtain the values in Table E1). Both alternatives are discussed in more detail in the following section.[44]

So conceptually, RCE traces the VaR curve shown in Figure 1, and then uses a simple measure of its presumed convexity to scale down the capital estimate. The goal is to scale the right amount so that on average, on the $Y$ axis (i.e. capital estimates), J.I. = $E[g(\hat{\beta})] - g(E[\hat{\beta}]) \approx 0$, or slightly above zero to be conservative. This is conceptually straightforward, but simulations of simulations (Steps 2 and 3) can be runtime prohibitive, depending on the sample size and number of UoMs for which capital must be estimated. In the implementation section below, I present a sampling method (actually, a perturbation method) that speeds this effort by orders of magnitude and provides even better stability than simple random sampling, especially for UoMs with smaller sample sizes.

[44] It should be noted here that because the first term of RCE is very close to capital based solely on, say, MLE, in a sense RCE can be viewed as an overlay, calibration, or adjustment to MLE-based capital. If they chose to swap this first term for LDA-based capital, banks currently using LDA would not have to change anything else in their framework to use RCE other than to apply RCE after Step 1 above. This change would alter the values of $c(sev,n)$ slightly, but this (re)calibration would not be onerous as it is identical to that described below: only with LDA-based capital as the first term of (3). The only disadvantage to this swap would be a slight increase in the variance of the capital estimate. Either way, this flexibility is a big advantage of RCE, especially for larger banks with more data, numerous UoMs, and large frameworks already implemented. Their use of RCE could merely be "on top of" their current framework: nothing else would need to change.

RCE Implemented

**Step 1**: Estimate LDA-based capital using the chosen method (e.g. MLE)

**Step 2**: Iso-Density Sampling – Use the severity parameter estimates from Step 1, treating them as reflecting the "true" data generating process, and invert their Fisher information to obtain their (asymptotic) variance-covariance matrix.[45] Then simply select 4 * *K* pairs of severity parameter estimates based on selected quantiles of the joint distribution of the severity parameters (those used in this paper correspond to the following percentiles: 1, 10, 25, 50, 75, 90, 99, so *K* = 7). Each severity parameter of the pair is incremented or decremented the same number of standard deviations away from the original estimates, in four directions tracing out two lines of severity parameter values as shown below in Figure 2 (the lines are orthogonal when the parameters are scaled by their standard deviations). In other words, taking the 99%tile as an example, i) both severity parameters are increased by the same number of standard deviations until the quantile corresponding to the 99%tile is reached; ii) both severity parameters are decreased by the same number of standard deviations until the quantile corresponding to the 99%tile is reached; iii) one severity parameter is increased while the other is decreased by the same number of standard deviations until the quantile corresponding to the 99%tile is reached; and iv) one severity parameter is decreased while the other is increased by the same number of standard deviations until the quantile corresponding to the 99%tile is reached. So we now have *K* = 7 * 4 = 28 pairs of severity parameter values. But we must also account for variation in $\lambda$, the frequency parameter, and so two values of $\lambda$ are used in this study: those corresponding to the 25%tile and the 75%tile of the Poisson distribution implied by the original estimate of $\lambda$. So now *K* = 28 * 2 = 56.

**Step 3**: Iso-Density Sampling – Using each of the *K* severity (and frequency) parameter estimates from Step 2 as defining the data generating process, generate via iso-density sampling *M* = 7 * 4 * 2 = 56 new severity (and frequency) parameter values for each set of estimates from Step 2, and now calculate their corresponding capital values. Then calculate the median of these *M* capital values to end up with *K* = 56 medians of capital.

**Step 4**: Using the *K* medians obtained from Step 3, calculate the median and calculate the weighted mean,[46] and multiply the median of medians by the ratio of the two (median over mean) raised to the power "*c*":

---

[45] Note that for many, if not most estimators used in this setting (e.g. M-class estimators), the joint distribution of the severity parameter estimates will be multivariate normal, and so the initial estimates taken together with the variance-covariance matrix completely define the estimated joint distribution.

[46] Because this is a weighted sampling, the mean is weighted by one minus the percentile associated with a particular iso-density multiplied by two times one minus that associated with the frequency percentile (since the frequency and severity parameters are assumed to be independent – see Ergashev, 2008, for more on this topic; weight = [1 – p-sev] * 2 * [1 – p-freq]). Technically the weighted median should be used alongside the weighted mean, but empirically the weighted median, which requires additional computational steps, always is identical to the unweighted median here due to the symmetry of the joint parameter distribution. And so for efficiency's sake, the unweighted median is used here.

$$\text{RCE} = \text{Median}\,(K \text{ capital medians}) * [\text{Median}\,(K \text{ capital medians}) \;/\; \text{Mean}\,(K \text{ capital medians})]^{c(sev,n)} \quad (3)$$

**FIGURE 2: Iso-density Sampling of the Joint Severity Parameter Distribution**

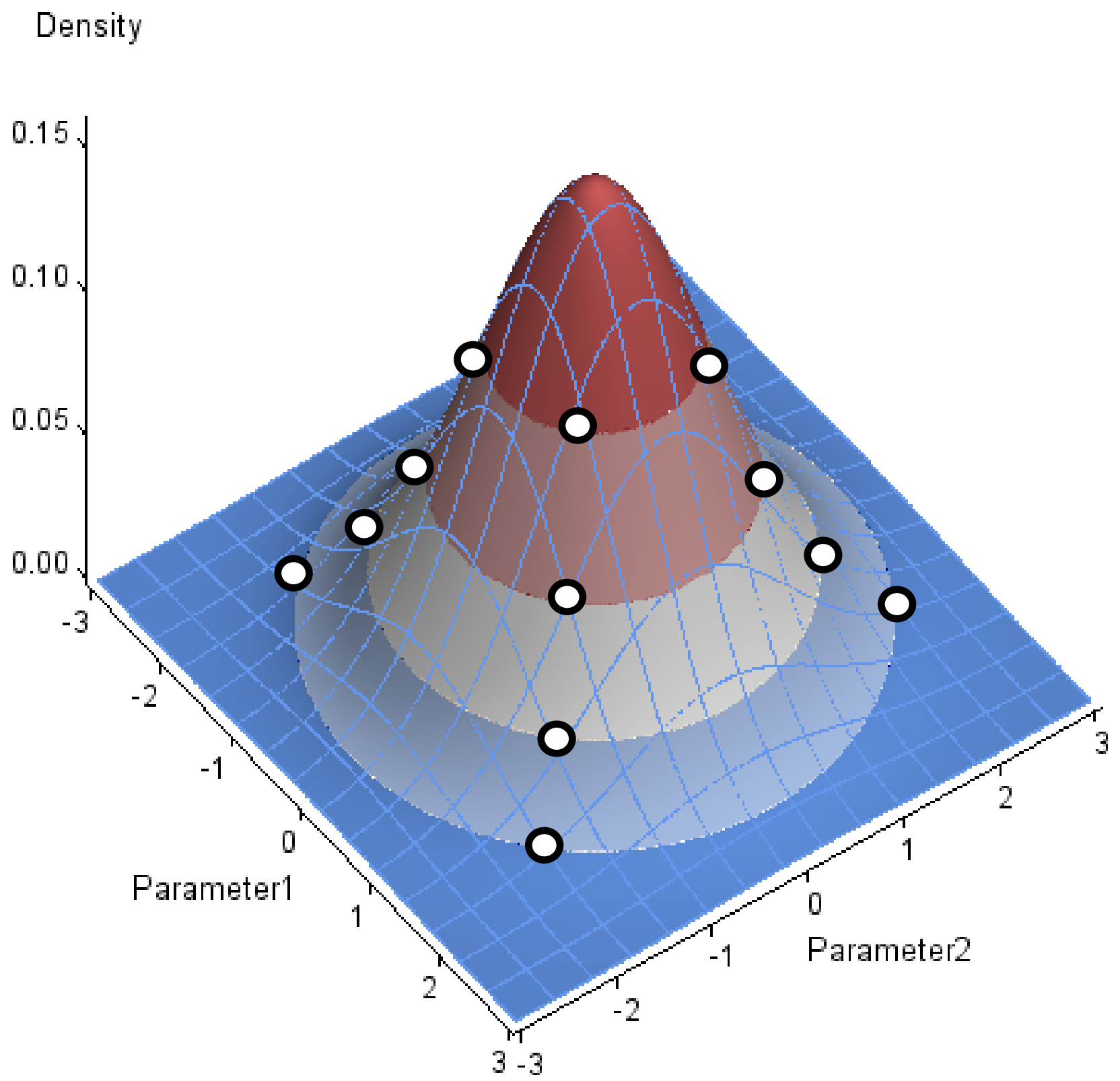

This is a rapid and stable way to systematically perturb parameters,[47] based on the joint (asymptotic) parameter distribution, to obtain a view of capital as a function of VaR. It also is quite accurate, arguably even more accurate for smaller samples than relying on simple random sampling, which for some of the data samples and some of the parameter values not uncommon in this setting can lead to truly enormous empirical variability. Asymptotically, in theory, both approaches are approximately equivalent as long as proper weighting is used when sampling via iso-densities. But in practice, simple random sampling in this setting can be i) extremely variable and unstable; ii) often more prone to enormous data outliers than theory would lead one to expect; and iii) often more prone to enormous estimate outliers because for many heavy-tailed severities, estimation of large parameter values simply is very difficult and algorithmic convergence is not always achieved. Even though iso-density "sampling" (perturbation) relies on an asymptotic result, it appears to not only be a much faster alternative, but also a more stable one in this setting, which is characterized by smallish samples and extremely skewed, heavy-tailed densities (not to mention heterogeneous loss data even within UoMs).

[47] Note that the point of iso-density "sampling" is *not* to draw a representative sample of the joint parameter distribution, but rather, to systematically perturb parameters for the purpose of tracing VaR as a function of the severity parameters.

To efficiently obtain the values of the severity parameters on a specified percentile ellipse,[48] one must utilize knowledge of the joint parameter distribution, or at least its approximation. If using, say, any M-class estimator to estimate severity parameters, we know the joint (asymptotic) distribution of the estimates is multivariate normal. With knowledge of the Fisher information of each,[49] therefore, we can use (4), where *x* is a *k*-dimensional vector of parameter values (incremented/decremented away from $\mu$ by the same number of standard deviations), $\mu$ is the known *k* -dimensional mean vector (the parameter estimates), $\sum$ is the known covariance matrix (the inverse of the Fisher information of the given severity), and $\chi_k^2\left(p\right)$ is the quantile function for probability *p* of the Chi-square distribution with *k* degrees of freedom.

$$\left(x-\mu\right)^T \Sigma^{-1}\left(x-\mu\right) \leq \chi_k^2\left(p\right) \tag{4}$$

In two-dimensional space, i.e. when $k = 2$, which is relevant for the widespread use of two-parameter severities in this setting, this defines the interior of an ellipse, which is a circle if there exists no dependence between the two severity parameters (if the joint parameter distribution is multivariate normal, a circle will be defined if the (Pearson's) correlation is zero). *x* represents the distance from the parameter estimates, $\mu$. Thus can one find the values of the severity parameters that provide a specified quantile of the joint distribution with (4). One can chose points on the ellipse that correspond to movement of each parameter the same number of standard deviations away from $\mu$ by using (5). Simply increment/decrement both parameters by *q* units of their respective standard deviations to obtain four pairs on the ellipse: increase both parameters by *q* standard deviations $\left(z_1 = z_2 = 1\right)$, decrease both parameters by *q* standard deviations $\left(z_1 = z_2 = -1\right)$, increase one while decreasing the other $\left(z_1 = 1, z_2 = -1\right)$, and decrease one while increasing the other $\left(z_1 = -1, z_2 = 1\right)$.

$$q\#stdev = \sqrt{\frac{\chi_k^2\left(p\right)\cdot\left(1+z_1 z_2 \rho_{1,2}\right)}{2}} \tag{5}$$

where $\sigma_1\left(\sigma_2\right) =$ stdev of parameter 1 (2), and $\rho_{1,2}$ is the correlation between the parameter estimates. (see Mayorov, 2014). Alternately, the eigenvalues and eigenvectors of $\Sigma^{-1}$ can be used to define the most extreme parameter values (smallest and largest) on the ellipses (corresponding to the largest/smallest eigenvalues) (see Johnson and Wichern, 2007), but this would likely change the values of *c*(*sev*,*n*) calculated in Appendix E, and (5) is arguably more straightforward.

---

[48] The specified percentile represents the percentage of the joint density within the ellipse. For severities with more than two parameters, and so dimensions higher than two, this is termed an ellipsoid.

[49] See Appendix D.

Other approaches to mitigating bias due to convexity, typically using bootstraps or exact bootstraps to shift the distribution of the estimator, simply do not appear to work in this setting either because the severity quantile that needs to be estimated is so extremely large (e.g. $[1 - (1-\alpha)/\lambda] = 0.99999$ for ECap assuming $\lambda=30$), or because this quantile is extrapolated so far "out-of-sample," or because VaR is the risk metric that must be used, or some combination of these reasons (see Kim and Hardy, 2007). Some that were tested here worked well for a particular severity for a very specific range of parameter values, but in the end all other options failed when applied across very different severities and very different sample sizes and very different parameter values. RCE was the only approach that reliably estimated capital unambiguously better than did MLE under the LDA framework,[50] across the wide range of conditions examined in this paper (see Simulation Study section below).

An important implementation note must be mentioned here: when calculating capital based on large severity parameter values, say, the 99%tile of the joint distribution in Step 3, based on 99%tile severity estimates generated in Step 2, based on (sometimes) already large estimate of severity parameters originally, sometimes capital becomes incalculable: in this example, the number simply is too large to estimate (in this study, this only occurred, and rarely, for the severities with the heaviest tails: TGPD and TLOGG). So we need to ensure that missing estimates do not cause bias: for example, that a scenario cannot occur whereby only the "decrease, decrease" arm of the iso-density sample in Figure 2 has no missing values. Therefore, if any capital values are incalculably large on an ellipse, the entire ellipse, and all ellipses "greater" than it, are discarded from the calculation. This ensures that the necessary exclusion of incalculably large capital numbers do not bias statistics calculated on the remaining values, which by definition are symmetric around the original estimates.

Finally, I address here how $c(sev, n)$ is defined and calculated. Table E1 in Appendix E presents values of $c(sev, n)$ by severity by sample size which were empirically determined via simulation studies. The simulation study simply generated 1,000 RCE capital estimates for a given sample size for a given severity for different values of $c$: the value of $c$ that came closest to being unbiased, with a slightly conservative leaning toward small positive bias, is the value of $c$ used. Sample sizes tested, for a ten year period, included average number of loss events = $\lambda$ = 15, 25, 50, 75, and 100 for samples of approximately $n \approx$ 150, 250, 500, 750, and 1,000 loss events.[51] This is a very wide range of sample sizes compared to those examined in the relevant literature (see Ergashev et al, 2014, Opdyke and Cavallo, 2012a and 2012b, and Joris, 2013), and it arguably covers the lion's share of sample sizes in practice, unfortunately with the exception of the very small UoM's. For all sample sizes in between, from 150 to 1,000, straightforward linear and non-linear interpolation is used, as shown in Figure E1 in Appendix E, and preliminary tests show this interpolation to be reasonably accurate.[52] The Results section describes in detail the effects of sample size (and severity selected) on RCE-based capital estimates.

[50] Again, "better" here means with greater capital accuracy, greater capital precision, and greater capital robustness.

[51] As described previously, a Poisson frequency distribution was assumed, as is widespread accepted practice in the industry. Sample sizes are approximate because they are a function of a random variable, $\lambda$. This is described in more detail below.

The second way to obtain and use values of $c(sev, n)$ is to simply conduct the above simulation study for a specific sample size and, say, three sets of severity parameters: the estimated pair (for a two-parameter severity), a pair at the 2.5%tile of the joint parameter distribution (obtained from (4)), and a pair at the 97.5%tile to provide a 95% joint confidence internal around the estimated values. If the same value of $c(sev, n)$ "works" for all three pairs of severity values,[53] thus appropriately taking into account severity parameter variability, then it is the right value for "*c*." As described in the Results section below, the distribution of RCE-based capital estimates was surprisingly robust to varying values of $c(sev, n)$. In other words, the same value of $c(sev, n)$ "worked" for very large changes in severity parameters (and capital), and very large changes in percentiles (e.g. RCap vs. ECap). So determining the value of $c(sev, n)$ empirically in this way, i.e. testing to make certain that the same value of $c(sev, n)$ holds for ±95% joint confidence interval (or a wider interval if deemed more appropriate), should properly account for the fact that our original severity parameter estimates are just that: inherently variable estimates of true and unobservable population values. All sample sizes beyond the range examined in this paper (i.e. $n < 150$ or $n > 1{,}000$) should make use of this approach, although for some severities, $c(sev, n)$ may vary by parameter values for small $n$. So caution is urged in the application of RCE to smaller samples than studied in this paper. Note again from Table 2 that for larger sample sizes beyond $n \approx 1{,}000$, all severities will exhibit much less bias because parameter variance is sufficiently small. However, RCE is extremely useful even in these cases in notably reducing capital variability and increasing model stability, as discussed further below.

Before addressing RCE runtimes below, I describe two more innovations, in addition to the efficient use of iso-density sampling, that are derived in this paper and that increase runtime speed by nearly an order of magnitude for one of the severities examined (a fourth innovation related to both runtime speed and extreme quantile approximation is presented in the next section). The two-parameter Truncated LogGamma distribution typically is parameterized in one of two ways: either with its second shape parameter, b, specified as b, or alternatively, 1/b. The latter is used throughout this paper. An analytic expression of the mean of the former is provided in Kim (2010),[54] but a corresponding result for the latter does not appear to have been derived in the literature, so this is done in Appendix D. Also, while the Fisher information of the Truncated LogGamma has been derived and used for operational risk capital estimation previously (see Opdyke and Cavallo, 2012a, and for the non-inverted parameterization of the Truncated LogGamma, see Zhou, 2013), both examples rely on computationally expensive numeric integration, so an analytic approximation is derived and presented in Appendix D that provides speed increases, for very precise approximations,[55] of between seven and ten times faster than that required for numeric integration.

---

[52] The non-linear interpolation is based on (6) presented in the next section.

[53] Here, "works" means that the three means of each of the three capital distributions of 1,000 RCE capital estimates all are very close to their respective "true" capital values.

Relying on these innovations, RCE runtimes are shown below in Table 3 by severity by sample size. All analyses presented in this paper were conducted using SAS® on a desktop computer (16GB RAM, 64-bit OS, CPU @ 3.40GHz). Note that because of the efficient use of the Fisher information and iso-density sampling, sample size has no effect on runtimes. Runtimes only vary by severity based on the complexity of the Fisher information and the capital calculation. Most importantly, note that to estimate capital for one UoM, implementing RCE as described above takes only a second or two, even for severities with the most complex Fisher informations (e.g. the Truncated LogGamma).

**TABLE 3: Runtime Speed of RCE by Severity (seconds)**

| | Non-Truncated | | Truncated (*H*=$10k) | |
|---|---|---|---|---|
| **Severity** | **Real Time** | **CPU Time** | **Real Time** | **CPU Time** |
| LogNormal | 0.14 | 0.14 | 1.10 | 1.10 |
| LogGamma | 1.13 | 1.12 | 2.96 | 2.94 |
| GPD | 0.21 | 0.18 | 1.35 | 1.35 |

From the above description of RCE, it should be clear that the general approach taken here to developing a capital estimator that is straightforward to understand, easy to implement, and *that works in practice* has been one of reliance on both appropriate theoretical results as well as practical empiricism. This is consistent with the late George Box's[56] approach to the development and use of robust methods. Box emphasized that "practical need often leads to theoretical development" (Box, 1984) and that "to obtain a useful procedure, one needs *both empiricism and theory*. But – more than that – one needs continuous iteration between them…" (Box and Luceño, 1998). This iteration not only guided the development of RCE specifically, and should guide its further development, but also should guide research on this topic area generally as increased focus is placed on the capital distribution. As operational risk modelers know, operational risk capital estimation is nothing if not highly resistant to idealized, textbook solutions that are mathematically convenient and neatly packaged. Keeping an open mind to pursuing approaches that appropriately balance important theoretical underpinnings with practical empirical, methodological, and regulatory constraints to arrive at workable, useable solutions for real world problems is absolutely necessary in this setting, and must drive the research agenda beyond strictly theoretical work that has limited use in practice.

[54] Technically, Kim (2010) provides the conditional tail expectation, which is different from the truncated mean, although the former simply needs to be multiplied by a constant to obtain the latter. See Appendix D for more details.

[55] The largest deviations from true capital when using this approximation in all the simulation studies conducted herein were a few thousand dollars when true capital was in the hundreds of millions, and sometimes billions of dollars.

[56] George Box is widely recognized as one of the fathers of modern, applied statistics. He often is cited as the source of the adage, "All models are wrong, but some models are useful."

RCE Range of Application: Severity Estimators

Although RCE can make use of most, if not all severity estimators used in this setting, it is implemented here essentially as an overlay or calibration to MLE for several reasons. Not only is MLE typically the fastest and easiest estimator to implement, but also it is most appropriate here for comparability purposes: we must hold all else constant when comparing RCE to the most widely used alternative, that is, MLE. Then the comparison is truly "apples-to-apples," because we know the only difference between the two capital distributions of RCE (when based on MLE) vs. MLE (used alone) is the use of RCE, and any observed differences are due only to RCE and cannot be the result of using some other estimator in Step 1 of RCE's implementation.

RCE Range of Application: Methods of Extreme Quantile Estimation

Another innovation presented in this paper is an improved method for approximating the extreme quantile of the aggregate loss distribution when the severity distribution can be characterized by infinite mean, *or close*.[57] As discussed above, the convolution of the frequency and severity distributions rarely yields a closed-form aggregate loss distribution from which VaR is easily estimated, but a number of methods for approximating VaR are widely used. These include mean-adjusted Single Loss Approximation (SLA, see Degen, 2010), Fast Fourier Transform (FFT, see Embrechts and Frei, 2009), Panjer Recursion (see Panjer, 1981, and Embrechts and Frei, 2009), extensive Monte Carlo simulation (see Opdyke and Cavallo, 2012a), numeric approximations (e.g. the Direct Method – see Kato, 2013), Indirect Estimation (see Sahay et al., 2007), and Closed-Form Approximations (see Hernandez et al., 2013). Extensive Monte Carlo simulation is the gold standard here, but it remains extremely computationally expensive because the quantiles being estimated are so large, so very large numbers of simulations are required to adequately represent the extreme empirical tail of the loss distribution. FFT is stable and faster than Panjer recursion, but mean-adjusted SLA is the most widely used method as it has the advantage of being a straightforward formula. As such it provides the fastest implementation, especially when testing or comparing estimators which typically requires many simulations. SLA also is widely accepted as being sufficiently accurate (see Hess, 2011; and Opdyke and Cavallo, 2012a). But it suffers from a serious implementation flaw: it contains divergent roots when, for severities that can have means approaching infinity (herein, for example, GPD, LogGamma, Truncated GPD, and Truncated LogGamma), the tail index approaches the value "one," either from above or below.[58] This is shown in Figure 3 for GPD, with reference to (2.a) and (2.c), below and above $\xi = 1$, respectively. Importantly, the parameter values do not need to be very close to "one" for the capital approximation to noticeably diverge from its true value,[59] so something must be done to address this when relying on data that is fitted to these distributions (or relying on data simulated from these distributions) to estimate capital. While the indirect method of Sahay et al. avoids this problem, it requires an additional loop for a root-finding algorithm, and so most likely is slower than a formula-based approach.

[57] "Or close" is emphasized here because any method that relies on simulation, as does RCE, will need to work under conditions of infinite mean upon encountering parameter values that are *close* to values that induce an infinite mean, because the simulations based on them will invariably generate some parameter values that correspond to infinite mean severities.

Both the MLE and RCE implementations of LDA in this paper make use of "ISLA" – Interpolated SLA – to avoid divergence in the capital estimate as tail index values approach one, yet still retain the speed advantages of a formula-based approximation. ISLA uses a straightforward nonlinear interpolation at predefined starting and ending points of the tail index values, as shown in (6) below. All notation corresponds to that used in (2.a,b,c).

Both the precision value (PRE = 1,000) and the root value (Root = 50) were sufficiently accurate in this setting: estimated capital via ISLA always was within ±1% of estimated capital based on extensive monte carlo simulation for all severities where infinite means are possible (herein, LogGamma, Truncated LogGamma, GPD, and Truncated GPD). The example of GPD is shown in Figure 3. Capital is calculated for $\xi = 0.8$ and $\xi = 1.2$ (using the estimated value for $\theta$), and if the estimated $\xi$ lies within this range, the interpolated capital value is used. This straightforward, if brute force method provides very accurate approximations to true capital, and importantly, is very fast computationally: ISLA is almost an order of magnitude faster than root-avoiding alternatives that require numeric integration (see Mignola and Opdyke, 2012). Of course, the start and endpoints for the interpolation must be determined for each severity, and sometimes can vary slightly based on

---

58 Verifying whether the Closed-Form Approximations of Hernandez et al. (2013) avoid this root divergence is beyond the scope of this paper.

59 This divergence holds regardless of sample size: it does not disappear asymptotically, that is, even if the number of losses in the sample approaches infinity. The divergence below $\xi = 1$ is due to the mean approaching $\infty$ (see 2.a), and divergence above $\xi = 1$ is due to the fact that $\Gamma(s)$ diverges as $s$ approaches zero from either direction, so $\Gamma(1 - 1/\xi)$ will diverge as $\xi$ approaches one (see 2.c).

**FIGURE 3: Correction for SLA Divergence at Root of $\xi = 1$ for GPD Severity ($\theta = 55{,}000$)**

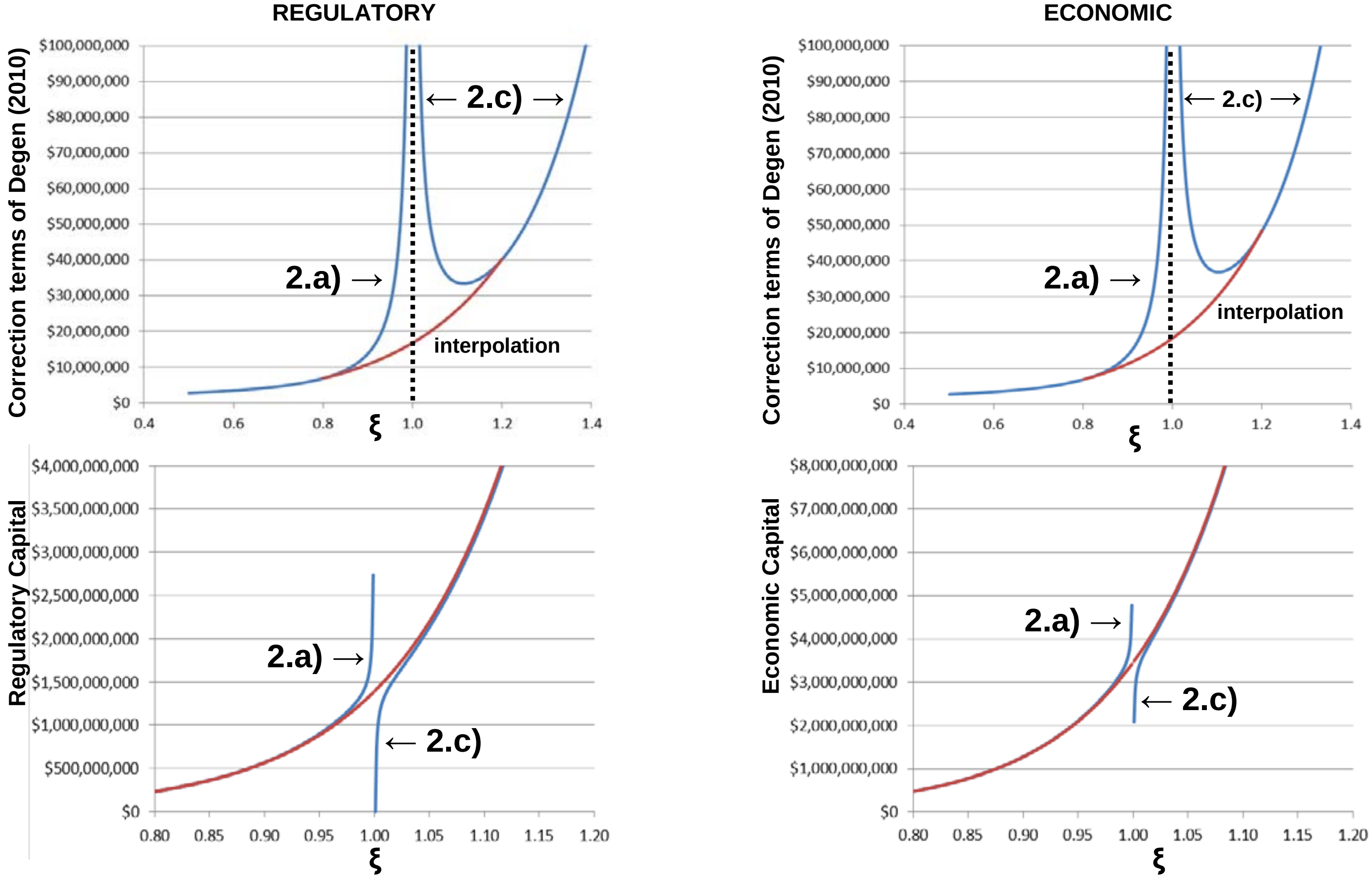

Low Value of Tail Parameter = LTP = $\xi_{LOW}$ (= 0.80 for GPD)

High Value of Tail Parameter = HTP = $\xi_{HIGH}$ (= 1.20 for GPD)

High Correction Term = HCT = $(1-\alpha)F^{-1}\left(1-\frac{1-\alpha}{\lambda};\xi_{HIGH}\right)\cdot\left(\frac{c_{\xi_{HIGH}}}{1-1/\xi_{HIGH}}\right)$ where $c_{\xi_{HIGH}} = (1-\xi_{HIGH})\frac{\Gamma^2(1-1/\xi_{HIGH})}{2\Gamma(1-2/\xi_{HIGH})}$

Low Correction Term = LCT = $\lambda\mu_{LOW}$ (where $\mu_{LOW}$ is based on $\xi_{LOW}$)

Precision = PRE = 1,000 ; Root = 50

Full Range Count = FRC = $(\xi_{HIGH} - \xi_{LOW})\cdot$ PRE

Count in Range = CIR = $(\xi - \xi_{LOW})\cdot$ PRE where $\xi$ is estimated from the sample

Difference Root Scale = DRS = $[\text{HCT}^{(1/\text{ROOT})} - \text{LCT}^{(1/\text{ROOT})}] / [\text{FRC} - 1]$

Interpolated Correction Term = ICT = $[\text{LCT}^{(1/\text{ROOT})} + \text{CIR}\cdot\text{DRS}]^{\text{ROOT}}$

Estimated Capital via ISLA = $\approx F^{-1}\left(1-\frac{1-\alpha}{\lambda}\right) + \text{ICT}$ (6)

the parameter values of the distribution.[60] But determining these values once, visually, is straightforward, as shown in Figure 3, and given the conditional nature of LDA capital estimation (i.e. it is conditioned upon the selection of a severity distribution first), this does not violate the ex ante nature of the estimation process.

**Simulation Study**

Framework

This simulation study compares i) LDA-based capital estimates relying solely on MLE for severity and frequency parameter estimation (the most common implementation of LDA) versus ii) LDA-based capital estimates generated by RCE implemented as described in the previous section (where MLE is used for the frequency and severity parameter estimation). One thousand samples of loss data, which are i.i.d. in the basecase, are simulated to generate one thousand capital estimates, and the characteristics of the two capital distributions are compared to each other, and to the true capital values. As described above, the three main criteria examined are capital accuracy (unbiasedness), capital precision (the spread of the capital distribution), and capital robustness (i.e. distribution characteristics under contaminated, non-i.i.d. data). Accuracy is determined by the capital bias: simply, are the expected values of the capital distributions close to, or far from the true capital values? Spread is determined by a number of descriptive statistics, including the standard deviation, inter-quartile range, coefficient of variation, 95% empirical confidence intervals, and RMSE (which also incorporates bias) of the two capital distributions. Other important distributional characteristics, such as the skewness and kurtosis, also are compared. Finally, robustness is determined by an MLE vs. RCE comparison of deviations from the two respective capital distributions under i.i.d. simulations when non-i.i.d. data simulations are generated, as described below.

It is very important to note that by both design and necessity, the results from this simulation study assume that the right model, i.e. the right severity (but of course, not static parameter values), is selected to estimate capital: the choice of severity in this setting has notoriously low statistical power, and is yet another source of enormous variance in capital estimation. However, severity selection is a distinct component of the capital estimation framework, and by design this study focuses on capital estimation conditional on (the right) severity selection, just as the LDA framework does. To compare capital estimators, by necessity all else must be held constant, so this study does not examine and include other sources of variance in LDA-based capital from other areas of the framework (such as severity selection or estimating dependence structure across UoMs).

[60] This is true for the LogGamma and Truncated LogGamma, but values of "b" do not approach "one" in this setting, so this is more a statistical coding precaution than an immediate concern for capital calculation, as it is for GPD (and Truncated GPD) as shown in Figure 3. Note, too, that only for the GPD does the tail index happen to equal $\xi$: for the LogGamma, for example, the tail index = 1/b.

*Sample Size*

The basecase sample size of losses ($n \approx 250$, corresponding to ten years of losses with $\lambda = 25$) is conservatively set a bit larger than that of many UoM's so that any observed effects of Jensen's inequality on the MLE-based capital numbers generally are understated relative to the "typical" UoM (because in relative terms, the effects almost always increase as sample sizes decrease). Table 2 above shows four additional simulations of the MLE-based capital estimates for smaller ($n \approx 150$) and larger ($n \approx 500$, 750, and 1,000) samples to demonstrate empirically the very strong role that the size of the variance of the severity parameter estimators, vis-à-vis sample size, plays in biasing capital estimates.[61]

*Severity (and Frequency) Distributions*

Operational risk losses are simulated based on a Poisson frequency distribution, and six of the most commonly used severity distributions:[62] the LogNormal, LogGamma, and the Generalized Pareto (GPD) distributions,[63] as well as the truncated versions of each with a truncation threshold of $H$ = \$10,000, arguably the most widely used data collection threshold. As described above, the use of truncated distributions is the most widely accepted method for addressing data collection thresholds, unlike some alternatives that have been pillared in regulatory review processes and in the literature (for example, the use of so-called "shifted" distributions – see Schevchenko, 2009, and Ergashev et al., 2014; but for a counter-argument supporting shifted distributions, see Cavallo et al., 2012).

As described above, the implementation of RCE efficiently utilizes the Fisher information, and analytic derivations exist for five of the six severities listed above (see Appendix D, which includes a) definitions of these severities, b) closed-form analytic expressions for their means for use in calculating capital, and c) their Fisher informations). For the Truncated LogGamma, a new analytic approximation of the Fisher information is derived in Appendix D, thus avoiding computationally costly numeric integration. This speeds computer runtime by almost an order of magnitude, so this approximation is used for this severity for both MLE and RCE-based capital estimates. Also derived in Appendix D is a closed-form analytic expression for the mean of the Truncated LogGamma severity,[64] which also decreases runtimes over the alternative requiring numeric integration.

---

[61] As mentioned previously, these findings are exactly consistent with other empirical results in the literature (for example, see Ergashev et al., 2014, and Opdyke and Cavallo, 2012a).

[62] See for example, Opdyke, 2013, Opdyke and Cavallo, 2012a, 2012b, Zhou et al., 2013, and Joris, 2013.

[63] These are three of the four parametric severity distributions listed in the most recent Interagency Guidance on Operational Risk AMA severity estimation (see OCC, 2014).

*Range of Parameter and Capital Values*

Severity parameter values for simulating capital were selected i) to reflect values commonly cited in the literature (see Opdyke and Cavallo, 2012a, 2012b; Joris, 2013, and Zhou, 2013); ii) to reflect a very wide range of capital values (from \$38m to over \$10.6b) to better represent the wide range of conditions under which capital inflation is slightly vs. notably vs. extremely large (sometimes well over 100% greater than true capital), and to fully test the behavior of RCE; and iii) to reflect different parameter values (of the same severity) while holding capital roughly constant to demonstrate the different effects of individual severity parameters on capital, and how these effects are exactly consistent with Jensen's inequality. The basecase is $\lambda = 25$ for ten years of data, or approximately $n \approx 250$. Parameter values generate (true) regulatory capital in the basecase that ranges from \$53m to over \$2.7b. While some may consider the upper end of this range large at the level of the UoM, the bulk of the range actually is fairly conservative for many large banks, and even many regional or mid-sized banks. For some of the larger sample sizes, however, such as $\lambda = 75$ and 100, even though the same parameter values are used as with the smaller sample sizes, the larger numbers of losses generate capital numbers that stretch the bounds of what would be seen in practice. Again, the purpose for this was to test the bounds of RCE, as well as to determine the point at which sample sizes increased enough to notably mitigate the upward bias of LDA-MLE-based capital. Otherwise, focus should be on the capital numbers corresponding to $\lambda = 25$, which despite their size, correspond to parameter values that are not uncommon in practice, especially for the larger banks.

*Method of Capital Calculation / Approximation*

As defined above, the ISLA method is used for approximating the VaR of the aggregate loss distribution to estimate capital, both for RCE and MLE. To calculate (2.a), analytic formulae for the means of all the severities used in this study are presented in Appendix D.

*Robustness*

Robustness of MLE vs. RCE is tested via "left tail," "right tail," and "both tail" contamination of the i.i.d. basecase sample. The left tail distribution is the same severity as the basecase, but with different parameter values that correspond to the lower 5%tile of the joint parameter distribution (where both basecase parameter values are decreased[65] by the same number of standard deviations via (5) to obtain these values). The same is true for the right tail distribution, but with parameter values that both are increased to correspond to the upper 95%tile. This conceptually is the equivalent of a multivariate 90% confidence interval, which is a very plausible, if not conservative reflection of a non-textbook, non-i.i.d. empirical reality (alternately, Johnson and Wichern, 2007, can be used to obtain parameter values on the same ellipse corresponding to the largest eigenvalues). Each contaminating distribution comprises 5%, on average, of the overall severity (and so the "both tails" case has 5% contamination from each tail, on average (because the percent of the distribution contaminated is stochastic)).

---

[64] Note that Kim (2010) presents an analytic expression for the mean of an alternate parameterization of the truncated LogGamma: his is based on a Gamma distribution with two shape parameters, rather than a shape parameter and an inverted (b=1/b) shape parameter, as is used in this paper.

While the above addresses violations of the "identical distribution" portion of the i.i.d. assumption, incorporating the effects of violations of the independence assumption on operational risk capital estimation has been largely ignored in this setting (the few exceptions include Guégan and Hassani, 2013, Umande, 2013, and Embrechts et al., 2013). This may be due, as least in part, to flexibility on this specific issue in the US Final Rule (2007).[66] While such testing is beyond the scope of this study, operational risk losses are very likely to be serially correlated (see Guégan and Hassani, 2013), and other empirical examinations of this issue generally show that its deleterious effects on statistical inference can be very material (see van Belle, 2008). So this is an issue that should continue to be further addressed in future research.

*RCE parameters*

Values of $c(sev, n)$ used are those listed in Table E1. Iso-densities used for sampling correspond to the percentiles listed above: 1, 10, 25, 50, 75, 90, and 99 for severity parameters, and 25 and 75 for the frequency parameter. RCE capital is based on severity and frequency parameters estimated using MLE. While RCE can be applied to capital estimated with almost any estimator, as mentioned above the use of MLE makes "all else equal" when comparing RCE capital to capital estimated via the most widely used estimator (MLE). In other words, the only differences must be attributed to RCE. And these differences will have the most direct relevance to the largest number of financial institutions because most currently are using only MLE.

---

[65] The exceptions to this are the LogGamma and Truncated LogGamma distributions. In this paper, they rely on a Gamma parameterization with a shape parameter and an inverted (b=1/b) shape parameter, which consequently means that smaller values of b correspond to larger capital estimates, so b is decreased where all other severity parameters are increased, and vice versa.

[66] See US Final Rule (2007), p.69318: "A bank's chosen unit of measure affects how it should account for dependence. Explicit assumptions regarding dependence across units of measure are always necessary to estimate operational risk exposure at the bank level. *However, explicit assumptions regarding dependence within units of measure are not necessary, and under many circumstances models assume statistical independence within each unit of measure. The use of only a few units of measure increases the need to ensure that dependence within units of measure is suitably reflected in the operational risk exposure estimate.*" (emphasis added).

Results

Results of the simulation study described above can be seen in Tables 4a,b and 5 below, and complete results can be found in Appendix F in Tables F4a,b-F11a,b. Based on the three criteria that arguably are the only ones that matter for assessing the effectiveness of an operational risk capital estimation framework – capital accuracy, capital precision, and capital robustness – the improvements provided by RCE over MLE are, respectively, nothing short of dramatic, very notable, and modest but consistent. Except for discussions relating either specifically to sample size or specifically to robustness, below I focus on Tables 4a,b, which is the basecase of $\lambda = 25$ with number of losses $n \approx 250$ (each of the tables, including Tables 4a,b as well as F4a,b-F8a,b in Appendix F, corresponds to a different sample size related to $\lambda = 25$, 15, 50, 75, and 100, respectively; and "a" and "b" indicate RCap and ECap, respectively. Additionally, Tables F9a,b-11a,b correspond to right-tail contamination, left-tail contamination, and both-tail contamination, respectively).

*Capital Accuracy*

Empirically, we can see that systemically positive bias in MLE-based capital estimates is driven by the three factors identified above: i) As shown earlier in Table 2, sample size drives parameter estimate variance which drives MLE-based capital bias: larger sample sizes are associated with smaller bias, all else equal (see Table 2, and Tables F4a,b-F8a,b). This is exactly consistent with the findings of Ergashev et al. (2014) and Opdyke and Cavallo (2012a). ii) Higher values of VaR, represented by ECap, always exhibit far more bias than lower values of VaR, represented by RCap, all else equal (see Table 4a vs. 4b). iii) Heavier tailed distributions exhibit more bias, all else equal (LogNormal is least heavy, GPD is heaviest, and truncated distributions exhibit more capital bias than their non-truncated counterparts, all else equal, although by design parameter values vary to some degree for the truncated and non-truncated versions of the same severity in this study to keep the overall range of capital estimates at comparable levels) (see Tables 4a,b).

To be more specific regarding iii), consistent with the results of Appendix A, Figure A1, we must note that for the same severity, capital is roughly the same in each pair of adjacent rows of Tables F4a,b-F8a,b, but MLE-based capital bias is consistently larger for the second row of the pair for the LogNormal distributions and GPD distributions. For the former, VaR is a convex function of both $\mu$ and $\sigma$, but the biasing effect of the shape parameter, $\sigma$, dominates over that of the scale parameter ($\log(\mu)$), so when $\sigma$ is the larger of the two for roughly the same level of capital, capital bias is larger. For the GPD distributions, VaR is a convex function of $\xi$, but a linear function of $\theta$, and so the latter induces no biasing effects at all on its own (although it is positively correlated with $\xi$). So when $\xi$ is the larger of the two for roughly the same level of capital, capital bias is larger. The LogGamma, on the other hand, has two shape parameters which are negatively correlated, and VaR is a convex function of both (see Appendix A, Figure A1). This is why the results are mixed for the LogGamma:

**TABLE 4a: RCE vs. MLE RCap Distributions – Bias and RMSE by Severity by Parameter Values ($millions)**

| Severity Dist. | Parm1 | Parm2 | True RCap | Mean MLE RCap | MLE Bias | MLE Bias% | Mean RCE RCap | RCE Bias | RCE Bias% | RMSE MLE RCap | RMSE RCE RCap | RMSE RCap RCE/MLE |
|---|---|---|---|---|---|---|---|---|---|---|---|---|
| | **μ** | **σ** | | | | | | | | | | |
| LogN | 10 | 2 | $63 | $67 | **$4** | **6.7%** | $63 | $0 | **0.5%** | $25 | $23 | **91.8%** |
| LogN | 7.7 | 2.55 | $53 | $59 | **$6** | **11.5%** | $54 | $1 | **1.5%** | $30 | $26 | **87.7%** |
| LogN | 10.4 | 2.5 | $649 | $720 | **$72** | **11.0%** | $658 | $9 | **1.4%** | $355 | $313 | **88.1%** |
| LogN | 9.27 | 2.77 | $603 | $686 | **$83** | **13.8%** | $614 | $12 | **2.0%** | $382 | $328 | **86.0%** |
| LogN | 10.75 | 2.7 | $2,012 | $2,275 | **$263** | **13.1%** | $2,048 | $37 | **1.8%** | $1,229 | $1,063 | **86.5%** |
| LogN | 9.63 | 2.97 | $1,893 | $2,198 | **$305** | **16.1%** | $1,939 | $46 | **2.4%** | $1,329 | $1,121 | **84.3%** |
| TLogN | 10.2 | 1.95 | $76 | $85 | **$9** | **11.5%** | $75 | -$1 | **-1.8%** | $52 | $41 | **79.4%** |
| TLogN | 9 | 2.2 | $76 | $96 | **$20** | **26.5%** | $75 | -$1 | **-1.4%** | $88 | $50 | **56.9%** |
| TLogN | 10.7 | 2.385 | $670 | $847 | **$177** | **26.4%** | $700 | $30 | **4.5%** | $665 | $469 | **70.5%** |
| TLogN | 9.4 | 2.65 | $643 | $894 | **$251** | **39.1%** | $628 | -$14 | **-2.2%** | $1,087 | $536 | **49.3%** |
| TLogN | 11 | 2.6 | $2,085 | $2,651 | **$566** | **27.1%** | $2,123 | $38 | **1.8%** | $2,568 | $1,771 | **69.0%** |
| TLogN | 10 | 2.8 | $1,956 | $2,743 | **$787** | **40.2%** | $1,965 | $9 | **0.5%** | $3,033 | $1,694 | **55.9%** |
| | **a** | **b** | | | | | | | | | | |
| Logg | 24 | 2.65 | $85 | $97 | **$12** | **13.6%** | $87 | $2 | **2.1%** | $62 | $53 | **86.0%** |
| Logg | 33 | 3.3 | $100 | $108 | **$8** | **8.5%** | $99 | $0 | **-0.4%** | $56 | $50 | **89.0%** |
| Logg | 25 | 2.5 | $444 | $513 | **$70** | **15.7%** | $455 | $11 | **2.5%** | $355 | $301 | **84.8%** |
| Logg | 34.5 | 3.15 | $448 | $497 | **$49** | **10.9%** | $452 | $4 | **0.8%** | $296 | $260 | **87.6%** |
| Logg | 25.25 | 2.45 | $766 | $906 | **$140** | **18.3%** | $799 | $32 | **4.2%** | $647 | $543 | **83.9%** |
| Logg | 34.7 | 3.07 | $818 | $930 | **$112** | **13.7%** | $841 | $23 | **2.8%** | $589 | $510 | **86.6%** |
| TLogg | 23.5 | 2.65 | $124 | $193 | **$70** | **56.1%** | $137 | $13 | **10.8%** | $273 | $181 | **66.4%** |
| TLogg | 33 | 3.3 | $130 | $174 | **$44** | **34.1%** | $130 | $0 | **-0.1%** | $173 | $93 | **53.9%** |
| TLogg | 24.5 | 2.5 | $495 | $794 | **$299** | **60.4%** | $516 | $20 | **4.1%** | $1,103 | $554 | **50.2%** |
| TLogg | 34.5 | 3.15 | $510 | $635 | **$125** | **24.5%** | $539 | $29 | **5.8%** | $544 | $397 | **73.1%** |
| TLogg | 24.75 | 2.45 | $801 | $1,305 | **$504** | **62.9%** | $848 | $47 | **5.9%** | $1,938 | $916 | **47.3%** |
| TLogg | 34.6 | 3.07 | $867 | $1,078 | **$211** | **24.3%** | $925 | $58 | **6.7%** | $927 | $709 | **76.5%** |
| | **ξ** | **θ** | | | | | | | | | | |
| GPD | 0.8 | 35,000 | $149 | $233 | **$85** | **56.9%** | $152 | $3 | **2.2%** | $295 | $167 | **56.7%** |
| GPD | 0.95 | 7,500 | $121 | $212 | **$91** | **75.6%** | $124 | $3 | **2.7%** | $311 | $156 | **50.3%** |
| GPD | 0.875 | 47,500 | $391 | $640 | **$249** | **63.7%** | $396 | $5 | **1.2%** | $870 | $466 | **53.6%** |
| GPD | 0.95 | 25,000 | $403 | $697 | **$295** | **73.2%** | $408 | $5 | **1.3%** | $1,019 | $513 | **50.3%** |
| GPD | 0.925 | 50,000 | $643 | $1,079 | **$436** | **67.8%** | $645 | $1 | **0.2%** | $1,535 | $792 | **51.6%** |
| GPD | 0.99 | 27,500 | $636 | $1,121 | **$486** | **76.4%** | $637 | $2 | **0.3%** | $1,698 | $828 | **48.8%** |
| TGPD | 0.775 | 33,500 | $141 | $214 | **$73** | **52.0%** | $144 | $3 | **2.2%** | $297 | $170 | **57.4%** |
| TGPD | 0.8 | 25,000 | $140 | $220 | **$80** | **56.9%** | $145 | $5 | **3.3%** | $315 | $179 | **56.8%** |
| TGPD | 0.8675 | 50,000 | $452 | $737 | **$285** | **63.0%** | $466 | $13 | **3.0%** | $1,062 | $576 | **54.3%** |
| TGPD | 0.91 | 31,000 | $451 | $761 | **$309** | **68.6%** | $463 | $12 | **2.7%** | $1,174 | $603 | **51.4%** |
| TGPD | 0.92 | 47,500 | $698 | $1,149 | **$451** | **64.7%** | $704 | $7 | **0.9%** | $1,668 | $888 | **53.2%** |
| TGPD | 0.95 | 35,000 | $717 | $1,206 | **$489** | **68.2%** | $715 | -$2 | **-0.2%** | $2,009 | $991 | **49.3%** |

*NOTE: #simulations = 1,000; λ = 25 for 10 years so n ~ 250; α=0.999

**TABLE 4b: RCE vs. MLE ECap Distributions – Bias and RMSE by Severity by Parameter Values ($millions)**

| Severity | | | | Mean | | | Mean | | | RMSE | RMSE | RMSE |
|---|---|---|---|---|---|---|---|---|---|---|---|---|
| Dist. | Parm1 | Parm2 | True | MLE | MLE | MLE | RCE | RCE | RCE | MLE | RCE | ECap |
| | μ | σ | ECap | ECap | Bias | Bias% | ECap | Bias | Bias% | ECap | ECap | RCE/MLE |
| LogN | 10 | 2 | $107 | $115 | **$8** | **7.8%** | $108 | $1 | **1.1%** | $47 | $43 | **91.3%** |
| LogN | 7.7 | 2.55 | $107 | $121 | **$14** | **13.2%** | $109 | $3 | **2.5%** | $66 | $57 | **87.0%** |
| LogN | 10.4 | 2.5 | $1,286 | $1,449 | **$163** | **12.7%** | $1,316 | $30 | **2.4%** | $769 | $673 | **87.4%** |
| LogN | 9.27 | 2.77 | $1,293 | $1,498 | **$205** | **15.8%** | $1,333 | $40 | **3.1%** | $898 | $764 | **85.2%** |
| LogN | 10.75 | 2.7 | $4,230 | $4,864 | **$634** | **15.0%** | $4,352 | $123 | **2.9%** | $2,828 | $2,425 | **85.8%** |
| LogN | 9.63 | 2.97 | $4,303 | $5,097 | **$794** | **18.5%** | $4,461 | $158 | **3.7%** | $3,321 | $2,769 | **83.4%** |
| TLogN | 10.2 | 1.95 | $126 | $144 | **$18** | **14.7%** | $124 | -$2 | **-1.3%** | $101 | $77 | **76.1%** |
| TLogN | 9 | 2.2 | $133 | $179 | **$46** | **35.0%** | $131 | -$2 | **-1.5%** | $202 | $97 | **48.0%** |
| TLogN | 10.7 | 2.385 | $1,267 | $1,678 | **$411** | **32.4%** | $1,338 | $71 | **5.6%** | $1,521 | $1,003 | **65.9%** |
| TLogN | 9.4 | 2.65 | $1,297 | $1,966 | **$669** | **51.6%** | $1,264 | -$33 | **-2.5%** | $2,910 | $1,192 | **41.0%** |
| TLogN | 11 | 2.6 | $4,208 | $5,639 | **$1,431** | **34.0%** | $4,337 | $129 | **3.1%** | $6,319 | $4,072 | **64.4%** |
| TLogN | 10 | 2.8 | $4,145 | $6,279 | **$2,134** | **51.5%** | $4,177 | $32 | **0.8%** | $8,119 | $3,972 | **48.9%** |
| | a | b | | | | | | | | | | |
| Logg | 24 | 2.65 | $192 | $225 | **$33** | **17.0%** | $199 | $7 | **3.4%** | $163 | $137 | **84.1%** |
| Logg | 33 | 3.3 | $203 | $225 | **$22** | **10.7%** | $204 | $1 | **0.4%** | $131 | $115 | **87.5%** |
| Logg | 25 | 2.5 | $1,064 | $1,272 | **$208** | **19.5%** | $1,105 | $42 | **3.9%** | $984 | $814 | **82.7%** |
| Logg | 34.5 | 3.15 | $960 | $1,090 | **$130** | **13.5%** | $978 | $18 | **1.8%** | $734 | $631 | **86.0%** |
| Logg | 25.25 | 2.45 | $1,877 | $2,300 | **$423** | **22.5%** | $1,986 | $109 | **5.8%** | $1,842 | $1,507 | **81.8%** |
| Logg | 34.7 | 3.07 | $1,794 | $2,097 | **$302** | **16.8%** | $1,869 | $74 | **4.1%** | $1,500 | $1,273 | **84.9%** |
| TLogg | 23.5 | 2.65 | $271 | $496 | **$225** | **83.1%** | $297 | $27 | **9.9%** | $903 | $530 | **58.6%** |
| TLogg | 33 | 3.3 | $261 | $382 | **$120** | **46.1%** | $247 | -$15 | **-5.6%** | $458 | $191 | **41.6%** |
| TLogg | 24.5 | 2.5 | $1,164 | $2,152 | **$988** | **84.9%** | $1,099 | -$65 | **-5.6%** | $3,620 | $1,370 | **37.8%** |
| TLogg | 34.5 | 3.15 | $1,086 | $1,437 | **$350** | **32.2%** | $1,158 | $72 | **6.6%** | $1,453 | $941 | **64.8%** |
| TLogg | 24.75 | 2.45 | $1,928 | $3,618 | **$1,689** | **87.6%** | $1,855 | -$74 | **-3.8%** | $6,604 | $2,269 | **34.4%** |
| TLogg | 34.6 | 3.07 | $1,892 | $2,493 | **$601** | **31.8%** | $2,050 | $158 | **8.4%** | $2,499 | $1,764 | **70.6%** |
| | ξ | θ | | | | | | | | | | |
| GPD | 0.8 | 35,000 | $382 | $696 | **$313** | **81.9%** | $396 | $13 | **3.5%** | $1,069 | $521 | **48.7%** |
| GPD | 0.95 | 7,500 | $375 | $785 | **$410** | **109.2%** | $390 | $14 | **3.8%** | $1,398 | $588 | **42.1%** |
| GPD | 0.875 | 47,500 | $1,106 | $2,123 | **$1,016** | **91.9%** | $1,130 | $24 | **2.2%** | $3,514 | $1,594 | **45.4%** |
| GPD | 0.95 | 25,000 | $1,251 | $2,576 | **$1,325** | **105.9%** | $1,279 | $28 | **2.2%** | $4,585 | $1,930 | **42.1%** |
| GPD | 0.925 | 50,000 | $1,938 | $3,835 | **$1,898** | **97.9%** | $1,955 | $17 | **0.9%** | $6,657 | $2,882 | **43.3%** |
| GPD | 0.99 | 27,500 | $2,076 | $4,375 | **$2,299** | **110.7%** | $2,095 | $19 | **0.9%** | $8,085 | $3,275 | **40.5%** |
| TGPD | 0.775 | 33,500 | $351 | $617 | **$266** | **75.7%** | $365 | $13 | **3.8%** | $1,109 | $543 | **48.9%** |
| TGPD | 0.8 | 25,000 | $361 | $660 | **$299** | **82.9%** | $379 | $18 | **5.0%** | $1,193 | $579 | **48.5%** |
| TGPD | 0.8675 | 50,000 | $1,267 | $2,432 | **$1,166** | **92.0%** | $1,327 | $61 | **4.8%** | $4,337 | $1,988 | **45.9%** |
| TGPD | 0.91 | 31,000 | $1,334 | $2,672 | **$1,338** | **100.4%** | $1,389 | $55 | **4.1%** | $5,165 | $2,203 | **42.6%** |
| TGPD | 0.92 | 47,500 | $2,088 | $4,048 | **$1,960** | **93.9%** | $2,129 | $41 | **2.0%** | $7,203 | $3,235 | **44.9%** |
| TGPD | 0.95 | 35,000 | $2,227 | $4,474 | **$2,246** | **100.8%** | $2,246 | $19 | **0.8%** | $9,606 | $3,873 | **40.3%** |

*NOTE: #simulations = 1,000; λ = 25 for 10 years so n ~ 250; α=0.9997

sometimes a larger value for "a" induces more bias at the same level of capital, and sometimes larger (smaller) values of "b" induce more bias at the same level of capital.

The bias of MLE-based capital estimates is material for, arguably, every result generated by this simulation study at $\lambda$ = 25. Even for the lowest capital bias in absolute terms in Table 4a – $4m corresponding to a true capital value of $63m under a (non-truncated) LogNormal severity – it is difficult to argue that pushing the button to implement RCE in under a second is not justified. And capital typically is subject to far greater inflation – sometimes even more than double true capital in relative terms, or well over $2b beyond true capital in absolute terms (see Table 4b). The larger numbers generally are more relevant to the larger banks, mainly because the parameter values of the severities typically are larger for them. But even for the smaller and medium-sized banks, it is important to remember that these numbers are *per UoM*. The cumulative effect of bias from every UoM is likely to be quite large, even for those banks not classified as the largest, and even after diversification benefits are taken into account. In contrast to MLE-based capital estimates, the accuracy of RCE is always within ±11% of true capital, except for when $\lambda$ = 15 where 11 of 72 RCE simulations (about 15%) deviated more than 11% from true capital (but were still much closer to true capital – see Tables F5a,b).

**TABLE 5: Summary of Capital Accuracy by Sample Size – MLE vs. RCE ($millions)**

| | ECap | | | | RCap | | | |
|---|---|---|---|---|---|---|---|---|
| | Mean of Absolute Bias | | Median of Absolute Bias | | Mean of Absolute Bias | | Median of Absolute Bias | |
| **λ =** | **RCE** | **MLE** | **RCE** | **MLE** | **RCE** | **MLE** | **RCE** | **MLE** |
| **15** | 7.8% | 92.6% | 2.6% | 82.3% | 5.9% | 61.6% | 1.6% | 58.1% |
| **25** | 3.4% | 53.1% | 3.3% | 40.6% | 2.4% | 38.1% | 2.0% | 30.6% |
| **50** | 2.8% | 25.7% | 2.7% | 17.7% | 2.0% | 19.4% | 1.9% | 14.3% |
| **75** | 1.2% | 15.5% | 0.8% | 10.7% | 0.8% | 11.9% | 0.5% | 8.7% |
| **100** | 0.9% | 11.3% | 0.5% | 7.9% | 0.5% | 8.7% | 0.4% | 6.1% |
| **15** | $61 | $825 | $18 | $502 | $21 | $228 | $5 | $154 |
| **25** | $45 | $727 | $29 | $410 | $14 | $209 | $8 | $133 |
| **50** | $69 | $617 | $52 | $320 | $20 | $182 | $15 | $109 |
| **75** | $40 | $526 | $14 | $250 | $11 | $157 | $3 | $80 |
| **100** | $32 | $485 | $15 | $223 | $7 | $142 | $5 | $73 |

A summary of capital accuracy from Tables F4a,b-F8a,b, in both relative and absolute terms, for MLE and RCE across sample sizes is shown in Table 5. Obviously this is heavily dependent on the severities used in this study (although they are the most widely used in practice), but it demonstrates how, even when MLE bias shrinks in relative terms as sample sizes increase, in absolute terms it remains quite large. RCE bias, in contrast, always is far smaller. It also is important to note that on average, RCE was slightly larger than true capital. This is important from the perspective of conservatism: any estimator should avoid even the appearance of benefiting from use merely to decrease required capital below what is consistent with regulatory intent (which, here, is estimation closest to the true capital numbers).

In sum, the inflation bias of MLE-based capital estimates can be enormous under conditions that are not uncommon, and that of RCE typically is small, and often de minimis: except for a few cases of smaller sample sizes (when $\lambda = 15$), RCE essentially and effectively eliminates capital bias due to Jensen's inequality.

*Capital Precision and Model Stability*

Model stability long has been cited by many as the most important and most difficult challenge for AMA operational risk capital estimation. Some of the major factors contributing to model (in)stability in this setting include the quality of the samples of loss event data (or lack thereof), notable data paucity, the arguably inherent heterogeneity of operational risk UoM definition, the (non-)robustness of the parameter estimators selected and used, low statistical power in the choice of severity, and critically, the size of the variance of the output of the model (here, the capital estimates). Achieving adequate precision in the model output under ideal, textbook data conditions unarguably is the first step in achieving broader model stability. Put differently, if under idealized conditions (e.g. perfectly homogenous i.i.d. loss event data) a model framework cannot generate capital estimates that are precise enough to use to make reliable inferences about the true values of capital, then there is no way that improving the other components of model stability will ever make a model "stable." The starting point for achieving and/or testing for model stability is testing the model when the data satisfies all, or nearly all of the assumptions required of the model (e.g. i.i.d. data). If a model fails this test, it will never achieve "model stability," and so this is the capital precision test applied here to RCE and standalone MLE.

Capital precision is measured by a number of descriptive statistics of the MLE-based and RCE-based capital distributions, including the standard deviation, the coefficient of variation, the RMSE (which also incorporates bias), the inter-quartile range, and empirical 95% confidence intervals. Although not statistics of "spread," relevant, too, are the skewness and kurtosis of these distributions. By every single one of these criteria, in every single simulation, RCE is more precise than MLE, and sometimes dramatically so (it also is less skewed and less kurtotic than MLE in all cases). This is a very strong result, although not unexpected because RCE capital is essentially scaled MLE capital, where the scaling factor, which is always less than one and greater than zero, is a function of the degree of apparent convexity of VaR for a given sample size and severity: the more convex is VaR for a particular estimate, the smaller the value of the scaling factor. Most notably, RMSE of RCE-based capital estimates is less than half that of MLE-based capital estimates in fully half of the ECap basecase simulations (see Table 4b). The numbers are similar for the standard deviations of the capital distributions of RCE vs. MLE, and even the inter-quartile range, which is less affected by MLE's extreme outliers than is RMSE or standard deviation, sometimes is less than half the value for RCE compared to MLE. Importantly, the empirical 95% confidence intervals, which are embarrassingly large for MLE, are much smaller for RCE, on average only two thirds the size across both the RCap and ECap results; and in one eighth of all cases, they are less than half the size of those of MLE. By any measure, RCE-based capital estimates are notably more precise than MLE-based capital estimates.

*Capital Robustness*

Tables F9a,b-F11a,b in Appendix F show the equivalent of the basecase Tables 4a,b for $\lambda = 25$ but with 5% right-tail contamination, 5% left-tail contamination, and 10% both-tail contamination, respectively, where contamination[67] comes from the same distribution with parameter values at either end of the 90% confidence interval of the joint parameter distribution, as described above. For all right-tail contamination simulations,[68] RCE capital still is much less biased than is MLE capital in absolute terms, but this is not what matters when assessing robustness since the capital distribution is now "wrong" – it no longer represents the original "true" capital numbers because it is contaminated. What matters is how much capital deviates from their original estimates under no contamination, i.e. under i.i.d. data samples. The average absolute deviation of MLE economic capital is 18.9% – more than twice as large as that of RCE economic capital (8.7%). Because ECap is subject to greater convexity farther out in the right tail, these numbers are larger than the respective RCap numbers – 11.8% and 6.4% – but this relative difference is still notable. That these differences are not even larger arguably is mostly a function of RCE's utilization of MLE as its capital estimator. While this was necessary for an apples-to-apples comparison to MLE that isolates the effects of using RCE, if RCE was used with a more robust severity estimator, these differences likely would be larger, ceteris paribus.

For all left-tail contamination simulations, RCE capital again is much less biased than MLE capital in absolute terms, but what matters is the extent to which capital deviates from the original estimates under no contamination, i.e. under i.i.d. data samples. For ECap, MLE deviates an average of 9.5% while RCE deviates an average of 6.5%. The respective numbers for RCap were 7.2% and 5.2%. Because we are dealing with the estimation of extremely large quantiles, it is not surprising that contamination in the other direction has less effect on the extent to which either estimator deviated from its i.i.d. values. But again, RCE deviated less.

[67] Note again that the %contamination is stochastic.

[68] Note that two Truncated LogGamma severities – those with the smallest values of b – required larger percentages of right-tail contamination to achieve some stability in their estimation. This is not surprising given that this parameter drives the extreme tail of the distribution, and estimation of such a heavy-tailed distribution is difficult even under i.i.d. conditions when data samples are not large; so under contaminated conditions, estimation difficulties are not uncommon. This was also true for one of the Truncated LogGamma distributions and one of the GPD distributions under left-tail contamination.

Finally, under both-tail contamination, for ECap MLE deviated an average of 6.3% while RCE deviated an average of 3.1%. The respective numbers for RCap were 4.1% and 2.1%.

In sum, while the increased robustness of RCE capital over MLE capital is consistent, is it not as notable as the strong capital precision gains provided by RCE or the dramatic capital accuracy gains provided by RCE over MLE. This is at least in part due to the reliance on MLE estimates in this study before the RCE capital adjustment is applied. Also, the largest robustness advantage of RCE over MLE occurred where robustness arguably matters most: under non-i.i.d. data in the extreme right tail of the aggregate loss distribution. Finally, it is worth noting that the deviations violated the expected order of the contaminated simulations more often for MLE capital than for RCE capital. In other words, for a given severity, in general one would expect a larger negative deviation under left-tail contamination, a smaller negative or a smaller positive deviation under both-tail contamination, and a large positive deviation under right-tail contamination. But due to random sampling error these comparisons "crossed over" each other, violating this order 9 of the 72 possible comparisons for MLE capital.[69] Given the conditions of the estimation exercise this is not surprising. But it is noteworthy that, in contrast, this happened only 3 times for RCE capital.[70] This is consistent with a more stable, less variable, and more robust capital estimator.

**Discussion**

"If you can't measure something, you can't understand it. If you can't understand it, you can't control it. If you can't control it, you can't improve it." - H. J. Harrington

I restate Harrington's measurement dictum here because operational risk capital modeling is all about measurement, yet it faces serious constraints and obstacles that span the empirical, the methodological, and the regulatory that make its measurement extremely challenging. As examined in this paper, these constraints often interact in material and complex ways: estimating an exceedingly large severity quantile (regulatory and methodological constraint) of medium- to heavy-tailed severities (empirical and regulatory constraint) under sample sizes that typically are fairly small and rarely "large" (empirical constraint) and subject to biasing effects apparently due to Jensen's inequality (methodological constraint) all converge in a perfect storm of estimation challenges that make what appears to be a fairly straightforward framework (LDA) arguably virtually unusable, as currently implemented across the industry, for operational risk mitigation purposes because it cannot effectively measure and estimate capital. One arrives at this conclusion not only based on the results provided in this paper, but also according to some of the most respected and established operational risk practitioners (see OR&R, 2013). Absent any one of the factors listed above, the deleterious effects on capital estimation either are notably mitigated or disappear altogether, but unfortunately their simultaneous occurrence in any given UoM is not uncommon, and arguably common.

[69] No results showed a "double" violation, that is, none showed estimated capital from a left-tail deviation that was larger than that from a right-tail deviation.

[70] Under a null hypothesis of 9 "cross-overs" out of 72 comparisons ($p$ = 0.125) and a binomial sample space, the probability of 3 or fewer cross-overs occurring is only about p = 0.016.

RCE was designed specifically to address these issues head-on. As shown above, RCE-based capital is dramatically more accurate, notably more precise, and modestly though consistently more robust than is MLE-based capital. Because RCE uniformly lowers capital requirements at the unit-of-measure level, it must do the same at the enterprise level, and likewise must increase capital stability from quarter to quarter, ceteris paribus. This decrease in capital, however, is not merely consistent with regulatory intent, but rather, is arguably *more* consistent with regulatory intent than most, if not all other implementations of LDA. We can say this simply because other implementations generate capital estimates that are systematically and materially biased upwards under conditions that are not uncommon. Regulatory intent cannot possibly support systemically and materially biased capital estimation but rather, unbiased capital estimation: in other words, capital estimates that, when based on multiple samples of loss event data, form a distribution with an *expected value* that is centered on true capital.[71] That regulators also would expect this unbiased capital estimator to be reasonably precise and robust to make reliable inferences about the true values of capital merely is consistent with their stated requirements for "credible … and verifiable processes … that most effectively enables it [the regulated bank/sifi] to quantify its exposure to operational risk" (see US Final Rule, 2007).

[71] One question raised about the capital distributions of LDA-based MLE vs. RCE is whether the median of the former is closer to true capital than is that of the latter, even though the mean of the latter is dramatically closer to true capital (which is to say RCE is essentially unbiased, whereas MLE often is very biased). An empirical examination of all the (i.i.d.) capital distributions generated in this study (360) shows this to be true: LDA-based MLE is consistently more "median-centered" than is RCE. On its face this might appear to be a comparative advantage of LDA-based MLE, but one must ask, "At what cost?" By every single measure of spread, RCE is consistently, if not dramatically more precise than is MLE: always. RCE also is systematically less skewed and less kurtotic than is MLE. So the cost of MLE's "median-centeredness" is an extremely long capital tail, which is exactly the thing all operational risk managers and analysts are trying to avoid the most. Also, we must question whether we are asking the right question: what is important is not whether a specific quantile (e.g. the median) of a distribution is closer to the true value of the estimate (here, capital), but rather, whether the entire distribution of a statistic is closer to the true value compared to that of another? RMSE is one metric, arguably the best one, that answers this question, and the RMSE of RCE is always notably, if not dramatically better (smaller) than that of MLE. Another way to answer address this issue is to find the cross-over points of the 72 baseline (i.e. $\lambda = 25$) capital distributions: that is, find the percentile at which the right tail of RCE-based capital becomes closer to true capital than does that of MLE-based capital. A percentile close to the median would indicate a very narrow range over which MLE arguably has any advantage. And the empirical answer is that all cross-over points are below the 62%tile, and over two thirds are below the 60%tile. So for the percentiles that matter (i.e. the right tail), RCE capital always is closer to true capital than is MLE capital.

## Conclusions

As we have seen above, RCE satisfies all the preferred criteria governing its development listed at the beginning of this paper. The typical consequences for large banks of using RCE rather than standalone MLE in their LDA frameworks will be: i) in many cases, notable reductions in capital, both for RCap and ECap, at both the UoM and enterprise levels; ii) in cases where capital inflation presumably due to Jensen's inequality is material, capital dramatically closer to "true" capital as defined by the LDA framework (and thus, capital estimates more consistent with and closer to that intended by regulators); iii) greater precision in these capital estimates, so capital will vary less and remain more stable from quarter to quarter, all else equal; and iv) more robustness to violations of the i.i.d. data assumption that plague even the most well-defined UoM's.

Smaller and medium-sized banks, whose severity parameter values typically will be smaller, all else equal, still should enjoy notable reductions in capital when RCE reduces bias across all UoM's, and they should enjoy the precision and robustness benefits of RCE as well. And to reemphasize yet again, all these reductions are *more* consistent with regulatory intent than are the MLE-based capital estimates under an LDA framework, so RCE unambiguously provides the proverbial "win-win." The capital estimator that systematically reduces the operational risk capital that banks and SIFIs must hold in reserve also is the one that gets to the "right" regulatory capital number. So there seems to be little to argue against the widespread use of RCE for AMA(LDA)-based operational risk capital estimation.

Areas of future research related to RCE include: i) Even though RCE dramatically mitigates capital inflation apparently due to Jensen's inequality, more needs to be done on the capital precision and capital robustness fronts. Where MLE-based capital variability is largest, RCE decreases variability the most: for example, for a GPD severity with $\xi = 0.99$ and $\theta = 27{,}500$, as shown on Table 4b, RMSE of MLE-based capital is about \$8.1b, but RMSE of RCE-based capital is only 39% of this value at \$3.3b. This is a large relative decrease, but in absolute terms \$3.3b is still a very large number (especially when true capital is \$2.1b!), and decreasing it further is vitally important for the capital planning that banks and SIFIs subject to these regulations must do. This may be one area, however, where methodological innovation may be limited by what many applied statisticians would call an ill-posed problem: that is, expecting to estimate severity quantiles associated with $p = 0.99999$ and higher with anything approaching a reasonable degree of precision. For example, Schevchenko (2011) uses a very straightforward calculation to show that under reasonable assumptions, even under i.i.d. loss data, sample size requirements for obtaining reasonable precision in the capital estimate are on the order of magnitude of a million or more loss events, or 50,000 to 100,000 years worth of loss data. Another way of stating this is that the variance associated with estimating extremely large quantiles – even taking any effects of Jensen's inequality out of the equation – inherently is extremely large, especially for the heavy-tailed severities that must be used in this setting. It may therefore take radical innovation and dramatic advances in applied statistics to even partially circumvent such a daunting challenge.

ii) A general mathematical proof of VaR's multidimensional surface convexity for extremely large quantile estimation for all the severities relevant to operational risk capital estimation would be desirable. However, it arguably is not strictly necessary in this setting since the number of severity distributions "allowed" and used in practice is quite finite, and the VaR of each at least can be checked for convexity in two straightforward ways: a) graphical checks for marginal convexity, as is done in Appendix A, Figure A1, and b) via a comparison of the mean of very straightforward capital simulations to true capital, which is required anyway to determine whether the size of the parameter estimates makes the capital bias material. Still, additional mathematical confirmation of these empirical findings would further validate them and more specifically define the conditions under which they hold.

iii) An analytic derivation of the values of $c(sev, n)$ as a function not only the severity distribution and the sample size, but also of the severity parameter values and the size of the quantile being estimated could be useful. If possible, this likely would improve RCE capital estimates under those few smaller-sample conditions where its bias is not negligible (see Tables F5a,b). And in Tables F6a,b, F7a,b, and F8a,b, we can see some differences in RCE accuracy by the size of the quantile being calculated, that is, by whether we look at RCap or ECap. Although RCE still vastly outperforms MLE in all these specific cases, it possibly could be improved if its value was based on an analytical derivation of its relationship with parameter values and quantile size.

Additionally, while a derivation of $c(sev, n)$ as an analytic function of sample size would be preferable to the empirical approach used in this paper, this is arguably more desirable than it is necessary because the range covered here spans most sample sizes where bias is material. This is true, also, of a derivation that holds across severity distributions because of the conditional nature of capital estimation under LDA: requiring knowledge of $c(sev, n)$ *after* the severity has been selected does not invalidate the ex ante nature of the capital estimation.

In the end, analytic derivations supporting empirical results always are preferable to empirical results alone, so such derivations would at the very least further validate the findings of this study and continue to rightly encourage focus of the research on operational risk capital estimation *on the capital distribution*, where it belongs. So at least for these purposes, such derivations are worth pursuing.

iv) As mentioned above, testing RCE's robustness based on violations of the independence assumption and not just the identically distributed assumption would be useful, especially since operational risk loss event data is likely to at least sometimes be serially correlated. Testing EVT-POT, spliced distributions, and kernel transformations for capital bias due to Jensen's inequality also could be important follow-ons to this paper.

I conclude by tying this research back to the broader operational risk setting with a recent quote from the head of a major US regulatory agency: "… it [operational risk] is currently at the top of the list of safety and soundness issues for the institutions we supervise. This is an extraordinary thing. Some of our most seasoned supervisors, people with 30 or more years of experience in some cases, tell me that this is the first time they have seen operational risk eclipse credit risk as a safety and soundness challenge. Rising operational risk concerns them, it concerns me, and it should concern you…" (Thomas J. Curry, Comptroller of the Currency, OCC, Before the Exchequer Club, May 16, 2012). From the perspective of operational risk capital estimation, the ominous tone here should trigger alarm based on the fact that the most widely used methods of estimating operational risk capital (such as MLE) under the most widely used framework (LDA) generate capital estimates that often are i) demonstrably, materially, and systematically inflated, and thus, utterly inconsistent with regulatory intent; ii) grossly imprecise by any measure; and iii) based on relatively fragile, mathematically convenient, idealized textbook assumptions that are consistently violated by real world data. This paper is the first to directly and comprehensively address these three issues by not only quantifying them, but also by i) identifying one of their major analytical sources, ii) specifying exactly the conditions under which they are material, and then iii) actually designing an approach to address, if not solve them. The broader objective, however, is to spur related research on these topics and focus more attention on the capital distribution. After all, capital estimation, not parameter estimation, is the endgame here, and as much time and resources must be dedicated to this as have been dedicated to research on severity parameter estimation if we are to make the existing framework more useable and useful in practice, not to mention more consistent with regulatory intent. Without this focus, the empirical evidence presented herein regarding MLE-based capital estimation under LDA transforms H. J. Harrington's aphorism on measurement and improvement into a dire warning that echoes Curry's concern. Simply put, we cannot improve business operations by effectively mitigating operational risk if the capital we estimate to represent it is not measured with reasonable accuracy, reasonable precision, or reasonable robustness.

In part because this paper began on a critical note, I would like to finish on a thought-provoking methodological note of optimism by placing RCE, and any similar capital-distribution-based research efforts, in a forward-looking context. The apparent effects of Jensen's inequality often are very damaging to capital estimation in this setting, and the approach taken by RCE is to control and eliminate them. But what if we could go a step further and actually exploit them? By this I mean, what if we could develop an estimator that not only mitigates or eliminates the biasing, imprecision, and non-robustness effects of convexity, but also becomes *less* rather than more biased, and *more* rather than less precise, and *more* rather than less robust, in the face of convexity? What if we could develop an estimator whose statistical properties actually *improve* under conditions of convexity? It is very telling that until the recent publication of Nicolas Nassim Taleb's book "Antifragile: Things that Gain from Disorder" (Taleb, 2014), there was not even a word to describe an estimator with such "antifragile" characteristics (note that "antifragility" is very distinct from "robustness").[72] But such an "antifragile" estimator would be enormously useful in this and many other risk settings where convexity is common, if not endemic, and plagues the effective use of a number of the most commonly used risk metrics (expected shortfall included), whether or not their users are aware of it (the need for this paper is a case in point). So regarding convexity and its deleterious effects on estimation generally, for capital and otherwise, we should aim high: first we must test for it, identify it, and measure it; then we must develop estimators like RCE to control it and in many ways, eliminate it; and finally, an ultimate goal would be to exploit it using "antifragile" estimators that actually *improve* when confronting it.

[72] For a mathematical treatment of antifragility, see Taleb and Douady (2013), and for an application to bank stress tests, see Taleb et al. (2012). The fragility heuristic (H) in both papers actually is conceptually similar to RCE in that both are measures of convexity based on perturbations of parameters: H measures the distance between the average of model results over a range of shocks and the model result of the average shock, while RCE is a scaling factor based on the ratio of the median to the mean of similar parameter perturbations. Both exploit Jensen's inequality to measure convexity: in the case of the fragility heuristic, to raise an alarm about it, and in the case of RCE, to eliminate it (or rather, to effectively mitigate its biasing effects on capital estimation).

## APPENDIX A

(note that the leftmost graphs in Figure A1, when scaled independently, exhibit notable convexity (except for GPD $\theta$, which is linear))

**FIGURE A1**: LogNormal Parameter $\boldsymbol{\mu}$ by Quantile by α, Parameter $\boldsymbol{\sigma}$ = 2, Threshold=$0

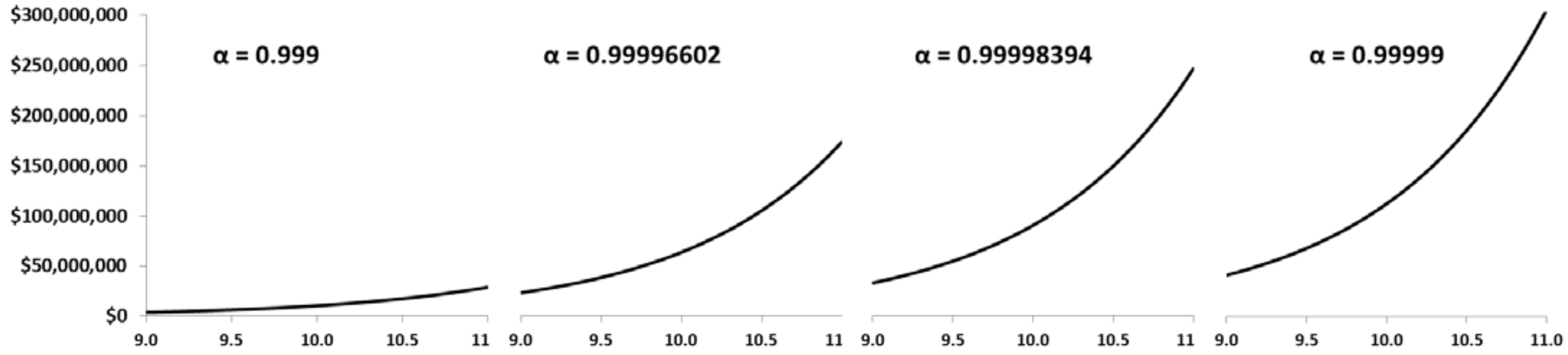

Parameter $\boldsymbol{\mu}$

LogNormal Parameter $\boldsymbol{\sigma}$ by Quantile by α, Parameter $\boldsymbol{\mu}$ = 10, Threshold=$0

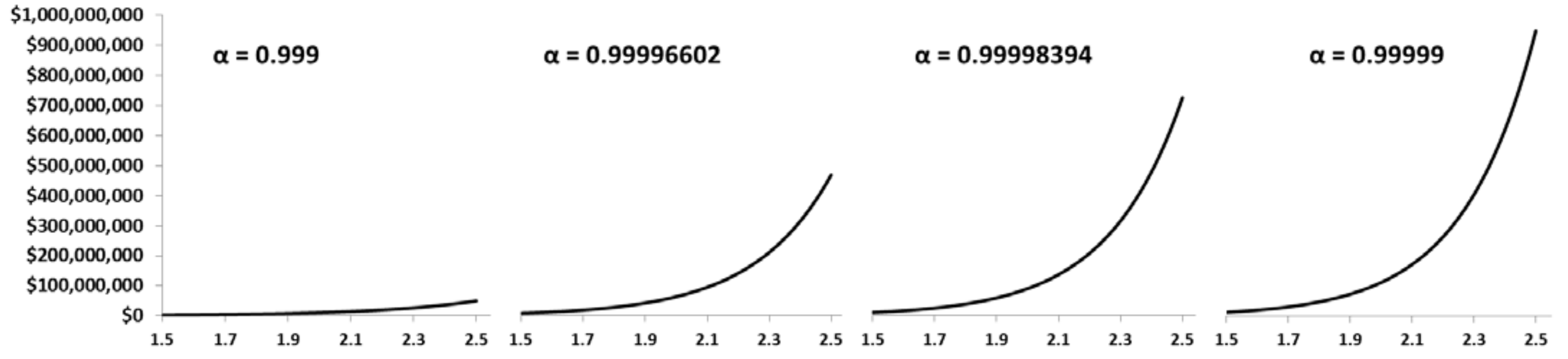

Parameter $\boldsymbol{\sigma}$

LogGamma Parameter ***a*** by Quantile by α, Parameter ***b*** = 2.65, Threshold=$0

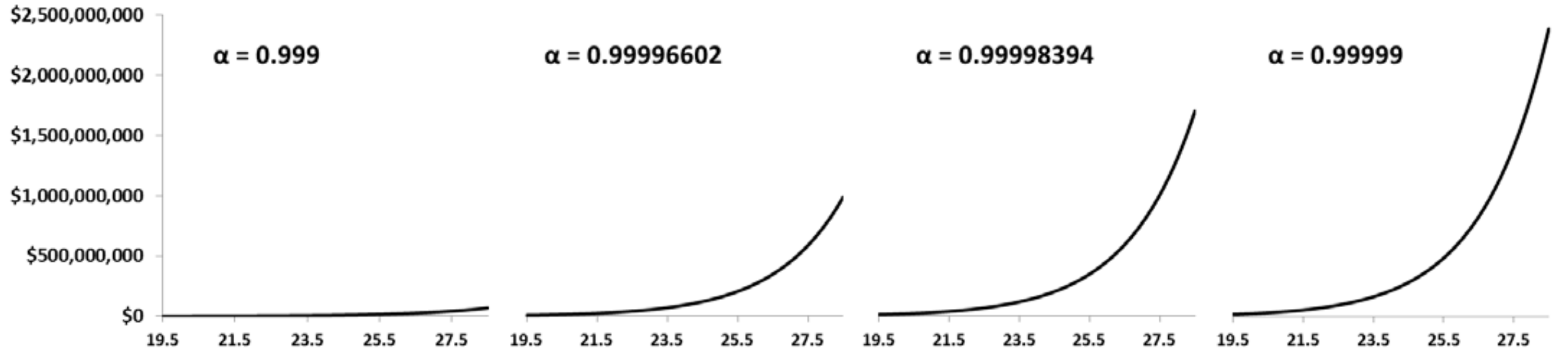

Parameter ***a***

LogGamma, Parameter ***b*** by Quantile by α, Parameter ***a*** = 24, Threshold=$0

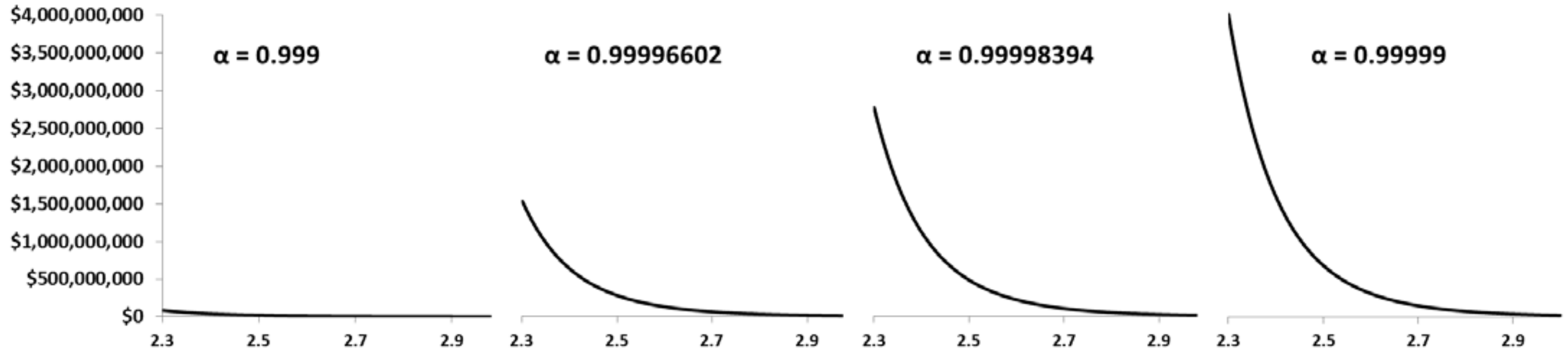

Parameter ***b***

GPD Parameter $\xi$ by Quantile by α, Parameter $\theta$ = 35,000, Threshold=$0

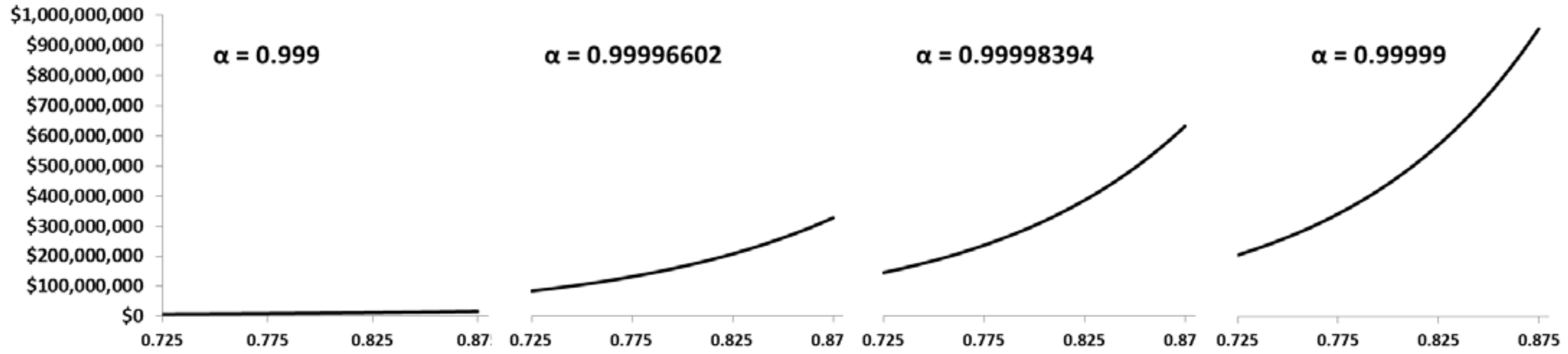

Parameter $\xi$

GPD, Parameter $\theta$ by Quantile by α, Parameter $\xi$ = 0.8, Threshold=$0

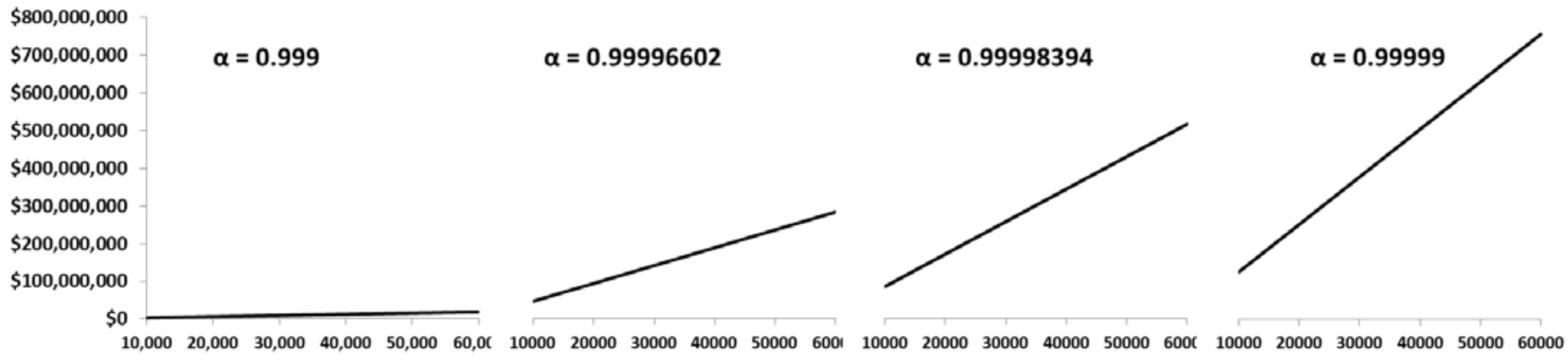

Parameter $\theta$

## APPENDIX B

**TABLE B1**: Regulatory and Economic Capital Distributions, 1,000 Simulations $\lambda$=25

| | GPD Severity ($\xi$ = 1.1, $\theta$ = 40,000) | | Normal Severity ($\mu$ = 500k, $\sigma$ = 1.5m) | |
|---|---|---|---|---|
| | RCap (α = 0.999) | ECap (α = 0.9997) | RCap (α = 0.999) | ECap (α = 0.9997) |
| True Capital | **$2,521,620,617** | **$9,432,295,763** | **$18,916,600** | **$19,336,006** |
| Mean | $4,606,994,975 | $20,895,168,520 | $18,890,719 | $19,310,226 |
| %Bias | 82.70% | 121.53% | -0.14% | -0.13% |
| RMSE | $7,809,076,769 | $43,461,854,111 | $2,583,208 | $2,584,629 |
| StdDev | $7,525,193,727 | $41,921,400,048 | $2,583,078 | $2,584,501 |
| IQR | $3,830,876,336 | $16,636,357,991 | $3,449,874 | $3,459,729 |
| 95%CIs | $22,811,712,755 | $117,827,228,494 | $9,948,476 | $9,963,910 |
| Skewness | 5.65 | 6.72 | 0.14 | 0.14 |
| Kurtosis | 48.80 | 66.47 | 0.07 | 0.06 |

(GPD Graphs $\xi$ = 1.1, $\theta$ = 40,000 are virtually identical to GPD graphs in Appendix A above)

**FIGURE B1**: Normal, Parameter $\mu$ by Quantile by α, Parameter $\sigma$ = 1,500,000, Threshold=$0

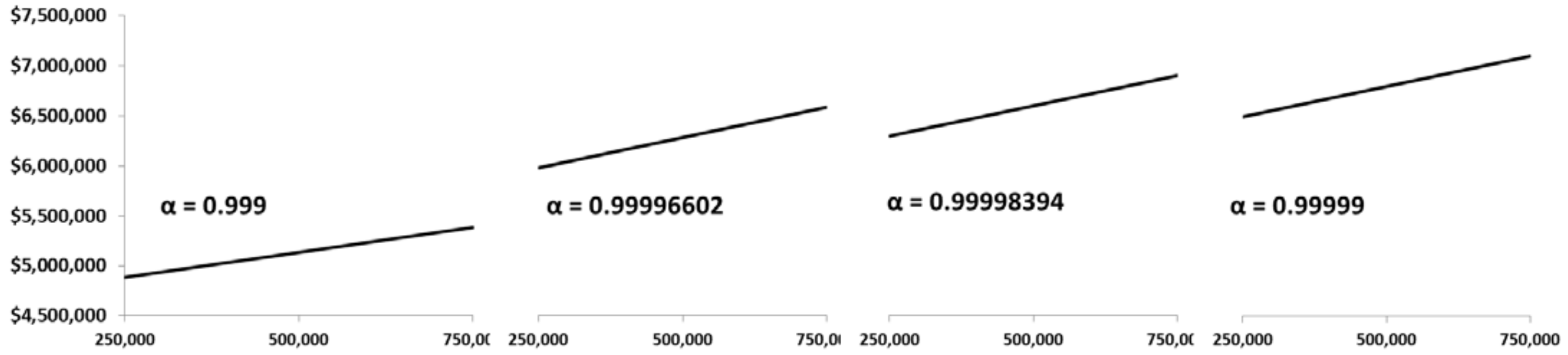

Parameter $\mu$

Normal, Parameter $\sigma$ by Quantile by α, Parameter $\mu$ = 500,000, Threshold=$0

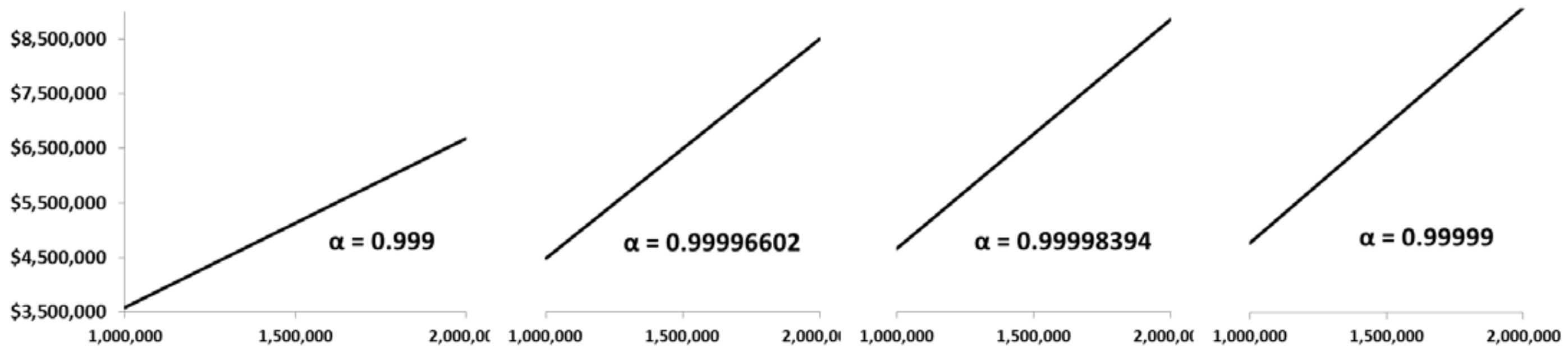

Parameter $\sigma$

## APPENDIX C

**TABLE C1: MLE-LDA for Economic and Regulatory Capital Estimation Varying λ Only***

| Severity Dist. | Parm1 | Parm2 | ECap Bias λ = 15 | ECap Bias λ = 25 | ECap Bias λ = 100 | RCap Bias λ = 15 | RCap Bias λ = 25 | RCap Bias λ = 100 |
|---|---|---|---|---|---|---|---|---|
| | **μ** | **σ** | | | | | | |
| LogN | 10 | 2 | -1.0% | -0.6% | -0.2% | -1.1% | -0.6% | -0.2% |
| LogN | 7.7 | 2.55 | -1.1% | -0.7% | -0.2% | -1.1% | -0.7% | -0.2% |
| LogN | 10.4 | 2.5 | -1.1% | -0.7% | -0.2% | -1.1% | -0.7% | -0.2% |
| LogN | 9.27 | 2.77 | -1.1% | -0.7% | -0.2% | -1.1% | -0.7% | -0.2% |
| LogN | 10.75 | 2.7 | -1.1% | -0.7% | -0.2% | -1.1% | -0.7% | -0.2% |
| LogN | 9.63 | 2.97 | -1.0% | -0.7% | -0.2% | -1.0% | -0.6% | -0.2% |
| TLogN | 10.2 | 1.95 | -1.0% | -0.6% | -0.2% | -1.0% | -0.6% | -0.2% |
| TLogN | 9 | 2.2 | -1.1% | -0.6% | -0.2% | -1.1% | -0.6% | -0.2% |
| TLogN | 10.7 | 2.385 | -1.1% | -0.7% | -0.2% | -1.1% | -0.7% | -0.2% |
| TLogN | 9.4 | 2.65 | -1.1% | -0.7% | -0.2% | -1.1% | -0.7% | -0.2% |
| TLogN | 11 | 2.6 | -1.1% | -0.7% | -0.2% | -1.1% | -0.7% | -0.2% |
| TLogN | 10 | 2.8 | -1.1% | -0.7% | -0.2% | -1.1% | -0.7% | -0.2% |
| | **a** | **b** | | | | | | |
| Logg | 24 | 2.65 | -1.0% | -0.6% | -0.2% | -1.0% | -0.6% | -0.2% |
| Logg | 33 | 3.3 | -1.0% | -0.6% | -0.2% | -1.0% | -0.6% | -0.2% |
| Logg | 25 | 2.5 | -0.9% | -0.6% | -0.2% | -0.9% | -0.6% | -0.2% |
| Logg | 34.5 | 3.15 | -1.0% | -0.6% | -0.2% | -1.0% | -0.6% | -0.2% |
| Logg | 25.25 | 2.45 | -0.9% | -0.6% | -0.2% | -0.9% | -0.6% | -0.2% |
| Logg | 34.7 | 3.07 | -1.0% | -0.6% | -0.2% | -1.0% | -0.6% | -0.2% |
| TLogg | 23.5 | 2.65 | -1.0% | -0.6% | -0.2% | -1.0% | -0.6% | -0.2% |
| TLogg | 33 | 3.3 | -1.0% | -0.6% | -0.2% | -1.0% | -0.6% | -0.2% |
| TLogg | 24.5 | 2.5 | -1.0% | -0.6% | -0.2% | -0.9% | -0.6% | -0.2% |
| TLogg | 34.5 | 3.15 | -1.0% | -0.6% | -0.2% | -1.0% | -0.6% | -0.2% |
| TLogg | 24.75 | 2.45 | -0.9% | -0.6% | -0.2% | -0.9% | -0.6% | -0.2% |
| TLogg | 34.6 | 3.07 | -1.0% | -0.6% | -0.2% | -1.0% | -0.6% | -0.2% |
| | **ξ** | **θ** | | | | | | |
| GPD | 0.8 | 35,000 | -0.8% | -0.5% | -0.2% | -0.8% | -0.5% | -0.2% |
| GPD | 0.95 | 7,500 | -0.4% | -0.3% | -0.2% | -0.4% | -0.3% | -0.2% |
| GPD | 0.875 | 47,500 | -0.6% | -0.4% | -0.2% | -0.6% | -0.4% | -0.2% |
| GPD | 0.95 | 25,000 | -0.4% | -0.3% | -0.2% | -0.4% | -0.3% | -0.2% |
| GPD | 0.925 | 50,000 | -0.5% | -0.3% | -0.2% | -0.5% | -0.3% | -0.2% |
| GPD | 0.99 | 27,500 | -0.3% | -0.2% | -0.1% | -0.3% | -0.2% | -0.1% |
| TGPD | 0.775 | 33,500 | -0.8% | -0.5% | -0.2% | -0.8% | -0.5% | -0.2% |
| TGPD | 0.8 | 25,000 | -0.8% | -0.5% | -0.2% | -0.8% | -0.5% | -0.2% |
| TGPD | 0.8675 | 50,000 | -0.6% | -0.4% | -0.2% | -0.6% | -0.4% | -0.2% |
| TGPD | 0.91 | 31,000 | -0.5% | -0.4% | -0.2% | -0.5% | -0.4% | -0.2% |
| TGPD | 0.92 | 47,500 | -0.5% | -0.3% | -0.2% | -0.5% | -0.3% | -0.2% |
| TGPD | 0.95 | 35,000 | -0.4% | -0.3% | -0.2% | -0.4% | -0.3% | -0.2% |

*NOTE: #simulations = 1,000; α=0.999 and 0.9997 for RCap and ECap, respectively.

**APPENDIX D**

**PDF, CDF, Mean, and Inverse of Fisher information for Six Severities**

**LogNormal PDF and CDF**:

$$f(x;\mu,\sigma)=\frac{1}{\sqrt{2\pi}\sigma x}e^{-\frac{1}{2}\left(\frac{\ln(x)-\mu}{\sigma}\right)^2} \text{ and } F(x;\mu,\sigma)=\frac{1}{2}\left[1+erf\left(\frac{\ln(x)-\mu}{\sqrt{2\sigma^2}}\right)\right] \text{ for } 0<x<\infty,\ 0<\sigma<\infty$$

**LogNormal Mean**:

$$E(X)=e^{\left(\mu+\sigma^2/2\right)}$$

**LogNormal Inverse of Fisher information**:

$$A(\theta)^{-1}=\begin{bmatrix}\sigma^2 & 0\\ 0 & \sigma^2/2\end{bmatrix}$$

**Truncated LogNormal PDF and CDF**:

$$g(x;\mu,\sigma)=\frac{f(x;\mu,\sigma)}{1-F(H;\mu,\sigma)} \text{ and } G(x;\mu,\sigma)=1-\frac{1-F(x;\mu,\sigma)}{1-F(H;\mu,\sigma)} \text{ for } H<x<\infty \text{ and } 0<\sigma<\infty$$

where $f(\ )$ is LogNormal PDF and $F(\ )$ is LogNormal CDF.

**Truncated LogNormal Mean**:

$$E(X)=e^{\mu+\sigma^2/2}\cdot\Phi\left(\frac{\mu+\sigma^2-\ln(H)}{\sigma}\right)\cdot\frac{1}{\left[1-F(H)\right]}$$ where $\Phi(\ )$ is the standard normal CDF.

**Truncated LogNormal Inverse of Fisher information**:

Let $u=\frac{\ln(H)-\mu}{\sigma}$, $j=\frac{-u^2/2}{\sqrt{2\pi}}$, $J=\frac{j}{1-\Phi(u)}$ where $\Phi$ is the CDFof the Standard Normal, and

$$INV=\frac{\sigma^2}{\left[2+J\cdot(J-u)\cdot\left(u\cdot(J-u)-3\right)\right]} \text{ then } A(\theta)^{-1}=INV\cdot\begin{bmatrix}2+J\cdot u\cdot\left(1-u\cdot(J-u)\right) & J\cdot\left(u\cdot(J-u)-1\right)\\ J\cdot\left(u\cdot(J-u)-1\right) & 1-\left(J\cdot(J-u)\right)\end{bmatrix}$$

from Roehr (2002) (Note that the first cell of this matrix as presented in Roehr (2002) contains a typo: this is corrected in the presentation above).

**Generalized Pareto Distribution (GPD) PDF and CDF**:

$$f(x;\xi,\theta)=\frac{1}{\theta}\left[1+\xi\frac{x}{\theta}\right]^{\left[-\frac{1}{\xi}-1\right]} \text{ and } F(x;\xi,\theta)=1-\left[1+\xi\frac{x}{\theta}\right]^{\left[-\frac{1}{\xi}\right]}$$

assuming $\xi\geq 0$, for $0\leq x<\infty;\ 0<\theta<\infty$

**GPD Mean**:

$$E(X)=\frac{\theta}{1-\xi} \text{ for } \xi<1 \ \left(=\infty \text{ for } \xi\geq 1\right)$$

**GPD Inverse of Fisher information**:

$$A(\theta)^{-1}=(1+\xi)\begin{bmatrix}1+\xi & -\theta\\ -\theta & 2\theta^2\end{bmatrix} \text{ from Smith (1987).}$$

**Truncated GPD PDF and CDF**:

$$g(x;\xi,\theta)=\frac{f(x;\xi,\theta)}{1-F(H;\xi,\theta)} \quad \text{and} \quad G(x;\xi,\theta)=1-\frac{1-F(x;\xi,\theta)}{1-F(H;\xi,\theta)}$$

assuming $\xi\geq 0$, for $H\leq x<\infty;\ 0<\theta<\infty$, where $f(\ )$ is GPD PDF and $F(\ )$ is GPD CDF.

**Truncated GPD Mean**:

$$E(X)=\frac{\theta}{\xi}\cdot\left(\frac{\left[1-F(H)\right]^{-\xi}}{1-\xi}-1\right) \text{ for } \xi<1 \ \left(=\infty \text{ for } \xi\geq 1\right)$$

As per Mayorov (2014), this also can be represented as $E(X)=\frac{H+\theta}{1-\xi}$ for $\xi<1$ $\left(=\infty \text{ for } \xi\geq 1\right)$.

**Truncated GPD Inverse of Fisher information**:

$$A(\theta)^{-1}=(1+\xi)\cdot\begin{bmatrix}(1+\xi) & -\theta\left(1+(1+2\xi)\left(\frac{H}{\theta}\right)\right)\\ -\theta\left(1+(1+2\xi)\left(\frac{H}{\theta}\right)\right) & \theta^2\left(2+2(1+2\xi)\left(\frac{H}{\theta}\right)+(1+\xi)(1+2\xi)\left(\frac{H}{\theta}\right)^2\right)\end{bmatrix} \text{ from Roehr (2002).}$$

**LogGamma PDF and CDF**:[73]

$$f(x;a,b)=\frac{b^a\left(\log(x)\right)^{(a-1)}}{\Gamma(a)x^{b+1}} \quad \text{and} \quad F(x;a,b)=\int_1^x\frac{b^a\left(\log(y)\right)^{(a-1)}}{\Gamma(a)y^{b+1}}dy \quad \text{for } 1\leq x<\infty;\ 0<a;\ 0<b$$ where $\Gamma(a)$

is the complete gamma function.The domain can be changed to $\mu\leq x<\infty$ if a location parameter, $\mu$, is added and $x-\mu$ is substituted for $x$ (so if $\mu$ = 1, the domain would range from zero and approach positive infinity).

**LogGamma Mean**:

$$E(X)=\left(\frac{b}{b-1}\right)^a \text{ for } b>1 \ \left(=\infty \text{ for } b\leq 1\right)$$

**LogGamma Inverse of Fisher information**:

[73] Note that this parameterization of the two-parameter LogGamma is the inverted parameterization (b = 1/b).

$$A(\theta)^{-1} = \frac{1}{\left(a/b^2\right)\cdot trigamma(a) - 1/b^2}\begin{bmatrix} a/b^2 & 1/b \\ 1/b & trigamma(a) \end{bmatrix}$$ from Opdyke and Cavallo (2012a).[74]

**Truncated LogGamma PDF and CDF**:

$$g(x;a,b) = \frac{f(x;a,b)}{1-F(H;a,b)} \quad \text{and} \quad G(x;a,b) = 1 - \frac{1-F(x;a,b)}{1-F(H;a,b)}$$

for $1 \le x < \infty;\ 0 < a;\ 0 < b$, where $f(\ )$ is LogGamma PDF and $F(\ )$ is LogGamma CDF.

**Truncated LogGamma Mean**:

$$E(X) = \left(\frac{b}{b-1}\right)^a \cdot \frac{1-J\left(\log(H)(b-1);a,1\right)}{\left[1-F(H)\right]}$$ for $b > 1$, $= \infty$ for $b \le 1$, where $J(\ )$ is the CDF of the Gamma distribution

Although Kim (2010) presents a derivation of the conditional (tail) mean for the LogGamma with direct parameterization (as opposed to LogGamma with inverted (1/b) parameterization as shown above and as used in this paper), the above mean for the Truncated LogGamma does not appear to have been presented in the literature previously and its derivation is shown below. It is known that for *J*( ) Gamma CDF,

$$1 - J(x;a,b=1) = \frac{\Gamma(a,x)}{\Gamma(a)}$$ where $\Gamma(a,x)$ is the upper incomplete gamma function and $\Gamma(\ )$ is the complete gamma function.

Also, the tail mean (i.e. mean beyond the threshold, *H*, as opposed to the mean of the truncated distribution) of the LogGamma is $TM(H) = \int_H^\infty y \cdot f(y;a,b)dy = \left(\frac{b}{b-1}\right)^a \cdot \frac{\Gamma\left(a, Log(H)(b-1)\right)}{\Gamma(a)}$

Therefore, because Mean of Truncated Distribution = *TM*(*H*) / (1-*F*(*H*)),

Mean of Truncated LogGamma = $\left(\frac{b}{b-1}\right)^a \cdot \left[1-J\left(\log(H)(b-1);a,1\right)\right] \cdot \frac{1}{1-F(H)}$

As per Mayorov (2014), this also can be represented as $\left(\frac{b}{b-1}\right)^a \cdot \frac{1-F(H;a,b-1)}{1-F(H;a,b)}$.

**Truncated LogGamma Inverse of Fisher information**:

$$A(\theta)^{-1} = \begin{bmatrix} A & B \\ B & D \end{bmatrix}^{-1}$$ where

$$A = trigamma(a) - \frac{\left[\int_{1^+}^{H} \ln(b) + \ln\left(\ln(x)\right) - digamma(a)\, f(x)dx\right]^2}{\left[1-F(H;a,b)\right]^2}$$

---

[74] The digamma and trigamma functions are the first and second order logarithmic derivatives of the complete gamma function: *digamma*(*z*) = *d*/*dz* ln [Γ(z)] and *trigamma*(*z*) = $d^2/dz^2$ ln [Γ(z)].

$$-\frac{\left[1-F\left(H;a,b\right)\right]\cdot\int_{1^+}^{H}\left[\ln\left(b\right)+\ln\left(\ln\left(x\right)\right)-digamma\left(a\right)\right]^2-trigamma\left(a\right)f\left(x\right)dx}{\left[1-F\left(H;a,b\right)\right]^2}$$

$$B=-\frac{1}{b}-\frac{\left[1-F\left(H;a,b\right)\right]\cdot\frac{1}{b}\cdot F\left(H;a,b\right)}{\left[1-F\left(H;a,b\right)\right]^2}$$

$$-\frac{\left[1-F\left(H;a,b\right)\right]\cdot\int_{1^+}^{H}\left[\ln\left(b\right)+\ln\left(\ln\left(x\right)\right)-digamma\left(a\right)\right]\cdot\left[\frac{a}{b}-\ln\left(x\right)\right]f\left(x\right)dx}{\left[1-F\left(H;a,b\right)\right]^2}$$

$$-\frac{\int_{1^+}^{H}\ln\left(b\right)+\ln\left(\ln\left(x\right)\right)-digamma\left(a\right)f\left(x\right)dx\cdot\int_{1^+}^{H}\left(\frac{a}{b}-\ln\left(x\right)\right)f\left(x\right)dx}{\left[1-F\left(H;a,b\right)\right]^2}$$

$$D=\frac{a}{b^2}-\frac{\left[\int_{1^+}^{H}\left(\frac{a}{b}-\ln\left(y\right)\right)f\left(x\right)dx\right]^2+\left[1-F\left(H;a,b\right)\right]\cdot\int_{1^+}^{H}\frac{a\left(a-1\right)}{b^2}-\frac{2a\ln\left(y\right)}{b}+\left[\ln\left(y\right)\right]^2f\left(x\right)dx}{\left[1-F\left(H;a,b\right)\right]^2}$$

from Opdyke and Cavallo (2012a).[75]

Zhou (2013) presents the equivalent of the above for the direct parameterization of the LogGamma,, but this, too, requires computationally expensive numeric integration.

Below I present an analytic approximation for the Fisher information of the Truncated LogGamma (inverted parameterization) that avoids numeric integration, and as a consequence, is over seven times faster to implement. For applications in this paper, the accuracy of the matrix terms, using $\eta$ = 0.001 (described below), ranges from eight to eleven decimal places, which is very sufficient for purposes of estimating operational risk capital (that is to say, no more than a few thousand dollars divergent from estimated capital that is hundreds of millions, or even billions of dollars).

$$A\left(\theta\right)^{-1}=\begin{bmatrix}A & B\\ B & D\end{bmatrix}^{-1}$$ where

[75] Note that the domain of the threshold, $H$, is $1 \le H < \infty$ instead of $0 \le H < \infty$ to avoid having to include a third location parameter for the LogGamma. The numeric effects of this mathematical convenience are de minimis, if even calculable. The same convention is used throughout the derivation of the analytic approximation for the Fisher information of the Truncated LogGamma, where the threshold is labeled "*t*" rather than *H*.

$$A=-\frac{MeijerG[\{\{\},\{1,1\}\},\{\{0,0,a\},\{\}\},bLog[t]]^2-2\Gamma[a,bLog[t]]MeijerG[\{\{\},\{1,1,1\}\},\{\{0,0,0,a\},\{\}\},bLog[t]]}{\Gamma[a,bLog[t]]^2}$$

$$B=\frac{-1+\frac{t^{-b}MeijerG[\{\{\},\{1+a,1+a\}\},\{\{a,a,2a\},\{\}\},bLog[t]]}{\Gamma[a,bLog[t]]^2}}{b}$$

$$D=t^{-3b}(1+(-1+a)t^bExpIntegralE[2-a,bLog[t]])\times$$

$$\times\frac{(-1+t^bExpIntegralE[1-a,bLog[t]](1-a+at^bExpIntegralE[1-a,bLog[t]]+bLog[t]))}{b^3ExpIntegralE[1-a,bLog[t]]^3Log[t]}$$

where

$\Gamma\left(a,x\right)$ is the upper incomplete gamma function,

$MeijerG[\{\{a1,...,an\},\{a\left(n+1\right),...,ap\}\},\{\{b1,...,bm\},\{b\left(m+1\right),...,bq\}\},z]=$

$$=\frac{1}{2\pi i}\int_{\gamma L}\frac{\prod_{j=1}^{m}\Gamma\left(b_j+s\right)\prod_{j=1}^{m}\Gamma\left(1-a_j-s\right)}{\prod_{j=n+1}^{p}\Gamma\left(a_j+s\right)\prod_{j=m+1}^{q}\Gamma\left(1-b_j-s\right)}x^{-s}ds$$

, and under certain conditions, in terms of generalized hypergeometric functions

$$=\sum_{h=1}^{m}\frac{\prod_{j=1}^{m}\Gamma\left(b_j-b_h\right)*\prod_{j=1}^{n}\Gamma\left(1+b_h-a_j\right)z^{b_h}}{\prod_{j=m+1}^{q}\Gamma\left(1+b_h-b_j\right)\prod_{j=n+1}^{p}\Gamma\left(a_j-b_h\right)}\times{}_pF_{q-1}\left(\begin{matrix}1+b_h-\mathrm{a}_\mathrm{p}\\ \left(1+b_h-\mathrm{a}_\mathrm{p}\right)^*\end{matrix}\middle|\left(-1\right)^{p-m-n}z\right)$$

and

$$ExpIntegralE[n,z]=\int_1^\infty\frac{e^{-zt}}{t^n}dt=z^{n-1}\Gamma\left(1-n,z\right)$$

Unfortunately, the MeijerG[ ] function converges too slowly, when used with values relevant to this setting, to be used by non-symbolic programming languages, that is, all statistical and computer programming languages.[76] In other words, the terms of the MeijerG[ ] become too small (i.e. less than 10E-16) to calculate for the computer chips on most modern computers, but we cannot discount these terms as they are not rapidly

[76] Mathematica is the major symbolic programming language that can correctly execute such calculations, but it is not as useable as the major statistical programming languages for nontrivial volumes of empirical statistical analyses, as are required for operational risk capital estimation. So these calculations must be put into terms that all major statistical programming languages, and most modern computer chips, can successfully calculate with sufficient precision.

divergent and notably affect the final result of the calculation. To circumvent this obstacle to practical usage, $A(\theta)^{-1}$ can be expanded as below:

$$A=\frac{1}{a^4\left[\Gamma\left(a,bLog[t]\right)\right]^2}\times$$

$$\times\left\{\left[-\left(GHG\left(\{a,a\},\{a+1,a+1\};-bLog[t]\right)\right)^2\right]\cdot\left(bLog[t]\right)^{2a}\right.$$

$$+2a\left(bLog[t]\right)^a\cdot\left[-\Gamma\left(a,bLog[t]\right)\cdot GHG\left(\{a,a,a\},\{a+1,a+1,a+1\};-bLog[t]\right)\right.$$

$$\left.+a\Gamma(a)\cdot GHG\left(\{a,a\},\{a+1,a+1\};-bLog[t]\right)\cdot\left(Log\left(bLog[t]\right)-digamma(a)\right)\right]$$

$$\left.+a^4\Gamma(a)\left[-\left(\Gamma(a)-\Gamma\left(a,bLog[t]\right)\right)\cdot\left(Log\left(bLog[t]\right)-digamma(a)\right)^2+\Gamma\left(a,bLog[t]\right)\cdot trigamma(a)\right]\right\}$$

$$B=\frac{1}{a^2b\left[\Gamma\left(a,bLog[t]\right)\right]^2}\times$$

$$\left\{t^{-b}\cdot GHG\left(\{a,a\},\{a+1,a+1\};-bLog[t]\right)\cdot\left(bLog[t]\right)^{2a}\right.$$

$$\left.-a^2\left(t^b\left[\Gamma\left(a,bLog[t]\right)\right]^2+\Gamma(a)\left(bLog[t]\right)^a\left(Log\left(bLog[t]\right)-digamma(a)\right)\right)\right\}$$

$$D=\frac{1}{b^3Log[t]\left[\Gamma\left(a,bLog[t]\right)\right]^3}\times\left\{\left[t^{-3b}\Gamma\left(a-1,bLog[t]\right)Log[t]+\left(bLog[t]\right)^a\right]\times\right.$$

$$\left.\times\left[at^{2b}\left[\Gamma\left(a,bLog[t]\right)\right]^2-\left(bLog[t]\right)^{2a}+t^{2b}\Gamma\left(a,bLog[t]\right)\left(bLog[t]\right)^{2a}\left(1-a+bLog[t]\right)\right]\right\}$$

which reduces to

$$D=\frac{a}{b^2}+\frac{t^{-b}\left(bLog[t]\right)^a\left(1-a+bLog[t]\right)}{b^2\Gamma\left(a,bLog[t]\right)}-\frac{t^{-2b}\left(bLog[t]\right)^{2a}}{b^2\left[\Gamma\left(a,bLog[t]\right)\right]^2}$$

where

$$GHG\left(\{a1,...,ap\},\{b1,...,bq\};z\right)=\sum_{n=0}^{\infty}\frac{(a1)_n\,...(ap)_n}{(b1)_n\,...(bq)_n}\frac{z^n}{n!},$$

where $(a)_n = a(a+1)(a+2)...(a+n-1),(a)_0 = 1$, is the generalized hypergeometric function.

Unfortunately, the generalized hypergeometric function suffers from the same usability challenge as does the MeijerG[ ] function: the terms become too small (i.e. less than 10E-16) to calculate for the computer chips on most modern computers, but these terms are material to the final result as they are not rapidly divergent. Fortunately, identities for the generalized hypergeometric function exist that contain terms similar to those found in the Fisher information above. For example,

$$GHG\left(\{a1,...,ap\},\{a1+1,...,ap+1\};z\right)=\sum_{k=1}^{q}\left[GHG\left(ak,ak+1;z\right)\cdot\prod_{j=1,\,j\neq k}^{q}\frac{a_j}{a_j-a_k}\right]$$

and

$$GHG\left(a,a+1;z\right)=\Gamma(a+1)(-z)^{-a}\left(1-\frac{\Gamma(a,-z)}{\Gamma(a)}\right)$$

The only problem with using these two identities to provide the generalized hypergeometric function in terms that can be readily calculated using most computer hardware and software (that is, non-symbolic programming languages) is that in our case, a1 = a2 = … = ap = a, so the last product term is undefined (i.e. $\frac{a_j}{a_j-a_j}=\frac{a_j}{0}\sim undefined$ ). Consequently, I use these identities to approximate the Fisher information by increasing and decreasing values for a1 and a2 and a3 by a small amount, $\eta$ (=0.001). This yields values of the elements of the Fisher information that deviate from true values by between eight and eleven decimal places, which as mentioned above, is very sufficient for purposes of estimating operational risk capital (that is to say, no more than a few thousand dollars divergent from estimated capital that is hundreds of millions, or even billions of dollars).

Also, we saw in deriving the mean of the truncated LogGamma severity above that the upper incomplete gamma function $\Gamma(s,x)=\Gamma(s)\cdot\left(1-J(x;s,b=1)\right)$ where *J*( ) is the CDF of the Gamma distribution, so this can be used to provide the upper incomplete gamma function in readily calculable terms.

Substituting both of the above term changes into the Fisher information, let

$a$ = parameter 1; $b$ = parameter 2; $t$ = threshold; $\eta = 0.001$ ; $adown = a - \eta$ ; $aup = a + \eta$ ; $z = -bLog\left[t\right]$

divide $a = diva = \dfrac{\Gamma\left(a+1\right)}{\left(-z\right)^{a}}$ ; divide $adown = divad = \dfrac{\Gamma\left(adown+1\right)}{\left(-z\right)^{adown}}$ ; divide $aup = divau = \dfrac{\Gamma\left(aup+1\right)}{\left(-z\right)^{aup}}$

so

$$GHG2 = divad \cdot J\left(-z;adown,1\right)\frac{aup}{aup-adown} + divau \cdot J\left(-z;aup,1\right)\frac{adown}{adown-aup}$$

$$GHG3 = divad \cdot J\left(-z;adown,1\right)\left(\frac{aup}{aup-adown}\right)\left(\frac{a}{a-adown}\right) + diva \cdot J\left(-z;a,1\right)\left(\frac{adown}{adown-a}\right)\left(\frac{aup}{aup-a}\right)$$

$$+divau \cdot J\left(-z;aup,1\right)\left(\frac{adown}{adown-aup}\right)\left(\frac{a}{a-adown}\right)$$ where $J(\,)$ is the CDF of the Gamma distribution.

$UIG$ = upper incomplete gamma function $= \Gamma\left(a,-z\right) = \Gamma\left(a\right)\left(1-J\left(-z;a,b=1\right)\right)$, then

$$A\left(\theta\right)^{-1} = \begin{bmatrix} A & B \\ B & D \end{bmatrix}^{-1}$$ where

$$A = \frac{1}{a^4 UIG^2} \times \Big\{ \left[-\left(GHG2\right)^2\right] \cdot \left(-z\right)^{2a} + 2a\left(-z\right)^{a} \cdot \left[-UIG \cdot GHG3 + a\Gamma\left(a\right) \cdot GHG2 \cdot \left(Log\left(-z\right) - digamma\left(a\right)\right)\right]$$
$$+a^4\Gamma\left(a\right)\left[-\left(\Gamma\left(a\right) - UIG\right) \cdot \left(Log\left(-z\right) - digamma\left(a\right)\right)^2 + UIG \cdot trigamma\left(a\right)\right]\Big\}$$

$$B = \frac{1}{a^2 bUIG^2} \times \left\{ t^{-b} \cdot GHG2 \cdot \left(-z\right)^{2a} - a^2\left(t^b UIG^2 + \Gamma\left(a\right)\left(-z\right)^{a}\left(Log\left(-z\right) - digamma\left(a\right)\right)\right)\right\}$$

$$D = \frac{a}{b^2} + \frac{t^{-b}\left(-z\right)^{a}\left(1-a-z\right)}{b^2 UIG} - \frac{t^{-2b}\left(-z\right)^{2a}}{b^2 UIG^2}$$

And now all terms use functions that are readily calculable using any computer and statistical and/or programming language. Note again that due to the use of $\eta$, this is an analytical approximation to the Fisher information for the Truncated LogGamma distribution (the first known to this author), but one that is sufficiently accurate for the purpose of estimating operational risk capital with deviations of, at most, a few thousand dollars for capital estimates of hundreds of millions, if not billions of dollars. And when compared to estimation that relies on numeric integration, the speed increases provided by this approximation approach an order of magnitude for the values used in this study.

## APPENDIX E

**TABLE E1**: Values of *c*(*sev*, *n*) by Severity by # of Loss Events

| Severity | 150 | 250 | 500 | 750 | 1000 | Root |
|---|---|---|---|---|---|---|
| LogN | 1.00 | 1.55 | 1.55 | 1.55 | 1.75 | 8 |
| TLogN | 1.20 | 1.70 | 1.80 | 1.80 | 1.80 | 8 |
| Logg | 1.00 | 1.00 | 1.00 | 1.00 | 0.30 | 3 |
| TLogg | 0.30 | 0.70 | 0.85 | 1.00 | 1.00 | 3 |
| GPD | 1.60 | 1.95 | 2.00 | 2.00 | 2.00 | 10 |
| TGPD | 1.50 | 1.85 | 2.00 | 2.10 | 2.10 | 10 |

**FIGURE E1**:

Values of *c*(*sev*, *n*) by Severity by # of Loss Events

(Linear and Non-Linear Interpolation via (5) with Roots Specified Above for Shaded Ranges)

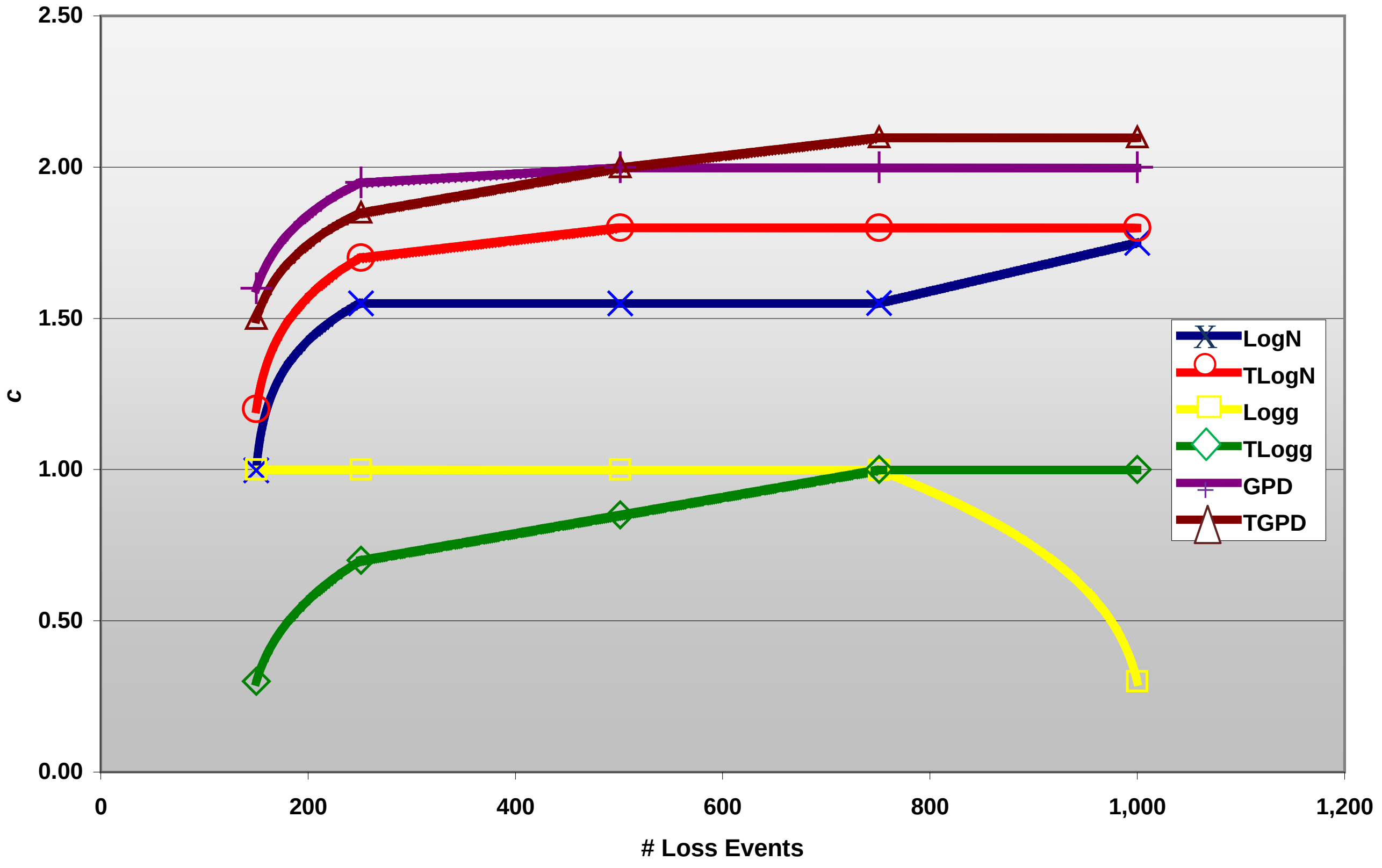

## APPENDIX F

**Available at [http:///www.risk.net/statis/about-the-journal-operational-risk](http:///www.risk.net/statis/about-the-journal-operational-risk) and at [http://www.DataMineit.com](http://www.DataMineit.com) and from the author upon request (JDOpdyke@DataMineit.com).**

**TABLE F4a**

**RCE vs. LDA for Regulatory Capital Estimation Under iid ($m, λ=25)***

| Severity Dist. | Parm1 | Parm2 | True RCap | Mean MLE RCap | MLE Bias | MLE Bias% | Mean RCE RCap | RCE Bias | RCE Bias% | RMSE MLE RCap | RMSE RCE RCap | RMSE RCap RCE/MLE | StdDev MLE RCap | StdDev RCE RCap | StdDev RCap RCE/MLE | 95%CIs MLE RCap | 95%CIs RCE RCap | 95%CIs RCap RCE/MLE | CV MLE | CV RCE | IQR MLE RCap | IQR RCE RCap | IQR RCap RCE/MLE | Skew MLE RCap | Skew RCE RCap | Kurtosis MLE RCap | Kurtosis RCE RCap |
|---|---|---|---|---|---|---|---|---|---|---|---|---|---|---|---|---|---|---|---|---|---|---|---|---|---|---|---|
| | **μ** | **σ** | | | | | | | | | | | | | | | | | | | | | | | | | |
| LogN | 10 | 2 | $63 | $67 | **$4** | **6.7%** | $63 | **$0** | **0.5%** | $25 | $23 | **91.8%** | $25 | $23 | 93.1% | $101 | $93 | **92.7%** | 0.373 | 0.368 | $30 | $29 | 94.5% | 1.11 | 1.09 | 1.95 | 1.86 |
| LogN | 7.7 | 2.55 | $53 | $59 | **$6** | **11.5%** | $54 | **$1** | **1.5%** | $30 | $26 | **87.7%** | $29 | $26 | 89.6% | $117 | $104 | **89.4%** | 0.494 | 0.487 | $33 | $30 | 91.4% | 1.48 | 1.44 | 3.58 | 3.36 |
| LogN | 10.4 | 2.5 | $649 | $720 | **$72** | **11.0%** | $658 | **$9** | **1.4%** | $355 | $313 | **88.1%** | $348 | $313 | 89.9% | $1,390 | $1,248 | **89.8%** | 0.483 | 0.476 | $399 | $366 | 91.7% | 1.44 | 1.41 | 3.41 | 3.20 |
| LogN | 9.27 | 2.77 | $603 | $686 | **$83** | **13.8%** | $614 | **$12** | **2.0%** | $382 | $328 | **86.0%** | $373 | $328 | 88.0% | $1,481 | $1,302 | **87.9%** | 0.543 | 0.534 | $414 | $372 | 89.9% | 1.64 | 1.59 | 4.43 | 4.12 |
| LogN | 10.75 | 2.7 | $2,012 | $2,275 | **$263** | **13.1%** | $2,048 | **$37** | **1.8%** | $1,229 | $1,063 | **86.5%** | $1,200 | $1,063 | 88.5% | $4,778 | $4,223 | **88.4%** | 0.528 | 0.519 | $1,345 | $1,215 | 90.4% | 1.59 | 1.55 | 4.15 | 3.87 |
| LogN | 9.63 | 2.97 | $1,893 | $2,198 | **$305** | **16.1%** | $1,939 | **$46** | **2.4%** | $1,329 | $1,121 | **84.3%** | $1,294 | $1,120 | 86.6% | $5,125 | $4,430 | **86.4%** | 0.589 | 0.578 | $1,401 | $1,240 | 88.5% | 1.79 | 1.73 | 5.33 | 4.90 |
| TLogN | 10.2 | 1.95 | $76 | $85 | **$9** | **11.5%** | $75 | **-$1** | **-1.8%** | $52 | $41 | **79.4%** | $51 | $41 | 80.5% | $179 | $146 | **81.6%** | 0.598 | 0.547 | $48 | $41 | 86.3% | 2.97 | 2.52 | 17.44 | 12.57 |
| TLogN | 9 | 2.2 | $76 | $96 | **$20** | **26.5%** | $75 | **-$1** | **-1.4%** | $88 | $50 | **56.9%** | $86 | $50 | 58.4% | $254 | $173 | **68.2%** | 0.899 | 0.673 | $68 | $50 | 73.8% | 4.89 | 2.49 | 41.21 | 10.26 |
| TLogN | 10.7 | 2.385 | $670 | $847 | **$177** | **26.4%** | $700 | **$30** | **4.5%** | $665 | $469 | **70.5%** | $641 | $468 | 73.0% | $2,288 | $1,755 | **76.7%** | 0.757 | 0.670 | $576 | $460 | 79.7% | 3.41 | 2.44 | 24.05 | 10.90 |
| TLogN | 9.4 | 2.65 | $643 | $894 | **$251** | **39.1%** | $628 | **-$14** | **-2.2%** | $1,087 | $536 | **49.3%** | $1,057 | $536 | 50.7% | $3,108 | $1,851 | **59.6%** | 1.183 | 0.853 | $659 | $470 | 71.3% | 5.39 | 3.19 | 43.88 | 17.16 |
| TLogN | 11 | 2.6 | $2,085 | $2,651 | **$566** | **27.1%** | $2,123 | **$38** | **1.8%** | $2,568 | $1,771 | **69.0%** | $2,504 | $1,771 | 70.7% | $8,801 | $6,308 | **71.7%** | 0.945 | 0.834 | $1,781 | $1,407 | 79.0% | 4.20 | 3.56 | 27.45 | 20.10 |
| TLogN | 10 | 2.8 | $1,956 | $2,743 | **$787** | **40.2%** | $1,965 | **$9** | **0.5%** | $3,033 | $1,694 | **55.9%** | $2,929 | $1,694 | 57.9% | $10,400 | $6,458 | **62.1%** | 1.068 | 0.862 | $2,150 | $1,469 | 68.3% | 3.83 | 2.63 | 21.70 | 9.88 |
| | **a** | **b** | | | | | | | | | | | | | | | | | | | | | | | | | |
| Logg | 24 | 2.65 | $85 | $97 | **$12** | **13.6%** | $87 | **$2** | **2.1%** | $62 | $53 | **86.0%** | $61 | $53 | 87.5% | $217 | $191 | **88.0%** | 0.628 | 0.612 | $63 | $56 | 88.9% | 2.67 | 2.57 | 13.66 | 12.74 |
| Logg | 33 | 3.3 | $100 | $108 | **$8** | **8.5%** | $99 | **$0** | **-0.4%** | $56 | $50 | **89.0%** | $56 | $50 | 90.0% | $210 | $190 | **90.2%** | 0.517 | 0.507 | $60 | $55 | 90.9% | 1.80 | 1.73 | 6.34 | 5.70 |
| Logg | 25 | 2.5 | $444 | $513 | **$70** | **15.7%** | $455 | **$11** | **2.5%** | $355 | $301 | **84.8%** | $348 | $301 | 86.5% | $1,241 | $1,069 | **86.2%** | 0.678 | 0.662 | $356 | $308 | 86.7% | 2.28 | 2.21 | 8.72 | 8.19 |
| Logg | 34.5 | 3.15 | $448 | $497 | **$49** | **10.9%** | $452 | **$4** | **0.8%** | $296 | $260 | **87.6%** | $292 | $260 | 88.8% | $1,032 | $920 | **89.1%** | 0.589 | 0.575 | $286 | $258 | 90.3% | 2.65 | 2.56 | 12.36 | 11.55 |
| Logg | 25.25 | 2.45 | $766 | $906 | **$140** | **18.3%** | $799 | **$32** | **4.2%** | $647 | $543 | **83.9%** | $632 | $542 | 85.8% | $2,277 | $1,909 | **83.8%** | 0.697 | 0.679 | $654 | $576 | 88.1% | 2.60 | 2.49 | 13.53 | 12.38 |
| Logg | 34.7 | 3.07 | $818 | $930 | **$112** | **13.7%** | $841 | **$23** | **2.8%** | $589 | $510 | **86.6%** | $578 | $510 | 88.1% | $2,043 | $1,807 | **88.5%** | 0.622 | 0.606 | $571 | $508 | 88.9% | 2.74 | 2.63 | 13.55 | 12.52 |
| TLogg | 23.5 | 2.65 | $124 | $193 | **$70** | **56.1%** | $137 | **$13** | **10.8%** | $273 | $181 | **66.4%** | $264 | $181 | 68.5% | $830 | $497 | **59.9%** | 1.365 | 1.317 | $151 | $100 | 66.6% | 5.49 | 6.34 | 46.55 | 62.21 |
| TLogg | 33 | 3.3 | $130 | $174 | **$44** | **34.1%** | $130 | **$0** | **-0.1%** | $173 | $93 | **53.9%** | $167 | $93 | 55.8% | $573 | $347 | **60.5%** | 0.964 | 0.721 | $128 | $91 | 70.9% | 3.69 | 2.53 | 20.94 | 10.23 |
| TLogg | 24.5 | 2.5 | $495 | $794 | **$299** | **60.4%** | $516 | **$20** | **4.1%** | $1,103 | $554 | **50.2%** | $1,062 | $554 | 52.1% | $3,610 | $1,957 | **54.2%** | 1.337 | 1.074 | $610 | $385 | 63.2% | 4.16 | 4.04 | 23.92 | 24.12 |
| TLogg | 34.5 | 3.15 | $510 | $635 | **$125** | **24.5%** | $539 | **$29** | **5.8%** | $544 | $397 | **73.1%** | $529 | $396 | 74.9% | $1,894 | $1,527 | **80.6%** | 0.834 | 0.736 | $495 | $411 | 83.1% | 3.73 | 2.52 | 29.80 | 12.36 |
| TLogg | 24.75 | 2.45 | $801 | $1,305 | **$504** | **62.9%** | $848 | **$47** | **5.9%** | $1,938 | $916 | **47.3%** | $1,871 | $915 | 48.9% | $6,376 | $2,809 | **44.0%** | 1.434 | 1.079 | $979 | $636 | 65.0% | 5.63 | 5.09 | 49.92 | 41.76 |
| TLogg | 34.6 | 3.07 | $867 | $1,078 | **$211** | **24.3%** | $925 | **$58** | **6.7%** | $927 | $709 | **76.5%** | $902 | $707 | 78.3% | $3,235 | $2,517 | **77.8%** | 0.838 | 0.764 | $771 | $645 | 83.7% | 3.04 | 2.61 | 15.19 | 11.16 |
| | **ξ** | **θ** | | | | | | | | | | | | | | | | | | | | | | | | | |
| GPD | 0.8 | 35,000 | $149 | $233 | **$85** | **56.9%** | $152 | **$3** | **2.2%** | $295 | $167 | **56.7%** | $282 | $167 | 59.2% | $973 | $579 | **59.5%** | 1.210 | 1.099 | $190 | $123 | 64.8% | 3.67 | 3.39 | 20.04 | 17.32 |
| GPD | 0.95 | 7,500 | $121 | $212 | **$91** | **75.6%** | $124 | **$3** | **2.7%** | $311 | $156 | **50.3%** | $297 | $156 | 52.6% | $971 | $529 | **54.4%** | 1.399 | 1.260 | $182 | $104 | 57.3% | 4.31 | 3.93 | 28.14 | 24.03 |
| GPD | 0.875 | 47,500 | $391 | $640 | **$249** | **63.7%** | $396 | **$5** | **1.2%** | $870 | $466 | **53.6%** | $834 | $466 | 55.9% | $2,753 | $1,555 | **56.5%** | 1.302 | 1.177 | $547 | $329 | 60.2% | 4.04 | 3.69 | 24.67 | 21.04 |
| GPD | 0.95 | 25,000 | $403 | $697 | **$295** | **73.2%** | $408 | **$5** | **1.3%** | $1,019 | $513 | **50.3%** | $976 | $513 | 52.6% | $3,139 | $1,750 | **55.8%** | 1.399 | 1.258 | $594 | $346 | 58.2% | 4.41 | 4.01 | 29.64 | 25.31 |
| GPD | 0.925 | 50,000 | $643 | $1,079 | **$436** | **67.8%** | $645 | **$1** | **0.2%** | $1,535 | $792 | **51.6%** | $1,472 | $792 | 53.8% | $4,845 | $2,715 | **56.0%** | 1.364 | 1.228 | $931 | $543 | 58.3% | 4.37 | 3.96 | 29.24 | 24.68 |
| GPD | 0.99 | 27,500 | $636 | $1,121 | **$486** | **76.4%** | $637 | **$2** | **0.3%** | $1,698 | $828 | **48.8%** | $1,627 | $828 | 50.9% | $5,131 | $2,781 | **54.2%** | 1.451 | 1.300 | $964 | $539 | 55.9% | 4.72 | 4.29 | 34.30 | 29.47 |
| TGPD | 0.775 | 33,500 | $141 | $214 | **$73** | **52.0%** | $144 | **$3** | **2.2%** | $297 | $170 | **57.4%** | $288 | $170 | 59.2% | $806 | $504 | **62.5%** | 1.341 | 1.181 | $178 | $118 | 65.9% | 7.53 | 6.40 | 102.72 | 77.40 |
| TGPD | 0.8 | 25,000 | $140 | $220 | **$80** | **56.9%** | $145 | **$5** | **3.3%** | $315 | $179 | **56.8%** | $305 | $179 | 58.7% | $992 | $591 | **59.5%** | 1.383 | 1.233 | $177 | $113 | 64.0% | 5.99 | 5.55 | 57.29 | 51.15 |
| TGPD | 0.8675 | 50,000 | $452 | $737 | **$285** | **63.0%** | $466 | **$13** | **3.0%** | $1,062 | $576 | **54.3%** | $1,023 | $576 | 56.3% | $3,309 | $1,935 | **58.5%** | 1.387 | 1.237 | $631 | $396 | 62.8% | 4.61 | 4.11 | 31.26 | 25.42 |
| TGPD | 0.91 | 31,000 | $451 | $761 | **$309** | **68.6%** | $463 | **$12** | **2.7%** | $1,174 | $603 | **51.4%** | $1,132 | $603 | 53.3% | $3,362 | $1,889 | **56.2%** | 1.489 | 1.302 | $646 | $392 | 60.7% | 5.34 | 4.52 | 43.26 | 31.44 |
| TGPD | 0.92 | 47,500 | $698 | $1,149 | **$451** | **64.7%** | $704 | **$7** | **0.9%** | $1,668 | $888 | **53.2%** | $1,606 | $888 | 55.3% | $5,192 | $2,946 | **56.7%** | 1.397 | 1.261 | $955 | $580 | 60.7% | 4.33 | 3.92 | 26.32 | 21.45 |
| TGPD | 0.95 | 35,000 | $717 | $1,206 | **$489** | **68.2%** | $715 | **-$2** | **-0.2%** | $2,009 | $991 | **49.3%** | $1,948 | $991 | 50.9% | $5,307 | $2,972 | **56.0%** | 1.615 | 1.386 | $1,023 | $613 | 60.0% | 7.57 | 6.14 | 98.27 | 67.33 |

*NOTE: #simulations = 1,000; λ = 25 for 10 years so n ~ 250; α=0.999

**TABLE F4b**

**RCE vs. LDA for Economic Capital Estimation Under iid ($m, λ=25)***

| Severity Dist. | Parm1 | Parm2 | True ECap | Mean MLE ECap | MLE Bias | MLE Bias% | Mean RCE ECap | RCE Bias | RCE Bias% | RMSE MLE ECap | RMSE RCE ECap | RMSE ECap RCE/MLE | StdDev MLE ECap | StdDev RCE ECap | StdDev ECap RCE/MLE | 95%CIs MLE ECap | 95%CIs RCE ECap | 95%CIs ECap RCE/MLE | CV MLE | CV RCE | IQR MLE ECap | IQR RCE ECap | IQR ECap RCE/MLE | Skew MLE ECap | Skew RCE ECap | Kurtosis MLE ECap | Kurtosis RCE ECap |
|---|---|---|---|---|---|---|---|---|---|---|---|---|---|---|---|---|---|---|---|---|---|---|---|---|---|---|---|
| | **μ** | **σ** | | | | | | | | | | | | | | | | | | | | | | | | | |
| LogN | 10 | 2 | $107 | $115 | **$8** | **7.8%** | $108 | **$1** | **1.1%** | $47 | $43 | **91.3%** | $46 | $43 | 92.7% | $186 | $172 | **92.3%** | 0.402 | 0.397 | $56 | $51 | 92.3% | 1.20 | 1.18 | 2.35 | 2.25 |
| LogN | 7.7 | 2.55 | $107 | $121 | **$14** | **13.2%** | $109 | **$3** | **2.5%** | $66 | $57 | **87.0%** | $64 | $57 | 89.0% | $255 | $227 | **88.9%** | 0.531 | 0.522 | $72 | $64 | 89.1% | 1.61 | 1.57 | 4.34 | 4.05 |
| LogN | 10.4 | 2.5 | $1,286 | $1,449 | **$163** | **12.7%** | $1,316 | **$30** | **2.4%** | $769 | $673 | **87.4%** | $752 | $672 | 89.4% | $2,998 | $2,675 | **89.2%** | 0.519 | 0.511 | $854 | $764 | 89.4% | 1.57 | 1.53 | 4.12 | 3.85 |
| LogN | 9.27 | 2.77 | $1,293 | $1,498 | **$205** | **15.8%** | $1,333 | **$40** | **3.1%** | $898 | $764 | **85.2%** | $874 | $763 | 87.4% | $3,465 | $3,024 | **87.3%** | 0.583 | 0.573 | $958 | $839 | 87.6% | 1.78 | 1.73 | 5.39 | 4.98 |
| LogN | 10.75 | 2.7 | $4,230 | $4,864 | **$634** | **15.0%** | $4,352 | **$123** | **2.9%** | $2,828 | $2,425 | **85.8%** | $2,756 | $2,422 | 87.9% | $10,945 | $9,609 | **87.8%** | 0.567 | 0.557 | $3,050 | $2,687 | 88.1% | 1.73 | 1.68 | 5.04 | 4.67 |
| LogN | 9.63 | 2.97 | $4,303 | $5,097 | **$794** | **18.5%** | $4,461 | **$158** | **3.7%** | $3,321 | $2,769 | **83.4%** | $3,224 | $2,765 | 85.8% | $12,725 | $10,912 | **85.8%** | 0.632 | 0.620 | $3,434 | $2,958 | 86.1% | 1.95 | 1.89 | 6.52 | 5.95 |
| TLogN | 10.2 | 1.95 | $126 | $144 | **$18** | **14.7%** | $124 | **-$2** | **-1.3%** | $101 | $77 | **76.1%** | $99 | $77 | 77.4% | $346 | $266 | **76.9%** | 0.687 | 0.618 | $88 | $73 | 83.7% | 3.49 | 2.89 | 23.76 | 16.25 |
| TLogN | 9 | 2.2 | $133 | $179 | **$46** | **35.0%** | $131 | **-$2** | **-1.5%** | $202 | $97 | **48.0%** | $197 | $97 | 49.4% | $537 | $335 | **62.3%** | 1.096 | 0.741 | $135 | $96 | 70.8% | 6.19 | 2.51 | 63.11 | 9.95 |
| TLogN | 10.7 | 2.385 | $1,267 | $1,678 | **$411** | **32.4%** | $1,338 | **$71** | **5.6%** | $1,521 | $1,003 | **65.9%** | $1,464 | $1,000 | 68.3% | $5,074 | $3,663 | **72.2%** | 0.873 | 0.747 | $1,211 | $924 | 76.3% | 4.33 | 2.81 | 38.66 | 14.78 |
| TLogN | 9.4 | 2.65 | $1,297 | $1,966 | **$669** | **51.6%** | $1,264 | **-$33** | **-2.5%** | $2,910 | $1,192 | **41.0%** | $2,832 | $1,192 | 42.1% | $7,535 | $4,089 | **54.3%** | 1.441 | 0.943 | $1,481 | $985 | 66.5% | 6.55 | 3.47 | 61.51 | 20.09 |
| TLogN | 11 | 2.6 | $4,208 | $5,639 | **$1,431** | **34.0%** | $4,337 | **$129** | **3.1%** | $6,319 | $4,072 | **64.4%** | $6,154 | $4,070 | 66.1% | $20,931 | $14,366 | **68.6%** | 1.091 | 0.938 | $3,951 | $2,976 | 75.3% | 4.86 | 3.98 | 36.00 | 24.79 |
| TLogN | 10 | 2.8 | $4,145 | $6,279 | **$2,134** | **51.5%** | $4,177 | **$32** | **0.8%** | $8,119 | $3,972 | **48.9%** | $7,833 | $3,972 | 50.7% | $26,918 | $15,066 | **56.0%** | 1.247 | 0.951 | $5,076 | $3,309 | 65.2% | 4.54 | 2.87 | 29.89 | 11.67 |
| | **a** | **b** | | | | | | | | | | | | | | | | | | | | | | | | | |
| Logg | 24 | 2.65 | $192 | $225 | **$33** | **17.0%** | $199 | **$7** | **3.4%** | $163 | $137 | **84.1%** | $159 | $137 | 85.7% | $557 | $478 | **85.9%** | 0.708 | 0.687 | $155 | $137 | 88.5% | 3.06 | 2.94 | 17.56 | 16.31 |
| Logg | 33 | 3.3 | $203 | $225 | **$22** | **10.7%** | $204 | **$1** | **0.4%** | $131 | $115 | **87.5%** | $129 | $115 | 88.7% | $492 | $436 | **88.6%** | 0.575 | 0.562 | $136 | $121 | 88.9% | 2.04 | 1.95 | 8.22 | 7.33 |
| Logg | 25 | 2.5 | $1,064 | $1,272 | **$208** | **19.5%** | $1,105 | **$42** | **3.9%** | $984 | $814 | **82.7%** | $962 | $813 | 84.5% | $3,424 | $2,849 | **83.2%** | 0.757 | 0.736 | $936 | $800 | 85.4% | 2.56 | 2.47 | 10.93 | 10.19 |
| Logg | 34.5 | 3.15 | $960 | $1,090 | **$130** | **13.5%** | $978 | **$18** | **1.8%** | $734 | $631 | **86.0%** | $722 | $631 | 87.3% | $2,505 | $2,207 | **88.1%** | 0.663 | 0.645 | $677 | $594 | 87.7% | 3.03 | 2.92 | 15.81 | 14.67 |
| Logg | 25.25 | 2.45 | $1,877 | $2,300 | **$423** | **22.5%** | $1,986 | **$109** | **5.8%** | $1,842 | $1,507 | **81.8%** | $1,793 | $1,503 | 83.8% | $6,365 | $5,300 | **83.3%** | 0.779 | 0.756 | $1,762 | $1,502 | 85.2% | 3.01 | 2.87 | 18.19 | 16.53 |
| Logg | 34.7 | 3.07 | $1,794 | $2,097 | **$302** | **16.8%** | $1,869 | **$74** | **4.1%** | $1,500 | $1,273 | **84.9%** | $1,470 | $1,271 | 86.5% | $5,113 | $4,437 | **86.8%** | 0.701 | 0.680 | $1,360 | $1,203 | 88.4% | 3.15 | 3.01 | 17.56 | 16.09 |
| TLogg | 23.5 | 2.65 | $271 | $496 | **$225** | **83.1%** | $297 | **$27** | **9.9%** | $903 | $530 | **58.6%** | $874 | $529 | 60.5% | $2,359 | $1,229 | **52.1%** | 1.764 | 1.778 | $385 | $217 | 56.4% | 7.20 | 8.71 | 78.04 | 112.48 |
| TLogg | 33 | 3.3 | $261 | $382 | **$120** | **46.1%** | $247 | **-$15** | **-5.6%** | $458 | $191 | **41.6%** | $442 | $190 | 42.9% | $1,462 | $707 | **48.4%** | 1.159 | 0.771 | $292 | $184 | 62.8% | 4.36 | 2.80 | 28.31 | 13.47 |
| TLogg | 24.5 | 2.5 | $1,164 | $2,152 | **$988** | **84.9%** | $1,099 | **-$65** | **-5.6%** | $3,620 | $1,370 | **37.8%** | $3,483 | $1,368 | 39.3% | $11,139 | $4,881 | **43.8%** | 1.618 | 1.245 | $1,629 | $782 | 48.0% | 4.86 | 4.88 | 31.98 | 34.33 |
| TLogg | 34.5 | 3.15 | $1,086 | $1,437 | **$350** | **32.2%** | $1,158 | **$72** | **6.6%** | $1,453 | $941 | **64.8%** | $1,410 | $938 | 66.5% | $4,792 | $3,638 | **75.9%** | 0.981 | 0.810 | $1,190 | $935 | 78.5% | 4.79 | 2.61 | 47.53 | 12.40 |
| TLogg | 24.75 | 2.45 | $1,928 | $3,618 | **$1,689** | **87.6%** | $1,855 | **-$74** | **-3.8%** | $6,604 | $2,269 | **34.4%** | $6,384 | $2,267 | 35.5% | $20,825 | $6,298 | **30.2%** | 1.765 | 1.223 | $2,710 | $1,406 | 51.9% | 6.84 | 6.34 | 72.80 | 61.09 |
| TLogg | 34.6 | 3.07 | $1,892 | $2,493 | **$601** | **31.8%** | $2,050 | **$158** | **8.4%** | $2,499 | $1,764 | **70.6%** | $2,425 | $1,757 | 72.4% | $8,627 | $6,151 | **71.3%** | 0.973 | 0.857 | $1,861 | $1,517 | 81.5% | 3.59 | 2.90 | 21.03 | 13.68 |
| | **ξ** | **θ** | | | | | | | | | | | | | | | | | | | | | | | | | |
| GPD | 0.8 | 35,000 | $382 | $696 | **$313** | **81.9%** | $396 | **$13** | **3.5%** | $1,069 | $521 | **48.7%** | $1,022 | $521 | 50.9% | $3,387 | $1,746 | **51.5%** | 1.469 | 1.315 | $577 | $332 | 57.5% | 4.38 | 3.97 | 27.86 | 23.37 |
| GPD | 0.95 | 7,500 | $375 | $785 | **$410** | **109.2%** | $390 | **$14** | **3.8%** | $1,398 | $588 | **42.1%** | $1,336 | $588 | 44.0% | $4,041 | $1,906 | **47.2%** | 1.701 | 1.509 | $657 | $330 | 50.2% | 5.16 | 4.69 | 39.16 | 33.55 |
| GPD | 0.875 | 47,500 | $1,106 | $2,123 | **$1,016** | **91.9%** | $1,130 | **$24** | **2.2%** | $3,514 | $1,594 | **45.4%** | $3,363 | $1,594 | 47.4% | $10,348 | $5,113 | **49.4%** | 1.585 | 1.410 | $1,818 | $960 | 52.8% | 4.87 | 4.41 | 34.76 | 29.38 |
| GPD | 0.95 | 25,000 | $1,251 | $2,576 | **$1,325** | **105.9%** | $1,279 | **$28** | **2.2%** | $4,585 | $1,930 | **42.1%** | $4,390 | $1,930 | 44.0% | $13,301 | $6,281 | **47.2%** | 1.704 | 1.509 | $2,182 | $1,099 | 50.4% | 5.30 | 4.81 | 41.36 | 35.48 |
| GPD | 0.925 | 50,000 | $1,938 | $3,835 | **$1,898** | **97.9%** | $1,955 | **$17** | **0.9%** | $6,657 | $2,882 | **43.3%** | $6,381 | $2,882 | 45.2% | $19,608 | $9,118 | **46.5%** | 1.664 | 1.475 | $3,298 | $1,681 | 51.0% | 5.28 | 4.76 | 41.18 | 34.90 |
| GPD | 0.99 | 27,500 | $2,076 | $4,375 | **$2,299** | **110.7%** | $2,095 | **$19** | **0.9%** | $8,085 | $3,275 | **40.5%** | $7,751 | $3,275 | 42.2% | $23,030 | $10,486 | **45.5%** | 1.772 | 1.563 | $3,686 | $1,797 | 48.8% | 5.68 | 5.15 | 47.71 | 41.14 |
| TGPD | 0.775 | 33,500 | $351 | $617 | **$266** | **75.7%** | $365 | **$13** | **3.8%** | $1,109 | $543 | **48.9%** | $1,077 | $543 | 50.4% | $2,654 | $1,446 | **54.5%** | 1.745 | 1.489 | $515 | $310 | 60.2% | 9.98 | 8.34 | 164.33 | 121.42 |
| TGPD | 0.8 | 25,000 | $361 | $660 | **$299** | **82.9%** | $379 | **$18** | **5.0%** | $1,193 | $579 | **48.5%** | $1,155 | $578 | 50.1% | $3,435 | $1,776 | **51.7%** | 1.749 | 1.526 | $525 | $307 | 58.6% | 7.36 | 6.79 | 81.65 | 72.55 |
| TGPD | 0.8675 | 50,000 | $1,267 | $2,432 | **$1,166** | **92.0%** | $1,327 | **$61** | **4.8%** | $4,337 | $1,988 | **45.9%** | $4,177 | $1,988 | 47.6% | $12,419 | $6,337 | **51.0%** | 1.717 | 1.498 | $2,065 | $1,144 | 55.4% | 5.54 | 4.89 | 43.68 | 35.05 |
| TGPD | 0.91 | 31,000 | $1,334 | $2,672 | **$1,338** | **100.4%** | $1,389 | **$55** | **4.1%** | $5,165 | $2,203 | **42.6%** | $4,989 | $2,202 | 44.1% | $13,302 | $6,344 | **47.7%** | 1.867 | 1.586 | $2,224 | $1,182 | 53.2% | 6.58 | 5.48 | 63.72 | 45.31 |
| TGPD | 0.92 | 47,500 | $2,088 | $4,048 | **$1,960** | **93.9%** | $2,129 | **$41** | **2.0%** | $7,203 | $3,235 | **44.9%** | $6,931 | $3,235 | 46.7% | $20,816 | $10,223 | **49.1%** | 1.712 | 1.519 | $3,384 | $1,765 | 52.2% | 5.13 | 4.60 | 36.26 | 28.88 |
| TGPD | 0.95 | 35,000 | $2,227 | $4,474 | **$2,246** | **100.8%** | $2,246 | **$19** | **0.8%** | $9,606 | $3,873 | **40.3%** | $9,339 | $3,873 | 41.5% | $22,320 | $10,320 | **46.2%** | 2.088 | 1.724 | $3,650 | $1,936 | 53.1% | 9.86 | 7.85 | 155.89 | 104.71 |

*NOTE: #simulations = 1,000; λ = 25 for 10 years so n ~ 250; α=0.9997

**TABLE F5a**
**RCE vs. LDA for Regulatory Capital Estimation Under iid ($m, λ=15)***

| Severity Dist. | Parm1 | Parm2 | True RCap | Mean MLE RCap | MLE Bias | MLE Bias% | Mean RCE RCap | RCE Bias | RCE Bias% | RMSE MLE RCap | RMSE RCE RCap | RMSE RCap RCE/MLE | StdDev MLE RCap | StdDev RCE RCap | StdDev RCap RCE/MLE | 95%CIs MLE RCap | 95%CIs RCE RCap | 95%CIs RCap RCE/MLE | CV MLE | CV RCE | IQR MLE RCap | IQR RCE RCap | IQR RCap RCE/MLE | Skew MLE RCap | Skew RCE RCap | Kurtosis MLE RCap | Kurtosis RCE RCap |
|---|---|---|---|---|---|---|---|---|---|---|---|---|---|---|---|---|---|---|---|---|---|---|---|---|---|---|---|
| | **μ** | **σ** | | | | | | | | | | | | | | | | | | | | | | | | | |
| LogN | 10 | 2 | $48 | $51 | **$3** | **6.0%** | $47 | **-$1** | **-1.6%** | $24 | $22 | **90.9%** | $24 | $22 | 91.5% | $92 | $85 | **91.8%** | 0.468 | 0.461 | $29 | $26 | 91.0% | 1.20 | 1.19 | 1.86 | 1.81 |
| LogN | 7.7 | 2.55 | $38 | $43 | **$5** | **11.9%** | $38 | **$0** | **-0.2%** | $27 | $23 | **86.3%** | $27 | $23 | 87.6% | $101 | $90 | **88.3%** | 0.617 | 0.606 | $30 | $26 | 87.4% | 1.59 | 1.57 | 3.44 | 3.33 |
| LogN | 10.4 | 2.5 | $473 | $527 | **$54** | **11.3%** | $471 | **-$2** | **-0.4%** | $322 | $279 | **86.8%** | $318 | $279 | 88.0% | $1,216 | $1,078 | **88.7%** | 0.603 | 0.593 | $358 | $314 | 87.7% | 1.55 | 1.53 | 3.27 | 3.17 |
| LogN | 9.27 | 2.77 | $426 | $489 | **$63** | **14.9%** | $428 | **$2** | **0.5%** | $337 | $285 | **84.4%** | $331 | $285 | 85.9% | $1,262 | $1,093 | **86.6%** | 0.678 | 0.665 | $359 | $308 | 85.8% | 1.75 | 1.72 | 4.22 | 4.08 |
| LogN | 10.75 | 2.7 | $1,434 | $1,633 | **$199** | **13.9%** | $1,438 | **$4** | **0.3%** | $1,093 | $929 | **85.0%** | $1,075 | $929 | 86.4% | $4,097 | $3,573 | **87.2%** | 0.658 | 0.646 | $1,177 | $1,015 | 86.3% | 1.70 | 1.67 | 3.96 | 3.83 |
| LogN | 9.63 | 2.97 | $1,306 | $1,540 | **$233** | **17.9%** | $1,324 | **$17** | **1.3%** | $1,154 | $952 | **82.5%** | $1,130 | $952 | 84.2% | $4,283 | $3,641 | **85.0%** | 0.734 | 0.719 | $1,184 | $997 | 84.3% | 1.90 | 1.86 | 5.03 | 4.84 |
| TLogN | 10.2 | 1.95 | $59 | $70 | **$11** | **18.9%** | $59 | **$0** | **0.3%** | $55 | $39 | **71.4%** | $54 | $39 | 72.9% | $185 | $143 | **77.5%** | 0.771 | 0.667 | $49 | $40 | 81.0% | 3.63 | 2.65 | 24.46 | 13.60 |
| TLogN | 9 | 2.2 | $57 | $87 | **$30** | **52.0%** | $56 | **-$1** | **-2.1%** | $173 | $44 | **25.4%** | $171 | $44 | 25.8% | $350 | $169 | **48.3%** | 1.955 | 0.782 | $64 | $45 | 70.0% | 17.77 | 2.05 | 433.88 | 5.66 |
| TLogN | 10.7 | 2.385 | $497 | $710 | **$213** | **42.9%** | $535 | **$39** | **7.8%** | $1,072 | $537 | **50.1%** | $1,050 | $535 | 51.0% | $2,468 | $1,661 | **67.3%** | 1.479 | 1.000 | $505 | $396 | 78.5% | 9.59 | 4.76 | 133.49 | 36.93 |
| TLogN | 9.4 | 2.65 | $465 | $764 | **$299** | **64.2%** | $457 | **-$8** | **-1.8%** | $1,116 | $434 | **38.9%** | $1,075 | $434 | 40.4% | $3,540 | $1,699 | **48.0%** | 1.408 | 0.950 | $611 | $386 | 63.2% | 4.75 | 3.00 | 32.25 | 14.29 |
| TLogN | 11 | 2.6 | $1,510 | $2,263 | **$753** | **49.9%** | $1,677 | **$167** | **11.1%** | $2,740 | $1,676 | **61.2%** | $2,635 | $1,668 | 63.3% | $9,381 | $5,847 | **62.3%** | 1.164 | 0.994 | $1,754 | $1,331 | 75.9% | 3.77 | 3.38 | 18.98 | 16.96 |
| TLogN | 10 | 2.8 | $1,389 | $2,652 | **$1,263** | **90.9%** | $1,507 | **$118** | **8.5%** | $5,725 | $1,743 | **30.4%** | $5,584 | $1,739 | 31.1% | $12,350 | $5,777 | **46.8%** | 2.105 | 1.154 | $1,924 | $1,239 | 64.4% | 10.36 | 3.82 | 167.66 | 21.93 |
| | **a** | **b** | | | | | | | | | | | | | | | | | | | | | | | | | |
| Logg | 24 | 2.65 | $59 | $73 | **$13** | **22.3%** | $60 | **$1** | **1.2%** | $60 | $46 | **76.6%** | $58 | $46 | 78.5% | $217 | $173 | **79.9%** | 0.802 | 0.761 | $52 | $42 | 81.3% | 2.86 | 2.68 | 12.85 | 11.31 |
| Logg | 33 | 3.3 | $72 | $85 | **$13** | **17.8%** | $73 | **$1** | **1.3%** | $61 | $49 | **80.6%** | $60 | $49 | 82.4% | $210 | $174 | **83.1%** | 0.707 | 0.677 | $59 | $50 | 84.4% | 2.69 | 2.57 | 12.04 | 11.24 |
| Logg | 25 | 2.5 | $301 | $380 | **$79** | **26.4%** | $306 | **$5** | **1.7%** | $344 | $253 | **73.4%** | $335 | $253 | 75.4% | $1,142 | $863 | **75.6%** | 0.881 | 0.826 | $277 | $221 | 79.8% | 3.39 | 3.07 | 18.59 | 15.01 |
| Logg | 34.5 | 3.15 | $317 | $369 | **$52** | **16.3%** | $312 | **-$5** | **-1.6%** | $262 | $209 | **79.9%** | $257 | $209 | 81.5% | $981 | $796 | **81.1%** | 0.697 | 0.670 | $270 | $224 | 82.9% | 2.06 | 2.00 | 5.94 | 5.66 |
| Logg | 25.25 | 2.45 | $515 | $659 | **$144** | **27.9%** | $523 | **$8** | **1.6%** | $607 | $440 | **72.6%** | $589 | $440 | 74.7% | $2,135 | $1,555 | **72.8%** | 0.894 | 0.841 | $490 | $384 | 78.3% | 3.04 | 2.86 | 14.03 | 12.53 |
| Logg | 34.7 | 3.07 | $575 | $685 | **$111** | **19.3%** | $574 | **$0** | **-0.1%** | $522 | $411 | **78.6%** | $511 | $411 | 80.5% | $1,806 | $1,463 | **81.0%** | 0.745 | 0.716 | $523 | $431 | 82.3% | 2.34 | 2.23 | 9.29 | 8.39 |
| TLogg | 23.5 | 2.65 | $87 | $232 | **$145** | **166.7%** | $150 | **$63** | **72.8%** | $1,121 | $372 | **33.2%** | $1,112 | $366 | 33.0% | $1,156 | $928 | **80.3%** | 4.789 | 2.436 | $145 | $98 | 67.3% | 24.42 | 8.44 | 688.38 | 95.43 |
| TLogg | 33 | 3.3 | $94 | $162 | **$68** | **72.7%** | $108 | **$14** | **15.4%** | $398 | $251 | **62.9%** | $392 | $250 | 63.8% | $726 | $399 | **55.0%** | 2.422 | 2.311 | $110 | $71 | 64.7% | 19.94 | 22.43 | 514.19 | 616.69 |
| TLogg | 24.5 | 2.5 | $339 | $712 | **$373** | **110.2%** | $408 | **$70** | **20.6%** | $1,547 | $825 | **53.3%** | $1,502 | $822 | 54.7% | $3,703 | $1,829 | **49.4%** | 2.109 | 2.014 | $522 | $282 | 54.0% | 7.01 | 7.66 | 63.01 | 78.35 |
| TLogg | 34.5 | 3.15 | $362 | $526 | **$164** | **45.3%** | $380 | **$18** | **5.0%** | $704 | $335 | **47.5%** | $685 | $334 | 48.8% | $2,027 | $1,189 | **58.6%** | 1.302 | 0.879 | $417 | $304 | 72.8% | 5.55 | 3.05 | 48.31 | 16.46 |
| TLogg | 24.75 | 2.45 | $543 | $1,097 | **$554** | **102.1%** | $586 | **$43** | **8.0%** | $2,098 | $999 | **47.6%** | $2,023 | $998 | 49.3% | $6,613 | $2,989 | **45.2%** | 1.845 | 1.703 | $802 | $415 | 51.7% | 5.41 | 5.80 | 39.73 | 47.30 |
| TLogg | 34.6 | 3.07 | $610 | $858 | **$248** | **40.7%** | $668 | **$58** | **9.5%** | $962 | $576 | **59.9%** | $929 | $573 | 61.7% | $3,157 | $2,027 | **64.2%** | 1.083 | 0.858 | $725 | $577 | 79.6% | 3.67 | 2.39 | 22.49 | 9.30 |
| | **ξ** | **θ** | | | | | | | | | | | | | | | | | | | | | | | | | |
| GPD | 0.8 | 35,000 | $99 | $178 | **$79** | **80.3%** | $101 | **$2** | **2.4%** | $257 | $123 | **47.6%** | $245 | $123 | 50.0% | $868 | $450 | **51.8%** | 1.378 | 1.215 | $165 | $94 | 57.0% | 3.30 | 2.89 | 14.01 | 10.60 |
| GPD | 0.95 | 7,500 | $74 | $155 | **$81** | **108.8%** | $76 | **$2** | **2.6%** | $255 | $103 | **40.6%** | $242 | $103 | 42.8% | $837 | $368 | **44.0%** | 1.558 | 1.356 | $139 | $70 | 50.5% | 3.68 | 3.13 | 17.78 | 12.60 |
| GPD | 0.875 | 47,500 | $250 | $477 | **$227** | **91.1%** | $253 | **$4** | **1.4%** | $725 | $320 | **44.2%** | $688 | $320 | 46.5% | $2,444 | $1,176 | **48.1%** | 1.443 | 1.265 | $443 | $237 | 53.6% | 3.47 | 2.97 | 15.94 | 11.37 |
| GPD | 0.95 | 25,000 | $248 | $509 | **$262** | **105.7%** | $251 | **$3** | **1.4%** | $832 | $339 | **40.7%** | $790 | $339 | 42.9% | $2,694 | $1,226 | **45.5%** | 1.551 | 1.350 | $458 | $233 | 50.9% | 3.68 | 3.13 | 18.11 | 12.67 |
| GPD | 0.925 | 50,000 | $401 | $761 | **$360** | **90.0%** | $386 | **-$14** | **-3.6%** | $1,245 | $528 | **42.4%** | $1,192 | $527 | 44.2% | $4,185 | $1,879 | **44.9%** | 1.566 | 1.366 | $703 | $365 | 51.9% | 3.70 | 3.14 | 18.55 | 12.95 |
| GPD | 0.99 | 27,500 | $383 | $802 | **$419** | **109.5%** | $381 | **-$2** | **-0.6%** | $1,348 | $527 | **39.1%** | $1,281 | $527 | 41.2% | $4,375 | $1,901 | **43.5%** | 1.596 | 1.385 | $722 | $355 | 49.2% | 3.90 | 3.28 | 20.94 | 14.37 |
| TGPD | 0.775 | 33,500 | $95 | $172 | **$77** | **81.6%** | $100 | **$5** | **5.4%** | $323 | $157 | **48.6%** | $313 | $157 | 50.0% | $909 | $451 | **49.6%** | 1.824 | 1.572 | $148 | $85 | 57.5% | 6.48 | 5.97 | 57.49 | 50.13 |
| TGPD | 0.8 | 25,000 | $93 | $159 | **$66** | **71.3%** | $92 | **-$1** | **-0.8%** | $250 | $122 | **48.8%** | $241 | $122 | 50.6% | $748 | $393 | **52.6%** | 1.513 | 1.322 | $149 | $84 | 56.6% | 5.16 | 4.59 | 41.59 | 34.13 |
| TGPD | 0.8675 | 50,000 | $290 | $583 | **$293** | **101.2%** | $315 | **$26** | **8.8%** | $964 | $427 | **44.3%** | $918 | $426 | 46.4% | $3,225 | $1,570 | **48.7%** | 1.574 | 1.351 | $527 | $283 | 53.7% | 4.74 | 4.00 | 31.27 | 22.49 |
| TGPD | 0.91 | 31,000 | $283 | $549 | **$266** | **93.8%** | $283 | **-$1** | **-0.2%** | $1,053 | $438 | **41.6%** | $1,019 | $438 | 43.0% | $2,953 | $1,375 | **46.6%** | 1.857 | 1.552 | $467 | $250 | 53.5% | 7.01 | 5.73 | 81.71 | 56.83 |
| TGPD | 0.92 | 47,500 | $436 | $940 | **$505** | **115.9%** | $483 | **$47** | **10.8%** | $1,567 | $657 | **41.9%** | $1,483 | $655 | 44.2% | $5,210 | $2,361 | **45.3%** | 1.578 | 1.358 | $912 | $480 | 52.6% | 4.18 | 3.60 | 22.83 | 17.62 |
| TGPD | 0.95 | 35,000 | $441 | $906 | **$465** | **105.6%** | $446 | **$5** | **1.2%** | $2,059 | $801 | **38.9%** | $2,005 | $801 | 39.9% | $4,830 | $2,241 | **46.4%** | 2.213 | 1.795 | $710 | $365 | 51.5% | 7.64 | 6.32 | 76.96 | 54.31 |

*NOTE: #simulations = 1,000; λ = 15 for 10 years so n ~ 150; α=0.999

**TABLE F5b**

**RCE vs. LDA for Economic Capital Estimation Under iid ($m, λ=15)***

| Severity Dist. | Parm1 | Parm2 | True ECap | Mean MLE ECap | MLE Bias | MLE Bias% | Mean RCE ECap | RCE Bias | RCE Bias% | RMSE MLE ECap | RMSE RCE ECap | RMSE ECap RCE/MLE | StdDev MLE ECap | StdDev RCE ECap | StdDev ECap RCE/MLE | 95%CIs MLE ECap | 95%CIs RCE ECap | 95%CIs ECap RCE/MLE | CV MLE | CV RCE | IQR MLE ECap | IQR RCE ECap | IQR ECap RCE/MLE | Skew MLE ECap | Skew RCE ECap | Kurtosis MLE ECap | Kurtosis RCE ECap |
|---|---|---|---|---|---|---|---|---|---|---|---|---|---|---|---|---|---|---|---|---|---|---|---|---|---|---|---|
| | μ | σ | | | | | | | | | | | | | | | | | | | | | | | | | |
| LogN | 10 | 2 | $84 | $90 | $6 | 7.3% | $83 | -$1 | -0.8% | $46 | $41 | 90.4% | $45 | $41 | 91.2% | $179 | $162 | 90.5% | 0.504 | 0.497 | $54 | $49 | 90.9% | 1.29 | 1.28 | 2.16 | 2.10 |
| LogN | 7.7 | 2.55 | $79 | $90 | $11 | 14.2% | $80 | $1 | 1.2% | $61 | $52 | 85.6% | $60 | $52 | 87.1% | $235 | $202 | 86.1% | 0.663 | 0.651 | $66 | $57 | 87.3% | 1.71 | 1.67 | 3.94 | 3.80 |
| LogN | 10.4 | 2.5 | $958 | $1,087 | $129 | 13.5% | $968 | $10 | 1.0% | $717 | $617 | 86.1% | $705 | $617 | 87.5% | $2,772 | $2,399 | 86.6% | 0.649 | 0.637 | $781 | $685 | 87.7% | 1.67 | 1.64 | 3.75 | 3.62 |
| LogN | 9.27 | 2.77 | $935 | $1,099 | $165 | 17.6% | $955 | $21 | 2.2% | $818 | $683 | 83.5% | $801 | $683 | 85.2% | $3,140 | $2,642 | 84.1% | 0.729 | 0.714 | $846 | $723 | 85.5% | 1.88 | 1.84 | 4.83 | 4.64 |
| LogN | 10.75 | 2.7 | $3,082 | $3,590 | $508 | 16.5% | $3,141 | $58 | 1.9% | $2,591 | $2,181 | 84.2% | $2,541 | $2,180 | 85.8% | $9,968 | $8,451 | 84.8% | 0.708 | 0.694 | $2,718 | $2,339 | 86.1% | 1.82 | 1.79 | 4.54 | 4.36 |
| LogN | 9.63 | 2.97 | $3,041 | $3,682 | $641 | 21.1% | $3,140 | $99 | 3.2% | $2,977 | $2,428 | 81.6% | $2,907 | $2,426 | 83.5% | $11,363 | $9,348 | 82.3% | 0.789 | 0.773 | $2,955 | $2,474 | 83.7% | 2.04 | 1.99 | 5.74 | 5.49 |
| TLogN | 10.2 | 1.95 | $99 | $124 | $24 | 24.6% | $101 | $2 | 1.7% | $115 | $76 | 66.4% | $113 | $76 | 67.9% | $362 | $279 | 77.1% | 0.911 | 0.758 | $92 | $73 | 79.3% | 4.51 | 3.13 | 35.70 | 18.58 |
| TLogN | 9 | 2.2 | $103 | $182 | $79 | 76.8% | $101 | -$2 | -2.0% | $543 | $88 | 16.2% | $537 | $88 | 16.3% | $805 | $332 | 41.2% | 2.951 | 0.869 | $128 | $85 | 66.6% | 22.07 | 2.27 | 592.72 | 7.09 |
| TLogN | 10.7 | 2.385 | $959 | $1,509 | $549 | 57.2% | $1,062 | $103 | 10.7% | $2,960 | $1,225 | 41.4% | $2,908 | $1,220 | 42.0% | $5,611 | $3,713 | 66.2% | 1.928 | 1.149 | $1,075 | $801 | 74.5% | 12.09 | 5.47 | 200.76 | 46.47 |
| TLogN | 9.4 | 2.65 | $959 | $1,801 | $842 | 87.8% | $943 | -$16 | -1.7% | $3,173 | $994 | 31.3% | $3,059 | $994 | 32.5% | $9,115 | $3,895 | 42.7% | 1.699 | 1.054 | $1,407 | $824 | 58.5% | 5.58 | 3.37 | 42.63 | 18.43 |
| TLogN | 11 | 2.6 | $3,113 | $5,093 | $1,980 | 63.6% | $3,551 | $438 | 14.1% | $7,129 | $3,969 | 55.7% | $6,848 | $3,945 | 57.6% | $23,464 | $13,431 | 57.2% | 1.345 | 1.111 | $3,941 | $2,864 | 72.7% | 4.16 | 3.66 | 22.33 | 19.17 |
| TLogN | 10 | 2.8 | $3,009 | $6,838 | $3,829 | 127.3% | $3,346 | $338 | 11.2% | $18,723 | $4,465 | 23.8% | $18,327 | $4,453 | 24.3% | $33,900 | $13,664 | 40.3% | 2.680 | 1.331 | $4,582 | $2,803 | 61.2% | 12.27 | 4.65 | 220.68 | 32.36 |
| | a | b | | | | | | | | | | | | | | | | | | | | | | | | | |
| Logg | 24 | 2.65 | $136 | $174 | $38 | 28.3% | $139 | $4 | 2.6% | $164 | $120 | 73.1% | $159 | $120 | 75.1% | $581 | $438 | 75.4% | 0.915 | 0.859 | $129 | $102 | 78.7% | 3.31 | 3.06 | 17.04 | 14.65 |
| Logg | 33 | 3.3 | $149 | $182 | $33 | 22.2% | $153 | $4 | 2.6% | $149 | $116 | 78.0% | $145 | $116 | 79.9% | $502 | $406 | 80.9% | 0.798 | 0.760 | $134 | $112 | 83.7% | 3.02 | 2.87 | 14.80 | 13.64 |
| Logg | 25 | 2.5 | $732 | $976 | $244 | 33.3% | $754 | $22 | 3.0% | $1,016 | $705 | 69.4% | $987 | $705 | 71.4% | $3,274 | $2,368 | 72.3% | 1.011 | 0.934 | $747 | $567 | 75.9% | 3.99 | 3.51 | 25.88 | 19.49 |
| Logg | 34.5 | 3.15 | $691 | $833 | $142 | 20.5% | $686 | -$5 | -0.7% | $663 | $510 | 77.0% | $647 | $510 | 78.9% | $2,401 | $1,942 | 80.9% | 0.777 | 0.744 | $633 | $517 | 81.7% | 2.26 | 2.18 | 7.08 | 6.65 |
| Logg | 25.25 | 2.45 | $1,281 | $1,733 | $452 | 35.2% | $1,319 | $38 | 3.0% | $1,811 | $1,244 | 68.7% | $1,753 | $1,244 | 70.9% | $6,125 | $4,363 | 71.2% | 1.012 | 0.943 | $1,349 | $1,016 | 75.3% | 3.36 | 3.14 | 16.76 | 14.89 |
| Logg | 34.7 | 3.07 | $1,280 | $1,589 | $309 | 24.2% | $1,294 | $14 | 1.1% | $1,357 | $1,027 | 75.7% | $1,321 | $1,027 | 77.7% | $4,567 | $3,618 | 79.2% | 0.831 | 0.794 | $1,279 | $1,020 | 79.7% | 2.60 | 2.47 | 11.36 | 10.18 |
| TLogg | 23.5 | 2.65 | $193 | $828 | $635 | 329.3% | $416 | $223 | 115.7% | $7,242 | $1,460 | 20.2% | $7,214 | $1,443 | 20.0% | $3,903 | $2,804 | 71.8% | 8.715 | 3.470 | $382 | $213 | 55.7% | 28.55 | 11.01 | 865.50 | 159.83 |
| TLogg | 33 | 3.3 | $192 | $404 | $212 | 110.5% | $220 | $29 | 14.9% | $1,465 | $795 | 54.3% | $1,450 | $795 | 54.8% | $2,039 | $897 | 44.0% | 3.588 | 3.605 | $257 | $143 | 55.8% | 23.68 | 26.41 | 659.71 | 777.87 |
| TLogg | 24.5 | 2.5 | $807 | $2,175 | $1,368 | 169.5% | $985 | $178 | 22.0% | $6,125 | $2,780 | 45.4% | $5,970 | $2,774 | 46.5% | $13,076 | $4,519 | 34.6% | 2.745 | 2.816 | $1,423 | $632 | 44.5% | 8.45 | 10.62 | 89.14 | 150.22 |
| TLogg | 34.5 | 3.15 | $783 | $1,278 | $495 | 63.3% | $789 | $6 | 0.8% | $2,135 | $759 | 35.6% | $2,077 | $759 | 36.5% | $5,585 | $2,599 | 46.5% | 1.625 | 0.962 | $1,015 | $642 | 63.2% | 6.84 | 3.75 | 70.02 | 26.23 |
| TLogg | 24.75 | 2.45 | $1,325 | $3,342 | $2,017 | 152.3% | $1,357 | $32 | 2.4% | $7,938 | $2,981 | 37.6% | $7,677 | $2,981 | 38.8% | $22,129 | $7,350 | 33.2% | 2.297 | 2.197 | $2,278 | $905 | 39.7% | 6.68 | 7.58 | 63.01 | 82.42 |
| TLogg | 34.6 | 3.07 | $1,352 | $2,095 | $744 | 55.0% | $1,458 | $106 | 7.8% | $2,805 | $1,352 | 48.2% | $2,704 | $1,348 | 49.8% | $8,667 | $4,701 | 54.2% | 1.291 | 0.925 | $1,797 | $1,288 | 71.7% | 4.65 | 2.62 | 36.54 | 11.51 |
| | ξ | θ | | | | | | | | | | | | | | | | | | | | | | | | | |
| GPD | 0.8 | 35,000 | $254 | $558 | $304 | 119.9% | $263 | $9 | 3.4% | $976 | $378 | 38.7% | $928 | $378 | 40.7% | $3,208 | $1,351 | 42.1% | 1.661 | 1.438 | $494 | $244 | 49.5% | 3.89 | 3.35 | 19.39 | 14.17 |
| GPD | 0.95 | 7,500 | $231 | $608 | $377 | 163.4% | $238 | $7 | 3.1% | $1,199 | $380 | 31.7% | $1,138 | $380 | 33.4% | $3,745 | $1,325 | 35.4% | 1.871 | 1.598 | $515 | $215 | 41.8% | 4.35 | 3.64 | 24.85 | 16.90 |
| GPD | 0.875 | 47,500 | $707 | $1,668 | $961 | 135.9% | $720 | $13 | 1.9% | $3,053 | $1,075 | 35.2% | $2,898 | $1,074 | 37.1% | $9,646 | $3,778 | 39.2% | 1.737 | 1.492 | $1,479 | $678 | 45.8% | 4.14 | 3.47 | 22.76 | 15.63 |
| GPD | 0.95 | 25,000 | $770 | $1,992 | $1,222 | 158.8% | $782 | $12 | 1.5% | $3,909 | $1,243 | 31.8% | $3,713 | $1,243 | 33.5% | $12,270 | $4,393 | 35.8% | 1.863 | 1.590 | $1,700 | $705 | 41.5% | 4.38 | 3.64 | 25.62 | 17.12 |
| GPD | 0.925 | 50,000 | $1,207 | $2,869 | $1,661 | 137.6% | $1,167 | -$41 | -3.4% | $5,633 | $1,874 | 33.3% | $5,382 | $1,873 | 34.8% | $17,924 | $6,552 | 36.6% | 1.876 | 1.606 | $2,498 | $1,066 | 42.7% | 4.44 | 3.68 | 26.65 | 17.79 |
| GPD | 0.99 | 27,500 | $1,252 | $3,315 | $2,063 | 164.9% | $1,241 | -$11 | -0.9% | $6,693 | $2,024 | 30.2% | $6,367 | $2,024 | 31.8% | $20,958 | $6,920 | 33.0% | 1.921 | 1.631 | $2,784 | $1,120 | 40.2% | 4.68 | 3.82 | 29.99 | 19.38 |
| TGPD | 0.775 | 33,500 | $236 | $538 | $302 | 127.8% | $257 | $21 | 9.0% | $1,295 | $508 | 39.3% | $1,259 | $508 | 40.3% | $3,192 | $1,326 | 41.5% | 2.341 | 1.974 | $443 | $216 | 48.7% | 7.68 | 7.08 | 77.92 | 67.73 |
| TGPD | 0.8 | 25,000 | $240 | $499 | $260 | 108.5% | $239 | $0 | 0.0% | $985 | $390 | 39.6% | $950 | $390 | 41.0% | $2,696 | $1,162 | 43.1% | 1.902 | 1.627 | $452 | $218 | 48.1% | 6.53 | 5.83 | 63.93 | 53.24 |
| TGPD | 0.8675 | 50,000 | $813 | $2,071 | $1,258 | 154.8% | $910 | $97 | 12.0% | $4,252 | $1,482 | 34.9% | $4,061 | $1,479 | 36.4% | $12,752 | $5,320 | 41.7% | 1.961 | 1.625 | $1,770 | $818 | 46.2% | 5.72 | 4.70 | 44.43 | 30.37 |
| TGPD | 0.91 | 31,000 | $837 | $2,065 | $1,228 | 146.7% | $846 | $9 | 1.0% | $5,096 | $1,638 | 32.1% | $4,945 | $1,638 | 33.1% | $12,708 | $4,708 | 37.1% | 2.394 | 1.936 | $1,629 | $727 | 44.6% | 9.16 | 7.50 | 132.86 | 94.59 |
| TGPD | 0.92 | 47,500 | $1,304 | $3,609 | $2,304 | 176.7% | $1,483 | $179 | 13.7% | $7,331 | $2,401 | 32.7% | $6,959 | $2,394 | 34.4% | $22,760 | $8,367 | 36.8% | 1.928 | 1.614 | $3,330 | $1,464 | 44.0% | 4.93 | 4.22 | 30.61 | 23.44 |
| TGPD | 0.95 | 35,000 | $1,371 | $3,681 | $2,310 | 168.6% | $1,416 | $45 | 3.3% | $10,884 | $3,185 | 29.3% | $10,636 | $3,184 | 29.9% | $21,357 | $8,114 | 38.0% | 2.889 | 2.249 | $2,580 | $1,100 | 42.6% | 9.09 | 7.37 | 104.00 | 70.72 |

*NOTE: #simulations = 1,000; λ = 15 for 10 years so n ~ 150; α=0.9997

**TABLE F6a**

**RCE vs. LDA for Regulatory Capital Estimation Under iid ($m, λ=50)***

| Severity Dist. | Parm1 | Parm2 | True RCap | Mean MLE RCap | MLE Bias | MLE Bias% | Mean RCE RCap | RCE Bias | RCE Bias% | RMSE MLE RCap | RMSE RCE RCap | RMSE RCap RCE/MLE | StdDev MLE RCap | StdDev RCE RCap | StdDev RCap RCE/MLE | 95%CIs MLE RCap | 95%CIs RCE RCap | 95%CIs RCap RCE/MLE | CV MLE | CV RCE | IQR MLE RCap | IQR RCE RCap | IQR RCap RCE/MLE | Skew MLE RCap | Skew RCE RCap | Kurtosis MLE RCap | Kurtosis RCE RCap |
|---|---|---|---|---|---|---|---|---|---|---|---|---|---|---|---|---|---|---|---|---|---|---|---|---|---|---|---|
| | **μ** | **σ** | | | | | | | | | | | | | | | | | | | | | | | | | |
| LogN | 10 | 2 | $90 | $92 | **$3** | **3.0%** | $89 | **$0** | **-0.2%** | $25 | $24 | **95.7%** | $25 | $24 | 96.3% | $97 | $93 | **96.0%** | 0.272 | 0.270 | $31 | $30 | 95.9% | 0.86 | 0.85 | 0.90 | 0.88 |
| LogN | 7.7 | 2.55 | $81 | $85 | **$4** | **5.4%** | $81 | **$0** | **0.3%** | $31 | $29 | **93.5%** | $31 | $29 | 94.4% | $118 | $112 | **94.5%** | 0.362 | 0.359 | $36 | $34 | 95.5% | 1.10 | 1.08 | 1.52 | 1.49 |
| LogN | 10.4 | 2.5 | $984 | $1,035 | **$51** | **5.1%** | $987 | **$3** | **0.3%** | $370 | $347 | **93.7%** | $367 | $347 | 94.6% | $1,402 | $1,327 | **94.7%** | 0.354 | 0.351 | $429 | $411 | 95.8% | 1.07 | 1.06 | 1.46 | 1.43 |
| LogN | 9.27 | 2.77 | $952 | $1,014 | **$62** | **6.5%** | $958 | **$6** | **0.6%** | $408 | $377 | **92.5%** | $403 | $377 | 93.6% | $1,534 | $1,437 | **93.7%** | 0.397 | 0.394 | $462 | $438 | 94.8% | 1.19 | 1.18 | 1.82 | 1.78 |
| LogN | 10.75 | 2.7 | $3,145 | $3,338 | **$194** | **6.2%** | $3,161 | **$16** | **0.5%** | $1,304 | $1,210 | **92.8%** | $1,289 | $1,210 | 93.8% | $4,914 | $4,617 | **94.0%** | 0.386 | 0.383 | $1,487 | $1,414 | 95.1% | 1.16 | 1.15 | 1.72 | 1.69 |
| LogN | 9.63 | 2.97 | $3,085 | $3,321 | **$236** | **7.7%** | $3,112 | **$27** | **0.9%** | $1,446 | $1,324 | **91.6%** | $1,427 | $1,324 | 92.8% | $5,412 | $5,027 | **92.9%** | 0.430 | 0.425 | $1,612 | $1,515 | 94.0% | 1.28 | 1.26 | 2.10 | 2.06 |
| TLogN | 10.2 | 1.95 | $108 | $117 | **$9** | **8.1%** | $109 | **$1** | **1.0%** | $49 | $44 | **88.3%** | $49 | $44 | 89.6% | $192 | $175 | **90.7%** | 0.416 | 0.399 | $54 | $49 | 92.3% | 1.64 | 1.54 | 4.25 | 3.70 |
| TLogN | 9 | 2.2 | $110 | $125 | **$15** | **13.9%** | $111 | **$1** | **1.3%** | $76 | $61 | **80.0%** | $75 | $61 | 81.6% | $272 | $224 | **82.6%** | 0.596 | 0.547 | $74 | $63 | 86.0% | 2.71 | 2.35 | 14.19 | 10.59 |
| TLogN | 10.7 | 2.385 | $996 | $1,120 | **$124** | **12.5%** | $1,014 | **$18** | **1.8%** | $620 | $527 | **84.9%** | $608 | $526 | 86.6% | $2,192 | $1,932 | **88.1%** | 0.543 | 0.519 | $615 | $547 | 88.9% | 2.06 | 1.97 | 7.57 | 6.98 |
| TLogN | 9.4 | 2.65 | $985 | $1,183 | **$197** | **20.0%** | $1,006 | **$21** | **2.1%** | $838 | $637 | **76.0%** | $815 | $637 | 78.1% | $3,008 | $2,384 | **79.3%** | 0.689 | 0.633 | $799 | $663 | 82.9% | 2.56 | 2.18 | 12.21 | 8.07 |
| TLogN | 11 | 2.6 | $3,195 | $3,667 | **$472** | **14.8%** | $3,270 | **$75** | **2.4%** | $2,201 | $1,834 | **83.3%** | $2,150 | $1,832 | 85.2% | $8,260 | $6,980 | **84.5%** | 0.586 | 0.560 | $2,178 | $1,910 | 87.7% | 2.29 | 2.15 | 9.06 | 8.02 |
| TLogN | 10 | 2.8 | $3,072 | $3,599 | **$527** | **17.1%** | $3,061 | **-$11** | **-0.4%** | $2,422 | $1,864 | **77.0%** | $2,364 | $1,864 | 78.9% | $8,573 | $6,851 | **79.9%** | 0.657 | 0.609 | $2,401 | $1,984 | 82.6% | 2.46 | 2.18 | 11.43 | 9.04 |
| | **a** | **b** | | | | | | | | | | | | | | | | | | | | | | | | | |
| Logg | 24 | 2.65 | $139 | $146 | **$8** | **5.6%** | $139 | **$0** | **0.0%** | $64 | $60 | **93.0%** | $64 | $60 | 93.7% | $246 | $233 | **94.5%** | 0.437 | 0.432 | $73 | $68 | 93.1% | 1.65 | 1.63 | 4.61 | 4.47 |
| Logg | 33 | 3.3 | $154 | $160 | **$6** | **3.6%** | $153 | **-$1** | **-0.8%** | $58 | $55 | **94.4%** | $58 | $55 | 94.8% | $223 | $213 | **95.3%** | 0.363 | 0.360 | $71 | $68 | 95.8% | 1.26 | 1.24 | 2.45 | 2.36 |
| Logg | 25 | 2.5 | $745 | $806 | **$62** | **8.3%** | $758 | **$13** | **1.8%** | $382 | $351 | **91.7%** | $377 | $350 | 92.8% | $1,423 | $1,327 | **93.2%** | 0.468 | 0.462 | $433 | $406 | 93.6% | 1.51 | 1.48 | 3.75 | 3.61 |
| Logg | 34.5 | 3.15 | $709 | $754 | **$45** | **6.3%** | $718 | **$9** | **1.2%** | $318 | $297 | **93.4%** | $315 | $297 | 94.3% | $1,188 | $1,118 | **94.1%** | 0.417 | 0.413 | $372 | $349 | 93.8% | 1.29 | 1.28 | 2.55 | 2.50 |
| Logg | 25.25 | 2.45 | $1,301 | $1,425 | **$124** | **9.5%** | $1,337 | **$36** | **2.7%** | $710 | $649 | **91.5%** | $699 | $648 | 92.8% | $2,698 | $2,517 | **93.3%** | 0.490 | 0.485 | $807 | $754 | 93.4% | 1.51 | 1.50 | 3.56 | 3.49 |
| Logg | 34.7 | 3.07 | $1,310 | $1,403 | **$93** | **7.1%** | $1,333 | **$23** | **1.7%** | $611 | $569 | **93.1%** | $604 | $569 | 94.1% | $2,194 | $2,061 | **94.0%** | 0.430 | 0.426 | $710 | $674 | 94.8% | 1.69 | 1.67 | 6.22 | 6.05 |
| TLogg | 23.5 | 2.65 | $199 | $262 | **$63** | **31.7%** | $204 | **$5** | **2.5%** | $310 | $182 | **58.7%** | $304 | $182 | 59.9% | $808 | $547 | **67.7%** | 1.161 | 0.894 | $182 | $136 | 74.8% | 8.04 | 5.32 | 110.67 | 49.66 |
| TLogg | 33 | 3.3 | $200 | $226 | **$26** | **13.2%** | $204 | **$4** | **2.0%** | $155 | $124 | **80.4%** | $152 | $124 | 81.6% | $503 | $439 | **87.3%** | 0.674 | 0.610 | $130 | $116 | 88.9% | 5.56 | 4.31 | 71.10 | 45.45 |
| TLogg | 24.5 | 2.5 | $823 | $1,035 | **$212** | **25.8%** | $860 | **$37** | **4.5%** | $901 | $634 | **70.3%** | $876 | $633 | 72.3% | $3,051 | $2,236 | **73.3%** | 0.846 | 0.736 | $719 | $590 | 82.0% | 3.53 | 2.77 | 20.37 | 12.59 |
| TLogg | 34.5 | 3.15 | $805 | $899 | **$93** | **11.6%** | $827 | **$21** | **2.6%** | $510 | $446 | **87.5%** | $502 | $446 | 88.9% | $1,941 | $1,745 | **89.9%** | 0.558 | 0.540 | $572 | $507 | 88.7% | 1.56 | 1.48 | 3.36 | 2.96 |
| TLogg | 24.75 | 2.45 | $1,348 | $1,664 | **$316** | **23.4%** | $1,400 | **$52** | **3.9%** | $1,455 | $1,048 | **72.0%** | $1,421 | $1,047 | 73.7% | $4,562 | $3,529 | **77.4%** | 0.854 | 0.748 | $1,238 | $1,005 | 81.1% | 4.21 | 3.10 | 32.51 | 17.37 |
| TLogg | 34.6 | 3.07 | $1,386 | $1,573 | **$188** | **13.6%** | $1,450 | **$65** | **4.7%** | $890 | $778 | **87.4%** | $870 | $775 | 89.1% | $3,272 | $2,914 | **89.1%** | 0.553 | 0.535 | $973 | $875 | 89.9% | 2.11 | 1.97 | 9.53 | 8.36 |
| | **ξ** | **θ** | | | | | | | | | | | | | | | | | | | | | | | | | |
| GPD | 0.8 | 35,000 | $260 | $339 | **$79** | **30.5%** | $268 | **$8** | **3.0%** | $304 | $220 | **72.4%** | $294 | $220 | 74.9% | $1,054 | $792 | **75.2%** | 0.866 | 0.822 | $276 | $213 | 77.1% | 2.90 | 2.73 | 12.83 | 11.32 |
| GPD | 0.95 | 7,500 | $234 | $327 | **$93** | **39.8%** | $243 | **$10** | **4.1%** | $339 | $230 | **67.8%** | $326 | $230 | 70.4% | $1,154 | $827 | **71.6%** | 0.998 | 0.944 | $276 | $204 | 74.0% | 3.33 | 3.10 | 16.90 | 14.51 |
| GPD | 0.875 | 47,500 | $719 | $970 | **$251** | **34.8%** | $744 | **$25** | **3.4%** | $939 | $658 | **70.1%** | $905 | $658 | 72.7% | $3,219 | $2,368 | **73.6%** | 0.933 | 0.885 | $808 | $600 | 74.2% | 3.13 | 2.94 | 14.95 | 13.16 |
| GPD | 0.95 | 25,000 | $779 | $1,088 | **$309** | **39.7%** | $810 | **$31** | **4.0%** | $1,129 | $765 | **67.8%** | $1,086 | $765 | 70.4% | $3,847 | $2,757 | **71.6%** | 0.998 | 0.944 | $918 | $681 | 74.2% | 3.34 | 3.10 | 16.97 | 14.56 |
| GPD | 0.925 | 50,000 | $1,223 | $1,681 | **$458** | **37.4%** | $1,265 | **$41** | **3.4%** | $1,701 | $1,168 | **68.7%** | $1,638 | $1,167 | 71.3% | $5,647 | $4,037 | **71.5%** | 0.974 | 0.923 | $1,411 | $1,040 | 73.7% | 3.30 | 3.08 | 16.65 | 14.48 |
| GPD | 0.99 | 27,500 | $1,264 | $1,790 | **$526** | **41.6%** | $1,311 | **$48** | **3.8%** | $1,917 | $1,275 | **66.5%** | $1,843 | $1,274 | 69.1% | $6,396 | $4,424 | **69.2%** | 1.030 | 0.971 | $1,530 | $1,119 | 73.1% | 3.49 | 3.22 | 18.80 | 15.93 |
| TGPD | 0.775 | 33,500 | $243 | $304 | **$61** | **25.3%** | $243 | **$0** | **-0.1%** | $249 | $183 | **73.5%** | $242 | $183 | 75.9% | $976 | $741 | **75.9%** | 0.794 | 0.755 | $224 | $177 | 79.2% | 2.42 | 2.30 | 8.75 | 7.88 |
| TGPD | 0.8 | 25,000 | $246 | $315 | **$70** | **28.3%** | $249 | **$3** | **1.2%** | $282 | $205 | **72.6%** | $273 | $205 | 74.9% | $984 | $745 | **75.7%** | 0.867 | 0.823 | $244 | $189 | 77.7% | 2.92 | 2.77 | 13.13 | 11.89 |
| TGPD | 0.8675 | 50,000 | $827 | $1,101 | **$274** | **33.1%** | $845 | **$18** | **2.2%** | $1,104 | $771 | **69.8%** | $1,069 | $770 | 72.1% | $3,440 | $2,530 | **73.6%** | 0.971 | 0.912 | $868 | $647 | 74.5% | 4.07 | 3.72 | 28.67 | 24.11 |
| TGPD | 0.91 | 31,000 | $850 | $1,140 | **$290** | **34.1%** | $861 | **$11** | **1.3%** | $1,091 | $750 | **68.7%** | $1,051 | $750 | 71.3% | $3,764 | $2,690 | **71.5%** | 0.922 | 0.870 | $915 | $672 | 73.5% | 3.53 | 3.19 | 25.80 | 21.07 |
| TGPD | 0.92 | 47,500 | $1,323 | $1,795 | **$472** | **35.7%** | $1,351 | **$28** | **2.1%** | $1,800 | $1,240 | **68.9%** | $1,736 | $1,240 | 71.4% | $6,365 | $4,514 | **70.9%** | 0.967 | 0.918 | $1,481 | $1,095 | 73.9% | 3.31 | 3.13 | 18.39 | 16.56 |
| TGPD | 0.95 | 35,000 | $1,387 | $1,928 | **$541** | **39.0%** | $1,428 | **$41** | **2.9%** | $2,245 | $1,487 | **66.2%** | $2,179 | $1,486 | 68.2% | $6,423 | $4,549 | **70.8%** | 1.130 | 1.041 | $1,592 | $1,168 | 73.3% | 6.65 | 5.70 | 83.48 | 63.04 |

*NOTE: #simulations = 1,000; λ = 50 for 10 years so n ~ 500; α=0.999

**TABLE F6b**

**RCE vs. LDA for Economic Capital Estimation Under iid ($m, λ=50)***

| Severity Dist. | Parm1 | Parm2 | True ECap | Mean MLE ECap | MLE Bias | MLE Bias% | Mean RCE ECap | RCE Bias | RCE Bias% | RMSE MLE ECap | RMSE RCE ECap | RMSE ECap RCE/MLE | StdDev MLE ECap | StdDev RCE ECap | StdDev ECap RCE/MLE | 95%CIs MLE ECap | 95%CIs RCE ECap | 95%CIs ECap RCE/MLE | CV MLE | CV RCE | IQR MLE ECap | IQR RCE ECap | IQR ECap RCE/MLE | Skew MLE ECap | Skew RCE ECap | Kurtosis MLE ECap | Kurtosis RCE ECap |
|---|---|---|---|---|---|---|---|---|---|---|---|---|---|---|---|---|---|---|---|---|---|---|---|---|---|---|---|
| | **μ** | **σ** | | | | | | | | | | | | | | | | | | | | | | | | | |
| LogN | 10 | 2 | $148 | $153 | **$5** | **3.5%** | $148 | **$0** | **0.1%** | $45 | $43 | **95.5%** | $45 | $43 | 96.1% | $173 | $166 | **95.8%** | 0.293 | 0.291 | $55 | $53 | 96.6% | 0.91 | 0.90 | 1.02 | 1.01 |
| LogN | 7.7 | 2.55 | $158 | $168 | **$10** | **6.2%** | $160 | **$1** | **0.8%** | $66 | $61 | **93.1%** | $65 | $61 | 94.1% | $249 | $234 | **94.0%** | 0.387 | 0.384 | $75 | $72 | 94.8% | 1.16 | 1.15 | 1.72 | 1.68 |
| LogN | 10.4 | 2.5 | $1,898 | $2,010 | **$112** | **5.9%** | $1,912 | **$15** | **0.8%** | $769 | $718 | **93.3%** | $761 | $718 | 94.3% | $2,917 | $2,747 | **94.2%** | 0.379 | 0.375 | $886 | $841 | 94.9% | 1.14 | 1.13 | 1.65 | 1.61 |
| LogN | 9.27 | 2.77 | $1,984 | $2,133 | **$149** | **7.5%** | $2,007 | **$23** | **1.2%** | $917 | $844 | **92.0%** | $905 | $844 | 93.2% | $3,453 | $3,214 | **93.1%** | 0.424 | 0.420 | $1,029 | $967 | 94.0% | 1.26 | 1.25 | 2.04 | 2.00 |
| LogN | 10.75 | 2.7 | $6,425 | $6,879 | **$454** | **7.1%** | $6,494 | **$69** | **1.1%** | $2,873 | $2,654 | **92.4%** | $2,837 | $2,654 | 93.5% | $10,840 | $10,120 | **93.4%** | 0.412 | 0.409 | $3,246 | $3,059 | 94.3% | 1.23 | 1.22 | 1.93 | 1.89 |
| LogN | 9.63 | 2.97 | $6,803 | $7,400 | **$596** | **8.8%** | $6,906 | **$103** | **1.5%** | $3,444 | $3,135 | **91.0%** | $3,392 | $3,133 | 92.4% | $12,900 | $11,891 | **92.2%** | 0.458 | 0.454 | $3,788 | $3,528 | 93.1% | 1.35 | 1.34 | 2.36 | 2.31 |
| TLogN | 10.2 | 1.95 | $173 | $190 | **$17** | **10.1%** | $176 | **$3** | **1.7%** | $91 | $79 | **86.6%** | $90 | $79 | 88.2% | $345 | $311 | **90.2%** | 0.471 | 0.449 | $97 | $86 | 88.7% | 1.84 | 1.72 | 5.41 | 4.64 |
| TLogN | 9 | 2.2 | $187 | $220 | **$33** | **17.7%** | $191 | **$4** | **2.1%** | $155 | $119 | **76.5%** | $152 | $119 | 78.3% | $548 | $434 | **79.2%** | 0.689 | 0.622 | $141 | $117 | 83.4% | 3.17 | 2.66 | 19.26 | 13.40 |
| TLogN | 10.7 | 2.385 | $1,833 | $2,110 | **$278** | **15.2%** | $1,881 | **$49** | **2.7%** | $1,303 | $1,081 | **83.0%** | $1,273 | $1,080 | 84.9% | $4,520 | $3,900 | **86.3%** | 0.603 | 0.574 | $1,261 | $1,085 | 86.0% | 2.26 | 2.15 | 8.95 | 8.21 |
| TLogN | 9.4 | 2.65 | $1,936 | $2,417 | **$481** | **24.8%** | $1,991 | **$56** | **2.9%** | $1,944 | $1,402 | **72.1%** | $1,883 | $1,401 | 74.4% | $6,771 | $5,241 | **77.4%** | 0.779 | 0.703 | $1,741 | $1,381 | 79.3% | 3.02 | 2.44 | 17.56 | 10.37 |
| TLogN | 11 | 2.6 | $6,271 | $7,382 | **$1,111** | **17.7%** | $6,471 | **$199** | **3.2%** | $4,923 | $3,998 | **81.2%** | $4,796 | $3,993 | 83.3% | $18,152 | $15,265 | **84.1%** | 0.650 | 0.617 | $4,638 | $3,996 | 86.2% | 2.54 | 2.37 | 11.14 | 9.70 |
| TLogN | 10 | 2.8 | $6,333 | $7,672 | **$1,339** | **21.1%** | $6,343 | **$9** | **0.1%** | $5,810 | $4,277 | **73.6%** | $5,653 | $4,277 | 75.7% | $20,252 | $15,400 | **76.0%** | 0.737 | 0.674 | $5,485 | $4,376 | 79.8% | 2.83 | 2.46 | 14.86 | 11.40 |
| | **a** | **b** | | | | | | | | | | | | | | | | | | | | | | | | | |
| Logg | 24 | 2.65 | $307 | $328 | **$22** | **7.0%** | $309 | **$2** | **0.7%** | $161 | $148 | **92.1%** | $159 | $148 | 92.9% | $616 | $570 | **92.6%** | 0.485 | 0.480 | $177 | $165 | 93.0% | 1.84 | 1.81 | 5.72 | 5.53 |
| Logg | 33 | 3.3 | $308 | $322 | **$14** | **4.6%** | $307 | **-$1** | **-0.4%** | $130 | $122 | **93.7%** | $129 | $122 | 94.2% | $495 | $469 | **94.7%** | 0.401 | 0.397 | $154 | $147 | 95.0% | 1.38 | 1.36 | 2.95 | 2.84 |
| Logg | 25 | 2.5 | $1,752 | $1,929 | **$177** | **10.1%** | $1,799 | **$47** | **2.7%** | $1,012 | $917 | **90.7%** | $996 | $916 | 92.0% | $3,713 | $3,395 | **91.5%** | 0.516 | 0.509 | $1,110 | $1,033 | 93.1% | 1.67 | 1.64 | 4.71 | 4.51 |
| Logg | 34.5 | 3.15 | $1,491 | $1,606 | **$115** | **7.7%** | $1,520 | **$29** | **2.0%** | $744 | $690 | **92.7%** | $735 | $689 | 93.7% | $2,764 | $2,591 | **93.7%** | 0.458 | 0.453 | $846 | $795 | 94.1% | 1.40 | 1.38 | 2.97 | 2.90 |
| Logg | 25.25 | 2.45 | $3,127 | $3,489 | **$362** | **11.6%** | $3,244 | **$118** | **3.8%** | $1,918 | $1,735 | **90.4%** | $1,883 | $1,731 | 91.9% | $7,221 | $6,717 | **93.0%** | 0.540 | 0.533 | $2,115 | $1,951 | 92.3% | 1.66 | 1.64 | 4.32 | 4.21 |
| Logg | 34.7 | 3.07 | $2,818 | $3,059 | **$241** | **8.6%** | $2,888 | **$70** | **2.5%** | $1,470 | $1,357 | **92.3%** | $1,450 | $1,355 | 93.4% | $5,230 | $4,902 | **93.7%** | 0.474 | 0.469 | $1,645 | $1,559 | 94.8% | 1.88 | 1.85 | 7.63 | 7.39 |
| TLogg | 23.5 | 2.65 | $427 | $618 | **$192** | **45.0%** | $427 | **$1** | **0.2%** | $954 | $437 | **45.8%** | $935 | $437 | 46.8% | $2,145 | $1,241 | **57.9%** | 1.511 | 1.023 | $451 | $299 | 66.3% | 10.07 | 5.66 | 160.42 | 50.99 |
| TLogg | 33 | 3.3 | $395 | $465 | **$70** | **17.7%** | $406 | **$12** | **3.0%** | $388 | $290 | **74.7%** | $382 | $289 | 75.9% | $1,181 | $984 | **83.4%** | 0.821 | 0.712 | $291 | $250 | 85.7% | 7.38 | 5.18 | 113.19 | 61.98 |
| TLogg | 24.5 | 2.5 | $1,900 | $2,550 | **$650** | **34.2%** | $1,987 | **$87** | **4.6%** | $2,642 | $1,628 | **61.6%** | $2,561 | $1,626 | 63.5% | $8,394 | $5,704 | **68.0%** | 1.004 | 0.818 | $1,877 | $1,462 | 77.9% | 4.21 | 2.94 | 28.24 | 13.90 |
| TLogg | 34.5 | 3.15 | $1,683 | $1,935 | **$252** | **15.0%** | $1,748 | **$65** | **3.9%** | $1,244 | $1,059 | **85.1%** | $1,218 | $1,057 | 86.8% | $4,685 | $4,112 | **87.8%** | 0.629 | 0.604 | $1,324 | $1,184 | 89.4% | 1.76 | 1.65 | 4.37 | 3.78 |
| TLogg | 24.75 | 2.45 | $3,186 | $4,181 | **$995** | **31.2%** | $3,333 | **$147** | **4.6%** | $4,386 | $2,800 | **63.8%** | $4,272 | $2,797 | 65.5% | $13,163 | $9,326 | **70.9%** | 1.022 | 0.839 | $3,282 | $2,555 | 77.9% | 5.29 | 3.33 | 49.11 | 19.31 |
| TLogg | 34.6 | 3.07 | $2,965 | $3,469 | **$504** | **17.0%** | $3,145 | **$180** | **6.1%** | $2,236 | $1,903 | **85.1%** | $2,178 | $1,895 | 87.0% | $8,111 | $7,066 | **87.1%** | 0.628 | 0.602 | $2,323 | $2,062 | 88.8% | 2.50 | 2.31 | 13.17 | 11.29 |
| | **ξ** | **θ** | | | | | | | | | | | | | | | | | | | | | | | | | |
| GPD | 0.8 | 35,000 | $667 | $944 | **$277** | **41.5%** | $697 | **$30** | **4.5%** | $1,013 | $678 | **67.0%** | $974 | $678 | 69.6% | $3,457 | $2,408 | **69.6%** | 1.032 | 0.972 | $822 | $594 | 72.3% | 3.44 | 3.20 | 17.63 | 15.28 |
| GPD | 0.95 | 7,500 | $726 | $1,117 | **$392** | **54.0%** | $767 | **$42** | **5.8%** | $1,384 | $854 | **61.7%** | $1,327 | $853 | 64.3% | $4,499 | $2,960 | **65.8%** | 1.188 | 1.112 | $990 | $672 | 67.9% | 3.98 | 3.66 | 23.72 | 20.04 |
| GPD | 0.875 | 47,500 | $2,031 | $2,992 | **$961** | **47.3%** | $2,132 | **$102** | **5.0%** | $3,459 | $2,227 | **64.4%** | $3,322 | $2,225 | 67.0% | $11,336 | $7,744 | **68.3%** | 1.110 | 1.043 | $2,654 | $1,846 | 69.5% | 3.71 | 3.44 | 20.60 | 17.70 |
| GPD | 0.95 | 25,000 | $2,418 | $3,720 | **$1,302** | **53.8%** | $2,555 | **$137** | **5.6%** | $4,610 | $2,846 | **61.7%** | $4,422 | $2,843 | 64.3% | $14,996 | $9,867 | **65.8%** | 1.189 | 1.113 | $3,292 | $2,230 | 67.7% | 3.99 | 3.67 | 23.80 | 20.11 |
| GPD | 0.925 | 50,000 | $3,681 | $5,550 | **$1,869** | **50.8%** | $3,861 | **$180** | **4.9%** | $6,704 | $4,203 | **62.7%** | $6,438 | $4,199 | 65.2% | $21,470 | $13,980 | **65.1%** | 1.160 | 1.088 | $4,936 | $3,365 | 68.2% | 3.93 | 3.63 | 23.30 | 19.79 |
| GPD | 0.99 | 27,500 | $4,124 | $6,457 | **$2,332** | **56.5%** | $4,346 | **$222** | **5.4%** | $8,272 | $4,993 | **60.4%** | $7,936 | $4,988 | 62.9% | $26,558 | $16,654 | **62.7%** | 1.229 | 1.148 | $5,761 | $3,871 | 67.2% | 4.19 | 3.84 | 26.50 | 22.21 |
| TGPD | 0.775 | 33,500 | $603 | $812 | **$209** | **34.7%** | $606 | **$3** | **0.6%** | $790 | $538 | **68.1%** | $762 | $538 | 70.6% | $3,019 | $2,130 | **70.6%** | 0.939 | 0.888 | $652 | $475 | 72.8% | 2.83 | 2.67 | 12.10 | 10.62 |
| TGPD | 0.8 | 25,000 | $629 | $874 | **$245** | **38.8%** | $644 | **$14** | **2.3%** | $935 | $627 | **67.1%** | $903 | $627 | 69.5% | $3,112 | $2,203 | **70.8%** | 1.033 | 0.974 | $715 | $525 | 73.5% | 3.45 | 3.25 | 18.04 | 16.12 |
| TGPD | 0.8675 | 50,000 | $2,313 | $3,366 | **$1,052** | **45.5%** | $2,397 | **$84** | **3.6%** | $4,111 | $2,623 | **63.8%** | $3,974 | $2,622 | 66.0% | $11,968 | $8,109 | **67.8%** | 1.181 | 1.094 | $2,823 | $1,968 | 69.7% | 5.06 | 4.53 | 42.28 | 34.44 |
| TGPD | 0.91 | 31,000 | $2,508 | $3,668 | **$1,161** | **46.3%** | $2,564 | **$56** | **2.2%** | $4,201 | $2,632 | **62.6%** | $4,037 | $2,631 | 65.2% | $13,671 | $9,199 | **67.3%** | 1.101 | 1.026 | $3,067 | $2,124 | 69.3% | 4.57 | 4.04 | 42.29 | 33.55 |
| TGPD | 0.92 | 47,500 | $3,953 | $5,875 | **$1,922** | **48.6%** | $4,088 | **$134** | **3.4%** | $7,004 | $4,409 | **63.0%** | $6,735 | $4,407 | 65.4% | $24,025 | $15,841 | **65.9%** | 1.146 | 1.078 | $5,115 | $3,486 | 68.1% | 3.95 | 3.70 | 25.63 | 22.80 |
| TGPD | 0.95 | 35,000 | $4,305 | $6,619 | **$2,313** | **53.7%** | $4,506 | **$201** | **4.7%** | $9,696 | $5,776 | **59.6%** | $9,416 | $5,772 | 61.3% | $25,151 | $15,986 | **63.6%** | 1.423 | 1.281 | $5,692 | $3,829 | 67.3% | 8.95 | 7.54 | 137.22 | 101.58 |

*NOTE: #simulations = 1,000; λ = 50 for 10 years so n ~ 500; α=0.9997

**TABLE F7a**

**RCE vs. LDA for Regulatory Capital Estimation Under iid ($m, λ=75)***

| Severity Dist. | Parm1 | Parm2 | True RCap | Mean MLE RCap | MLE Bias | MLE Bias% | Mean RCE RCap | RCE Bias | RCE Bias% | RMSE MLE RCap | RMSE RCE RCap | RMSE RCap RCE/MLE | StdDev MLE RCap | StdDev RCE RCap | StdDev RCap RCE/MLE | 95%CIs MLE RCap | 95%CIs RCE RCap | 95%CIs RCap RCE/MLE | CV MLE | CV RCE | IQR MLE RCap | IQR RCE RCap | IQR RCap RCE/MLE | Skew MLE RCap | Skew RCE RCap | Kurtosis MLE RCap | Kurtosis RCE RCap |
|---|---|---|---|---|---|---|---|---|---|---|---|---|---|---|---|---|---|---|---|---|---|---|---|---|---|---|---|
| | μ | σ | | | | | | | | | | | | | | | | | | | | | | | | | |
| LogN | 10 | 2 | $110 | $112 | $2 | 1.5% | $110 | -$1 | -0.5% | $25 | $24 | 97.4% | $25 | $24 | 97.6% | $96 | $93 | 97.5% | 0.222 | 0.221 | $33 | $32 | 98.8% | 0.75 | 0.74 | 1.63 | 1.61 |
| LogN | 7.7 | 2.55 | $103 | $106 | $3 | 3.0% | $103 | $0 | -0.3% | $32 | $30 | 95.9% | $32 | $30 | 96.4% | $120 | $115 | 95.8% | 0.297 | 0.296 | $40 | $39 | 97.0% | 1.04 | 1.03 | 2.88 | 2.83 |
| LogN | 10.4 | 2.5 | $1,250 | $1,285 | $35 | 2.8% | $1,246 | -$4 | -0.3% | $375 | $360 | 96.1% | $373 | $360 | 96.5% | $1,422 | $1,365 | 95.9% | 0.291 | 0.289 | $477 | $464 | 97.3% | 1.01 | 1.00 | 2.75 | 2.70 |
| LogN | 9.27 | 2.77 | $1,236 | $1,281 | $45 | 3.7% | $1,234 | -$2 | -0.1% | $420 | $400 | 95.2% | $418 | $400 | 95.8% | $1,582 | $1,508 | 95.4% | 0.326 | 0.324 | $528 | $508 | 96.2% | 1.16 | 1.15 | 3.52 | 3.45 |
| LogN | 10.75 | 2.7 | $4,059 | $4,198 | $139 | 3.4% | $4,051 | -$7 | -0.2% | $1,338 | $1,277 | 95.5% | $1,330 | $1,277 | 96.0% | $5,045 | $4,819 | 95.5% | 0.317 | 0.315 | $1,685 | $1,626 | 96.5% | 1.12 | 1.11 | 3.31 | 3.24 |
| LogN | 9.63 | 2.97 | $4,074 | $4,252 | $177 | 4.4% | $4,075 | $1 | 0.0% | $1,509 | $1,427 | 94.6% | $1,498 | $1,427 | 95.3% | $5,646 | $5,355 | 94.8% | 0.352 | 0.350 | $1,874 | $1,791 | 95.6% | 1.28 | 1.26 | 4.19 | 4.10 |
| TLogN | 10.2 | 1.95 | $133 | $138 | $5 | 3.6% | $132 | -$1 | -0.8% | $45 | $42 | 92.9% | $45 | $42 | 93.4% | $174 | $160 | 92.4% | 0.323 | 0.315 | $55 | $51 | 93.6% | 1.14 | 1.09 | 2.22 | 2.04 |
| TLogN | 9 | 2.2 | $136 | $146 | $10 | 7.3% | $136 | -$1 | -0.5% | $65 | $56 | 87.1% | $64 | $56 | 88.2% | $241 | $212 | 88.1% | 0.437 | 0.416 | $73 | $66 | 89.6% | 1.56 | 1.45 | 3.63 | 3.00 |
| TLogN | 10.7 | 2.385 | $1,251 | $1,327 | $76 | 6.0% | $1,244 | -$7 | -0.6% | $543 | $491 | 90.5% | $538 | $491 | 91.3% | $2,168 | $1,988 | 91.7% | 0.405 | 0.395 | $616 | $569 | 92.3% | 1.28 | 1.24 | 2.18 | 2.03 |
| TLogN | 9.4 | 2.65 | $1,259 | $1,434 | $175 | 13.9% | $1,290 | $31 | 2.4% | $798 | $666 | 83.4% | $779 | $665 | 85.4% | $2,716 | $2,337 | 86.1% | 0.543 | 0.515 | $862 | $749 | 86.9% | 2.03 | 1.89 | 7.26 | 6.35 |
| TLogN | 11 | 2.6 | $4,079 | $4,456 | $377 | 9.2% | $4,130 | $51 | 1.3% | $2,049 | $1,814 | 88.6% | $2,014 | $1,814 | 90.1% | $7,870 | $7,053 | 89.6% | 0.452 | 0.439 | $2,323 | $2,120 | 91.3% | 1.62 | 1.57 | 4.15 | 3.88 |
| TLogN | 10 | 2.8 | $3,978 | $4,505 | $527 | 13.2% | $4,046 | $68 | 1.7% | $2,606 | $2,194 | 84.2% | $2,552 | $2,193 | 85.9% | $9,400 | $8,172 | 86.9% | 0.567 | 0.542 | $2,836 | $2,487 | 87.7% | 2.05 | 1.94 | 7.51 | 6.72 |
| | a | b | | | | | | | | | | | | | | | | | | | | | | | | | |
| Logg | 24 | 2.65 | $184 | $192 | $8 | 4.4% | $185 | $1 | 0.8% | $68 | $64 | 95.2% | $67 | $64 | 95.9% | $256 | $246 | 96.0% | 0.350 | 0.348 | $84 | $80 | 95.5% | 1.06 | 1.06 | 1.89 | 1.88 |
| Logg | 33 | 3.3 | $198 | $205 | $6 | 3.2% | $199 | $1 | 0.3% | $64 | $61 | 96.1% | $63 | $61 | 96.6% | $236 | $228 | 96.4% | 0.310 | 0.308 | $82 | $80 | 96.9% | 0.99 | 0.98 | 1.69 | 1.64 |
| Logg | 25 | 2.5 | $1,004 | $1,062 | $58 | 5.8% | $1,019 | $16 | 1.5% | $427 | $404 | 94.6% | $424 | $404 | 95.4% | $1,616 | $1,543 | 95.5% | 0.399 | 0.396 | $516 | $497 | 96.3% | 1.16 | 1.15 | 2.00 | 1.98 |
| Logg | 34.5 | 3.15 | $925 | $962 | $37 | 4.0% | $932 | $7 | 0.7% | $316 | $302 | 95.6% | $313 | $302 | 96.3% | $1,232 | $1,183 | 96.0% | 0.326 | 0.324 | $401 | $390 | 97.2% | 0.93 | 0.93 | 1.63 | 1.60 |
| Logg | 25.25 | 2.45 | $1,765 | $1,857 | $92 | 5.2% | $1,780 | $15 | 0.8% | $760 | $718 | 94.4% | $754 | $718 | 95.1% | $2,946 | $2,796 | 94.9% | 0.406 | 0.403 | $948 | $902 | 95.1% | 1.15 | 1.14 | 1.95 | 1.91 |
| Logg | 34.7 | 3.07 | $1,720 | $1,777 | $58 | 3.3% | $1,719 | -$1 | -0.1% | $615 | $588 | 95.6% | $612 | $588 | 96.1% | $2,424 | $2,340 | 96.5% | 0.344 | 0.342 | $766 | $739 | 96.4% | 1.07 | 1.06 | 1.59 | 1.58 |
| TLogg | 23.5 | 2.65 | $261 | $299 | $38 | 14.6% | $258 | -$3 | -1.0% | $211 | $162 | 76.6% | $208 | $162 | 77.9% | $723 | $587 | 81.2% | 0.694 | 0.625 | $196 | $165 | 84.1% | 2.73 | 2.39 | 11.50 | 9.10 |
| TLogg | 33 | 3.3 | $257 | $277 | $20 | 7.7% | $257 | $0 | 0.1% | $134 | $119 | 88.3% | $133 | $119 | 89.3% | $489 | $439 | 89.8% | 0.481 | 0.462 | $146 | $132 | 90.7% | 1.87 | 1.72 | 7.04 | 5.75 |
| TLogg | 24.5 | 2.5 | $1,104 | $1,290 | $186 | 16.9% | $1,138 | $34 | 3.1% | $920 | $740 | 80.4% | $901 | $739 | 82.0% | $3,334 | $2,762 | 82.9% | 0.699 | 0.649 | $860 | $737 | 85.6% | 2.38 | 2.09 | 8.63 | 6.25 |
| TLogg | 34.5 | 3.15 | $1,049 | $1,130 | $81 | 7.7% | $1,062 | $13 | 1.2% | $506 | $459 | 90.6% | $500 | $459 | 91.8% | $1,873 | $1,740 | 92.9% | 0.443 | 0.432 | $593 | $552 | 93.0% | 1.34 | 1.30 | 2.80 | 2.60 |
| TLogg | 24.75 | 2.45 | $1,820 | $2,111 | $291 | 16.0% | $1,879 | $59 | 3.2% | $1,390 | $1,153 | 83.0% | $1,359 | $1,151 | 84.7% | $5,320 | $4,437 | 83.4% | 0.644 | 0.613 | $1,365 | $1,196 | 87.7% | 1.98 | 1.88 | 5.86 | 5.30 |
| TLogg | 34.6 | 3.07 | $1,817 | $1,967 | $150 | 8.3% | $1,852 | $36 | 2.0% | $893 | $812 | 91.0% | $880 | $812 | 92.2% | $3,313 | $3,053 | 92.2% | 0.448 | 0.438 | $1,034 | $952 | 92.1% | 1.51 | 1.45 | 4.01 | 3.65 |
| | ξ | θ | | | | | | | | | | | | | | | | | | | | | | | | | |
| GPD | 0.8 | 35,000 | $361 | $424 | $63 | 17.6% | $360 | $0 | -0.1% | $278 | $223 | 80.4% | $271 | $223 | 82.6% | $1,035 | $843 | 81.4% | 0.638 | 0.620 | $313 | $262 | 83.8% | 1.75 | 1.71 | 4.25 | 4.05 |
| GPD | 0.95 | 7,500 | $344 | $423 | $79 | 23.0% | $345 | $1 | 0.4% | $318 | $244 | 76.8% | $308 | $244 | 79.3% | $1,170 | $923 | 78.9% | 0.728 | 0.706 | $332 | $265 | 79.8% | 2.01 | 1.96 | 5.93 | 5.65 |
| GPD | 0.875 | 47,500 | $1,027 | $1,232 | $206 | 20.0% | $1,027 | $0 | 0.0% | $867 | $681 | 78.6% | $842 | $681 | 80.9% | $3,180 | $2,555 | 80.3% | 0.683 | 0.663 | $932 | $764 | 82.0% | 1.88 | 1.83 | 5.07 | 4.82 |
| GPD | 0.95 | 25,000 | $1,146 | $1,407 | $261 | 22.8% | $1,149 | $3 | 0.2% | $1,057 | $812 | 76.8% | $1,024 | $812 | 79.3% | $3,900 | $3,077 | 78.9% | 0.728 | 0.706 | $1,103 | $881 | 79.9% | 2.02 | 1.97 | 5.99 | 5.71 |
| GPD | 0.925 | 50,000 | $1,782 | $2,170 | $388 | 21.8% | $1,784 | $2 | 0.1% | $1,595 | $1,235 | 77.4% | $1,547 | $1,235 | 79.8% | $5,867 | $4,647 | 79.2% | 0.713 | 0.692 | $1,679 | $1,344 | 80.1% | 1.97 | 1.93 | 5.69 | 5.41 |
| GPD | 0.99 | 27,500 | $1,889 | $2,347 | $458 | 24.3% | $1,896 | $7 | 0.4% | $1,822 | $1,384 | 75.9% | $1,763 | $1,384 | 78.5% | $6,780 | $5,254 | 77.5% | 0.751 | 0.730 | $1,862 | $1,467 | 78.8% | 2.10 | 2.05 | 6.56 | 6.25 |
| TGPD | 0.775 | 33,500 | $334 | $393 | $59 | 17.7% | $333 | $0 | -0.1% | $276 | $220 | 80.0% | $269 | $220 | 81.9% | $994 | $814 | 81.9% | 0.685 | 0.661 | $270 | $226 | 83.9% | 2.50 | 2.42 | 10.67 | 9.98 |
| TGPD | 0.8 | 25,000 | $341 | $408 | $67 | 19.6% | $343 | $2 | 0.7% | $301 | $238 | 79.0% | $293 | $238 | 81.0% | $1,039 | $844 | 81.3% | 0.720 | 0.693 | $280 | $232 | 83.1% | 2.89 | 2.71 | 17.85 | 15.44 |
| TGPD | 0.8675 | 50,000 | $1,178 | $1,421 | $243 | 20.6% | $1,174 | -$4 | -0.3% | $1,129 | $877 | 77.7% | $1,102 | $877 | 79.6% | $3,765 | $3,012 | 80.0% | 0.776 | 0.747 | $994 | $806 | 81.1% | 3.69 | 3.56 | 27.76 | 26.34 |
| TGPD | 0.91 | 31,000 | $1,231 | $1,516 | $285 | 23.2% | $1,236 | $6 | 0.5% | $1,193 | $913 | 76.5% | $1,159 | $912 | 78.8% | $4,039 | $3,183 | 78.8% | 0.764 | 0.738 | $1,157 | $927 | 80.1% | 2.45 | 2.36 | 10.07 | 9.29 |
| TGPD | 0.92 | 47,500 | $1,923 | $2,383 | $461 | 24.0% | $1,940 | $17 | 0.9% | $1,969 | $1,504 | 76.4% | $1,915 | $1,504 | 78.6% | $6,545 | $5,202 | 79.5% | 0.803 | 0.776 | $1,733 | $1,377 | 79.5% | 3.12 | 3.03 | 19.22 | 18.38 |
| TGPD | 0.95 | 35,000 | $2,041 | $2,542 | $501 | 24.6% | $2,050 | $9 | 0.4% | $2,066 | $1,562 | 75.6% | $2,004 | $1,562 | 77.9% | $7,470 | $5,817 | 77.9% | 0.789 | 0.762 | $1,911 | $1,507 | 78.8% | 2.34 | 2.26 | 8.34 | 7.85 |

*NOTE: #simulations = 1,000; λ = 75 for 10 years so n ~ 750; α=0.999

**TABLE F7b**

**RCE vs. LDA for Economic Capital Estimation Under iid ($m, λ=75)***

| Severity Dist. | Parm1 | Parm2 | True ECap | Mean MLE ECap | MLE Bias | MLE Bias% | Mean RCE ECap | RCE Bias | RCE Bias% | RMSE MLE ECap | RMSE RCE ECap | RMSE ECap RCE/MLE | StdDev MLE ECap | StdDev RCE ECap | StdDev ECap RCE/MLE | 95%CIs MLE ECap | 95%CIs RCE ECap | 95%CIs ECap RCE/MLE | CV MLE | CV RCE | IQR MLE ECap | IQR RCE ECap | IQR ECap RCE/MLE | Skew MLE ECap | Skew RCE ECap | Kurtosis MLE ECap | Kurtosis RCE ECap |
|---|---|---|---|---|---|---|---|---|---|---|---|---|---|---|---|---|---|---|---|---|---|---|---|---|---|---|---|
| | μ | σ | | | | | | | | | | | | | | | | | | | | | | | | | |
| LogN | 10 | 2 | $179 | $182 | $3 | 1.8% | $178 | -$1 | -0.4% | $44 | $42 | 97.2% | $44 | $42 | 97.5% | $166 | $162 | 97.4% | 0.240 | 0.238 | $57 | $56 | 97.6% | 0.82 | 0.81 | 1.93 | 1.90 |
| LogN | 7.7 | 2.55 | $199 | $206 | $7 | 3.4% | $199 | $0 | -0.1% | $66 | $63 | 95.6% | $65 | $63 | 96.1% | $246 | $236 | 96.0% | 0.317 | 0.315 | $83 | $80 | 96.3% | 1.13 | 1.12 | 3.41 | 3.35 |
| LogN | 10.4 | 2.5 | $2,372 | $2,449 | $77 | 3.2% | $2,370 | -$2 | -0.1% | $763 | $731 | 95.8% | $759 | $731 | 96.2% | $2,869 | $2,757 | 96.1% | 0.310 | 0.308 | $968 | $934 | 96.4% | 1.10 | 1.09 | 3.25 | 3.19 |
| LogN | 9.27 | 2.77 | $2,534 | $2,641 | $107 | 4.2% | $2,537 | $3 | 0.1% | $923 | $876 | 94.9% | $917 | $876 | 95.5% | $3,444 | $3,284 | 95.4% | 0.347 | 0.345 | $1,153 | $1,102 | 95.6% | 1.27 | 1.25 | 4.18 | 4.09 |
| LogN | 10.75 | 2.7 | $8,160 | $8,482 | $322 | 3.9% | $8,165 | $5 | 0.1% | $2,881 | $2,740 | 95.1% | $2,863 | $2,740 | 95.7% | $10,772 | $10,292 | 95.6% | 0.338 | 0.336 | $3,613 | $3,462 | 95.8% | 1.22 | 1.21 | 3.92 | 3.84 |
| LogN | 9.63 | 2.97 | $8,838 | $9,278 | $441 | 5.0% | $8,864 | $26 | 0.3% | $3,508 | $3,303 | 94.2% | $3,480 | $3,303 | 94.9% | $13,001 | $12,320 | 94.8% | 0.375 | 0.373 | $4,320 | $4,105 | 95.0% | 1.40 | 1.38 | 4.99 | 4.88 |
| TLogN | 10.2 | 1.95 | $209 | $218 | $10 | 4.6% | $207 | -$1 | -0.6% | $80 | $73 | 91.8% | $79 | $73 | 92.5% | $306 | $284 | 92.7% | 0.364 | 0.354 | $95 | $89 | 93.4% | 1.28 | 1.22 | 2.78 | 2.55 |
| TLogN | 9 | 2.2 | $228 | $250 | $22 | 9.5% | $228 | $0 | 0.0% | $127 | $108 | 84.9% | $125 | $108 | 86.2% | $460 | $403 | 87.5% | 0.498 | 0.471 | $138 | $121 | 87.7% | 1.75 | 1.61 | 4.77 | 3.81 |
| TLogN | 10.7 | 2.385 | $2,266 | $2,433 | $167 | 7.4% | $2,259 | -$7 | -0.3% | $1,098 | $979 | 89.2% | $1,085 | $979 | 90.2% | $4,365 | $3,960 | 90.7% | 0.446 | 0.433 | $1,233 | $1,121 | 90.9% | 1.39 | 1.35 | 2.58 | 2.40 |
| TLogN | 9.4 | 2.65 | $2,435 | $2,849 | $414 | 17.0% | $2,513 | $77 | 3.2% | $1,781 | $1,440 | 80.9% | $1,733 | $1,438 | 83.0% | $5,938 | $4,994 | 84.1% | 0.608 | 0.572 | $1,827 | $1,577 | 86.3% | 2.30 | 2.11 | 9.09 | 7.76 |
| TLogN | 11 | 2.6 | $7,882 | $8,751 | $869 | 11.0% | $8,024 | $142 | 1.8% | $4,434 | $3,865 | 87.2% | $4,348 | $3,862 | 88.8% | $17,033 | $15,067 | 88.5% | 0.497 | 0.481 | $4,824 | $4,397 | 91.1% | 1.77 | 1.70 | 4.92 | 4.55 |
| TLogN | 10 | 2.8 | $8,072 | $9,372 | $1,300 | 16.1% | $8,266 | $194 | 2.4% | $6,039 | $4,945 | 81.9% | $5,897 | $4,941 | 83.8% | $21,344 | $18,081 | 84.7% | 0.629 | 0.598 | $6,260 | $5,420 | 86.6% | 2.28 | 2.14 | 9.32 | 8.18 |
| | a | b | | | | | | | | | | | | | | | | | | | | | | | | | |
| Logg | 24 | 2.65 | $401 | $423 | $22 | 5.4% | $406 | $5 | 1.2% | $164 | $155 | 94.6% | $163 | $155 | 95.4% | $618 | $590 | 95.5% | 0.384 | 0.381 | $201 | $192 | 95.5% | 1.17 | 1.16 | 2.26 | 2.24 |
| Logg | 33 | 3.3 | $392 | $407 | $16 | 4.0% | $394 | $3 | 0.6% | $139 | $133 | 95.6% | $139 | $133 | 96.2% | $512 | $497 | 97.0% | 0.340 | 0.338 | $178 | $171 | 96.0% | 1.09 | 1.08 | 2.05 | 1.99 |
| Logg | 25 | 2.5 | $2,337 | $2,501 | $164 | 7.0% | $2,389 | $52 | 2.2% | $1,105 | $1,037 | 93.9% | $1,092 | $1,036 | 94.8% | $4,134 | $3,945 | 95.4% | 0.437 | 0.434 | $1,299 | $1,241 | 95.5% | 1.26 | 1.25 | 2.40 | 2.36 |
| Logg | 34.5 | 3.15 | $1,922 | $2,015 | $93 | 4.8% | $1,943 | $22 | 1.1% | $723 | $688 | 95.1% | $717 | $688 | 95.9% | $2,800 | $2,687 | 95.9% | 0.356 | 0.354 | $919 | $879 | 95.6% | 1.03 | 1.02 | 1.99 | 1.96 |
| Logg | 25.25 | 2.45 | $4,198 | $4,467 | $268 | 6.4% | $4,258 | $60 | 1.4% | $2,003 | $1,878 | 93.7% | $1,985 | $1,877 | 94.5% | $7,800 | $7,332 | 94.0% | 0.444 | 0.441 | $2,419 | $2,293 | 94.8% | 1.25 | 1.24 | 2.34 | 2.29 |
| Logg | 34.7 | 3.07 | $3,657 | $3,808 | $152 | 4.1% | $3,667 | $10 | 0.3% | $1,443 | $1,372 | 95.1% | $1,435 | $1,372 | 95.6% | $5,710 | $5,468 | 95.8% | 0.377 | 0.374 | $1,782 | $1,704 | 95.7% | 1.15 | 1.14 | 1.85 | 1.83 |
| TLogg | 23.5 | 2.65 | $555 | $666 | $111 | 20.1% | $547 | -$8 | -1.4% | $558 | $389 | 69.8% | $546 | $389 | 71.3% | $1,827 | $1,392 | 76.2% | 0.820 | 0.711 | $476 | $377 | 79.3% | 3.16 | 2.64 | 14.87 | 10.80 |
| TLogg | 33 | 3.3 | $501 | $553 | $52 | 10.3% | $505 | $4 | 0.8% | $310 | $266 | 85.9% | $306 | $266 | 87.1% | $1,111 | $981 | 88.2% | 0.553 | 0.527 | $318 | $285 | 89.7% | 2.21 | 1.99 | 10.14 | 7.90 |
| TLogg | 24.5 | 2.5 | $2,523 | $3,089 | $566 | 22.4% | $2,632 | $109 | 4.3% | $2,570 | $1,944 | 75.6% | $2,507 | $1,941 | 77.4% | $9,207 | $7,231 | 78.5% | 0.812 | 0.738 | $2,184 | $1,852 | 84.8% | 2.81 | 2.34 | 12.31 | 7.90 |
| TLogg | 34.5 | 3.15 | $2,167 | $2,380 | $213 | 9.8% | $2,210 | $42 | 2.0% | $1,202 | $1,070 | 89.0% | $1,183 | $1,069 | 90.4% | $4,360 | $4,012 | 92.0% | 0.497 | 0.484 | $1,365 | $1,263 | 92.5% | 1.52 | 1.46 | 3.67 | 3.38 |
| TLogg | 24.75 | 2.45 | $4,259 | $5,138 | $879 | 20.6% | $4,444 | $185 | 4.3% | $3,867 | $3,073 | 79.5% | $3,766 | $3,067 | 81.4% | $14,868 | $11,839 | 79.6% | 0.733 | 0.690 | $3,562 | $3,002 | 84.3% | 2.23 | 2.09 | 7.40 | 6.56 |
| TLogg | 34.6 | 3.07 | $3,843 | $4,239 | $396 | 10.3% | $3,948 | $104 | 2.7% | $2,161 | $1,933 | 89.4% | $2,125 | $1,930 | 90.8% | $7,924 | $7,205 | 90.9% | 0.501 | 0.489 | $2,456 | $2,218 | 90.3% | 1.71 | 1.64 | 5.25 | 4.73 |
| | ξ | θ | | | | | | | | | | | | | | | | | | | | | | | | | |
| GPD | 0.8 | 35,000 | $924 | $1,139 | $215 | 23.3% | $927 | $3 | 0.4% | $868 | $664 | 76.4% | $841 | $664 | 78.9% | $3,207 | $2,494 | 77.8% | 0.738 | 0.716 | $920 | $734 | 79.9% | 1.99 | 1.94 | 5.58 | 5.28 |
| GPD | 0.95 | 7,500 | $1,067 | $1,389 | $322 | 30.2% | $1,076 | $10 | 0.9% | $1,207 | $872 | 72.2% | $1,163 | $872 | 75.0% | $4,372 | $3,251 | 74.4% | 0.837 | 0.810 | $1,165 | $896 | 76.9% | 2.30 | 2.23 | 7.80 | 7.34 |
| GPD | 0.875 | 47,500 | $2,897 | $3,665 | $767 | 26.5% | $2,912 | $14 | 0.5% | $2,987 | $2,221 | 74.3% | $2,887 | $2,221 | 76.9% | $10,866 | $8,327 | 76.6% | 0.788 | 0.763 | $3,005 | $2,362 | 78.6% | 2.15 | 2.09 | 6.66 | 6.27 |
| GPD | 0.95 | 25,000 | $3,556 | $4,622 | $1,066 | 30.0% | $3,582 | $26 | 0.7% | $4,015 | $2,902 | 72.3% | $3,870 | $2,902 | 75.0% | $14,574 | $10,836 | 74.4% | 0.837 | 0.810 | $3,870 | $2,985 | 77.2% | 2.31 | 2.24 | 7.87 | 7.40 |
| GPD | 0.925 | 50,000 | $5,358 | $6,898 | $1,539 | 28.7% | $5,391 | $33 | 0.6% | $5,867 | $4,281 | 73.0% | $5,661 | $4,281 | 75.6% | $21,129 | $15,859 | 75.1% | 0.821 | 0.794 | $5,732 | $4,409 | 76.9% | 2.26 | 2.19 | 7.47 | 7.03 |
| GPD | 0.99 | 27,500 | $6,163 | $8,127 | $1,964 | 31.9% | $6,214 | $51 | 0.8% | $7,291 | $5,192 | 71.2% | $7,021 | $5,192 | 73.9% | $26,759 | $19,663 | 73.5% | 0.864 | 0.836 | $6,967 | $5,190 | 74.5% | 2.40 | 2.33 | 8.64 | 8.10 |
| TGPD | 0.775 | 33,500 | $826 | $1,024 | $198 | 23.9% | $830 | $4 | 0.4% | $851 | $645 | 75.8% | $828 | $645 | 78.0% | $2,988 | $2,326 | 77.8% | 0.808 | 0.777 | $780 | $613 | 78.7% | 2.95 | 2.83 | 14.66 | 13.57 |
| TGPD | 0.8 | 25,000 | $872 | $1,103 | $231 | 26.5% | $885 | $13 | 1.5% | $968 | $722 | 74.6% | $940 | $722 | 76.8% | $3,242 | $2,506 | 77.3% | 0.852 | 0.816 | $825 | $655 | 79.4% | 3.62 | 3.34 | 27.96 | 23.58 |
| TGPD | 0.8675 | 50,000 | $3,291 | $4,203 | $912 | 27.7% | $3,302 | $11 | 0.3% | $3,988 | $2,921 | 73.3% | $3,882 | $2,921 | 75.3% | $12,841 | $9,740 | 75.9% | 0.924 | 0.885 | $3,175 | $2,433 | 76.6% | 4.54 | 4.37 | 40.78 | 38.39 |
| TGPD | 0.91 | 31,000 | $3,628 | $4,751 | $1,122 | 30.9% | $3,672 | $44 | 1.2% | $4,379 | $3,144 | 71.8% | $4,233 | $3,144 | 74.3% | $14,211 | $10,598 | 74.6% | 0.891 | 0.856 | $3,889 | $2,951 | 75.9% | 2.91 | 2.77 | 14.08 | 12.78 |
| TGPD | 0.92 | 47,500 | $5,743 | $7,581 | $1,837 | 32.0% | $5,844 | $101 | 1.8% | $7,385 | $5,298 | 71.7% | $7,153 | $5,297 | 74.1% | $23,816 | $17,844 | 74.9% | 0.944 | 0.906 | $5,940 | $4,474 | 75.3% | 3.76 | 3.62 | 27.63 | 26.07 |
| TGPD | 0.95 | 35,000 | $6,330 | $8,404 | $2,073 | 32.8% | $6,406 | $75 | 1.2% | $7,964 | $5,636 | 70.8% | $7,689 | $5,635 | 73.3% | $28,569 | $20,769 | 72.7% | 0.915 | 0.880 | $6,704 | $5,060 | 75.5% | 2.71 | 2.61 | 11.34 | 10.46 |

*NOTE: #simulations = 1,000; λ = 75 for 10 years so n ~ 750; α=0.9997

**TABLE F8a**

**RCE vs. LDA for Regulatory Capital Estimation Under iid ($m, λ=100)***

| Severity Dist. | Parm1 | Parm2 | True RCap | Mean MLE RCap | MLE Bias | MLE Bias% | Mean RCE RCap | RCE Bias | RCE Bias% | RMSE MLE RCap | RMSE RCE RCap | RMSE RCap RCE/MLE | StdDev MLE RCap | StdDev RCE RCap | StdDev RCap RCE/MLE | 95%CIs MLE RCap | 95%CIs RCE RCap | 95%CIs RCap RCE/MLE | CV MLE | CV RCE | IQR MLE RCap | IQR RCE RCap | IQR RCap RCE/MLE | Skew MLE RCap | Skew RCE RCap | Kurtosis MLE RCap | Kurtosis RCE RCap |
|---|---|---|---|---|---|---|---|---|---|---|---|---|---|---|---|---|---|---|---|---|---|---|---|---|---|---|---|
| | **μ** | **σ** | | | | | | | | | | | | | | | | | | | | | | | | | |
| LogN | 10 | 2 | $128 | $130 | **$2** | **1.5%** | $128 | **$0** | **-0.1%** | $25 | $25 | **97.7%** | $25 | $25 | 97.9% | $100 | $98 | **98.2%** | 0.195 | 0.194 | $32 | $32 | 98.4% | 0.75 | 0.74 | 1.07 | 1.07 |
| LogN | 7.7 | 2.55 | $122 | $126 | **$3** | **2.8%** | $123 | **$0** | **0.1%** | $33 | $32 | **96.4%** | $33 | $32 | 96.9% | $130 | $126 | **96.7%** | 0.263 | 0.262 | $41 | $40 | 97.6% | 0.96 | 0.95 | 1.69 | 1.69 |
| LogN | 10.4 | 2.5 | $1,478 | $1,518 | **$40** | **2.7%** | $1,479 | **$1** | **0.1%** | $392 | $378 | **96.5%** | $390 | $378 | 97.0% | $1,534 | $1,485 | **96.8%** | 0.257 | 0.256 | $484 | $473 | 97.7% | 0.94 | 0.93 | 1.63 | 1.62 |
| LogN | 9.27 | 2.77 | $1,483 | $1,534 | **$51** | **3.4%** | $1,487 | **$3** | **0.2%** | $446 | $427 | **95.8%** | $443 | $427 | 96.4% | $1,740 | $1,674 | **96.2%** | 0.289 | 0.287 | $543 | $526 | 97.0% | 1.04 | 1.04 | 1.99 | 1.98 |
| LogN | 10.75 | 2.7 | $4,852 | $5,010 | **$157** | **3.2%** | $4,861 | **$9** | **0.2%** | $1,414 | $1,357 | **96.0%** | $1,405 | $1,357 | 96.6% | $5,522 | $5,322 | **96.4%** | 0.280 | 0.279 | $1,727 | $1,679 | 97.2% | 1.01 | 1.01 | 1.89 | 1.88 |
| LogN | 9.63 | 2.97 | $4,949 | $5,148 | **$200** | **4.0%** | $4,966 | **$18** | **0.4%** | $1,619 | $1,542 | **95.2%** | $1,606 | $1,542 | 96.0% | $6,303 | $6,037 | **95.8%** | 0.312 | 0.310 | $1,953 | $1,883 | 96.4% | 1.12 | 1.12 | 2.28 | 2.27 |
| TLogN | 10.2 | 1.95 | $155 | $159 | **$4** | **2.9%** | $154 | **-$1** | **-0.4%** | $43 | $41 | **94.5%** | $43 | $41 | 95.0% | $165 | $158 | **95.3%** | 0.270 | 0.265 | $53 | $50 | 95.7% | 1.06 | 1.03 | 1.68 | 1.58 |
| TLogN | 9 | 2.2 | $159 | $168 | **$8** | **5.3%** | $158 | **-$1** | **-0.4%** | $63 | $57 | **90.4%** | $63 | $57 | 91.2% | $250 | $226 | **90.5%** | 0.374 | 0.361 | $71 | $66 | 92.1% | 1.27 | 1.23 | 1.82 | 1.67 |
| TLogN | 10.7 | 2.385 | $1,470 | $1,547 | **$77** | **5.2%** | $1,472 | **$2** | **0.2%** | $573 | $529 | **92.4%** | $568 | $529 | 93.3% | $2,204 | $2,053 | **93.1%** | 0.367 | 0.360 | $683 | $645 | 94.3% | 1.28 | 1.25 | 2.19 | 2.10 |
| TLogN | 9.4 | 2.65 | $1,496 | $1,621 | **$125** | **8.4%** | $1,498 | **$2** | **0.1%** | $742 | $652 | **87.8%** | $731 | $652 | 89.1% | $2,809 | $2,538 | **90.3%** | 0.451 | 0.435 | $843 | $763 | 90.5% | 1.62 | 1.54 | 4.21 | 3.78 |
| TLogN | 11 | 2.6 | $4,842 | $5,114 | **$272** | **5.6%** | $4,829 | **-$12** | **-0.3%** | $1,963 | $1,801 | **91.8%** | $1,944 | $1,801 | 92.7% | $7,232 | $6,704 | **92.7%** | 0.380 | 0.373 | $2,431 | $2,250 | 92.6% | 1.07 | 1.05 | 1.27 | 1.21 |
| TLogN | 10 | 2.8 | $4,767 | $5,184 | **$418** | **8.8%** | $4,781 | **$14** | **0.3%** | $2,601 | $2,291 | **88.1%** | $2,567 | $2,291 | 89.2% | $9,703 | $8,673 | **89.4%** | 0.495 | 0.479 | $2,923 | $2,641 | 90.4% | 1.94 | 1.84 | 7.94 | 7.01 |
| | **a** | **b** | | | | | | | | | | | | | | | | | | | | | | | | | |
| Logg | 24 | 2.65 | $224 | $226 | **$3** | **1.1%** | $224 | **$0** | **0.0%** | $68 | $67 | **98.6%** | $68 | $67 | 98.6% | $269 | $265 | **98.6%** | 0.302 | 0.302 | $90 | $88 | 98.2% | 0.80 | 0.80 | 0.62 | 0.62 |
| Logg | 33 | 3.3 | $237 | $238 | **$1** | **0.4%** | $236 | **-$1** | **-0.6%** | $62 | $61 | **98.9%** | $62 | $61 | 98.9% | $241 | $238 | **98.8%** | 0.259 | 0.259 | $81 | $80 | 98.9% | 0.70 | 0.70 | 0.51 | 0.50 |
| Logg | 25 | 2.5 | $1,238 | $1,254 | **$16** | **1.3%** | $1,238 | **$0** | **0.0%** | $401 | $395 | **98.5%** | $401 | $395 | 98.5% | $1,539 | $1,521 | **98.8%** | 0.319 | 0.319 | $518 | $509 | 98.2% | 0.93 | 0.93 | 1.01 | 1.01 |
| Logg | 34.5 | 3.15 | $1,116 | $1,122 | **$7** | **0.6%** | $1,110 | **-$5** | **-0.5%** | $314 | $310 | **98.8%** | $314 | $310 | 98.8% | $1,189 | $1,175 | **98.8%** | 0.280 | 0.279 | $401 | $397 | 98.9% | 0.83 | 0.83 | 1.10 | 1.10 |
| Logg | 25.25 | 2.45 | $2,188 | $2,223 | **$35** | **1.6%** | $2,193 | **$5** | **0.2%** | $749 | $737 | **98.4%** | $748 | $737 | 98.5% | $2,830 | $2,794 | **98.7%** | 0.336 | 0.336 | $945 | $930 | 98.5% | 1.01 | 1.01 | 1.55 | 1.55 |
| Logg | 34.7 | 3.07 | $2,083 | $2,092 | **$9** | **0.4%** | $2,069 | **-$14** | **-0.7%** | $596 | $589 | **98.7%** | $596 | $588 | 98.7% | $2,225 | $2,197 | **98.7%** | 0.285 | 0.284 | $748 | $736 | 98.4% | 0.90 | 0.90 | 1.26 | 1.25 |
| TLogg | 23.5 | 2.65 | $317 | $360 | **$43** | **13.5%** | $324 | **$7** | **2.1%** | $244 | $197 | **80.6%** | $240 | $197 | 81.9% | $798 | $672 | **84.2%** | 0.668 | 0.608 | $219 | $191 | 87.2% | 4.40 | 3.52 | 41.30 | 26.74 |
| TLogg | 33 | 3.3 | $307 | $327 | **$20** | **6.6%** | $310 | **$3** | **0.9%** | $147 | $134 | **91.2%** | $145 | $134 | 92.0% | $577 | $531 | **92.0%** | 0.444 | 0.431 | $162 | $150 | 92.9% | 1.63 | 1.58 | 3.99 | 3.71 |
| TLogg | 24.5 | 2.5 | $1,357 | $1,492 | **$135** | **9.9%** | $1,366 | **$9** | **0.6%** | $872 | $757 | **86.9%** | $862 | $757 | 87.9% | $3,102 | $2,732 | **88.1%** | 0.577 | 0.555 | $907 | $817 | 90.0% | 2.03 | 1.92 | 6.67 | 5.93 |
| TLogg | 34.5 | 3.15 | $1,264 | $1,325 | **$61** | **4.8%** | $1,265 | **$1** | **0.0%** | $529 | $493 | **93.3%** | $525 | $493 | 93.9% | $2,018 | $1,900 | **94.2%** | 0.397 | 0.390 | $637 | $601 | 94.4% | 1.39 | 1.37 | 3.38 | 3.27 |
| TLogg | 24.75 | 2.45 | $2,249 | $2,471 | **$222** | **9.9%** | $2,270 | **$21** | **0.9%** | $1,446 | $1,266 | **87.5%** | $1,429 | $1,266 | 88.6% | $5,816 | $5,118 | **88.0%** | 0.578 | 0.558 | $1,526 | $1,369 | 89.7% | 1.96 | 1.88 | 5.89 | 5.35 |
| TLogg | 34.6 | 3.07 | $2,198 | $2,294 | **$95** | **4.3%** | $2,192 | **-$6** | **-0.3%** | $907 | $849 | **93.6%** | $902 | $849 | 94.1% | $3,487 | $3,273 | **93.9%** | 0.393 | 0.387 | $1,099 | $1,038 | 94.4% | 1.25 | 1.23 | 2.11 | 2.05 |
| | **ξ** | **θ** | | | | | | | | | | | | | | | | | | | | | | | | | |
| GPD | 0.8 | 35,000 | $455 | $519 | **$64** | **14.0%** | $457 | **$2** | **0.3%** | $315 | $266 | **84.2%** | $309 | $266 | 86.0% | $1,132 | $982 | **86.8%** | 0.596 | 0.582 | $336 | $292 | 86.9% | 2.01 | 1.97 | 6.63 | 6.35 |
| GPD | 0.95 | 7,500 | $452 | $534 | **$82** | **18.2%** | $456 | **$4** | **0.8%** | $374 | $304 | **81.2%** | $365 | $304 | 83.2% | $1,264 | $1,064 | **84.1%** | 0.683 | 0.667 | $378 | $316 | 83.7% | 2.33 | 2.28 | 9.05 | 8.69 |
| GPD | 0.875 | 47,500 | $1,322 | $1,536 | **$213** | **16.1%** | $1,331 | **$9** | **0.7%** | $1,006 | $832 | **82.7%** | $983 | $832 | 84.6% | $3,491 | $2,992 | **85.7%** | 0.640 | 0.625 | $1,045 | $895 | 85.6% | 2.15 | 2.10 | 7.68 | 7.34 |
| GPD | 0.95 | 25,000 | $1,507 | $1,784 | **$277** | **18.3%** | $1,521 | **$14** | **0.9%** | $1,250 | $1,015 | **81.2%** | $1,219 | $1,014 | 83.2% | $4,214 | $3,546 | **84.1%** | 0.683 | 0.667 | $1,261 | $1,058 | 83.9% | 2.32 | 2.27 | 8.98 | 8.62 |
| GPD | 0.925 | 50,000 | $2,327 | $2,730 | **$403** | **17.3%** | $2,341 | **$14** | **0.6%** | $1,866 | $1,525 | **81.7%** | $1,822 | $1,525 | 83.7% | $6,360 | $5,351 | **84.1%** | 0.667 | 0.651 | $1,900 | $1,606 | 84.5% | 2.28 | 2.22 | 8.68 | 8.31 |
| GPD | 0.99 | 27,500 | $2,512 | $2,997 | **$484** | **19.3%** | $2,534 | **$22** | **0.9%** | $2,168 | $1,745 | **80.5%** | $2,113 | $1,745 | 82.6% | $7,218 | $6,034 | **83.6%** | 0.705 | 0.689 | $2,153 | $1,786 | 83.0% | 2.43 | 2.38 | 9.94 | 9.59 |
| TGPD | 0.775 | 33,500 | $418 | $479 | **$60** | **14.4%** | $421 | **$2** | **0.6%** | $289 | $242 | **83.7%** | $283 | $242 | 85.6% | $1,019 | $871 | **85.5%** | 0.591 | 0.576 | $312 | $276 | 88.3% | 2.15 | 2.09 | 8.19 | 7.75 |
| TGPD | 0.8 | 25,000 | $430 | $499 | **$69** | **16.0%** | $436 | **$6** | **1.4%** | $297 | $246 | **83.0%** | $289 | $246 | 85.3% | $1,123 | $961 | **85.6%** | 0.579 | 0.565 | $325 | $278 | 85.7% | 1.71 | 1.66 | 4.97 | 4.61 |
| TGPD | 0.8675 | 50,000 | $1,514 | $1,753 | **$239** | **15.8%** | $1,511 | **-$3** | **-0.2%** | $1,098 | $901 | **82.1%** | $1,071 | $901 | 84.1% | $4,063 | $3,418 | **84.1%** | 0.611 | 0.596 | $1,189 | $1,007 | 84.7% | 1.76 | 1.71 | 4.52 | 4.28 |
| TGPD | 0.91 | 31,000 | $1,600 | $1,885 | **$285** | **17.8%** | $1,609 | **$9** | **0.6%** | $1,207 | $978 | **81.0%** | $1,173 | $978 | 83.4% | $4,254 | $3,544 | **83.3%** | 0.622 | 0.608 | $1,361 | $1,138 | 83.6% | 1.65 | 1.61 | 4.76 | 4.46 |
| TGPD | 0.92 | 47,500 | $2,508 | $2,937 | **$429** | **17.1%** | $2,501 | **-$6** | **-0.2%** | $1,982 | $1,604 | **80.9%** | $1,935 | $1,604 | 82.9% | $6,927 | $5,740 | **82.9%** | 0.659 | 0.641 | $2,025 | $1,683 | 83.1% | 2.22 | 2.16 | 8.73 | 8.24 |
| TGPD | 0.95 | 35,000 | $2,684 | $3,196 | **$512** | **19.1%** | $2,703 | **$19** | **0.7%** | $2,240 | $1,796 | **80.2%** | $2,181 | $1,796 | 82.4% | $8,066 | $6,654 | **82.5%** | 0.682 | 0.665 | $2,389 | $1,991 | 83.3% | 1.95 | 1.89 | 5.31 | 5.03 |

*NOTE: #simulations = 1,000; λ = 100 for 10 years so n ~ 1,000; α=0.999

**TABLE F8b**

**RCE vs. LDA for Economic Capital Estimation Under iid ($m, λ=100)***

| Severity Dist. | Parm1 | Parm2 | True ECap | Mean MLE ECap | MLE Bias | MLE Bias% | Mean RCE ECap | RCE Bias | RCE Bias% | RMSE MLE ECap | RMSE RCE ECap | RMSE ECap RCE/MLE | StdDev MLE ECap | StdDev RCE ECap | StdDev ECap RCE/MLE | 95%CIs MLE ECap | 95%CIs RCE ECap | 95%CIs ECap RCE/MLE | CV MLE | CV RCE | IQR MLE ECap | IQR RCE ECap | IQR ECap RCE/MLE | Skew MLE ECap | Skew RCE ECap | Kurtosis MLE ECap | Kurtosis RCE ECap |
|---|---|---|---|---|---|---|---|---|---|---|---|---|---|---|---|---|---|---|---|---|---|---|---|---|---|---|---|
| | μ | σ | | | | | | | | | | | | | | | | | | | | | | | | | |
| LogN | 10 | 2 | $204 | $208 | $4 | 1.8% | $204 | $0 | 0.0% | $44 | $43 | 97.5% | $44 | $43 | 97.8% | $173 | $169 | 97.6% | 0.211 | 0.210 | $56 | $55 | 98.0% | 0.79 | 0.79 | 1.19 | 1.19 |
| LogN | 7.7 | 2.55 | $233 | $241 | $8 | 3.3% | $234 | $1 | 0.3% | $68 | $65 | 96.1% | $67 | $65 | 96.7% | $266 | $257 | 96.5% | 0.280 | 0.279 | $83 | $81 | 97.4% | 1.01 | 1.01 | 1.88 | 1.87 |
| LogN | 10.4 | 2.5 | $2,773 | $2,860 | $86 | 3.1% | $2,781 | $7 | 0.3% | $789 | $759 | 96.2% | $784 | $759 | 96.8% | $3,098 | $2,992 | 96.6% | 0.274 | 0.273 | $967 | $943 | 97.5% | 0.99 | 0.99 | 1.81 | 1.80 |
| LogN | 9.27 | 2.77 | $3,008 | $3,126 | $118 | 3.9% | $3,021 | $13 | 0.4% | $968 | $924 | 95.4% | $961 | $924 | 96.2% | $3,787 | $3,635 | 96.0% | 0.307 | 0.306 | $1,171 | $1,133 | 96.7% | 1.10 | 1.10 | 2.21 | 2.20 |
| LogN | 10.75 | 2.7 | $9,647 | $10,005 | $358 | 3.7% | $9,685 | $38 | 0.4% | $3,011 | $2,880 | 95.6% | $2,990 | $2,880 | 96.3% | $11,789 | $11,334 | 96.1% | 0.299 | 0.297 | $3,655 | $3,543 | 96.9% | 1.07 | 1.07 | 2.10 | 2.09 |
| LogN | 9.63 | 2.97 | $10,612 | $11,100 | $488 | 4.6% | $10,675 | $63 | 0.6% | $3,716 | $3,524 | 94.8% | $3,683 | $3,523 | 95.7% | $14,489 | $13,836 | 95.5% | 0.332 | 0.330 | $4,442 | $4,273 | 96.2% | 1.19 | 1.18 | 2.54 | 2.52 |
| TLogN | 10.2 | 1.95 | $239 | $247 | $9 | 3.7% | $238 | -$1 | -0.2% | $76 | $71 | 93.7% | $75 | $71 | 94.3% | $287 | $272 | 94.5% | 0.305 | 0.299 | $92 | $87 | 94.5% | 1.17 | 1.13 | 2.10 | 1.97 |
| TLogN | 9 | 2.2 | $263 | $281 | $18 | 6.9% | $263 | $0 | -0.1% | $121 | $107 | 88.8% | $120 | $107 | 89.9% | $481 | $429 | 89.3% | 0.425 | 0.409 | $136 | $122 | 90.0% | 1.38 | 1.33 | 2.21 | 2.01 |
| TLogN | 10.7 | 2.385 | $2,631 | $2,799 | $168 | 6.4% | $2,645 | $14 | 0.5% | $1,143 | $1,045 | 91.5% | $1,131 | $1,045 | 92.5% | $4,407 | $4,064 | 92.2% | 0.404 | 0.395 | $1,339 | $1,249 | 93.3% | 1.38 | 1.34 | 2.56 | 2.44 |
| TLogN | 9.4 | 2.65 | $2,861 | $3,156 | $295 | 10.3% | $2,876 | $15 | 0.5% | $1,611 | $1,385 | 86.0% | $1,584 | $1,385 | 87.4% | $6,059 | $5,313 | 87.7% | 0.502 | 0.482 | $1,767 | $1,583 | 89.6% | 1.79 | 1.70 | 5.20 | 4.61 |
| TLogN | 11 | 2.6 | $9,252 | $9,877 | $625 | 6.8% | $9,254 | $1 | 0.0% | $4,136 | $3,752 | 90.7% | $4,088 | $3,752 | 91.8% | $15,245 | $13,984 | 91.7% | 0.414 | 0.406 | $5,101 | $4,697 | 92.1% | 1.15 | 1.12 | 1.52 | 1.44 |
| TLogN | 10 | 2.8 | $9,568 | $10,598 | $1,031 | 10.8% | $9,644 | $77 | 0.8% | $5,901 | $5,092 | 86.3% | $5,810 | $5,091 | 87.6% | $21,708 | $19,140 | 88.2% | 0.548 | 0.528 | $6,394 | $5,671 | 88.7% | 2.20 | 2.06 | 10.32 | 8.94 |
| | a | b | | | | | | | | | | | | | | | | | | | | | | | | | |
| Logg | 24 | 2.65 | $485 | $494 | $8 | 1.7% | $487 | $2 | 0.4% | $164 | $161 | 98.4% | $163 | $161 | 98.5% | $643 | $632 | 98.4% | 0.331 | 0.330 | $210 | $207 | 98.5% | 0.87 | 0.87 | 0.77 | 0.77 |
| Logg | 33 | 3.3 | $464 | $468 | $3 | 0.7% | $463 | -$1 | -0.3% | $133 | $131 | 98.8% | $133 | $131 | 98.8% | $520 | $513 | 98.7% | 0.284 | 0.284 | $172 | $170 | 98.8% | 0.77 | 0.76 | 0.64 | 0.63 |
| Logg | 25 | 2.5 | $2,862 | $2,915 | $53 | 1.9% | $2,875 | $13 | 0.4% | $1,019 | $1,002 | 98.3% | $1,018 | $1,002 | 98.4% | $3,916 | $3,871 | 98.9% | 0.349 | 0.348 | $1,296 | $1,276 | 98.4% | 1.01 | 1.01 | 1.23 | 1.23 |
| Logg | 34.5 | 3.15 | $2,298 | $2,321 | $23 | 1.0% | $2,295 | -$3 | -0.1% | $710 | $701 | 98.7% | $710 | $701 | 98.7% | $2,686 | $2,647 | 98.6% | 0.306 | 0.305 | $896 | $884 | 98.6% | 0.91 | 0.91 | 1.35 | 1.34 |
| Logg | 25.25 | 2.45 | $5,166 | $5,282 | $116 | 2.2% | $5,205 | $40 | 0.8% | $1,943 | $1,909 | 98.2% | $1,940 | $1,908 | 98.4% | $7,331 | $7,209 | 98.3% | 0.367 | 0.367 | $2,420 | $2,397 | 99.1% | 1.10 | 1.10 | 1.89 | 1.88 |
| Logg | 34.7 | 3.07 | $4,393 | $4,430 | $37 | 0.8% | $4,378 | -$15 | -0.3% | $1,379 | $1,360 | 98.6% | $1,379 | $1,360 | 98.7% | $5,159 | $5,082 | 98.5% | 0.311 | 0.311 | $1,725 | $1,693 | 98.1% | 0.98 | 0.97 | 1.45 | 1.44 |
| TLogg | 23.5 | 2.65 | $668 | $792 | $123 | 18.5% | $689 | $21 | 3.1% | $655 | $490 | 74.8% | $643 | $489 | 76.1% | $2,039 | $1,640 | 80.4% | 0.812 | 0.710 | $523 | $446 | 85.3% | 5.72 | 4.15 | 65.49 | 35.78 |
| TLogg | 33 | 3.3 | $593 | $646 | $52 | 8.8% | $604 | $11 | 1.8% | $333 | $298 | 89.5% | $329 | $298 | 90.5% | $1,301 | $1,181 | 90.7% | 0.509 | 0.493 | $354 | $328 | 92.6% | 1.83 | 1.76 | 4.97 | 4.58 |
| TLogg | 24.5 | 2.5 | $3,080 | $3,491 | $411 | 13.3% | $3,129 | $49 | 1.6% | $2,343 | $1,968 | 84.0% | $2,307 | $1,968 | 85.3% | $8,199 | $7,028 | 85.7% | 0.661 | 0.629 | $2,302 | $2,020 | 87.7% | 2.34 | 2.18 | 8.77 | 7.60 |
| TLogg | 34.5 | 3.15 | $2,590 | $2,752 | $162 | 6.3% | $2,605 | $15 | 0.6% | $1,233 | $1,136 | 92.1% | $1,222 | $1,136 | 92.9% | $4,645 | $4,350 | 93.7% | 0.444 | 0.436 | $1,454 | $1,356 | 93.3% | 1.55 | 1.52 | 4.21 | 4.06 |
| TLogg | 24.75 | 2.45 | $5,225 | $5,914 | $689 | 13.2% | $5,327 | $102 | 2.0% | $3,953 | $3,361 | 85.0% | $3,892 | $3,360 | 86.3% | $15,824 | $13,649 | 86.3% | 0.658 | 0.631 | $3,985 | $3,505 | 88.0% | 2.22 | 2.10 | 7.50 | 6.70 |
| TLogg | 34.6 | 3.07 | $4,614 | $4,875 | $261 | 5.7% | $4,622 | $8 | 0.2% | $2,151 | $1,991 | 92.6% | $2,135 | $1,991 | 93.3% | $8,274 | $7,695 | 93.0% | 0.438 | 0.431 | $2,538 | $2,382 | 93.9% | 1.37 | 1.35 | 2.53 | 2.44 |
| | ξ | θ | | | | | | | | | | | | | | | | | | | | | | | | | |
| GPD | 0.8 | 35,000 | $1,164 | $1,380 | $216 | 18.6% | $1,175 | $11 | 1.0% | $978 | $792 | 81.0% | $954 | $792 | 83.0% | $3,443 | $2,892 | 84.0% | 0.691 | 0.674 | $978 | $824 | 84.3% | 2.32 | 2.26 | 8.70 | 8.29 |
| GPD | 0.95 | 7,500 | $1,402 | $1,737 | $335 | 23.9% | $1,425 | $22 | 1.6% | $1,412 | $1,094 | 77.5% | $1,372 | $1,094 | 79.8% | $4,654 | $3,740 | 80.4% | 0.789 | 0.768 | $1,330 | $1,076 | 80.9% | 2.70 | 2.62 | 11.94 | 11.35 |
| GPD | 0.875 | 47,500 | $3,728 | $4,522 | $794 | 21.3% | $3,780 | $52 | 1.4% | $3,440 | $2,725 | 79.2% | $3,347 | $2,725 | 81.4% | $11,683 | $9,629 | 82.4% | 0.740 | 0.721 | $3,384 | $2,784 | 82.3% | 2.49 | 2.42 | 10.11 | 9.61 |
| GPD | 0.95 | 25,000 | $4,674 | $5,800 | $1,125 | 24.1% | $4,756 | $81 | 1.7% | $4,713 | $3,652 | 77.5% | $4,576 | $3,651 | 79.8% | $15,512 | $12,466 | 80.4% | 0.789 | 0.768 | $4,440 | $3,600 | 81.1% | 2.69 | 2.61 | 11.86 | 11.28 |
| GPD | 0.925 | 50,000 | $6,994 | $8,588 | $1,594 | 22.8% | $7,088 | $94 | 1.3% | $6,812 | $5,320 | 78.1% | $6,623 | $5,320 | 80.3% | $22,671 | $18,364 | 81.0% | 0.771 | 0.751 | $6,545 | $5,336 | 81.5% | 2.64 | 2.56 | 11.46 | 10.89 |
| GPD | 0.99 | 27,500 | $8,195 | $10,265 | $2,070 | 25.3% | $8,329 | $134 | 1.6% | $8,610 | $6,596 | 76.6% | $8,357 | $6,594 | 78.9% | $27,783 | $22,204 | 79.9% | 0.814 | 0.792 | $7,991 | $6,364 | 79.6% | 2.82 | 2.74 | 13.17 | 12.52 |
| TGPD | 0.775 | 33,500 | $1,034 | $1,231 | $197 | 19.1% | $1,046 | $12 | 1.2% | $874 | $703 | 80.4% | $851 | $703 | 82.5% | $3,018 | $2,491 | 82.5% | 0.691 | 0.672 | $890 | $745 | 83.7% | 2.52 | 2.44 | 10.98 | 10.33 |
| TGPD | 0.8 | 25,000 | $1,098 | $1,328 | $230 | 20.9% | $1,121 | $23 | 2.1% | $918 | $731 | 79.6% | $889 | $730 | 82.2% | $3,423 | $2,821 | 82.4% | 0.669 | 0.651 | $959 | $795 | 82.9% | 2.01 | 1.93 | 7.04 | 6.46 |
| TGPD | 0.8675 | 50,000 | $4,226 | $5,102 | $876 | 20.7% | $4,236 | $10 | 0.2% | $3,687 | $2,896 | 78.5% | $3,581 | $2,896 | 80.9% | $13,519 | $11,043 | 81.7% | 0.702 | 0.684 | $3,772 | $3,064 | 81.2% | 2.01 | 1.95 | 5.89 | 5.55 |
| TGPD | 0.91 | 31,000 | $4,716 | $5,810 | $1,094 | 23.2% | $4,767 | $51 | 1.1% | $4,275 | $3,304 | 77.3% | $4,132 | $3,304 | 80.0% | $14,834 | $11,848 | 79.9% | 0.711 | 0.693 | $4,582 | $3,694 | 80.6% | 1.94 | 1.87 | 6.81 | 6.31 |
| TGPD | 0.92 | 47,500 | $7,486 | $9,172 | $1,686 | 22.5% | $7,504 | $19 | 0.2% | $7,190 | $5,548 | 77.2% | $6,989 | $5,548 | 79.4% | $24,710 | $19,648 | 79.5% | 0.762 | 0.739 | $6,800 | $5,470 | 80.4% | 2.61 | 2.52 | 12.06 | 11.26 |
| TGPD | 0.95 | 35,000 | $8,322 | $10,408 | $2,086 | 25.1% | $8,441 | $119 | 1.4% | $8,429 | $6,431 | 76.3% | $8,166 | $6,430 | 78.7% | $30,104 | $23,589 | 78.4% | 0.785 | 0.762 | $8,365 | $6,690 | 80.0% | 2.23 | 2.16 | 6.92 | 6.51 |

*NOTE: #simulations = 1,000; λ = 100 for 10 years so n ~ 1,000; α=0.9997

**TABLE F9a**

**RCE vs. LDA for Regulatory Capital Estimation Under 5% Right-Tail Contamination ($m, λ=25)***

| Severity Dist. | Parm1 | Parm2 | True RCap | Mean MLE RCap | MLE Bias | MLE Bias% | Mean RCE RCap | RCE Bias | RCE Bias% | RMSE MLE RCap | RMSE RCE RCap | RMSE RCap RCE/MLE | StdDev MLE RCap | StdDev RCE RCap | StdDev RCap RCE/MLE | 95%CIs MLE RCap | 95%CIs RCE RCap | 95%CIs RCap RCE/MLE | CV MLE | CV RCE | IQR MLE RCap | IQR RCE RCap | IQR RCap RCE/MLE | Skew MLE RCap | Skew RCE RCap | Kurtosis MLE RCap | Kurtosis RCE RCap |
|---|---|---|---|---|---|---|---|---|---|---|---|---|---|---|---|---|---|---|---|---|---|---|---|---|---|---|---|
| | μ | σ | | | | | | | | | | | | | | | | | | | | | | | | | |
| LogN | 10 | 2 | $63 | $91 | $28 | 44.6% | $80 | $17 | 26.7% | $63 | $48 | 76.5% | $56 | $45 | 80.1% | $203 | $169 | 83.3% | 0.620 | 0.567 | $57 | $49 | 85.6% | 2.31 | 2.01 | 8.95 | 6.91 |
| LogN | 7.7 | 2.55 | $53 | $62 | $9 | 17.1% | $56 | $3 | 6.5% | $34 | $29 | 86.9% | $32 | $29 | 89.6% | $122 | $111 | 91.2% | 0.523 | 0.515 | $36 | $32 | 89.8% | 2.08 | 2.04 | 11.87 | 11.41 |
| LogN | 10.4 | 2.5 | $649 | $755 | $106 | 16.4% | $689 | $41 | 6.3% | $400 | $349 | 87.3% | $386 | $347 | 89.9% | $1,452 | $1,330 | 91.6% | 0.511 | 0.503 | $427 | $385 | 90.2% | 2.01 | 1.97 | 11.06 | 10.64 |
| LogN | 9.27 | 2.77 | $603 | $724 | $121 | 20.1% | $648 | $45 | 7.5% | $436 | $371 | 85.2% | $418 | $368 | 88.0% | $1,554 | $1,395 | 89.8% | 0.578 | 0.568 | $446 | $394 | 88.2% | 2.43 | 2.37 | 16.15 | 15.40 |
| LogN | 10.75 | 2.7 | $2,012 | $2,396 | $384 | 19.1% | $2,157 | $145 | 7.2% | $1,397 | $1,198 | 85.8% | $1,343 | $1,189 | 88.5% | $5,008 | $4,518 | 90.2% | 0.560 | 0.551 | $1,447 | $1,285 | 88.7% | 2.32 | 2.26 | 14.66 | 14.01 |
| LogN | 9.63 | 2.97 | $1,893 | $2,330 | $437 | 23.1% | $2,055 | $162 | 8.5% | $1,532 | $1,281 | 83.6% | $1,469 | $1,271 | 86.5% | $5,393 | $4,767 | 88.4% | 0.630 | 0.618 | $1,518 | $1,316 | 86.7% | 2.80 | 2.72 | 21.20 | 20.07 |
| TLogN | 10.2 | 1.95 | $76 | $91 | $15 | 19.2% | $80 | $3 | 4.4% | $58 | $45 | 77.8% | $56 | $45 | 80.1% | $203 | $169 | 83.3% | 0.620 | 0.567 | $57 | $49 | 85.6% | 2.31 | 2.01 | 8.95 | 6.91 |
| TLogN | 9 | 2.2 | $76 | $103 | $27 | 35.9% | $79 | $4 | 4.9% | $102 | $59 | 57.8% | $98 | $59 | 59.9% | $360 | $224 | 62.3% | 0.953 | 0.739 | $68 | $51 | 75.6% | 3.99 | 2.75 | 23.67 | 11.16 |
| TLogN | 10.7 | 2.385 | $670 | $882 | $212 | 31.7% | $728 | $58 | 8.7% | $719 | $513 | 71.4% | $687 | $510 | 74.2% | $2,476 | $1,865 | 75.3% | 0.779 | 0.700 | $616 | $488 | 79.2% | 3.16 | 2.46 | 20.95 | 11.69 |
| TLogN | 9.4 | 2.65 | $643 | $970 | $327 | 50.9% | $675 | $33 | 5.1% | $1,192 | $581 | 48.7% | $1,146 | $580 | 50.6% | $3,759 | $2,123 | 56.5% | 1.182 | 0.859 | $709 | $494 | 69.6% | 5.23 | 2.83 | 45.49 | 12.85 |
| TLogN | 11 | 2.6 | $2,085 | $2,965 | $880 | 42.2% | $2,338 | $253 | 12.1% | $3,229 | $2,085 | 64.6% | $3,107 | $2,069 | 66.6% | $9,520 | $6,968 | 73.2% | 1.048 | 0.885 | $2,076 | $1,614 | 77.8% | 4.91 | 3.75 | 37.57 | 22.17 |
| TLogN | 10 | 2.8 | $1,956 | $2,897 | $941 | 48.1% | $2,058 | $103 | 5.3% | $3,457 | $1,856 | 53.7% | $3,326 | $1,853 | 55.7% | $10,530 | $6,562 | 62.3% | 1.148 | 0.900 | $2,262 | $1,579 | 69.8% | 4.99 | 3.30 | 40.71 | 19.17 |
| | a | b | | | | | | | | | | | | | | | | | | | | | | | | | |
| Logg | 24 | 2.65 | $85 | $102 | $17 | 19.8% | $92 | $6 | 7.6% | $69 | $59 | 85.4% | $67 | $59 | 87.5% | $257 | $227 | 88.2% | 0.658 | 0.641 | $67 | $60 | 88.9% | 2.61 | 2.53 | 11.90 | 11.23 |
| Logg | 33 | 3.3 | $100 | $114 | $14 | 14.2% | $104 | $5 | 4.9% | $62 | $54 | 87.9% | $60 | $54 | 89.9% | $221 | $198 | 89.6% | 0.531 | 0.520 | $64 | $58 | 91.1% | 1.86 | 1.79 | 5.58 | 5.09 |
| Logg | 25 | 2.5 | $444 | $535 | $92 | 20.7% | $474 | $30 | 6.8% | $362 | $303 | 83.8% | $350 | $302 | 86.1% | $1,356 | $1,159 | 85.4% | 0.654 | 0.637 | $346 | $302 | 87.4% | 2.01 | 1.92 | 6.22 | 5.54 |
| Logg | 34.5 | 3.15 | $448 | $516 | $68 | 15.2% | $469 | $21 | 4.7% | $300 | $261 | 87.0% | $292 | $260 | 89.0% | $1,113 | $992 | 89.2% | 0.567 | 0.555 | $313 | $281 | 90.0% | 1.88 | 1.84 | 5.15 | 4.99 |
| Logg | 25.25 | 2.45 | $766 | $933 | $167 | 21.7% | $821 | $55 | 7.2% | $697 | $580 | 83.3% | $677 | $578 | 85.4% | $2,340 | $2,027 | 86.6% | 0.725 | 0.704 | $607 | $525 | 86.4% | 3.11 | 2.98 | 15.62 | 14.47 |
| Logg | 34.7 | 3.07 | $818 | $947 | $129 | 15.8% | $857 | $39 | 4.8% | $572 | $495 | 86.5% | $557 | $493 | 88.5% | $2,107 | $1,862 | 88.4% | 0.588 | 0.575 | $601 | $536 | 89.1% | 2.06 | 2.00 | 7.22 | 6.80 |
| TLogg | 23.5 | 2.65 | $124 | $237 | $113 | 91.5% | $151 | $28 | 22.3% | $906 | $790 | 87.1% | $899 | $789 | 87.8% | $932 | $471 | 50.5% | 3.790 | 5.211 | $173 | $95 | 54.6% | 27.17 | 29.88 | 811.09 | 925.45 |
| TLogg | 33 | 3.3 | $130 | $180 | $51 | 39.3% | $132 | $2 | 1.9% | $202 | $96 | 47.6% | $195 | $96 | 49.2% | $539 | $323 | 60.0% | 1.081 | 0.727 | $134 | $95 | 70.5% | 4.71 | 2.85 | 31.29 | 14.03 |
| TLogg** | 24.5 | 2.5 | $495 | $1,158 | $663 | 133.8% | $629 | $134 | 27.0% | $9,870 | $4,828 | 48.9% | $9,847 | $4,826 | 49.0% | $3,331 | $1,456 | 43.7% | 8.504 | 7.673 | $699 | $366 | 52.3% | 31.14 | 31.34 | 979.37 | 988.09 |
| TLogg | 34.5 | 3.15 | $510 | $671 | $162 | 31.8% | $560 | $50 | 9.8% | $661 | $464 | 70.2% | $641 | $461 | 72.0% | $2,171 | $1,620 | 74.6% | 0.954 | 0.824 | $499 | $414 | 83.0% | 3.86 | 3.03 | 22.38 | 14.12 |
| TLogg## | 24.75 | 2.45 | $801 | $1,547 | $746 | 93.1% | $896 | $95 | 11.8% | $2,251 | $747 | 33.2% | $2,124 | $741 | 34.9% | $6,347 | $2,705 | 42.6% | 1.373 | 0.827 | $1,264 | $748 | 59.1% | 6.69 | 2.82 | 78.91 | 14.85 |
| TLogg | 34.6 | 3.07 | $867 | $1,123 | $256 | 29.6% | $956 | $89 | 10.3% | $1,030 | $777 | 75.4% | $998 | $772 | 77.4% | $3,409 | $2,702 | 79.2% | 0.889 | 0.808 | $799 | $664 | 83.1% | 3.52 | 3.11 | 19.50 | 15.69 |
| | ξ | θ | | | | | | | | | | | | | | | | | | | | | | | | | |
| GPD | 0.8 | 35,000 | $149 | $248 | $99 | 66.5% | $160 | $12 | 7.9% | $323 | $178 | 55.2% | $307 | $178 | 57.8% | $1,025 | $600 | 58.6% | 1.242 | 1.108 | $216 | $139 | 64.7% | 3.87 | 3.40 | 21.19 | 16.68 |
| GPD | 0.95 | 7,500 | $121 | $228 | $107 | 88.8% | $132 | $11 | 9.4% | $353 | $171 | 48.4% | $336 | $170 | 50.7% | $1,059 | $561 | 52.9% | 1.475 | 1.289 | $204 | $121 | 59.2% | 4.73 | 4.05 | 32.21 | 24.20 |
| GPD | 0.875 | 47,500 | $391 | $679 | $287 | 73.5% | $417 | $26 | 6.8% | $937 | $487 | 52.0% | $891 | $487 | 54.6% | $3,052 | $1,726 | 56.6% | 1.314 | 1.166 | $595 | $368 | 61.7% | 4.02 | 3.48 | 22.93 | 17.52 |
| GPD | 0.95 | 25,000 | $403 | $742 | $340 | 84.4% | $432 | $29 | 7.2% | $1,105 | $538 | 48.7% | $1,051 | $538 | 51.2% | $3,521 | $1,868 | 53.1% | 1.416 | 1.245 | $660 | $392 | 59.4% | 4.31 | 3.71 | 26.17 | 19.95 |
| GPD | 0.925 | 50,000 | $643 | $1,149 | $505 | 78.6% | $682 | $38 | 6.0% | $1,666 | $832 | 49.9% | $1,588 | $831 | 52.3% | $5,261 | $2,926 | 55.6% | 1.382 | 1.219 | $1,042 | $633 | 60.7% | 4.29 | 3.69 | 26.10 | 19.77 |
| GPD | 0.99 | 27,500 | $636 | $1,202 | $566 | 89.1% | $678 | $42 | 6.6% | $1,859 | $875 | 47.1% | $1,770 | $874 | 49.4% | $5,455 | $3,006 | 55.1% | 1.473 | 1.289 | $1,100 | $630 | 57.2% | 4.53 | 3.90 | 28.95 | 22.21 |
| TGPD | 0.775 | 33,500 | $141 | $218 | $77 | 54.4% | $147 | $6 | 4.0% | $260 | $153 | 58.8% | $248 | $153 | 61.5% | $912 | $565 | 62.0% | 1.139 | 1.040 | $180 | $119 | 65.9% | 3.24 | 3.04 | 14.73 | 13.25 |
| TGPD | 0.8 | 25,000 | $140 | $222 | $82 | 58.1% | $146 | $6 | 4.2% | $296 | $166 | 56.1% | $284 | $166 | 58.3% | $938 | $562 | 60.0% | 1.281 | 1.132 | $189 | $125 | 66.1% | 5.68 | 4.71 | 56.76 | 38.75 |
| TGPD | 0.8675 | 50,000 | $452 | $766 | $314 | 69.5% | $484 | $32 | 7.0% | $1,006 | $548 | 54.5% | $956 | $547 | 57.3% | $3,231 | $1,959 | 60.6% | 1.247 | 1.131 | $662 | $415 | 62.6% | 3.30 | 2.98 | 14.60 | 11.67 |
| TGPD | 0.91 | 31,000 | $451 | $789 | $338 | 74.9% | $482 | $31 | 6.8% | $1,147 | $599 | 52.3% | $1,096 | $599 | 54.6% | $3,504 | $1,957 | 55.9% | 1.388 | 1.242 | $655 | $402 | 61.4% | 4.22 | 3.81 | 24.90 | 20.60 |
| TGPD | 0.92 | 47,500 | $698 | $1,223 | $526 | 75.4% | $741 | $44 | 6.3% | $1,960 | $1,008 | 51.4% | $1,888 | $1,007 | 53.3% | $5,529 | $3,130 | 56.6% | 1.543 | 1.358 | $993 | $616 | 62.0% | 5.96 | 5.19 | 55.30 | 42.24 |
| TGPD | 0.95 | 35,000 | $717 | $1,364 | $647 | 90.3% | $802 | $85 | 11.8% | $2,479 | $1,219 | 49.2% | $2,393 | $1,216 | 50.8% | $6,397 | $3,594 | 56.2% | 1.754 | 1.516 | $1,019 | $605 | 59.4% | 7.15 | 6.10 | 79.49 | 59.77 |

*NOTE: #simulations = 1,000; λ = 25 for 10 years so n ~ 250; α=0.999

** 15% contamination used instead of 5%.

## 25% contamination used instead of 5%.

**TABLE F9b**

**RCE vs. LDA for Economic Capital Estimation Under 5% Right-Tail Contamination ($m, λ = 25)***

| Severity Dist. | Parm1 | Parm2 | True ECap | Mean MLE ECap | MLE Bias | MLE Bias% | Mean RCE ECap | RCE Bias | RCE Bias% | RMSE MLE ECap | RMSE RCE ECap | RMSE ECap RCE/MLE | StdDev MLE ECap | StdDev RCE ECap | StdDev ECap RCE/MLE | 95%CIs MLE ECap | 95%CIs RCE ECap | 95%CIs ECap RCE/MLE | CV MLE | CV RCE | IQR MLE ECap | IQR RCE ECap | IQR ECap RCE/MLE | Skew MLE ECap | Skew RCE ECap | Kurtosis MLE ECap | Kurtosis RCE ECap |
|---|---|---|---|---|---|---|---|---|---|---|---|---|---|---|---|---|---|---|---|---|---|---|---|---|---|---|---|
| | μ | σ | | | | | | | | | | | | | | | | | | | | | | | | | |
| LogN | 10 | 2 | $107 | $155 | **$48** | **45.2%** | $133 | **$26** | **24.1%** | $120 | $88 | **73.6%** | $110 | $85 | 77.0% | $391 | $316 | **80.9%** | 0.708 | 0.638 | $104 | $87 | 83.3% | 2.65 | 2.25 | 11.58 | 8.56 |
| LogN | 7.7 | 2.55 | $107 | $127 | **$20** | **19.1%** | $115 | **$8** | **7.8%** | $75 | $64 | **86.4%** | $72 | $64 | 89.1% | $268 | $239 | **89.3%** | 0.565 | 0.556 | $78 | $70 | 89.9% | 2.39 | 2.34 | 15.86 | 15.22 |
| LogN | 10.4 | 2.5 | $1,286 | $1,523 | **$237** | **18.4%** | $1,383 | **$97** | **7.5%** | $872 | $757 | **86.8%** | $840 | $751 | 89.4% | $3,146 | $2,819 | **89.6%** | 0.551 | 0.543 | $924 | $833 | 90.2% | 2.30 | 2.26 | 14.72 | 14.15 |
| LogN | 9.27 | 2.77 | $1,293 | $1,584 | **$292** | **22.6%** | $1,410 | **$117** | **9.0%** | $1,033 | $874 | **84.6%** | $991 | $866 | 87.4% | $3,653 | $3,201 | **87.6%** | 0.625 | 0.614 | $1,039 | $920 | 88.5% | 2.83 | 2.75 | 21.87 | 20.83 |
| LogN | 10.75 | 2.7 | $4,230 | $5,137 | **$907** | **21.4%** | $4,595 | **$366** | **8.6%** | $3,241 | $2,760 | **85.2%** | $3,111 | $2,735 | 87.9% | $11,523 | $10,160 | **88.2%** | 0.606 | 0.595 | $3,306 | $2,941 | 88.9% | 2.68 | 2.62 | 19.76 | 18.87 |
| LogN | 9.63 | 2.97 | $4,303 | $5,421 | **$1,118** | **26.0%** | $4,743 | **$440** | **10.2%** | $3,869 | $3,206 | **82.9%** | $3,703 | $3,176 | 85.7% | $13,459 | $11,586 | **86.1%** | 0.683 | 0.670 | $3,737 | $3,254 | 87.1% | 3.29 | 3.19 | 29.05 | 27.46 |
| TLogN | 10.2 | 1.95 | $126 | $155 | **$30** | **23.6%** | $133 | **$7** | **5.6%** | $114 | $85 | **74.6%** | $110 | $85 | 77.0% | $391 | $316 | **80.9%** | 0.708 | 0.638 | $104 | $87 | 83.3% | 2.65 | 2.25 | 11.58 | 8.56 |
| TLogN | 9 | 2.2 | $133 | $195 | **$62** | **46.6%** | $140 | **$7** | **5.5%** | $232 | $116 | **50.1%** | $223 | $116 | 51.9% | $777 | $429 | **55.2%** | 1.145 | 0.826 | $134 | $96 | 71.5% | 4.78 | 2.94 | 34.11 | 12.70 |
| TLogN | 10.7 | 2.385 | $1,267 | $1,753 | **$486** | **38.3%** | $1,397 | **$130** | **10.3%** | $1,629 | $1,096 | **67.3%** | $1,555 | $1,089 | 70.0% | $5,510 | $3,940 | **71.5%** | 0.887 | 0.779 | $1,292 | $1,001 | 77.4% | 3.88 | 2.83 | 32.01 | 15.71 |
| TLogN | 9.4 | 2.65 | $1,297 | $2,153 | **$856** | **66.0%** | $1,369 | **$72** | **5.5%** | $3,199 | $1,303 | **40.7%** | $3,083 | $1,301 | 42.2% | $9,216 | $4,795 | **52.0%** | 1.432 | 0.950 | $1,591 | $1,036 | 65.1% | 6.66 | 3.06 | 72.84 | 14.99 |
| TLogN | 11 | 2.6 | $4,208 | $6,408 | **$2,200** | **52.3%** | $4,822 | **$614** | **14.6%** | $8,217 | $4,863 | **59.2%** | $7,916 | $4,824 | 60.9% | $22,384 | $15,736 | **70.3%** | 1.235 | 1.000 | $4,589 | $3,488 | 76.0% | 5.84 | 4.22 | 51.84 | 27.39 |
| TLogN | 10 | 2.8 | $4,145 | $6,678 | **$2,533** | **61.1%** | $4,391 | **$245** | **5.9%** | $9,445 | $4,384 | **46.4%** | $9,099 | $4,377 | 48.1% | $27,395 | $15,340 | **56.0%** | 1.362 | 0.997 | $5,303 | $3,523 | 66.4% | 6.06 | 3.64 | 57.58 | 23.07 |
| | a | b | | | | | | | | | | | | | | | | | | | | | | | | | |
| Logg | 24 | 2.65 | $192 | $238 | **$46** | **23.9%** | $211 | **$18** | **9.5%** | $182 | $152 | **83.6%** | $176 | $151 | 85.8% | $666 | $573 | **86.1%** | 0.740 | 0.718 | $168 | $146 | 86.9% | 2.95 | 2.85 | 14.86 | 13.93 |
| Logg | 33 | 3.3 | $203 | $238 | **$35** | **17.0%** | $216 | **$12** | **6.2%** | $145 | $125 | **86.5%** | $141 | $125 | 88.6% | $508 | $453 | **89.1%** | 0.593 | 0.579 | $142 | $127 | 89.3% | 2.10 | 2.01 | 7.19 | 6.52 |
| Logg | 25 | 2.5 | $1,064 | $1,329 | **$266** | **25.0%** | $1,154 | **$91** | **8.5%** | $1,005 | $821 | **81.7%** | $969 | $816 | 84.2% | $3,706 | $3,130 | **84.5%** | 0.729 | 0.707 | $915 | $791 | 86.5% | 2.27 | 2.15 | 7.97 | 7.03 |
| Logg | 34.5 | 3.15 | $960 | $1,135 | **$175** | **18.2%** | $1,017 | **$57** | **6.0%** | $735 | $629 | **85.5%** | $714 | $626 | 87.6% | $2,697 | $2,386 | **88.5%** | 0.630 | 0.615 | $738 | $647 | 87.7% | 2.06 | 2.01 | 6.18 | 5.94 |
| Logg | 25.25 | 2.45 | $1,877 | $2,372 | **$495** | **26.3%** | $2,046 | **$169** | **9.0%** | $2,004 | $1,626 | **81.1%** | $1,942 | $1,617 | 83.3% | $6,583 | $5,586 | **84.9%** | 0.819 | 0.790 | $1,635 | $1,387 | 84.8% | 3.52 | 3.36 | 19.42 | 17.86 |
| Logg | 34.7 | 3.07 | $1,794 | $2,135 | **$341** | **19.0%** | $1,905 | **$110** | **6.2%** | $1,439 | $1,221 | **84.8%** | $1,398 | $1,216 | 86.9% | $5,220 | $4,546 | **87.1%** | 0.655 | 0.638 | $1,421 | $1,273 | 89.6% | 2.31 | 2.24 | 9.01 | 8.40 |
| TLogg | 23.5 | 2.65 | $271 | $703 | **$432** | **159.7%** | **$394** | **$124** | **45.7%** | $5,152 | $4,436 | **86.1%** | $5,133 | $4,434 | 86.4% | $2,834 | $1,065 | **37.6%** | 7.302 | 11.241 | $444 | $190 | 42.8% | 30.11 | 31.23 | 935.40 | 983.28 |
| TLogg | 33 | 3.3 | $261 | $400 | **$139** | **53.2%** | **$248** | **-$13** | **-5.0%** | $554 | $189 | **34.1%** | $536 | $188 | 35.1% | $1,337 | $630 | **47.1%** | 1.340 | 0.759 | $302 | $186 | 61.6% | 5.57 | 3.07 | 42.29 | 16.94 |
| TLogg** | 24.5 | 2.5 | $1,164 | $4,748 | **$3,584** | **307.9%** | $1,978 | **$814** | **70.0%** | $78,844 | $33,228 | **42.1%** | $78,762 | $33,218 | 42.2% | $10,042 | $2,896 | **28.8%** | 16.587 | 16.790 | $1,912 | $711 | 37.2% | 31.54 | 31.60 | 996.57 | 998.89 |
| TLogg | 34.5 | 3.15 | $1,086 | $1,537 | **$451** | **41.5%** | **$1,205** | **$119** | **10.9%** | $1,796 | $1,105 | **61.5%** | $1,739 | $1,099 | 63.2% | $5,537 | $3,848 | **69.5%** | 1.131 | 0.912 | $1,194 | $937 | 78.5% | 4.52 | 3.21 | 29.62 | 15.57 |
| TLogg## | 24.75 | 2.45 | $1,928 | $4,297 | **$2,369** | **122.9%** | $1,863 | **-$65** | **-3.4%** | $7,864 | $1,579 | **20.1%** | $7,499 | $1,578 | 21.0% | $19,566 | $5,575 | **28.5%** | 1.745 | 0.847 | $3,550 | $1,562 | 44.0% | 9.02 | 3.07 | 132.89 | 18.36 |
| TLogg | 34.6 | 3.07 | $1,892 | $2,607 | **$715** | **37.8%** | **$2,121** | **$229** | **12.1%** | $2,793 | $1,942 | **69.5%** | $2,700 | $1,928 | 71.4% | $9,079 | $6,624 | **73.0%** | 1.036 | 0.909 | $1,936 | $1,561 | 80.6% | 4.05 | 3.44 | 25.19 | 18.98 |
| | ξ | θ | | | | | | | | | | | | | | | | | | | | | | | | | |
| GPD | 0.8 | 35,000 | $382 | $745 | **$362** | **94.7%** | $421 | **$38** | **10.0%** | $1,189 | $559 | **47.0%** | $1,133 | $558 | 49.3% | $3,487 | $1,792 | **51.4%** | 1.522 | 1.327 | $653 | $376 | 57.5% | 4.63 | 3.99 | 29.42 | 22.45 |
| GPD | 0.95 | 7,500 | $375 | $855 | **$480** | **127.8%** | $419 | **$43** | **11.5%** | $1,627 | $650 | **40.0%** | $1,555 | $649 | 41.7% | $4,591 | $2,027 | **44.1%** | 1.818 | 1.549 | $734 | $383 | 52.2% | 5.64 | 4.78 | 43.61 | 32.59 |
| GPD | 0.875 | 47,500 | $1,106 | $2,268 | **$1,162** | **105.0%** | $1,199 | **$92** | **8.4%** | $3,836 | $1,675 | **43.7%** | $3,655 | $1,672 | 45.8% | $11,760 | $5,646 | **48.0%** | 1.612 | 1.395 | $1,965 | $1,076 | 54.8% | 4.88 | 4.16 | 32.90 | 24.65 |
| GPD | 0.95 | 25,000 | $1,251 | $2,767 | **$1,516** | **121.2%** | $1,361 | **$110** | **8.8%** | $5,045 | $2,033 | **40.3%** | $4,812 | $2,030 | 42.2% | $14,885 | $6,672 | **44.8%** | 1.739 | 1.492 | $2,385 | $1,242 | 52.1% | 5.20 | 4.43 | 36.82 | 27.73 |
| GPD | 0.925 | 50,000 | $1,938 | $4,121 | **$2,183** | **112.7%** | $2,080 | **$142** | **7.3%** | $7,342 | $3,043 | **41.4%** | $7,010 | $3,039 | 43.4% | $21,907 | $10,180 | **46.5%** | 1.701 | 1.461 | $3,650 | $1,930 | 52.9% | 5.20 | 4.42 | 36.98 | 27.76 |
| GPD | 0.99 | 27,500 | $2,076 | $4,743 | **$2,667** | **128.5%** | $2,243 | **$167** | **8.0%** | $8,997 | $3,470 | **38.6%** | $8,592 | $3,466 | 40.3% | $24,631 | $11,151 | **45.3%** | 1.811 | 1.545 | $4,195 | $2,119 | 50.5% | 5.45 | 4.64 | 40.30 | 30.48 |
| TGPD | 0.775 | 33,500 | $351 | $623 | **$272** | **77.3%** | $370 | **$19** | **5.3%** | $900 | $459 | **51.0%** | $858 | $459 | 53.5% | $3,044 | $1,623 | **53.3%** | 1.378 | 1.241 | $524 | $315 | 60.1% | 3.83 | 3.56 | 20.66 | 18.32 |
| TGPD | 0.8 | 25,000 | $361 | $662 | **$301** | **83.4%** | $381 | **$20** | **5.5%** | $1,113 | $528 | **47.4%** | $1,072 | $528 | 49.2% | $3,265 | $1,682 | **51.5%** | 1.620 | 1.386 | $573 | $336 | 58.5% | 7.48 | 5.90 | 93.28 | 58.49 |
| TGPD | 0.8675 | 50,000 | $1,267 | $2,517 | **$1,250** | **98.7%** | $1,376 | **$110** | **8.7%** | $3,962 | $1,837 | **46.4%** | $3,759 | $1,834 | 48.8% | $12,050 | $6,352 | **52.7%** | 1.494 | 1.332 | $2,143 | $1,201 | 56.0% | 3.86 | 3.42 | 20.09 | 15.44 |
| TGPD | 0.91 | 31,000 | $1,334 | $2,766 | **$1,433** | **107.4%** | $1,445 | **$111** | **8.3%** | $4,904 | $2,154 | **43.9%** | $4,690 | $2,152 | 45.9% | $13,991 | $6,589 | **47.1%** | 1.695 | 1.489 | $2,259 | $1,224 | 54.2% | 4.99 | 4.49 | 33.95 | 28.04 |
| TGPD | 0.92 | 47,500 | $2,088 | $4,377 | **$2,289** | **109.6%** | $2,264 | **$176** | **8.4%** | $8,814 | $3,785 | **42.9%** | $8,511 | $3,781 | 44.4% | $22,898 | $10,671 | **46.6%** | 1.944 | 1.670 | $3,446 | $1,887 | 54.8% | 7.34 | 6.27 | 81.48 | 59.87 |
| TGPD | 0.95 | 35,000 | $2,227 | $5,195 | **$2,967** | **133.2%** | $2,574 | **$347** | **15.6%** | $12,102 | $4,880 | **40.3%** | $11,733 | $4,868 | 41.5% | $27,632 | $13,122 | **47.5%** | 2.258 | 1.891 | $3,692 | $1,894 | 51.3% | 8.99 | 7.58 | 119.45 | 88.69 |

*NOTE: #simulations = 1,000; λ = 25 for 10 years so n ~ 250; α=0.9997

** 15% contamination used instead of 5%.

## 25% contamination used instead of 5%.

**TABLE F10a**

**RCE vs. LDA for Regulatory Capital Estimation Under 5% Left-Tail Contamination ($m, λ=25)***

| Severity Dist. | Parm1 | Parm2 | True RCap | Mean MLE RCap | MLE Bias | MLE Bias% | Mean RCE RCap | RCE Bias | RCE Bias% | RMSE MLE RCap | RMSE RCE RCap | RMSE RCap RCE/MLE | StdDev MLE RCap | StdDev RCE RCap | StdDev RCap RCE/MLE | 95%CIs MLE RCap | 95%CIs RCE RCap | 95%CIs RCap RCE/MLE | CV MLE | CV RCE | IQR MLE RCap | IQR RCE RCap | IQR RCap RCE/MLE | Skew MLE RCap | Skew RCE RCap | Kurtosis MLE RCap | Kurtosis RCE RCap |
|---|---|---|---|---|---|---|---|---|---|---|---|---|---|---|---|---|---|---|---|---|---|---|---|---|---|---|---|
| | μ | σ | | | | | | | | | | | | | | | | | | | | | | | | | |
| LogN | 10 | 2 | $63 | $64 | $1 | 2.1% | $60 | -$2 | -3.8% | $25 | $23 | 93.5% | $25 | $23 | 93.2% | $93 | $87 | 93.7% | 0.388 | 0.384 | $30 | $28 | 93.3% | 1.46 | 1.44 | 6.13 | 5.98 |
| LogN | 7.7 | 2.55 | $53 | $56 | $3 | 5.5% | $51 | -$2 | -3.9% | $29 | $26 | 89.6% | $29 | $26 | 89.7% | $106 | $96 | 90.2% | 0.521 | 0.513 | $33 | $30 | 90.5% | 2.23 | 2.18 | 14.22 | 13.63 |
| LogN | 10.4 | 2.5 | $649 | $682 | $33 | 5.1% | $623 | -$25 | -3.9% | $348 | $313 | 89.9% | $347 | $312 | 90.0% | $1,266 | $1,147 | 90.6% | 0.509 | 0.501 | $391 | $355 | 90.8% | 2.15 | 2.10 | 13.21 | 12.69 |
| LogN | 9.27 | 2.77 | $603 | $646 | $43 | 7.2% | $579 | -$23 | -3.9% | $375 | $329 | 87.8% | $372 | $328 | 88.2% | $1,337 | $1,186 | 88.7% | 0.577 | 0.567 | $403 | $359 | 89.1% | 2.63 | 2.56 | 19.50 | 18.56 |
| LogN | 10.75 | 2.7 | $2,012 | $2,145 | $133 | 6.6% | $1,933 | -$78 | -3.9% | $1,206 | $1,065 | 88.4% | $1,198 | $1,062 | 88.7% | $4,322 | $3,855 | 89.2% | 0.559 | 0.549 | $1,312 | $1,174 | 89.5% | 2.49 | 2.43 | 17.65 | 16.84 |
| LogN | 9.63 | 2.97 | $1,893 | $2,062 | $169 | 8.9% | $1,821 | -$72 | -3.8% | $1,309 | $1,127 | 86.1% | $1,298 | $1,125 | 86.7% | $4,592 | $4,009 | 87.3% | 0.630 | 0.618 | $1,358 | $1,192 | 87.8% | 3.05 | 2.96 | 25.77 | 24.35 |
| TLogN | 10.2 | 1.95 | $76 | $85 | $9 | 11.6% | $75 | -$2 | -2.0% | $53 | $42 | 79.4% | $52 | $42 | 80.5% | $189 | $153 | 80.8% | 0.612 | 0.561 | $53 | $45 | 84.5% | 2.62 | 2.28 | 14.74 | 11.21 |
| TLogN | 9 | 2.2 | $76 | $98 | $22 | 28.8% | $74 | -$1 | -1.9% | $101 | $53 | 52.4% | $99 | $53 | 53.6% | $276 | $178 | 64.7% | 1.015 | 0.715 | $69 | $51 | 73.9% | 5.32 | 2.87 | 43.35 | 14.59 |
| TLogN | 10.7 | 2.385 | $670 | $826 | $157 | 23.4% | $680 | $10 | 1.5% | $738 | $524 | 71.0% | $721 | $524 | 72.6% | $2,304 | $1,788 | 77.6% | 0.873 | 0.771 | $580 | $460 | 79.3% | 3.89 | 3.12 | 25.93 | 16.51 |
| TLogN | 9.4 | 2.65 | $643 | $857 | $214 | 33.4% | $612 | -$31 | -4.8% | $926 | $520 | 56.2% | $900 | $519 | 57.6% | $3,166 | $1,899 | 60.0% | 1.051 | 0.848 | $645 | $452 | 70.1% | 3.89 | 3.00 | 25.33 | 15.87 |
| TLogN | 11 | 2.6 | $2,085 | $2,737 | $652 | 31.3% | $2,154 | $69 | 3.3% | $3,254 | $2,040 | 62.7% | $3,188 | $2,039 | 64.0% | $9,160 | $6,699 | 73.1% | 1.165 | 0.947 | $2,051 | $1,597 | 77.9% | 7.47 | 5.09 | 92.10 | 45.06 |
| TLogN | 10 | 2.8 | $1,956 | $2,652 | $697 | 35.6% | $1,889 | -$67 | -3.4% | $3,317 | $1,831 | 55.2% | $3,242 | $1,830 | 56.4% | $9,669 | $6,401 | 66.2% | 1.222 | 0.969 | $1,980 | $1,425 | 72.0% | 5.70 | 3.96 | 52.65 | 27.05 |
| | a | b | | | | | | | | | | | | | | | | | | | | | | | | | |
| Logg | 24 | 2.65 | $85 | $97 | $11 | 13.0% | $87 | $1 | 1.5% | $65 | $56 | 86.1% | $64 | $56 | 87.3% | $226 | $200 | 88.4% | 0.664 | 0.645 | $59 | $52 | 89.0% | 2.86 | 2.75 | 13.53 | 12.63 |
| Logg | 33 | 3.3 | $100 | $109 | $9 | 9.2% | $100 | $0 | 0.3% | $57 | $51 | 88.8% | $56 | $51 | 90.0% | $216 | $193 | 89.3% | 0.519 | 0.509 | $59 | $54 | 90.8% | 1.65 | 1.60 | 3.92 | 3.70 |
| Logg | 25 | 2.5 | $444 | $505 | $62 | 13.9% | $447 | $4 | 0.9% | $336 | $285 | 84.8% | $330 | $285 | 86.2% | $1,261 | $1,092 | 86.6% | 0.653 | 0.636 | $323 | $284 | 87.8% | 1.99 | 1.91 | 5.85 | 5.28 |
| Logg | 34.5 | 3.15 | $448 | $493 | $45 | 10.0% | $448 | $0 | 0.0% | $291 | $257 | 88.2% | $288 | $257 | 89.2% | $1,057 | $942 | 89.1% | 0.584 | 0.573 | $296 | $265 | 89.7% | 2.42 | 2.40 | 11.32 | 11.29 |
| Logg | 25.25 | 2.45 | $766 | $876 | $109 | 14.3% | $771 | $5 | 0.7% | $630 | $531 | 84.3% | $620 | $531 | 85.6% | $2,250 | $1,961 | 87.2% | 0.709 | 0.689 | $570 | $496 | 87.1% | 3.02 | 2.92 | 15.40 | 14.53 |
| Logg | 34.7 | 3.07 | $818 | $902 | $85 | 10.3% | $817 | -$1 | -0.1% | $541 | $473 | 87.5% | $535 | $473 | 88.5% | $2,049 | $1,803 | 88.0% | 0.592 | 0.579 | $560 | $492 | 87.9% | 2.11 | 2.07 | 7.12 | 6.84 |
| TLogg | 23.5 | 2.65 | $124 | $222 | $98 | 79.5% | $142 | $18 | 14.7% | $591 | $490 | 83.0% | $582 | $490 | 84.2% | $959 | $474 | 49.4% | 2.619 | 3.450 | $143 | $83 | 58.3% | 17.83 | 23.52 | 411.12 | 641.75 |
| TLogg | 33 | 3.3 | $130 | $168 | $39 | 29.9% | $125 | -$5 | -3.8% | $180 | $93 | 51.6% | $176 | $93 | 52.8% | $519 | $304 | 58.5% | 1.044 | 0.744 | $132 | $91 | 68.9% | 5.87 | 3.92 | 64.96 | 36.53 |
| TLogg | 24.5 | 2.5 | $495 | $806 | $311 | 62.8% | $425 | -$70 | -14.2% | $1,380 | $447 | 32.4% | $1,344 | $441 | 32.8% | $3,194 | $1,238 | 38.8% | 1.667 | 1.038 | $636 | $319 | 50.2% | 8.06 | 5.89 | 96.75 | 60.93 |
| TLogg | 34.5 | 3.15 | $510 | $630 | $120 | 23.6% | $525 | $15 | 3.0% | $641 | $446 | 69.5% | $630 | $446 | 70.7% | $1,972 | $1,516 | 76.9% | 1.000 | 0.849 | $455 | $375 | 82.3% | 5.95 | 5.06 | 59.69 | 49.38 |
| TLogg## | 24.75 | 2.45 | $801 | $1,244 | $443 | 55.3% | $684 | -$117 | -14.6% | $2,250 | $653 | 29.0% | $2,206 | $643 | 29.1% | $5,169 | $1,858 | 36.0% | 1.774 | 0.939 | $964 | $503 | 52.2% | 12.16 | 5.60 | 227.64 | 58.78 |
| TLogg | 34.6 | 3.07 | $867 | $1,074 | $208 | 24.0% | $912 | $45 | 5.2% | $920 | $687 | 74.7% | $896 | $685 | 76.5% | $3,141 | $2,470 | 78.6% | 0.834 | 0.751 | $824 | $678 | 82.3% | 2.94 | 2.47 | 13.34 | 9.91 |
| | ξ | θ | | | | | | | | | | | | | | | | | | | | | | | | | |
| GPD | 0.8 | 35,000 | $149 | $216 | $67 | 45.0% | $141 | -$8 | -5.2% | $275 | $156 | 56.6% | $267 | $156 | 58.3% | $886 | $545 | 61.5% | 1.237 | 1.103 | $177 | $113 | 63.6% | 3.92 | 3.47 | 22.03 | 17.52 |
| GPD | 0.95 | 7,500 | $121 | $195 | $75 | 61.7% | $114 | -$6 | -5.3% | $296 | $147 | 49.7% | $287 | $147 | 51.3% | $891 | $489 | 54.8% | 1.468 | 1.286 | $159 | $95 | 59.6% | 4.67 | 4.04 | 30.71 | 23.59 |
| GPD | 0.875 | 47,500 | $391 | $590 | $199 | 51.0% | $367 | -$24 | -6.2% | $804 | $430 | 53.5% | $779 | $429 | 55.2% | $2,605 | $1,496 | 57.4% | 1.319 | 1.171 | $494 | $303 | 61.4% | 4.14 | 3.62 | 24.70 | 19.15 |
| GPD | 0.95 | 25,000 | $403 | $650 | $247 | 61.5% | $381 | -$22 | -5.4% | $987 | $491 | 49.7% | $955 | $490 | 51.3% | $2,972 | $1,628 | 54.8% | 1.469 | 1.287 | $533 | $317 | 59.5% | 4.68 | 4.05 | 30.79 | 23.68 |
| GPD | 0.925 | 50,000 | $643 | $993 | $350 | 54.4% | $595 | -$48 | -7.5% | $1,422 | $730 | 51.3% | $1,378 | $728 | 52.8% | $4,498 | $2,519 | 56.0% | 1.387 | 1.223 | $842 | $506 | 60.1% | 4.41 | 3.81 | 27.97 | 21.39 |
| GPD | 0.99 | 27,500 | $636 | $1,036 | $401 | 63.0% | $590 | -$45 | -7.1% | $1,580 | $764 | 48.4% | $1,529 | $763 | 49.9% | $4,825 | $2,569 | 53.3% | 1.475 | 1.292 | $857 | $493 | 57.5% | 4.62 | 4.00 | 30.51 | 23.59 |
| TGPD | 0.775 | 33,500 | $141 | $206 | $65 | 45.8% | $138 | -$3 | -2.0% | $272 | $159 | 58.4% | $264 | $159 | 60.2% | $894 | $570 | 63.8% | 1.285 | 1.149 | $173 | $115 | 66.6% | 4.35 | 3.78 | 29.45 | 22.10 |
| TGPD** | 0.8 | 25,000 | $140 | $189 | $49 | 34.9% | $126 | -$14 | -10.1% | $236 | $140 | 59.3% | $231 | $139 | 60.3% | $742 | $459 | 61.8% | 1.222 | 1.105 | $155 | $101 | 65.2% | 4.06 | 3.77 | 23.69 | 20.84 |
| TGPD | 0.8675 | 50,000 | $452 | $676 | $224 | 49.6% | $429 | -$23 | -5.1% | $1,014 | $558 | 55.0% | $989 | $557 | 56.4% | $3,353 | $1,938 | 57.8% | 1.462 | 1.299 | $537 | $337 | 62.8% | 5.69 | 5.00 | 51.06 | 39.99 |
| TGPD | 0.91 | 31,000 | $451 | $678 | $227 | 50.3% | $419 | -$32 | -7.1% | $936 | $502 | 53.7% | $908 | $501 | 55.2% | $3,022 | $1,785 | 59.1% | 1.338 | 1.196 | $548 | $338 | 61.6% | 3.97 | 3.62 | 20.76 | 17.57 |
| TGPD | 0.92 | 47,500 | $698 | $1,129 | $432 | 61.9% | $686 | -$12 | -1.7% | $1,943 | $1,003 | 51.6% | $1,894 | $1,003 | 53.0% | $5,193 | $2,987 | 57.5% | 1.677 | 1.462 | $913 | $555 | 60.8% | 7.77 | 6.69 | 101.10 | 78.70 |
| TGPD | 0.95 | 35,000 | $717 | $1,128 | $411 | 57.3% | $675 | -$42 | -5.9% | $1,795 | $916 | 51.1% | $1,747 | $915 | 52.4% | $5,160 | $2,905 | 56.3% | 1.549 | 1.357 | $919 | $545 | 59.3% | 5.98 | 5.12 | 55.57 | 41.61 |

*NOTE: #simulations = 1,000; λ = 25 for 10 years so n ~ 250; α=0.999

** 10% contamination instead of 5%.

## 25% contamination used instead of 5%.

**TABLE F10b**

**RCE vs. LDA for Economic Capital Estimation Under 5% Left-Tail Contamination ($m, λ = 25)***

| Severity Dist. | Parm1 | Parm2 | True ECap | Mean MLE ECap | MLE Bias | MLE Bias% | Mean RCE ECap | RCE Bias | RCE Bias% | RMSE MLE ECap | RMSE RCE ECap | RMSE ECap RCE/MLE | StdDev MLE ECap | StdDev RCE ECap | StdDev ECap RCE/MLE | 95%CIs MLE ECap | 95%CIs RCE ECap | 95%CIs ECap RCE/MLE | CV MLE | CV RCE | IQR MLE ECap | IQR RCE ECap | IQR ECap RCE/MLE | Skew MLE ECap | Skew RCE ECap | Kurtosis MLE ECap | Kurtosis RCE ECap |
|---|---|---|---|---|---|---|---|---|---|---|---|---|---|---|---|---|---|---|---|---|---|---|---|---|---|---|---|
| | μ | σ | | | | | | | | | | | | | | | | | | | | | | | | | |
| LogN | 10 | 2 | $107 | $110 | $3 | 2.8% | $103 | -$4 | -3.5% | $46 | $43 | 93.0% | $46 | $43 | 92.8% | $171 | $158 | 92.3% | 0.420 | 0.415 | $55 | $51 | 92.6% | 1.65 | 1.63 | 7.97 | 7.78 |
| LogN | 7.7 | 2.55 | $107 | $114 | $7 | 6.7% | $103 | -$4 | -3.3% | $64 | $57 | 88.8% | $64 | $57 | 89.2% | $230 | $204 | 88.4% | 0.563 | 0.554 | $70 | $62 | 88.3% | 2.58 | 2.52 | 19.12 | 18.32 |
| LogN | 10.4 | 2.5 | $1,286 | $1,367 | $81 | 6.3% | $1,243 | -$43 | -3.3% | $755 | $674 | 89.2% | $751 | $672 | 89.5% | $2,711 | $2,406 | 88.7% | 0.550 | 0.541 | $833 | $739 | 88.7% | 2.48 | 2.43 | 17.71 | 17.00 |
| LogN | 9.27 | 2.77 | $1,293 | $1,405 | $112 | 8.7% | $1,252 | -$41 | -3.1% | $885 | $769 | 86.9% | $878 | $768 | 87.5% | $3,103 | $2,689 | 86.7% | 0.625 | 0.613 | $929 | $806 | 86.8% | 3.08 | 2.99 | 26.58 | 25.27 |
| LogN | 10.75 | 2.7 | $4,230 | $4,570 | $340 | 8.0% | $4,094 | -$135 | -3.2% | $2,784 | $2,436 | 87.5% | $2,763 | $2,433 | 88.0% | $9,824 | $8,569 | 87.2% | 0.605 | 0.594 | $2,963 | $2,586 | 87.3% | 2.91 | 2.84 | 23.96 | 22.83 |
| LogN | 9.63 | 2.97 | $4,303 | $4,765 | $462 | 10.7% | $4,177 | -$126 | -2.9% | $3,290 | $2,800 | 85.1% | $3,257 | $2,797 | 85.9% | $11,312 | $9,624 | 85.1% | 0.684 | 0.670 | $3,314 | $2,827 | 85.3% | 3.62 | 3.50 | 35.53 | 33.52 |
| TLogN | 10.2 | 1.95 | $126 | $145 | $19 | 15.3% | $124 | -$2 | -1.3% | $103 | $79 | 76.1% | $101 | $79 | 77.4% | $358 | $286 | 79.8% | 0.701 | 0.634 | $97 | $80 | 82.9% | 3.08 | 2.61 | 20.27 | 14.76 |
| TLogN | 9 | 2.2 | $133 | $185 | $52 | 39.3% | $130 | -$3 | -2.2% | $239 | $102 | 42.5% | $233 | $102 | 43.5% | $599 | $348 | 58.1% | 1.259 | 0.780 | $139 | $97 | 69.6% | 6.44 | 2.94 | 59.90 | 15.85 |
| TLogN | 10.7 | 2.385 | $1,267 | $1,644 | $377 | 29.7% | $1,303 | $36 | 2.9% | $1,702 | $1,131 | 66.5% | $1,659 | $1,131 | 68.1% | $5,182 | $3,732 | 72.0% | 1.009 | 0.867 | $1,213 | $930 | 76.7% | 4.64 | 3.54 | 36.27 | 20.95 |
| TLogN | 9.4 | 2.65 | $1,297 | $1,869 | $572 | 44.1% | $1,232 | -$65 | -5.0% | $2,354 | $1,158 | 49.2% | $2,283 | $1,156 | 50.6% | $7,926 | $4,178 | 52.7% | 1.222 | 0.939 | $1,442 | $959 | 66.5% | 4.46 | 3.29 | 32.62 | 19.13 |
| TLogN | 11 | 2.6 | $4,208 | $5,910 | $1,702 | 40.4% | $4,433 | $224 | 5.3% | $8,521 | $4,806 | 56.4% | $8,349 | $4,800 | 57.5% | $20,903 | $15,066 | 72.1% | 1.413 | 1.083 | $4,522 | $3,443 | 76.1% | 9.22 | 5.89 | 132.02 | 58.19 |
| TLogN | 10 | 2.8 | $4,145 | $6,099 | $1,954 | 47.1% | $4,021 | -$124 | -3.0% | $9,060 | $4,343 | 47.9% | $8,846 | $4,342 | 49.1% | $24,136 | $14,824 | 61.4% | 1.450 | 1.080 | $4,625 | $3,102 | 67.1% | 6.78 | 4.38 | 70.89 | 32.61 |
| | a | b | | | | | | | | | | | | | | | | | | | | | | | | | |
| Logg | 24 | 2.65 | $192 | $224 | $32 | 16.5% | $198 | $6 | 2.9% | $171 | $144 | 84.1% | $168 | $144 | 85.6% | $569 | $495 | 87.0% | 0.750 | 0.726 | $144 | $128 | 88.9% | 3.22 | 3.10 | 16.65 | 15.51 |
| Logg | 33 | 3.3 | $203 | $227 | $24 | 11.7% | $206 | $3 | 1.3% | $133 | $116 | 87.3% | $131 | $116 | 88.7% | $494 | $442 | 89.4% | 0.578 | 0.565 | $132 | $119 | 90.5% | 1.83 | 1.77 | 4.88 | 4.57 |
| Logg | 25 | 2.5 | $1,064 | $1,252 | $188 | 17.7% | $1,088 | $24 | 2.3% | $931 | $768 | 82.6% | $911 | $768 | 84.3% | $3,405 | $2,898 | 85.1% | 0.728 | 0.706 | $852 | $741 | 87.0% | 2.23 | 2.13 | 7.44 | 6.65 |
| Logg | 34.5 | 3.15 | $960 | $1,082 | $122 | 12.7% | $971 | $11 | 1.1% | $716 | $620 | 86.5% | $706 | $620 | 87.8% | $2,570 | $2,248 | 87.5% | 0.652 | 0.638 | $694 | $610 | 87.8% | 2.72 | 2.68 | 14.04 | 13.89 |
| Logg | 25.25 | 2.45 | $1,877 | $2,218 | $341 | 18.2% | $1,916 | $39 | 2.1% | $1,802 | $1,479 | 82.1% | $1,770 | $1,479 | 83.6% | $6,253 | $5,258 | 84.1% | 0.798 | 0.772 | $1,519 | $1,295 | 85.3% | 3.43 | 3.30 | 19.41 | 18.17 |
| Logg | 34.7 | 3.07 | $1,794 | $2,031 | $237 | 13.2% | $1,812 | $18 | 1.0% | $1,360 | $1,166 | 85.7% | $1,339 | $1,165 | 87.0% | $5,075 | $4,428 | 87.2% | 0.659 | 0.643 | $1,319 | $1,173 | 88.9% | 2.34 | 2.28 | 8.65 | 8.23 |
| TLogg | 23.5 | 2.65 | $271 | $630 | $359 | 132.8% | $344 | $73 | 27.1% | $2,787 | $2,374 | 85.2% | $2,764 | $2,373 | 85.9% | $2,925 | $1,142 | 39.0% | 4.387 | 6.897 | $360 | $171 | 47.7% | 22.70 | 27.83 | 599.55 | 828.72 |
| TLogg | 33 | 3.3 | $261 | $371 | $110 | 41.9% | $234 | -$27 | -10.5% | $497 | $187 | 37.7% | $484 | $185 | 38.3% | $1,327 | $579 | 43.7% | 1.307 | 0.793 | $296 | $177 | 59.9% | 7.68 | 4.67 | 104.18 | 52.17 |
| TLogg | 24.5 | 2.5 | $1,164 | $2,232 | $1,068 | 91.8% | $824 | -$340 | -29.2% | $4,992 | $1,085 | 21.7% | $4,877 | $1,030 | 21.1% | $9,990 | $2,694 | 27.0% | 2.185 | 1.250 | $1,677 | $618 | 36.8% | 9.71 | 9.50 | 129.78 | 154.87 |
| TLogg | 34.5 | 3.15 | $1,086 | $1,436 | $350 | 32.2% | $1,124 | $37 | 3.4% | $1,776 | $1,073 | 60.4% | $1,741 | $1,072 | 61.6% | $5,098 | $3,455 | 67.8% | 1.212 | 0.954 | $1,097 | $846 | 77.1% | 6.99 | 5.76 | 76.69 | 62.48 |
| TLogg## | 24.75 | 2.45 | $1,928 | $3,467 | $1,539 | 79.8% | $1,381 | -$547 | -28.4% | $8,408 | $1,462 | 17.4% | $8,266 | $1,356 | 16.4% | $15,646 | $4,054 | 25.9% | 2.384 | 0.982 | $2,651 | $1,017 | 38.4% | 15.16 | 5.53 | 323.75 | 53.12 |
| TLogg | 34.6 | 3.07 | $1,892 | $2,488 | $596 | 31.5% | $2,016 | $124 | 6.6% | $2,483 | $1,690 | 68.1% | $2,410 | $1,685 | 69.9% | $8,178 | $6,061 | 74.1% | 0.969 | 0.836 | $1,977 | $1,572 | 79.5% | 3.46 | 2.73 | 17.98 | 12.13 |
| | ξ | θ | | | | | | | | | | | | | | | | | | | | | | | | | |
| GPD | 0.8 | 35,000 | $382 | $637 | $254 | 66.5% | $364 | -$19 | -4.9% | $1,001 | $483 | 48.2% | $968 | $482 | 49.8% | $3,064 | $1,646 | 53.7% | 1.520 | 1.326 | $536 | $304 | 56.7% | 4.73 | 4.10 | 31.39 | 24.13 |
| GPD | 0.95 | 7,500 | $375 | $718 | $343 | 91.3% | $356 | -$19 | -5.1% | $1,346 | $552 | 41.0% | $1,301 | $552 | 42.4% | $3,716 | $1,718 | 46.2% | 1.812 | 1.549 | $577 | $296 | 51.4% | 5.58 | 4.78 | 41.94 | 32.11 |
| GPD | 0.875 | 47,500 | $1,106 | $1,939 | $833 | 75.3% | $1,037 | -$69 | -6.3% | $3,256 | $1,460 | 44.8% | $3,147 | $1,458 | 46.3% | $9,942 | $4,819 | 48.5% | 1.623 | 1.406 | $1,626 | $867 | 53.3% | 5.05 | 4.33 | 35.94 | 27.15 |
| GPD | 0.95 | 25,000 | $1,251 | $2,390 | $1,139 | 91.0% | $1,186 | -$66 | -5.2% | $4,483 | $1,839 | 41.0% | $4,336 | $1,838 | 42.4% | $12,387 | $5,726 | 46.2% | 1.815 | 1.550 | $1,924 | $988 | 51.3% | 5.58 | 4.79 | 42.02 | 32.20 |
| GPD | 0.925 | 50,000 | $1,938 | $3,500 | $1,563 | 80.7% | $1,787 | -$150 | -7.8% | $6,195 | $2,636 | 42.6% | $5,994 | $2,632 | 43.9% | $18,064 | $8,640 | 47.8% | 1.712 | 1.473 | $2,948 | $1,533 | 52.0% | 5.37 | 4.59 | 40.26 | 30.46 |
| GPD | 0.99 | 27,500 | $2,076 | $4,014 | $1,938 | 93.4% | $1,923 | -$153 | -7.4% | $7,557 | $2,993 | 39.6% | $7,304 | $2,989 | 40.9% | $21,356 | $9,449 | 44.2% | 1.820 | 1.554 | $3,265 | $1,623 | 49.7% | 5.60 | 4.78 | 43.16 | 32.90 |
| TGPD | 0.775 | 33,500 | $351 | $591 | $240 | 68.2% | $349 | -$2 | -0.6% | $971 | $486 | 50.0% | $941 | $486 | 51.6% | $3,024 | $1,629 | 53.9% | 1.592 | 1.391 | $497 | $298 | 60.0% | 5.39 | 4.53 | 43.82 | 31.38 |
| TGPD** | 0.8 | 25,000 | $361 | $553 | $192 | 53.2% | $323 | -$38 | -10.5% | $855 | $434 | 50.8% | $833 | $433 | 51.9% | $2,475 | $1,336 | 54.0% | 1.507 | 1.339 | $461 | $270 | 58.7% | 4.86 | 4.45 | 33.19 | 28.30 |
| TGPD | 0.8675 | 50,000 | $1,267 | $2,207 | $940 | 74.2% | $1,211 | -$55 | -4.4% | $4,158 | $1,932 | 46.5% | $4,050 | $1,931 | 47.7% | $12,655 | $6,284 | 49.7% | 1.835 | 1.594 | $1,730 | $955 | 55.2% | 6.99 | 6.09 | 74.85 | 58.29 |
| TGPD | 0.91 | 31,000 | $1,334 | $2,322 | $988 | 74.1% | $1,231 | -$103 | -7.7% | $3,918 | $1,763 | 45.0% | $3,791 | $1,760 | 46.4% | $11,971 | $6,065 | 50.7% | 1.633 | 1.430 | $1,901 | $1,023 | 53.8% | 4.62 | 4.16 | 27.53 | 22.73 |
| TGPD | 0.92 | 47,500 | $2,088 | $4,035 | $1,947 | 93.2% | $2,091 | $3 | 0.1% | $8,964 | $3,843 | 42.9% | $8,750 | $3,843 | 43.9% | $20,662 | $10,129 | 49.0% | 2.169 | 1.838 | $3,159 | $1,688 | 53.4% | 10.01 | 8.64 | 156.71 | 124.03 |
| TGPD | 0.95 | 35,000 | $2,227 | $4,133 | $1,906 | 85.5% | $2,101 | -$126 | -5.7% | $8,286 | $3,492 | 42.1% | $8,064 | $3,490 | 43.3% | $21,228 | $10,201 | 48.1% | 1.951 | 1.661 | $3,294 | $1,706 | 51.8% | 7.38 | 6.22 | 81.17 | 59.19 |

*NOTE: #simulations = 1,000; λ = 25 for 10 years so n ~ 250; α=0.9997

** 10% contamination instead of 5%.

## 25% contamination used instead of 5%.

**TABLE F11a**

**RCE vs. LDA for Regulatory Capital Estimation Under 5% Right, 5%Left-Tail Contamination ($m, λ=25)***

| Severity Dist. | Parm1 | Parm2 | True RCap | Mean MLE RCap | MLE Bias | MLE Bias% | Mean RCE RCap | RCE Bias | RCE Bias% | RMSE MLE RCap | RMSE RCE RCap | RMSE RCap RCE/MLE | StdDev MLE RCap | StdDev RCE RCap | StdDev RCap RCE/MLE | 95%CIs MLE RCap | 95%CIs RCE RCap | 95%CIs RCap RCE/MLE | CV MLE | CV RCE | IQR MLE RCap | IQR RCE RCap | IQR RCap RCE/MLE | Skew MLE RCap | Skew RCE RCap | Kurtosis MLE RCap | Kurtosis RCE RCap |
|---|---|---|---|---|---|---|---|---|---|---|---|---|---|---|---|---|---|---|---|---|---|---|---|---|---|---|---|
| | **μ** | **σ** | | | | | | | | | | | | | | | | | | | | | | | | | |
| LogN | 10 | 2 | $63 | $67 | **$4** | **6.7%** | $63 | **$0** | **0.6%** | $26 | $24 | **91.9%** | $26 | $24 | 93.1% | $99 | $93 | **93.6%** | 0.390 | 0.385 | $31 | $29 | 92.3% | 1.46 | 1.44 | 6.08 | 5.93 |
| LogN | 7.7 | 2.55 | $53 | $59 | **$6** | **11.7%** | $54 | **$1** | **1.7%** | $32 | $28 | **87.9%** | $31 | $28 | 89.6% | $114 | $103 | **90.0%** | 0.523 | 0.515 | $35 | $31 | 89.5% | 2.22 | 2.18 | 14.12 | 13.53 |
| LogN | 10.4 | 2.5 | $649 | $721 | **$73** | **11.2%** | $659 | **$10** | **1.6%** | $375 | $331 | **88.3%** | $368 | $331 | 90.0% | $1,365 | $1,232 | **90.3%** | 0.510 | 0.503 | $416 | $374 | 89.8% | 2.14 | 2.10 | 13.12 | 12.60 |
| LogN | 9.27 | 2.77 | $603 | $688 | **$85** | **14.1%** | $616 | **$14** | **2.3%** | $407 | $351 | **86.2%** | $398 | $350 | 88.1% | $1,452 | $1,285 | **88.5%** | 0.579 | 0.569 | $431 | $381 | 88.4% | 2.62 | 2.55 | 19.37 | 18.42 |
| LogN | 10.75 | 2.7 | $2,012 | $2,280 | **$269** | **13.3%** | $2,054 | **$42** | **2.1%** | $1,306 | $1,133 | **86.8%** | $1,278 | $1,133 | 88.6% | $4,685 | $4,168 | **88.9%** | 0.561 | 0.551 | $1,400 | $1,242 | 88.8% | 2.49 | 2.43 | 17.53 | 16.72 |
| LogN | 9.63 | 2.97 | $1,893 | $2,207 | **$313** | **16.5%** | $1,947 | **$54** | **2.8%** | $1,428 | $1,208 | **84.6%** | $1,394 | $1,207 | 86.6% | $5,017 | $4,369 | **87.1%** | 0.632 | 0.620 | $1,456 | $1,269 | 87.2% | 3.04 | 2.95 | 25.61 | 24.18 |
| TLogN | 10.2 | 1.95 | $76 | $89 | **$13** | **17.2%** | $78 | **$2** | **2.7%** | $60 | $47 | **77.8%** | $59 | $47 | 79.6% | $226 | $178 | **78.7%** | 0.656 | 0.596 | $55 | $47 | 84.7% | 2.83 | 2.51 | 13.49 | 10.61 |
| TLogN | 9 | 2.2 | $76 | $102 | **$26** | **34.7%** | $78 | **$2** | **2.9%** | $104 | $57 | **55.1%** | $101 | $57 | 56.9% | $320 | $206 | **64.3%** | 0.986 | 0.734 | $72 | $54 | 74.9% | 4.64 | 2.71 | 37.03 | 12.00 |
| TLogN | 10.7 | 2.385 | $670 | $863 | **$193** | **28.9%** | $709 | **$39** | **5.8%** | $802 | $557 | **69.5%** | $778 | $555 | 71.4% | $2,463 | $1,880 | **76.3%** | 0.901 | 0.784 | $587 | $472 | 80.4% | 4.86 | 3.68 | 43.70 | 24.26 |
| TLogN | 9.4 | 2.65 | $643 | $959 | **$316** | **49.2%** | $652 | **$9** | **1.4%** | $1,454 | $605 | **41.6%** | $1,419 | $605 | 42.6% | $3,583 | $2,140 | **59.7%** | 1.480 | 0.928 | $668 | $481 | 72.1% | 8.37 | 3.75 | 102.69 | 22.80 |
| TLogN | 11 | 2.6 | $2,085 | $2,818 | **$733** | **35.1%** | $2,231 | **$146** | **7.0%** | $3,125 | $2,041 | **65.3%** | $3,037 | $2,036 | 67.0% | $8,555 | $6,329 | **74.0%** | 1.078 | 0.912 | $1,948 | $1,542 | 79.1% | 6.34 | 4.91 | 66.54 | 42.98 |
| TLogN | 10 | 2.8 | $1,956 | $2,844 | **$888** | **45.4%** | $2,008 | **$53** | **2.7%** | $3,429 | $1,866 | **54.4%** | $3,312 | $1,866 | 56.3% | $11,047 | $6,805 | **61.6%** | 1.165 | 0.929 | $2,233 | $1,578 | 70.7% | 4.38 | 2.98 | 30.22 | 12.87 |
| | **a** | **b** | | | | | | | | | | | | | | | | | | | | | | | | | |
| Logg | 24 | 2.65 | $85 | $100 | **$15** | **17.6%** | $90 | **$5** | **5.6%** | $70 | $60 | **85.6%** | $69 | $60 | 87.4% | $248 | $218 | **87.9%** | 0.683 | 0.664 | $65 | $57 | 88.1% | 2.94 | 2.84 | 14.19 | 13.33 |
| Logg | 33 | 3.3 | $100 | $112 | **$12** | **12.3%** | $103 | **$3** | **3.1%** | $59 | $52 | **88.2%** | $58 | $52 | 90.0% | $219 | $196 | **89.4%** | 0.521 | 0.510 | $62 | $57 | 91.8% | 1.66 | 1.61 | 4.00 | 3.79 |
| Logg | 25 | 2.5 | $444 | $524 | **$80** | **18.1%** | $464 | **$20** | **4.6%** | $356 | $299 | **84.1%** | $347 | $299 | 86.2% | $1,326 | $1,139 | **85.9%** | 0.661 | 0.644 | $334 | $297 | 89.1% | 2.05 | 1.97 | 6.21 | 5.61 |
| Logg | 34.5 | 3.15 | $448 | $506 | **$58** | **13.0%** | $460 | **$12** | **2.7%** | $299 | $262 | **87.6%** | $293 | $261 | 89.2% | $1,118 | $984 | **88.0%** | 0.579 | 0.568 | $310 | $270 | 87.2% | 2.28 | 2.26 | 10.08 | 10.08 |
| Logg | 25.25 | 2.45 | $766 | $916 | **$149** | **19.5%** | $806 | **$40** | **5.2%** | $704 | $589 | **83.6%** | $688 | $587 | 85.4% | $2,332 | $2,017 | **86.5%** | 0.751 | 0.729 | $613 | $540 | 88.1% | 3.57 | 3.43 | 21.15 | 19.67 |
| Logg | 34.7 | 3.07 | $818 | $928 | **$110** | **13.4%** | $840 | **$22** | **2.6%** | $559 | $485 | **86.9%** | $548 | $485 | 88.5% | $2,068 | $1,816 | **87.8%** | 0.590 | 0.578 | $549 | $494 | 89.9% | 2.18 | 2.13 | 7.60 | 7.27 |
| TLogg | 23.5 | 2.65 | $124 | $230 | **$106** | **85.5%** | $145 | **$21** | **17.0%** | $543 | $445 | **81.9%** | $533 | $444 | 83.4% | $938 | $530 | **56.5%** | 2.319 | 3.067 | $156 | $86 | 55.5% | 16.99 | 23.30 | 400.47 | 645.03 |
| TLogg | 33 | 3.3 | $130 | $178 | **$49** | **37.6%** | $130 | **$1** | **0.7%** | $244 | $118 | **48.5%** | $239 | $118 | 49.5% | $573 | $320 | **55.8%** | 1.338 | 0.905 | $121 | $90 | 74.4% | 9.61 | 6.94 | 142.23 | 87.47 |
| TLogg | 24.5 | 2.5 | $495 | $872 | **$377** | **76.2%** | $457 | **-$38** | **-7.6%** | $1,581 | $537 | **34.0%** | $1,536 | $536 | 34.9% | $3,573 | $1,485 | **41.6%** | 1.760 | 1.171 | $692 | $328 | 47.4% | 7.61 | 6.54 | 80.18 | 66.51 |
| TLogg | 34.5 | 3.15 | $510 | $658 | **$148** | **29.0%** | $552 | **$42** | **8.3%** | $613 | $454 | **74.2%** | $595 | $452 | 76.1% | $2,053 | $1,589 | **77.4%** | 0.904 | 0.820 | $518 | $423 | 81.8% | 3.53 | 3.20 | 21.50 | 18.18 |
| TLogg | 24.75 | 2.45 | $801 | $1,336 | **$535** | **66.8%** | $752 | **-$49** | **-6.2%** | $2,541 | $938 | **36.9%** | $2,484 | $936 | 37.7% | $5,528 | $2,139 | **38.7%** | 1.859 | 1.246 | $1,012 | $542 | 53.6% | 11.34 | 11.46 | 187.34 | 210.41 |
| TLogg | 34.6 | 3.07 | $867 | $1,097 | **$231** | **26.6%** | $936 | **$70** | **8.0%** | $906 | $696 | **76.9%** | $876 | $693 | 79.1% | $3,270 | $2,611 | **79.9%** | 0.798 | 0.740 | $813 | $676 | 83.1% | 2.41 | 2.23 | 8.50 | 7.56 |
| | **ξ** | **θ** | | | | | | | | | | | | | | | | | | | | | | | | | |
| GPD | 0.8 | 35,000 | $149 | $235 | **$87** | **58.3%** | $153 | **$4** | **3.0%** | $303 | $169 | **55.8%** | $291 | $169 | 58.2% | $999 | $606 | **60.7%** | 1.235 | 1.105 | $199 | $127 | 64.0% | 3.78 | 3.36 | 20.19 | 16.18 |
| GPD | 0.95 | 7,500 | $121 | $215 | **$94** | **78.1%** | $125 | **$4** | **3.7%** | $331 | $162 | **49.0%** | $317 | $162 | 51.1% | $1,005 | $545 | **54.2%** | 1.473 | 1.293 | $181 | $104 | 57.6% | 4.69 | 4.05 | 31.66 | 23.99 |
| GPD | 0.875 | 47,500 | $391 | $643 | **$252** | **64.5%** | $397 | **$6** | **1.6%** | $880 | $464 | **52.7%** | $843 | $464 | 55.0% | $2,918 | $1,594 | **54.6%** | 1.310 | 1.167 | $542 | $334 | 61.6% | 3.96 | 3.48 | 22.28 | 17.37 |
| GPD | 0.95 | 25,000 | $403 | $712 | **$309** | **76.8%** | $414 | **$12** | **2.9%** | $1,096 | $536 | **48.9%** | $1,051 | $536 | 51.0% | $3,350 | $1,806 | **53.9%** | 1.476 | 1.294 | $603 | $347 | 57.6% | 4.74 | 4.10 | 32.37 | 24.63 |
| GPD | 0.925 | 50,000 | $643 | $1,082 | **$439** | **68.2%** | $645 | **$2** | **0.2%** | $1,554 | $786 | **50.6%** | $1,491 | $786 | 52.7% | $4,946 | $2,709 | **54.8%** | 1.377 | 1.219 | $933 | $552 | 59.1% | 4.26 | 3.72 | 25.82 | 19.97 |
| GPD | 0.99 | 27,500 | $636 | $1,133 | **$497** | **78.3%** | $642 | **$6** | **1.0%** | $1,731 | $825 | **47.7%** | $1,657 | $825 | 49.8% | $5,462 | $2,788 | **51.0%** | 1.463 | 1.286 | $992 | $556 | 56.0% | 4.49 | 3.91 | 28.47 | 22.24 |
| TGPD | 0.775 | 33,500 | $141 | $219 | **$78** | **55.3%** | $147 | **$5** | **3.9%** | $286 | $165 | **57.6%** | $276 | $165 | 59.8% | $880 | $585 | **66.4%** | 1.258 | 1.125 | $179 | $118 | 65.9% | 4.09 | 3.70 | 24.33 | 20.59 |
| TGPD | 0.8 | 25,000 | $140 | $221 | **$80** | **57.2%** | $145 | **$4** | **3.2%** | $307 | $172 | **56.2%** | $296 | $172 | 58.2% | $934 | $557 | **59.7%** | 1.342 | 1.189 | $178 | $115 | 64.8% | 4.50 | 4.01 | 28.12 | 22.80 |
| TGPD | 0.8675 | 50,000 | $452 | $770 | **$318** | **70.3%** | $482 | **$30** | **6.6%** | $1,341 | $698 | **52.0%** | $1,302 | $697 | 53.5% | $3,510 | $2,061 | **58.7%** | 1.692 | 1.447 | $628 | $392 | 62.5% | 10.81 | 8.52 | 199.20 | 134.44 |
| TGPD | 0.91 | 31,000 | $451 | $759 | **$307** | **68.1%** | $464 | **$12** | **2.7%** | $1,126 | $581 | **51.6%** | $1,083 | $581 | 53.7% | $3,188 | $1,862 | **58.4%** | 1.428 | 1.254 | $672 | $410 | 61.1% | 5.23 | 4.38 | 42.92 | 30.28 |
| TGPD | 0.92 | 47,500 | $698 | $1,138 | **$441** | **63.1%** | $698 | **$0** | **0.0%** | $1,565 | $835 | **53.3%** | $1,501 | $835 | 55.6% | $4,788 | $2,799 | **58.5%** | 1.319 | 1.196 | $963 | $583 | 60.5% | 3.88 | 3.55 | 22.39 | 18.89 |
| TGPD | 0.95 | 35,000 | $717 | $1,268 | **$551** | **76.9%** | $753 | **$36** | **5.0%** | $1,938 | $986 | **50.9%** | $1,858 | $986 | 53.0% | $6,149 | $3,326 | **54.1%** | 1.465 | 1.309 | $1,035 | $614 | 59.4% | 4.57 | 4.04 | 31.88 | 24.64 |

*NOTE: #simulations = 1,000; λ = 25 for 10 years so n ~ 250; α=0.999

**TABLE F11b**

**RCE vs. LDA for Economic Capital Estimation Under 5% Right, 5%Left-Tail Contamination ($m, λ=25)***

| Severity Dist. | Parm1 | Parm2 | True ECap | Mean MLE ECap | MLE Bias | MLE Bias% | Mean RCE ECap | RCE Bias | RCE Bias% | RMSE MLE ECap | RMSE RCE ECap | RMSE ECap RCE/MLE | StdDev MLE ECap | StdDev RCE ECap | StdDev ECap RCE/MLE | 95%CIs MLE ECap | 95%CIs RCE ECap | 95%CIs ECap RCE/MLE | CV MLE | CV RCE | IQR MLE ECap | IQR RCE ECap | IQR ECap RCE/MLE | Skew MLE ECap | Skew RCE ECap | Kurtosis MLE ECap | Kurtosis RCE ECap |
|---|---|---|---|---|---|---|---|---|---|---|---|---|---|---|---|---|---|---|---|---|---|---|---|---|---|---|---|
| | μ | σ | | | | | | | | | | | | | | | | | | | | | | | | | |
| LogN | 10 | 2 | $107 | $115 | $8 | 7.8% | $108 | $1 | 1.2% | $49 | $45 | 91.5% | $49 | $45 | 92.8% | $181 | $168 | 92.6% | 0.422 | 0.417 | $57 | $54 | 93.5% | 1.65 | 1.62 | 7.91 | 7.72 |
| LogN | 7.7 | 2.55 | $107 | $121 | $14 | 13.4% | $109 | $3 | 2.7% | $70 | $61 | 87.3% | $68 | $61 | 89.1% | $247 | $220 | 89.2% | 0.565 | 0.556 | $74 | $67 | 90.5% | 2.58 | 2.52 | 19.00 | 18.19 |
| LogN | 10.4 | 2.5 | $1,286 | $1,451 | $165 | 12.8% | $1,319 | $33 | 2.6% | $817 | $717 | 87.7% | $800 | $716 | 89.5% | $2,902 | $2,598 | 89.5% | 0.551 | 0.543 | $876 | $796 | 90.8% | 2.47 | 2.42 | 17.60 | 16.88 |
| LogN | 9.27 | 2.77 | $1,293 | $1,502 | $209 | 16.2% | $1,337 | $44 | 3.4% | $964 | $824 | 85.5% | $941 | $823 | 87.4% | $3,343 | $2,930 | 87.7% | 0.627 | 0.615 | $981 | $874 | 89.1% | 3.07 | 2.99 | 26.42 | 25.10 |
| LogN | 10.75 | 2.7 | $4,230 | $4,876 | $646 | 15.3% | $4,365 | $135 | 3.2% | $3,027 | $2,605 | 86.1% | $2,958 | $2,602 | 88.0% | $10,567 | $9,315 | 88.2% | 0.607 | 0.596 | $3,126 | $2,800 | 89.6% | 2.90 | 2.83 | 23.82 | 22.68 |
| LogN | 9.63 | 2.97 | $4,303 | $5,118 | $815 | 18.9% | $4,482 | $179 | 4.2% | $3,603 | $3,017 | 83.7% | $3,510 | $3,011 | 85.8% | $12,246 | $10,555 | 86.2% | 0.686 | 0.672 | $3,515 | $3,081 | 87.6% | 3.61 | 3.49 | 35.34 | 33.32 |
| TLogN | 10.2 | 1.95 | $126 | $153 | $27 | 21.7% | $130 | $5 | 3.9% | $119 | $88 | 74.5% | $116 | $88 | 76.4% | $445 | $335 | 75.2% | 0.756 | 0.677 | $99 | $83 | 83.7% | 3.23 | 2.81 | 17.49 | 13.18 |
| TLogN | 9 | 2.2 | $133 | $194 | $61 | 45.8% | $137 | $4 | 3.3% | $240 | $111 | 46.3% | $232 | $111 | 47.8% | $694 | $398 | 57.3% | 1.199 | 0.808 | $140 | $101 | 72.4% | 5.76 | 2.70 | 56.19 | 11.14 |
| TLogN | 10.7 | 2.385 | $1,267 | $1,722 | $455 | 35.9% | $1,362 | $95 | 7.5% | $1,874 | $1,212 | 64.7% | $1,817 | $1,208 | 66.5% | $5,385 | $3,983 | 74.0% | 1.055 | 0.887 | $1,243 | $961 | 77.3% | 5.98 | 4.21 | 64.76 | 31.17 |
| TLogN | 9.4 | 2.65 | $1,297 | $2,154 | $857 | 66.0% | $1,321 | $24 | 1.9% | $4,206 | $1,364 | 32.4% | $4,117 | $1,364 | 33.1% | $8,757 | $4,740 | 54.1% | 1.912 | 1.032 | $1,498 | $1,010 | 67.4% | 10.69 | 4.06 | 156.41 | 26.36 |
| TLogN | 11 | 2.6 | $4,208 | $6,059 | $1,850 | 44.0% | $4,589 | $381 | 9.1% | $7,956 | $4,782 | 60.1% | $7,738 | $4,766 | 61.6% | $19,963 | $14,205 | 71.2% | 1.277 | 1.039 | $4,282 | $3,313 | 77.4% | 7.53 | 5.67 | 88.00 | 54.80 |
| TLogN | 10 | 2.8 | $4,145 | $6,579 | $2,434 | 58.7% | $4,289 | $144 | 3.5% | $9,318 | $4,406 | 47.3% | $8,995 | $4,404 | 49.0% | $28,334 | $15,565 | 54.9% | 1.367 | 1.027 | $5,206 | $3,509 | 67.4% | 5.23 | 3.22 | 42.69 | 14.97 |
| | a | b | | | | | | | | | | | | | | | | | | | | | | | | | |
| Logg | 24 | 2.65 | $192 | $234 | $42 | 21.6% | $207 | $14 | 7.4% | $185 | $155 | 83.7% | $180 | $154 | 85.6% | $650 | $550 | 84.7% | 0.770 | 0.746 | $158 | $139 | 87.9% | 3.28 | 3.16 | 17.25 | 16.10 |
| Logg | 33 | 3.3 | $203 | $234 | $31 | 15.0% | $212 | $9 | 4.3% | $139 | $120 | 86.7% | $135 | $120 | 88.6% | $509 | $450 | 88.4% | 0.579 | 0.566 | $137 | $124 | 90.4% | 1.84 | 1.79 | 4.97 | 4.67 |
| Logg | 25 | 2.5 | $1,064 | $1,301 | $238 | 22.3% | $1,130 | $66 | 6.2% | $989 | $811 | 82.0% | $960 | $808 | 84.2% | $3,588 | $3,070 | 85.6% | 0.738 | 0.715 | $874 | $760 | 86.9% | 2.29 | 2.19 | 7.83 | 6.99 |
| Logg | 34.5 | 3.15 | $960 | $1,114 | $153 | 16.0% | $999 | $39 | 4.0% | $735 | $632 | 86.0% | $719 | $631 | 87.8% | $2,722 | $2,369 | 87.0% | 0.646 | 0.632 | $717 | $632 | 88.1% | 2.56 | 2.53 | 12.51 | 12.41 |
| Logg | 25.25 | 2.45 | $1,877 | $2,329 | $452 | 24.1% | $2,009 | $132 | 7.0% | $2,037 | $1,657 | 81.4% | $1,986 | $1,652 | 83.2% | $6,463 | $5,481 | 84.8% | 0.853 | 0.822 | $1,653 | $1,410 | 85.3% | 4.05 | 3.87 | 26.26 | 24.23 |
| Logg | 34.7 | 3.07 | $1,794 | $2,090 | $296 | 16.5% | $1,864 | $70 | 3.9% | $1,407 | $1,198 | 85.2% | $1,375 | $1,196 | 87.0% | $5,191 | $4,514 | 87.0% | 0.658 | 0.642 | $1,343 | $1,187 | 88.4% | 2.42 | 2.35 | 9.27 | 8.79 |
| TLogg | 23.5 | 2.65 | $271 | $639 | $368 | 136.1% | $340 | $69 | 25.6% | $2,465 | $2,102 | 85.3% | $2,437 | $2,101 | 86.2% | $2,863 | $1,280 | 44.7% | 3.813 | 6.180 | $396 | $178 | 44.8% | 23.14 | 28.29 | 638.93 | 855.18 |
| TLogg | 33 | 3.3 | $261 | $400 | $139 | 53.3% | $247 | -$14 | -5.5% | $742 | $256 | 34.4% | $729 | $255 | 35.0% | $1,466 | $611 | 41.7% | 1.821 | 1.034 | $277 | $178 | 64.4% | 12.49 | 8.97 | 218.19 | 133.34 |
| TLogg | 24.5 | 2.5 | $1,164 | $2,455 | $1,291 | 110.9% | $898 | -$267 | -22.9% | $5,805 | $1,286 | 22.1% | $5,659 | $1,258 | 22.2% | $10,711 | $3,249 | 30.3% | 2.305 | 1.401 | $1,887 | $618 | 32.8% | 9.04 | 8.17 | 108.19 | 96.40 |
| TLogg | 34.5 | 3.15 | $1,086 | $1,498 | $412 | 37.9% | $1,190 | $104 | 9.6% | $1,627 | $1,097 | 67.4% | $1,574 | $1,092 | 69.4% | $5,400 | $3,767 | 69.8% | 1.051 | 0.918 | $1,241 | $964 | 77.7% | 4.11 | 3.57 | 28.75 | 22.34 |
| TLogg | 24.75 | 2.45 | $1,928 | $3,815 | $1,887 | 97.9% | $1,573 | -$356 | -18.4% | $10,543 | $2,702 | 25.6% | $10,373 | $2,678 | 25.8% | $17,336 | $4,406 | 25.4% | 2.719 | 1.703 | $2,733 | $1,143 | 41.8% | 15.79 | 18.01 | 328.59 | 435.49 |
| TLogg | 34.6 | 3.07 | $1,892 | $2,537 | $645 | 34.1% | $2,074 | $182 | 9.6% | $2,399 | $1,719 | 71.7% | $2,311 | $1,710 | 74.0% | $8,524 | $6,382 | 74.9% | 0.911 | 0.824 | $1,968 | $1,564 | 79.5% | 2.71 | 2.44 | 10.59 | 9.12 |
| | ξ | θ | | | | | | | | | | | | | | | | | | | | | | | | | |
| GPD | 0.8 | 35,000 | $382 | $704 | $321 | 83.9% | $399 | $17 | 4.4% | $1,109 | $528 | 47.6% | $1,061 | $528 | 49.7% | $3,502 | $1,831 | 52.3% | 1.508 | 1.323 | $597 | $340 | 57.0% | 4.51 | 3.93 | 28.12 | 21.78 |
| GPD | 0.95 | 7,500 | $375 | $801 | $426 | 113.4% | $394 | $19 | 5.1% | $1,514 | $613 | 40.5% | $1,452 | $613 | 42.2% | $4,326 | $1,989 | 46.0% | 1.813 | 1.554 | $660 | $330 | 49.9% | 5.58 | 4.77 | 43.03 | 32.29 |
| GPD | 0.875 | 47,500 | $1,106 | $2,137 | $1,031 | 93.2% | $1,135 | $29 | 2.6% | $3,579 | $1,586 | 44.3% | $3,427 | $1,585 | 46.3% | $11,343 | $5,256 | 46.3% | 1.604 | 1.397 | $1,842 | $974 | 52.9% | 4.79 | 4.14 | 31.97 | 24.38 |
| GPD | 0.95 | 25,000 | $1,251 | $2,650 | $1,398 | 111.8% | $1,304 | $53 | 4.2% | $5,018 | $2,031 | 40.5% | $4,819 | $2,031 | 42.1% | $14,421 | $6,412 | 44.5% | 1.819 | 1.557 | $2,166 | $1,099 | 50.7% | 5.64 | 4.83 | 43.87 | 33.08 |
| GPD | 0.925 | 50,000 | $1,938 | $3,853 | $1,916 | 98.9% | $1,955 | $18 | 0.9% | $6,795 | $2,860 | 42.1% | $6,519 | $2,860 | 43.9% | $19,904 | $9,416 | 47.3% | 1.692 | 1.463 | $3,273 | $1,688 | 51.6% | 5.15 | 4.44 | 36.70 | 27.99 |
| GPD | 0.99 | 27,500 | $2,076 | $4,437 | $2,361 | 113.7% | $2,111 | $35 | 1.7% | $8,311 | $3,257 | 39.2% | $7,968 | $3,257 | 40.9% | $23,923 | $10,500 | 43.9% | 1.796 | 1.543 | $3,767 | $1,816 | 48.2% | 5.39 | 4.64 | 39.80 | 30.52 |
| TGPD | 0.775 | 33,500 | $351 | $634 | $283 | 80.5% | $372 | $21 | 5.8% | $1,022 | $506 | 49.5% | $982 | $505 | 51.5% | $2,935 | $1,677 | 57.1% | 1.549 | 1.359 | $529 | $309 | 58.5% | 4.86 | 4.35 | 33.45 | 27.94 |
| TGPD | 0.8 | 25,000 | $361 | $663 | $302 | 83.7% | $379 | $18 | 5.0% | $1,147 | $548 | 47.7% | $1,106 | $547 | 49.5% | $3,266 | $1,656 | 50.7% | 1.669 | 1.444 | $532 | $307 | 57.8% | 5.34 | 4.68 | 38.41 | 30.31 |
| TGPD | 0.8675 | 50,000 | $1,267 | $2,582 | $1,316 | 103.9% | $1,390 | $123 | 9.7% | $6,016 | $2,577 | 42.8% | $5,871 | $2,574 | 43.8% | $13,331 | $6,710 | 50.3% | 2.274 | 1.852 | $2,084 | $1,130 | 54.3% | 14.74 | 11.50 | 323.63 | 219.01 |
| TGPD | 0.91 | 31,000 | $1,334 | $2,651 | $1,317 | 98.8% | $1,384 | $50 | 3.8% | $4,916 | $2,106 | 42.8% | $4,736 | $2,105 | 44.4% | $12,491 | $6,249 | 50.0% | 1.787 | 1.521 | $2,333 | $1,224 | 52.5% | 6.60 | 5.39 | 65.93 | 45.09 |
| TGPD | 0.92 | 47,500 | $2,088 | $3,987 | $1,899 | 90.9% | $2,099 | $11 | 0.5% | $6,618 | $2,984 | 45.1% | $6,340 | $2,984 | 47.1% | $18,525 | $9,553 | 51.6% | 1.590 | 1.422 | $3,366 | $1,782 | 52.9% | 4.61 | 4.21 | 31.11 | 26.22 |
| TGPD | 0.95 | 35,000 | $2,227 | $4,713 | $2,486 | 111.6% | $2,377 | $150 | 6.7% | $8,807 | $3,742 | 42.5% | $8,449 | $3,739 | 44.3% | $26,750 | $11,880 | 44.4% | 1.793 | 1.573 | $3,800 | $1,960 | 51.6% | 5.55 | 4.83 | 46.92 | 35.21 |

*NOTE: #simulations = 1,000; λ = 25 for 10 years so n ~ 250; α=0.9997